\documentclass[twocolumn]{aastex631}
\usepackage{tabularx}
\usepackage{comment}
\usepackage{here}

\shorttitle{Full SED Analysis of $z\sim0.5\text{--}6$ Lensed Galaxies Detected with Millimeter Observations}
\shortauthors{Uematsu et al.}

\begin{document}

\title{ALMA Lensing Cluster Survey: Full SED Analysis of $z\sim0.5\text{--}6$ Lensed Galaxies Detected with Millimeter Observations}

\correspondingauthor{Ryosuke Uematsu}
\email{uematsu@kusastro.kyoto-u.ac.jp}

\author[0000-0001-6653-779X]{Ryosuke Uematsu}
\affiliation{Department of Astronomy, Kyoto University, Kyoto 606-8502, Japan}

\author[0000-0001-7821-6715]{Yoshihiro Ueda}
\affiliation{Department of Astronomy, Kyoto University, Kyoto 606-8502, Japan}

\author[0000-0002-4052-2394]{Kotaro Kohno}
\affiliation{Institute of Astronomy, Graduate School of Science, The University of Tokyo, 2-21-1 Osawa, Mitaka, Tokyo 181-0015, Japan}
\affiliation{Research Center for the Early Universe, School of Science, The University of Tokyo, 7-3-1 Hongo, Bunkyo-ku, Tokyo 113-0033, Japan}

\author[0000-0002-3531-7863]{Yoshiki Toba}
\affiliation{National Astronomical Observatory of Japan, 2-21-1 Osawa, Mitaka, Tokyo 181-8588, Japan}
\affiliation{Academia Sinica Institute of Astronomy and Astrophysics, 11F of Astronomy-Mathematics Building, AS/NTU, No.1, Section 4, Roosevelt Road, Taipei 10617, Taiwan}
\affiliation{Research Center for Space and Cosmic Evolution, Ehime University, 2-5 Bunkyo-cho, Matsuyama, Ehime 790-8577, Japan}

\author[0000-0002-9754-3081]{Satoshi Yamada}
\affiliation{Institute of Physical and Chemical Research (RIKEN), 2-1 Hirosawa, Wako, Saitama 351-0198}

\author[0000-0003-3037-257X]{Ian Smail}
\affiliation{Centre for Extragalactic Astronomy, Department of Physics, Durham University, South Road, Durham DH1 3LE, UK}

\author[0000-0003-1937-0573]{Hideki Umehata}
\affiliation{Institute for Advanced Research, Nagoya University, Furocho, Chikusa, Nagoya 464-8602, Japan}
\affiliation{Department of Physics, Nagoya University, Furo-cho, Chikusa-ku, Nagoya 464-8601, Japan}

\author[0000-0001-7201-5066]{Seiji Fujimoto}
\affiliation{Department of Astronomy, The University of Texas at Austin, Austin, TX, USA}
\affiliation{Cosmic Dawn Center (DAWN), Denmark}
\affiliation{Niels Bohr Institute, University of Copenhagen, Copenhagen N, Denmark}

\author[0000-0001-6469-8725]{Bunyo Hatsukade}
\affiliation{Institute of Astronomy, Graduate School of Science, The University of Tokyo, 2-21-1 Osawa, Mitaka, Tokyo 181-0015, Japan}

\author[0000-0003-3139-2724]{Yiping Ao}
\affiliation{Purple Mountain Observatory and Key Laboratory for Radio Astronomy, Chinese Academy of Sciences, Nanjing, China}

\author[0000-0002-8686-8737]{Franz Erik Bauer}
\affiliation{Instituto de Astrof{\'{\i}}sica, Facultad de F{\'{i}}sica, Pontificia Universidad Cat{\'{o}}lica de Chile, Campus San Joaquín, Av. Vicuña Mackenna 4860, Macul Santiago, Chile, 7820436} 
\affiliation{Centro de Astroingenier{\'{\i}}a, Facultad de F{\'{i}}sica, Pontificia Universidad Cat{\'{o}}lica de Chile, Campus San Joaquín, Av. Vicuña Mackenna 4860, Macul Santiago, Chile, 7820436} 
\affiliation{Millennium Institute of Astrophysics, Nuncio Monse{\~{n}}or S{\'{o}}tero Sanz 100, Of 104, Providencia, Santiago, Chile} 
\affiliation{Space Science Institute, 4750 Walnut Street, Suite 205, Boulder, Colorado 80301} 

\author[0000-0003-2680-005X]{Gabriel Brammer}
\affiliation{Cosmic Dawn Center (DAWN), Denmark}
\affiliation{Niels Bohr Institute, University of Copenhagen, Copenhagen N, Denmark}

\author[0000-0003-0348-2917]{Miroslava Dessauges-Zavadsky}
\affiliation{Department of Astronomy, University of Geneva, 51 chemin Pegasi, 1290 Versoix, Switzerland}

\author[0000-0002-8726-7685]{Daniel Espada}
\affiliation{Departamento de F\'{i}sica Te\'{o}rica y del Cosmos, Campus de Fuentenueva, Edificio Mecenas, Universidad de Granada, E-18071, Granada, Spain}
\affiliation{Instituto Carlos I de F\'{i}sica Te\'{o}rica y Computacional, Facultad de Ciencias, E-18071, Granada, Spain}

\author[0000-0002-3405-5646]{Jean-Baptiste Jolly}
\affiliation{Max-Planck-Institut für Extraterrestrische Physik (MPE), Giessenbachstraße 1, D-85748 Garching, Germany}

\author[0000-0002-6610-2048]{Anton M. Koekemoer}
\affiliation{Space Telescope Science Institute, 3700 San Martin Dr., Baltimore, MD 21218, USA}

\author[0000-0002-5588-9156]{Vasily Kokorev}
\affiliation{Kapteyn Astronomical Institute, University of Groningen, P.O. Box 800, 9700AV Groningen, The Netherlands}
\affiliation{Cosmic Dawn Center (DAWN), Denmark}
\affiliation{Niels Bohr Institute, University of Copenhagen, Copenhagen N, Denmark}

\author[0000-0002-4872-2294]{Georgios E. Magdis}
\affiliation{Cosmic Dawn Center (DAWN), Denmark}
\affiliation{DTU-Space, Technical University of Denmark, Elektrovej 327, 2800, Kgs. Lyngby, Denmark}
\affiliation{Niels Bohr Institute, University of Copenhagen, Copenhagen N, Denmark}
\affiliation{Millennium Institute of Astrophysics, Nuncio Monse{\~{n}}or S{\'{o}}tero Sanz 100, Of 104, Providencia, Santiago, Chile}

\author[0000-0003-3484-399X]{Masamune Oguri}
\affiliation{Center for Frontier Science, Chiba University, 1-33 Yayoi-cho, Inage-ku, Chiba 263-8522, Japan}
\affiliation{Department of Physics, Graduate School of Science, Chiba University, 1-33 Yayoi-Cho, Inage-Ku, Chiba 263-8522, Japan}

\author[0000-0002-4622-6617]{Fengwu Sun}
\affiliation{Steward Observatory, University of Arizona, 933 N. Cherry Avenue, Tucson, 85721, USA}

\begin{abstract}

Sub/millimeter galaxies are a key population for the study of galaxy evolution because the majority of star formation at high redshifts occurred in galaxies deeply embedded in dust. To search for this population, we have performed an extensive survey with ALMA, called the ALMA Lensing Cluster Survey (ALCS). This survey covers 133 arcmin$^2$ area and securely detects 180 sources at $z\sim0.5\text{--}6$ with a flux limit of ${\sim}0.2$ mJy at 1.2 mm \citep{2023arXiv230301658F}. Here we report the results of multi-wavelength spectral energy distribution (SED) analysis of the whole ALCS sample, utilizing the observed-frame UV to millimeter photometry. We find that the majority of the ALCS sources lie on the star-forming main sequence, with a smaller fraction showing intense starburst activities. The ALCS sample contains high infrared-excess sources ($\mathrm{IRX}=\log (L_{\mathrm{dust}}/L_{\mathrm{UV}})>1$), including two extremely dust-obscured galaxies ($\mathrm{IRX}>5$). We also confirm that the ALCS sample probes a broader range in lower dust mass than conventional SMG samples in the same redshift range. We identify six heavily obscured AGN candidates that are not detected in the archival Chandra data in addition to the three X-ray AGNs reported by \citet{2023ApJ...945..121U}. The inferred AGN luminosity density shows a possible excess at $z=2\text{--}3$ compared with that determined from X-ray surveys below 10 keV.

\end{abstract}


\section{Introduction} \label{sec:intro}

The evolutionary scenario of galaxies and the supermassive black holes (SMBHs) at their centers is one of the most perplexing mysteries in modern astronomy. The tight bulge-mass-to-SMBH-mass correlation in the local universe (e.g., \citealt{1998AJ....115.2285M,2003ApJ...589L..21M,2013ARA&A..51..511K}) suggests the co-evolution of galaxies and SMBHs. This idea is also supported by the similarity between the cosmological evolution of star-formation density and active galactic nucleus (AGN) luminosity density, which both reached a peak around $z\sim2$ \citep{2003ApJ...598..886U,2014ARA&A..52..415M}. However, the evolutionary scenario of an individual system is still unclear. High-redshift dusty star-forming galaxies (DSFGs) are a key population to elucidate the co-evolution scenario, because the majority of star formation at high redshifts occurred in galaxies deeply enshrouded by dust. 

Sub/millimeter observations are a powerful tool to study DSFGs in the high-redshift universe ($z\geq3$). DSFGs are characterized by prominent far-infrared (FIR) emission from dust heated by stars. At high redshifts, the peak of dust emission is shifted to the sub/millimeter band, so that the sensitivity required to detect a galaxy with a given FIR luminosity is nearly constant across $z=1\text{--}8$ in the millimeter band (\citealt{2002PhR...369..111B,2014PhR...541...45C} for reviews). Hence, sub/millimeter observations can efficiently search for DSFGs in the high-redshift universe (often referred to as submillimeter galaxies; SMGs).

The Atacama Large Millimeter/submillimeter Array (ALMA) achieves excellent sensitivity and angular resolution in the millimeter band, and has been used to study SMGs. Here we introduce some examples of representative ALMA surveys that are used as the comparison samples in this paper (see also \citealt{2020RSOS....700556H} for a recent review). The ALMA LABOCA E-CDFS Submillimeter Surveys (ALESS; \citealt{2012Msngr.149...40S,2012MNRAS.427.1066S,2013ApJ...768...91H,2013MNRAS.432....2K,2013ApJ...778..179W,2014ApJ...780..115D,2014MNRAS.438.1267S,2014ApJ...788..125S,2014MNRAS.442..577T,2015ApJ...799..194C,2015ApJ...806..110D,2017ApJ...840...78D}) is an ALMA cycle 0 program to follow up ${\sim}100$ SMGs that were previously detected by the single-dish LABOCA 870 $\mu$m survey in the extended Chandra Deep Field South (LESS; \citealt{2009ApJ...707.1201W}). It was able to separate unresolved multiple sources that were confused in the single-dish survey, and successfully yielded a reliable catalog of SMGs. In order to search for rare populations and to improve statistics, \citet{2019MNRAS.487.4648S} performed ALMA imaging follow-up of the SCUBA-2 UDS survey (AS2UDS) in cycles 1, 3, 4, and 5. This survey observed 716 SMGs in the UKIDSS Ultra Deep Survey (UDS) field that were previously detected by the SCUBA-2 Cosmology Legacy Survey (S2CLS: \citealt{2017MNRAS.465.1789G}), and yielded the largest homogeneously-selected SMG sample to date (see also \citealt{2018ApJ...860..161S,2018ApJ...861..100C,2019MNRAS.490.4956G,2020MNRAS.492.4927K,2020MNRAS.494.3828D,2020ApJ...903..138A,2021MNRAS.502.3426S}). To search for fainter (hence more numerous) sources, \citet{2017MNRAS.466..861D} performed an ALMA deep survey in the GOODS-S/Hubble Ultra Deep Field (HUDF; \citealt{2006AJ....132.1729B}). It covered an area of ${\sim}$4.5 arcmin$^2$ and detected 16 secure sources with a flux limit of 0.12 mJy at 1.3 mm. The ALMA twenty-Six Arcmin$^2$ survey of GOODS-S At One-millimeter (ASAGAO; \citealt{2018ApJ...853...24U,2018ApJ...861....7F,2018PASJ...70..105H,2019ApJ...878...73Y,2020IAUS..352..239H,2020PASJ...72...69Y}) was designed to fill the gap between the ALESS and HUDF surveys in the area versus sensitivity space. The ASAGAO survey covered an area of 26 arcmin$^2$ and detected 25 sources at ${\geq}5.0\sigma$ and 45 sources at ${\geq}4.5\sigma$ (1.2 mm).\footnote{The GOODS-S area is also covered by an extensive survey at 1.1m called GOODS-ALMA \citep{2018A&A...620A.152F,2020A&A...642A.155Z,2020A&A...643A..30F,2020A&A...643A..53F,2022A&A...658A..43G,2022A&A...659A.196G,2023A&A...672A.191C}.} To simultaneously search for blank-field CO line and dust continuum, the large program called ALMA Spectroscopic Survey in the HUDF  (ASPECS) was performed in cycle 4 \citep{2016ApJ...833...67W,2016ApJ...833...68A,2016ApJ...833...69D,2016ApJ...833...70D,2016ApJ...833...71A,2016ApJ...833...72B,2016ApJ...833...73C,2019ApJ...882..136A,2019ApJ...882..137P,2019ApJ...882..138D,2019ApJ...882..140B,2019ApJ...887...37U,2020ApJ...891..135P,2020ApJ...892...66M,2020ApJ...897...91G,2020ApJ...901...79A,2020ApJ...902..109B,2020ApJ...902..110D,2020ApJ...902..113I,2021ApJ...912...67U}. This survey covered ${\sim}4$ arcmin$^2$ and detected 35 secure sources with a flux limit of ${\sim}0.04$ mJy at 1.2 mm.

In ALMA cycle 6, our team performed a large survey in 33 lensing clusters called ALMA Lensing Cluster Survey (ALCS; \citealt{2021ApJ...908..146C,2021ApJ...911...99F,2021A&A...652A.128J,2021MNRAS.505.4838L,2022ApJ...932...77S,2022ApJS..263...38K,2023ApJ...945..121U,2023arXiv230301658F,2023arXiv230515126K}), in order to detect faint sources by utilizing the lensing effect. The ALMA observations were designed to target areas which were covered by the major HST treasury programs: Cluster Lensing And Supernova Survey with Hubble (CLASH; \citealt{2012ApJS..199...25P}), Hubble Frontier Fields (HFFs; \citealt{2017ApJ...837...97L}), Beyond Ultra-deep Frontier Fields And Legacy Observations (BUFFALO; \citealt{2020ApJS..247...64S}), and Reionization Lensing Cluster Survey (RELICS; \citealt{2019ApJ...884...85C}). The total area covered is 133 arcmin$^2$ (primary beam sensitivity in the mosaic maps of $>30$\%), with 180 sources securely detected down to an image-plane flux limit of ${\sim}0.2$ mJy at 1.2 mm and a source plane flux limit of ${\sim}7\,\mu$Jy, which is one of the deepest surveys after lensing correction (see Section~\ref{subsection:alma} for details). 

In this paper, we report the results of multi-component spectral energy distribution (SED) analysis for the whole ALCS sample, utilizing the UV-to-millimeter photometry obtained by ALMA, Hubble Space Telescope (HST), Spitzer Space Telescope (Spitzer), and Herschel Space Observatory (Herschel). In Section~\ref{section:observation}, we summarize the observations and data reduction. In Section~\ref{section:analysis}, we explain the SED analysis and comparison samples. In Section~\ref{section:result}, we show the results of the SED analysis. In Section~\ref{section:discussion}, we discuss the dust properties and the nature of AGNs in our sample. Finally, we provide a summary and conclusions in Section~\ref{section:summary}. Throughout the paper, we assume a flat universe with $H_0=70.4\ \mathrm{km\ s^{-1}\ Mpc^{-1}}$ and $\Omega_M=0.272$ \citep{2011ApJS..192...18K}. The Chabrier initial mass function (IMF, \citealt{2003PASP..115..763C}) is adopted. If not specifically mentioned, errors correspond to $1\sigma$ confidence.

\section{Observations and data reduction} \label{section:observation}

\subsection{Observations and source detection with ALMA} \label{subsection:alma}

The 1.2 mm continuum source catalog was built by \citet{2023arXiv230301658F}. Here we briefly explain the observations and the data reduction. The ALCS fields were observed in ALMA Band 6 between 2018 December and 2019 December through the program 2018.1.00035.L (PI: K. Kohno; \citealt{2023arXiv230515126K}). These observations covered 250.0--257.5 GHz and 265.0--272.5 GHz with a total bandwidth of 15 GHz. The synthesized beam size is ${\sim}1$\arcsec. All the ALMA data were calibrated and reduced with the Common Astronomy Software Applications package (CASA). The existing ALMA data from 2013.1.00999S and 2015.1.01425.S (PI: Bauer; \citealt{2017A&A...597A..41G}) were also combined for five Hubble Frontier Fields. The source detection was performed with the \texttt{SExtractor} in native (full width at half maximum, $\mathrm{FWHM}\sim 1$\arcsec) and tapered maps ($\mathrm{FWHM}\sim 2$\arcsec) with natural weighting. First, in a blind search, 141 sources were detected with a signal-to-noise ratio (SNR) over 5.0 in the natural maps or over 4.5 in the tapered maps, which corresponds to a false-detection rate of ${<}$1\%. Second, in a prior-based approach, 39 sources are detected with $5.0\geq \mathrm{SNR}\geq 4.0$ (a false-detection rate of ${<}$50\%) in the natural maps associated with Spitzer/IRAC channel 2 ($4.5\,\mu \mathrm{m}$) detection ($\mathrm{SNR}\geq 5.0$). Hereafter, we refer to the former as the MAIN sample and the latter as the SECONDARY sample.

\subsection{HST, Spitzer, and Herschel counterparts}

The ALCS fields were observed with HST, Spitzer, and Herschel previously. The images and photometric catalogs of optical and near-infrared bands were built by \citet{2022ApJS..263...38K} by reprocessing the best-available archival data from HST and Spitzer/IRAC channels 1 and 2. \citet{2022ApJ...932...77S} produced the far-infrared images and catalogs, utilizing the best-available data of Herschel/PACS and SPIRE. The quality of the multi-wavelength photometry is summarized in Section~\ref{subsection:sample}. We note that the photometric data of Spitzer/IRAC channel 2 is sometimes inconsistent with the other photometries, which may be due to remaining blending effects or oversubtraction of nearby sources. For this reason, the photometric data of Spitzer/IRAC channel 2 in M0416-ID120, M1206-ID60, ACT0102-ID11, M1115-ID34, and ACT0102-ID224 are not used in the SED modeling.\footnote{We note that these sources were not detected in the Spitzer/IRAC channel 3, 4 and Spitzer/MIPS channel 1 in the SEIP catalog.}

In addition to these catalogs, we utilize the photometric data of Spitzer/IRAC channels 3 (5.8 $\mu m$) and 4 (8.0 $\mu m$), and those of Spitzer/MIPS channel 1 (24 $\mu m$), taken from the Spitzer Enhanced Imaging Products (SEIP; \citealt{https://doi.org/10.26131/irsa433}). Those sources are cross-matched with the ALMA source positions within 1\farcs5. Since IRAC and MIPS have large beam sizes (full-width half maximum of 1\farcs8, 1\farcs9, and 5\farcs9 at 5.8 $\mu$m, 8.0 $\mu$m, and 24 $\mu$m, respectively), the measured photometric data can be contaminated by nearby sources. To evaluate this effect, we compare the flux densities of IRAC channel 2 extracted from the SEIP ($f_{\mathrm{SEIP}}$) with those in the catalog by \citet{2022ApJS..263...38K} ($f_{\mathrm{ALCS}}$), who de-blended the photometric data utilizing the HST source positions. If $0.5 < f_{\mathrm{ALCS}}/f_{\mathrm{SEIP}} < 2.0$, we use the photometric data from the SEIP for our SED modeling. If $f_{\mathrm{ALCS}}/f_{\mathrm{SEIP}} < 0.5$, we treat the SEIP data as upper limits. In the case of $f_{\mathrm{ALCS}}/f_{\mathrm{SEIP}} > 2.0$, we conservatively do not utilize the data from the SEIP, considering the possibility of mismatching with the ALMA source. 

\subsection{Chandra observations}

\citet{2023ApJ...945..121U} detected three AGNs among the lensed ALCS sources, utilizing the archived data of Chandra; hereafter, we refer to these AGNs as ``ALCS-XAGNs". In this paper, we derive the 3$\sigma$ upperbounds\footnote{In this paper, we refer to the upper edge of a confidence interval as ``upperbound'' (see \citealt{2010ApJ...719..900K} for more details).} of the X-ray luminosities for the rest of the sources. First, we reprocessed all the data obtained with Chandra by 2021, following the standard procedures with the Chandra interactive analysis of observations software (CIAO v4.12) and the calibration database (CALDB v4.9.1). Next, we combined the products by using \texttt{merge\_obs}. Then, we ran \texttt{srcflux} to the combined data and obtained the $3\sigma$ upperbounds of the 0.5--7 keV count rates for each source. For the source and background regions, we employed circular regions with radii of 1.5\arcsec\ and annuli with inner and outer radii of 2\arcsec\ and 4\arcsec, respectively, after masking obvious X-ray point sources. Then, we converted the count rates to the intrinsic X-ray luminosities, assuming typical X-ray spectra of AGNs. Since there are no Chandra observations in RXC0032 and RXC0600, the X-ray luminosity upperbounds in those fields are not given. We excluded the brightest cluster galaxies (BCGs) from the catalog because of the difficulty in separating the X-ray point sources from the diffuse X-ray emission of the galaxy clusters. Details of the X-ray spectral models and the results are summarized in Appendix~\ref{appendix:xray}.

\subsection{Data quality, sample selection, and magnification}\label{subsection:sample}

We classify the ALCS sources into eight groups based on the quality of the multi-wavelength photometry. The criteria are summarized in Table~\ref{table:quality}. In this classification, BCGs (11 sources\footnote{We follow the classification in \citet{2022ApJ...932...77S}.}; A383-ID40, M0159-ID46, M0429-ID19, M1115-ID36, M1206-ID58, M1423-ID50, M1931-ID41/42, R0032-ID162, R1347-ID75, R2129-ID20), a possible mismatched source (R0600-ID67), and a possible contaminated source (R0600-ID164) are excluded. A370-ID31 is also excluded because it is spectroscopically confirmed as a member galaxy of the foreground cluster (Abell 370). We note that this classification is independent of the one that is based on the source detection (either blind [MAIN] or prior-based [SECONDARY]) in the ALMA band (see Section~\ref{subsection:alma}). 

\begin{deluxetable}{ccccc}
 \tablewidth{\linewidth}
 \tablecaption{Classification of the ALCS sources\label{table:quality}}
 \tablehead{
 \colhead{\hspace{8pt}Tier\hspace{8pt} } &\colhead{\hspace{8pt}HST\hspace{8pt} } &\colhead{\hspace{8pt}Spitzer\hspace{8pt} } &\colhead{\hspace{8pt}Herschel\hspace{8pt} } &\colhead{\hspace{8pt}N\hspace{8pt} }
 }
 \colnumbers
 \startdata
 1 & $\checkmark$ & $\checkmark$ & $\checkmark$ & 99 (79)\\
 2 & $\checkmark$ & $\checkmark$ & \nodata & 28 (16)\\
 3 & \nodata & $\checkmark$ & $\checkmark$ & 21 (20)\\
 4 & \nodata & $\checkmark$ & \nodata & 10 (7)\\
 5 & $\checkmark$ & \nodata & $\checkmark$ & 3 (3)\\
 6 & $\checkmark$ & \nodata & \nodata & 1 (0)\\
 7 & \nodata & \nodata & $\checkmark$ & 2 (2)\\
 8 & \nodata & \nodata & \nodata & 2 (2)\\
 \enddata
 \tablecomments{
 (1) Class. (2)--(4) The quality of the data. If a source is detected above 2$\sigma$ in at least one band, the corresponding column is checked by the symbol ($\checkmark$). (5) The number of ALCS sources that meet the criteria. The number in parentheses shows the number of sources that belong to the MAIN sample.\\
 In this classification, BCGs (11 sources), a possible mismatched source, a possible contaminated source, and a spectroscopically-confirmed member galaxy of a cluster are excluded. The multiply-imaged sources are counted individually.
 }
\end{deluxetable}

\begin{figure}[H]
 \epsscale{1.1}
 \plotone{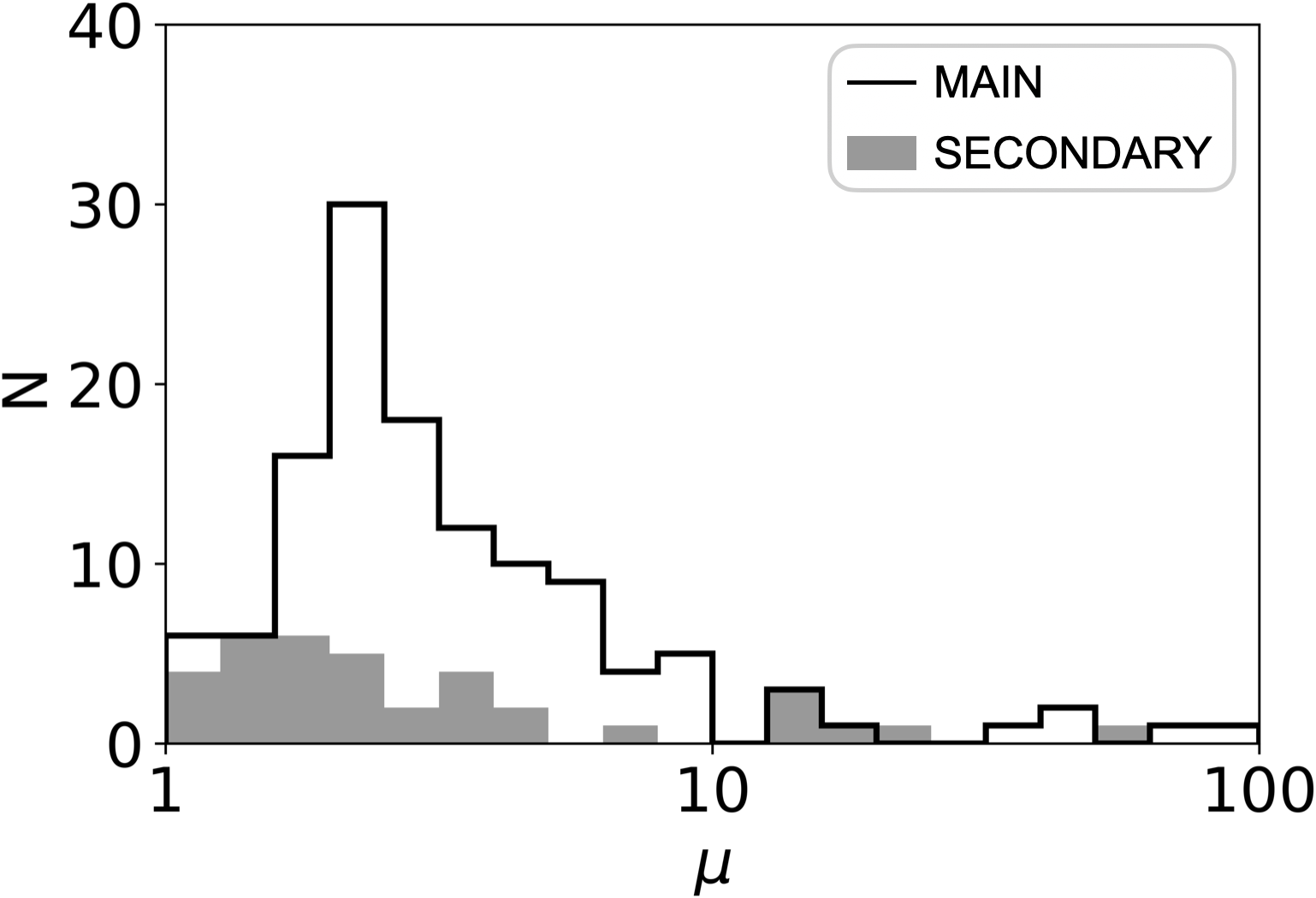}
 \caption{Distribution of the magnification factor in the ALCS tier-1/2/3/4/5/6 samples. One source belonging to the SECONDARY sample lies off the plot with $\mu=0.15$ ($z=0.63\pm0.10$). The multiply-imaged sources are counted individually. \label{figure:mag}}
\end{figure}

The magnification factors are extracted from \citet{2023arXiv230301658F}, where the lens model of \texttt{GLAFIC} \citep{2010PASJ...62.1017O} is adopted as a fiducial. Figure~\ref{figure:mag} shows the histogram of the magnification factors in the tier-1/2/3/4/5/6 samples. Here we note that the uncertainties in the lens models and magnification factors are not considered in this paper. Such uncertainties can affect the physical quantities of each source, however, they do not change the qualitative discussion. We spectroscopically confirm six multiply-imaged sources (ACT0102-ID118/215/224; $z=4.32$, MACS0417-ID46/58/121; $z=3.65$, MACS0553-ID133/190/249; $z=1.14$, M1206-ID27/55/60/61; $z=1.04$, R0032-ID32/53/55/57/58; $z=3.63$, R0032-ID208/281/304; $z=2.98$). We also regard R0032-ID127/131/198 and ACT0102-ID223/294 as multiple images of single sources at $z=2.391$ and $z=4.0\pm0.5$, based on the source positions, the lens models, and the similarities in their SEDs. Moreover, A2537-ID24/66 and R0032-ID63/81 are also treated as multiple images at $z=3.2$ and $z=3.0$, which are used to construct the lens models in the literature (see \citealt{2023arXiv230301658F} for details).

\section{Analysis} \label{section:analysis}

\subsection{SED modeling with \texttt{CIGALE}} \label{section:sedanalysis}

We conduct multi-component SED modeling to the UV to millimeter photometry of the ALCS tier-1/2/3/4/5/6 samples, except for the three ALCS-XAGNs, whose SEDs are already analyzed by \citet{2023ApJ...945..121U}. The tier-7/8 samples are not analyzed in this paper because it is challenging to derive the physical properties from their limited photometric data. Although these sources are located within the coverages of HST and Spitzer, they are too faint to be detected with these observatories. This suggests that those sources are heavily obscured galaxies in high-redshift universe. We use \texttt{CIGALE} (Code Investigating GALaxy Emission; \citealt{2019A&A...622A.103B,2022ApJ...927..192Y}), incorporating modifications to the dust emission model. \texttt{CIGALE} assumes energy balance between UV/optical absorption and infrared re-emission, which enables us to self-consistently model the SED of each galaxy. Moreover, \texttt{CIGALE} can simultaneously treat AGN and galaxy components, which enables us to assess the contribution of AGN in the SED.

The SED modules assumed in this study are described in the following subsections. The free parameters are summarized in Appendix~\ref{appendix:freeparameter}. In the SED modeling, the multiply-imaged sources are treated individually. The redshifts are fixed at the values reported in \citet{2023arXiv230301658F} (69/180 sources have spectroscopic redshifts). The uncertainties of the photometric redshifts are not considered in the main part of this paper. The redshift uncertainties and their impact on the estimation of the physical properties are discussed in Appendix~\ref{appendix:redshift}, where we treat photometric redshifts utilizing \texttt{MAGPHYS} \citep{2008MNRAS.388.1595D,2015ApJ...806..110D,2019ApJ...882...61B,2020ApJ...888..108B}.

\subsubsection{Stellar emission, dust attenuation, and nebular emission}

In our \texttt{CIGALE} SED modeling, we employ a delayed star-formation history (SFH), assuming an additional burst with exponential decay. The simple stellar population (SSP) is modeled with the stellar templates of \citet{2003MNRAS.344.1000B}, assuming the \citet{2003PASP..115..763C} initial mass function. The dust attenuation is modeled with the Calzetti starburst attenuation law \citep{2000ApJ...533..682C}, where we allow slightly steeper or flatter curves than the original one (see Appendix~\ref{appendix:freeparameter} for the parameter range). We also add nebulae emission according to \citet{2011MNRAS.415.2920I}.

\subsubsection{AGN emission} \label{subsubsection:AGNemission}

We employ the Skirtor model \citep{2012MNRAS.420.2756S,2016MNRAS.458.2288S} for the optical and infrared emission from an AGN. The Skirtor model assumes anisotropic radiation from an accretion disk and a dusty torus composed of two-phase matter. We also add a polar dust emission modeled by a single graybody, where we assume an emissivity index of 1.6, temperature of 150 K, and opacity of unity at 200 $\mu$m. To identify whether a galaxy hosts an AGN or not, we utilize the Bayesian Information Criterion (BIC; \citealt{1978AnSta...6..461S}). The BIC is calculated as $\mathrm{BIC}=\chi^2+k\times\ln(n)$, where $\chi^2$ is the non-reduced chi-square, $k$ is the number of degrees of freedom (dof), and $n$ is the number of the photometric data points. If the addition of an AGN component reduces BIC by more than 10 ($\Delta\mathrm{BIC}=\mathrm{BIC_{noAGN}}-\mathrm{BIC_{AGN}}>10$), we consider that the AGN component is needed for the SED (see \citealt{2020ApJ...899...35T}). For the tier-2/3/4/5/6 samples, the AGN component is always excluded, because it is challenging to identify the presence of an AGN from the limited photometry of those samples.

\subsubsection{Dust emission from host galaxy}

Recent studies have shown that some SMGs and local Ultra Luminous Infrared Galaxies (ULIRGs) exhibit higher dust temperatures compared to typical local galaxies \citep{2014MNRAS.440..942C,2018MNRAS.475.2097C,2021ApJ...919...30D}. Currently, the physically-based dust emission models implemented in \texttt{CIGALE} cannot properly account for these populations due to the limited range of dust temperatures. This limitation in the higher temperature regimes may result in an overestimation of AGN contribution, particularly that of polar dust emission from the AGN, or an underestimation of infrared luminosity from star-formation activity. Therefore, we develop a new dust emission model by extending the parameter range of the Themis model \citep{2017A&A...602A..46J}, which is the latest dust-emission model implemented in \texttt{CIGALE}.

\begin{figure}[htb]
\epsscale{1.1}
\plotone{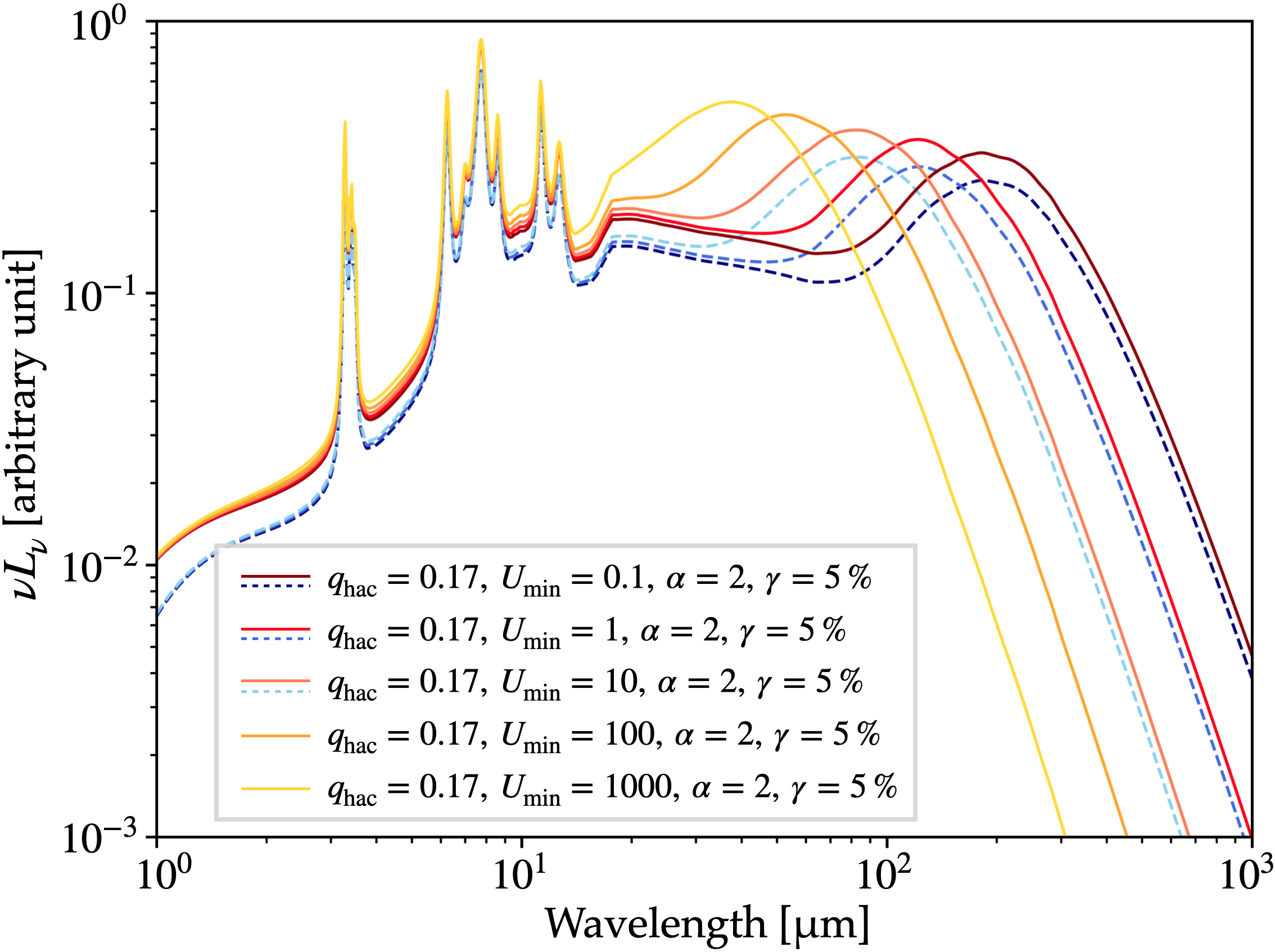}
\caption{Example of SEDs in the Themis (dashed lines) and the EThemis model (solid lines) spanning a wide range in dust temperature. All the SEDs in this figure are normalized by the total luminosity, while a small systematic offset is imposed in the Themis model to improve visibility. \label{figure:ethemis}}
\end{figure}

We use the DustEM code \citep{2011A&A...525A.103C} to extend the Themis model. Dust temperature is characterized by the minimum radiation field illuminating the interstellar dust ($U_{\mathrm{min}}$). In the current version of the Themis model, $U_{\mathrm{min}}$ is limited to 0.1--80, which corresponds to 12--39 K.\footnote{Here, we use $T_{\mathrm{dust}}$ to refer to the intrinsic temperature of interstellar dust illuminated with $U=U_{\mathrm{min}}$ ($T_{\mathrm{dust}}=18.3\ U_{\mathrm{min}}^{1/5.79}\ \mathrm{K}$; \citealt{2019A&A...624A..80N}).} We extend it to $U_{\mathrm{min}}=0.1\text{--}6309$ ($10^{-1}\text{--}10^{3.8}$) with a step size of 0.2 in logarithmic space. This parameter range corresponds to $T_{\mathrm{dust}}=12\text{--}83\,\mathrm{K}$ ($T_{\mathrm{dust}}^{\mathrm{peak}}=20\text{--}141\,\mathrm{K}$; see Section~\ref{subsection:Tchara}), which is almost equivalent to the likelihood distribution of dust temperature in the ALESS sample (20--80 K; \citealt{2015ApJ...806..110D}). We confirm that the extended model successfully reproduces the original Themis model by covering high dust temperature cases. Figure~\ref{figure:ethemis} compares the original Themis model and our new model, which we hereafter call EThemis (Extended Themis).

\subsection{Estimation of physical properties} \label{subsection:physicalproperty}

For the estimation of physical properties, we calculate the likelihood-weighted mean of all the models on the grid as a Bayesian estimation, where we treat the following parameters in a logarithmic space:
\begin{itemize}
\setlength{\parskip}{0cm}
\setlength{\itemsep}{0.1cm}
\item Star-formation rate (SFR)\footnote{In this paper, the term ``SFR" refers to the average star-formation rate in the last 10 Myr.}
\item Stellar mass ($M_*$)
\item Dust luminosity ($L_{\mathrm{dust}}$) and mass ($M_{\mathrm{dust}}$)\footnote{In this paper, we use the terms ``dust luminosity" and ``dust mass" to describe those of interstellar dust (i.e., not including those in the AGN torus).}
\end{itemize}
while the following parameters are treated in a linear space:
\begin{itemize}
\setlength{\parskip}{0cm}
\setlength{\itemsep}{0.1cm}
\item Dust temperature ($T_{\mathrm{dust}}$)
\item Color excess of stellar continuum attenuation ($E(B-V)$)
\item Infrared excess ($\mathrm{IRX}=\log (L_{\mathrm{dust}}/L_{\mathrm{UV}})$)
\item Power-law index of observed UV slope ($\beta$)\footnote{The power-law index of observed UV slope is measured in the same way as \citet{1994ApJ...429..582C}.}
\item Power-law index to modify the attenuation slope ($\delta$)\footnote{The definition of $\delta$ is described in Equation~(8) in \citet{2019A&A...622A.103B}.}
\item Bolometric AGN luminosity ($L_{\mathrm{AGN,\,bol}}$)
\end{itemize}
This means that we calculate the likelihood-weighted mean of the marginalized PDF with a log-uniform prior for the former parameters and with a uniform prior for the latter, respectively. The specific star-formation rates (sSFR = SFR / stellar mass) are derived from the stellar masses and the SFRs, where the confidence intervals of the sSFRs are calculated by propagating the uncertainties associated with the stellar masses and the SFRs. We correct all the quantitative variables for the lensing effect by dividing the observed values by the magnification factors.

\subsection{Characteristic dust temperature} \label{subsection:Tchara}

\begin{figure*}[t]
 \epsscale{1.17}
 \plotone{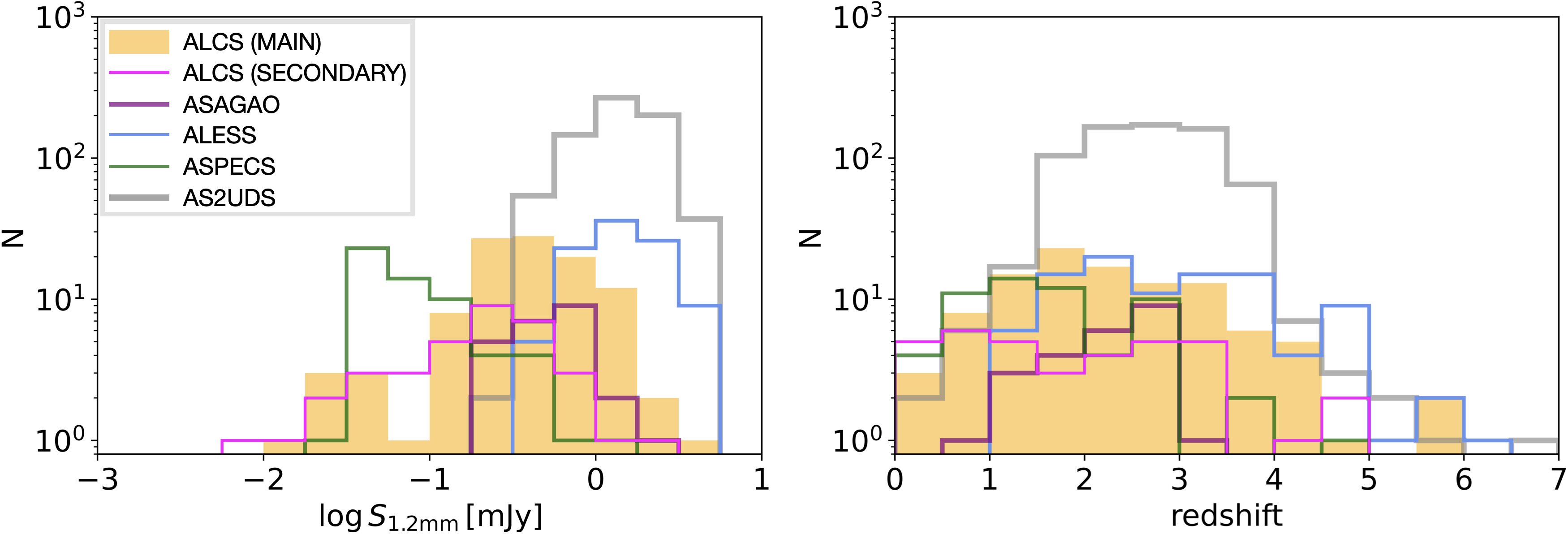}
 \caption{(Left) Flux-density distribution of sources detected in major ALMA surveys. (Right) Their redshift distribution. In these histograms, only the ALCS tier-1/2/3/4/5/6 samples are counted for the ALCS sample. For each set of multiple images, we include only one image which is selected with priorities given to a higher tier sample and to an image with a smaller magnification factor. The flux densities of the ALCS sources are corrected for lensing magnification. \label{figure:sample}}
\end{figure*}

The definition of the term ``dust temperature'' varies from study to study. For example, \citet{2019A&A...624A..80N} employ the term to denote the intrinsic temperature of interstellar dust illuminated by the minimum radiation field ($U=U_{\mathrm{min}}$), which is the same definition employed in our EThemis model. By contrast, \citet{2022ApJ...932...77S} define the term as the luminosity-weighted average of the intrinsic temperature of two optically thin graybodys. Furthermore, the measured dust temperature is also highly dependent on the assumptions of models, e.g., the assumption of dust opacity \citep{2012MNRAS.425.3094C,2020MNRAS.494.3828D,2021ApJ...919...30D}. Therefore, to properly compare different studies, the dust temperature in each study should be converted according to a common definition. One way is to use the inverse ``peak wavelength" temperature ($T_{\mathrm{dust}}^{\mathrm{peak}}$), which is calculated from the peak wavelength of the far-infrared dust emission using the Wien's displacement law\footnote{Since far-infrared photometry is often given in the dimension of ``Jy'', we utilize the peak wavelength in the dimension of ``Jy'' ($\mathrm{erg\,s^{-1}\,cm^{-2}\,Hz^{-1}}$) to calculate the characteristic dust temperature.}:
\begin{equation}
T_{\mathrm{dust}}^{\mathrm{peak}}\,[\mathrm{K}]=5099/\lambda^{\mathrm{peak}}\,[\mu\mathrm{m}]
\end{equation}
Since this value directly reflects the shape of the dust SED, it is less affected by the assumptions of models and hence is useful for comparing different studies. Therefore, we use the inverse peak-wavelength temperature to compare different samples; hereafter we refer to this temperature as ``characteristic dust temperature''. However, it should be noted here that the characteristic dust temperature does not denote the intrinsic temperature of dust, and one should be careful when interpreting this value.

\subsection{Comparison samples}

We discuss our results in comparison with previous ALMA surveys (e.g., ASAGAO, ASPECS, ALESS, AS2UDS, and ALPINE). The flux densities and the redshifts of the ASAGAO, ASPECS, ALESS, and AS2UDS samples are extracted from \citet{2020PASJ...72...69Y}, \citet{2020ApJ...901...79A}, \citet{2013ApJ...768...91H}, and \citet{2020MNRAS.494.3828D}, respectively. The other physical properties discussed in this paper are taken from \citet{2020PASJ...72...69Y}, \citet{2020ApJ...901...79A}, \citet{2015ApJ...806..110D}, and \citet{2020MNRAS.494.3828D}, respectively. We also utilize the results from \citet{2021ApJ...919...30D}, which discussed the dust properties of the well-sampled subset of the ALESS sample utilizing 2mm observation with ALMA. In addition, the physical properties of the ALPINE sample are extracted from \citet{2022A&A...663A..50B}.

Figure~\ref{figure:sample} shows the histograms of 1.2 mm flux densities.\footnote{The flux densities of the ALESS and AS2UDS samples are converted from those at 870 $\mu$m, assuming an optically thin graybody with an emissivity index of 1.8 and an intrinsic temperature of 35 K (e.g., \citealt{2008MNRAS.384.1597C,2011A&A...536A..25P}).} We confirm that the ALCS basically fills the gap between ALESS/AS2UDS and ASPECS, although the ALCS detects especially faint sources that were not detected in the previous surveys thanks to the lensing effect. The right panel of Figure~\ref{figure:sample} shows the histograms of redshifts. We see that the ALCS sample contains mainly high-redshift ($z>1$) galaxies like the other sub/millimeter selected samples (ALESS, ASAGAO, ASPECS, AS2UDS), reflecting the effect of negative $K$-correction.

The SEDs of the ALESS, ASAGAO, ASPECS, and AS2UDS samples were analyzed with \texttt{MAGPHYS} \citep{2008MNRAS.388.1595D,2015ApJ...806..110D}, while those of the ALPINE and ALCS samples are analyzed with \texttt{CIGALE}. To check the systematic difference between these codes, we also analyze the SEDs of the ALCS sample with \texttt{MAGPHYS}. This results are summarized in Appendix~\ref{appendix:magphys}. Accordingly, we impose a systematic offset of --0.37 dex in the stellar mass of the AS2UDS, ASAGAO, ALESS, and ASPECS samples. The SFRs, dust luminosities, and dust masses are not corrected, because we did not find obvious systematic differences for these parameters. Moreover, we use the characteristic dust temperature to compare the dust temperature among different studies. The conversion methods of dust temperature among different definitions are summarized in Appendix~\ref{appendix:temperature}.

\section{Results} \label{section:result}

\begin{figure*}[t]
 \epsscale{0.6}
 \plotone{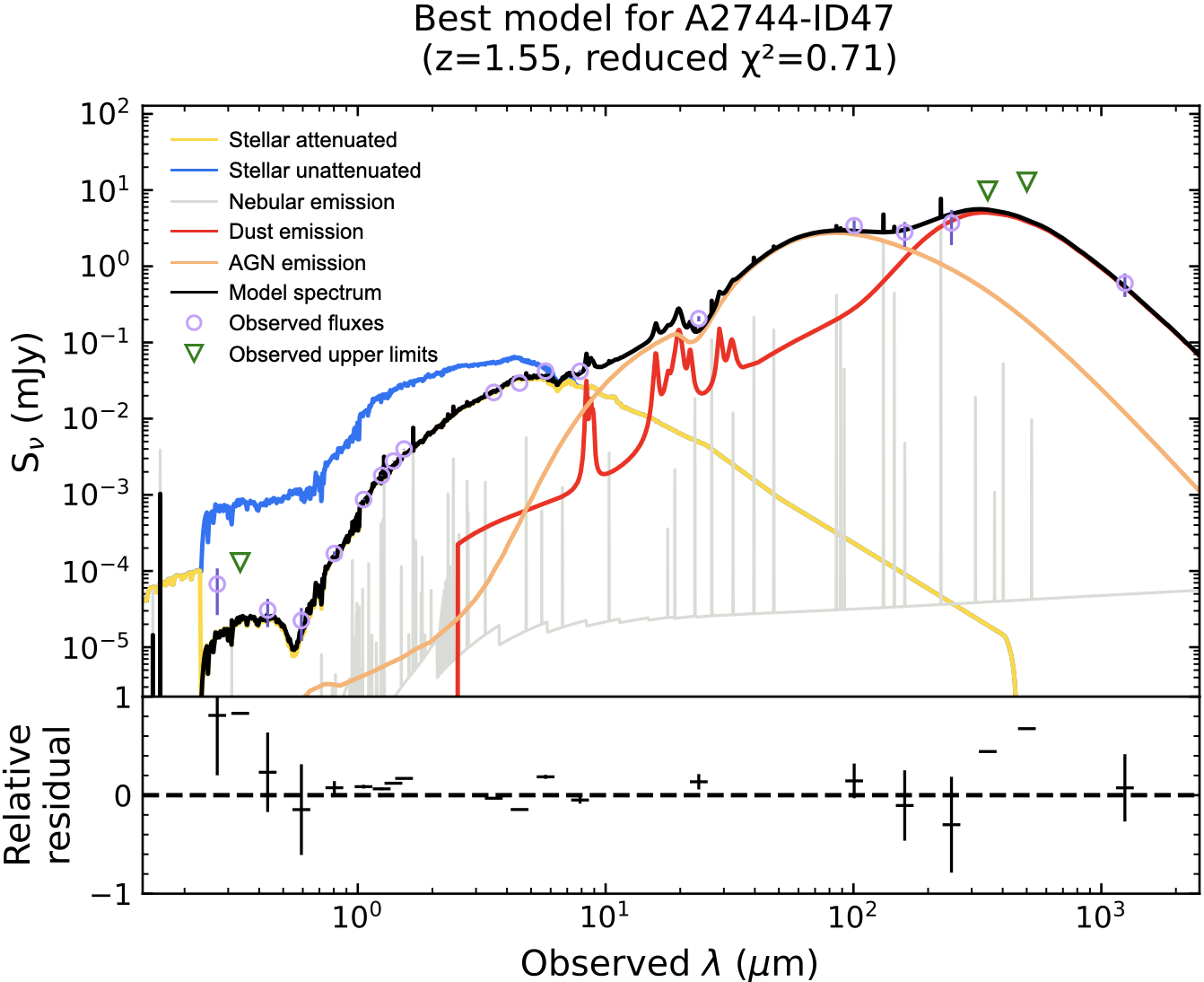}
 \caption{The SED and the best-fit model of A2744-ID47. Lensing magnification is NOT corrected in this figure. The black solid line represents the composite spectrum of the galaxy. The yellow line illustrates the stellar emission attenuated by interstellar dust. The blue line depicts the unattenuated stellar emission for reference purpose. The orange line corresponds to the emission from an AGN. The red line shows the infrared emission from interstellar dust. The gray line denotes the nebulae emission. The observed data points are represented by purple circles, accompanied by 1$\sigma$ error bars. The bottom panel displays the relative residuals. \label{figure:example}}
\end{figure*}

We successfully reproduce the SEDs of the ALCS tier-1/2/3/4/5/6 samples with a fit quality of $\chi^2/\mathrm{dof}<5$. Figure~\ref{figure:example} shows the example SED of the ALCS sample. The fit quality is shown in Figure~\ref{figure:chi2}. All the SEDs and the best-fit models are summarized in Appendix~\ref{appendix:freeparameter}. 

\begin{figure}[h]
 \epsscale{1.0}
 \plotone{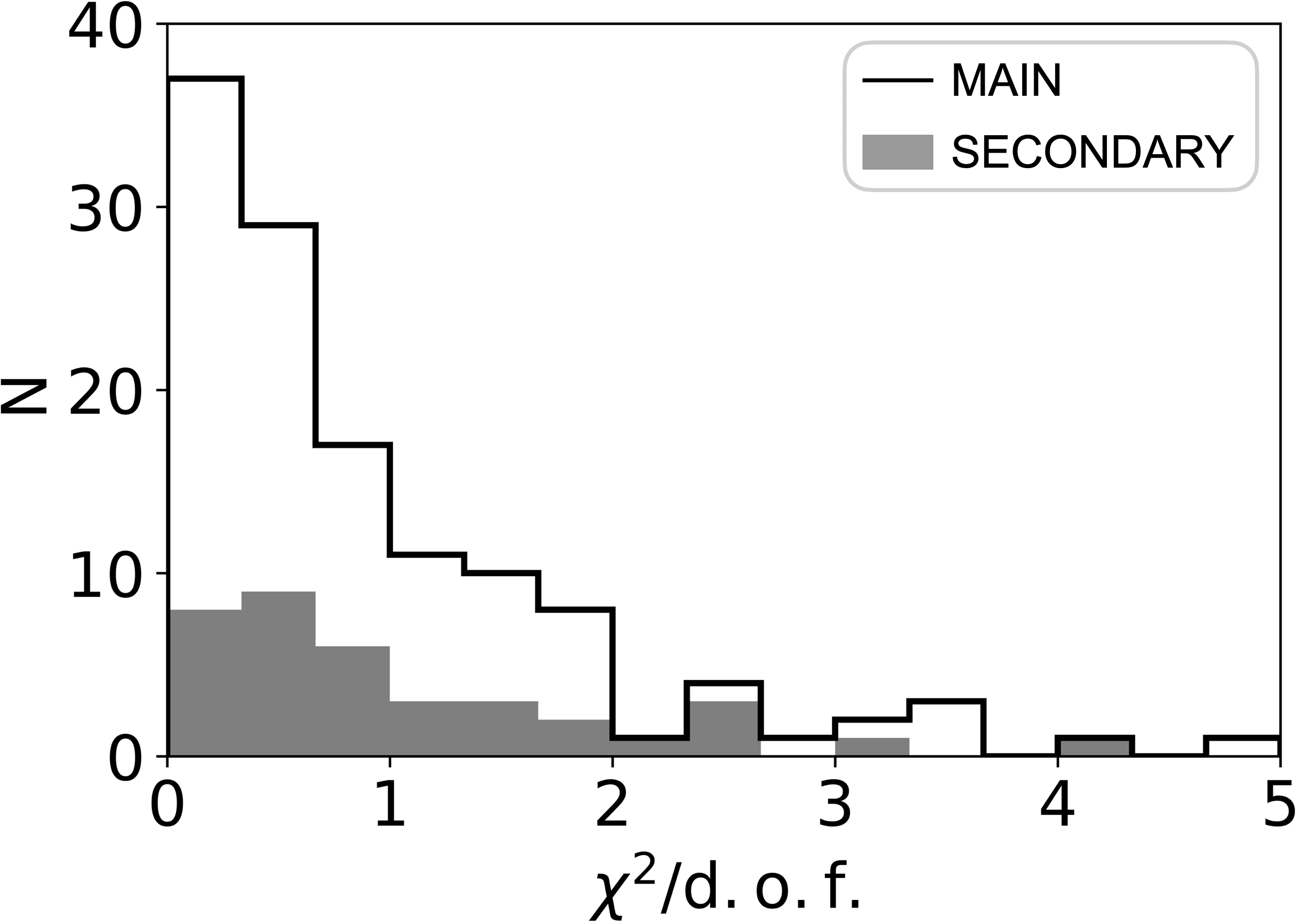}
 \caption{Histogram of fit quality of the ALCS tier-1/2/3/4/5/6 samples. The multiply-imaged sources are counted individually. \label{figure:chi2}}
\end{figure}

\subsection{Overall properties of ALCS sample}

\begin{figure*}[h]
 \epsscale{1.17}
 \plotone{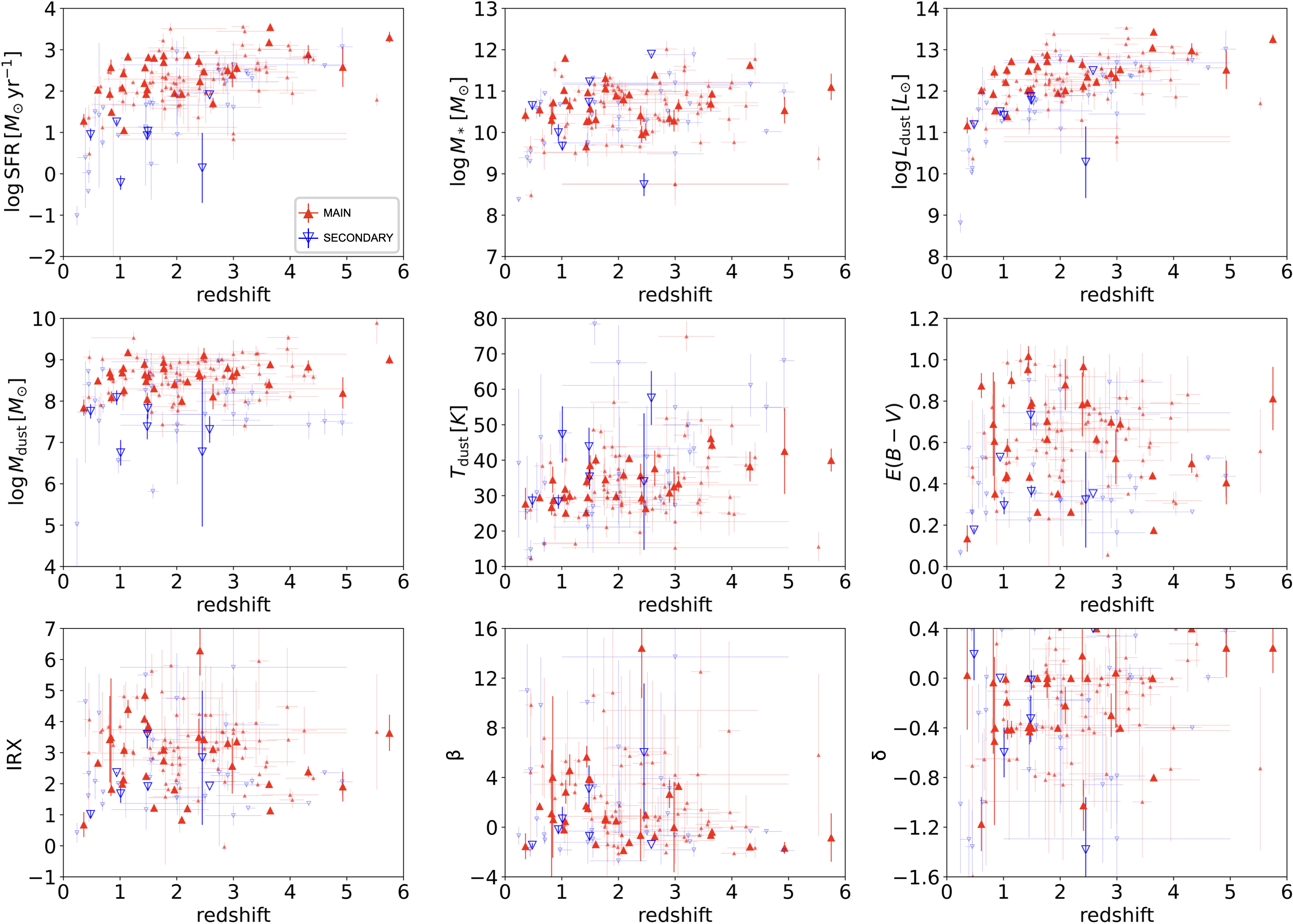}
 \caption{Distribution of the ALCS tier-1/2/3/4/5/6 samples in nine parameter spaces. The physical properties of ALCS-XAGNs are extracted from \citet{2023ApJ...945..121U}. The filled red triangles represent the MAIN sample and the blue open triangles the SECONDARY sample. The large symbols show the galaxies that have spectroscopic redshifts, while the small symbols show those with photometric redshifts. For each set of multiple images, we include only one image, which is selected with a priority given to a higher tier sample and to an image with a smaller magnification factor, whose uncertainty is expected to be smaller. All the physical quantities in this figure are NOT corrected for lensing magnification. \label{figure:distribution}}
\end{figure*}

Figure~\ref{figure:distribution} shows the distribution of the ALCS sample across nine parameter spaces. We confirm that the MAIN sample comprises galaxies with higher dust luminosities, dust mass, and SFRs. We also confirm that the SECONDARY sample tends to contain galaxies with lower $E(B-V)$ and lower IRX. These differences can be explained by the selection effects. Since the MAIN sample is selected by higher significance in the ALMA band, it contains galaxies that are brighter in the millimeter band. This results in the higher dust luminosities, dust mass, and SFRs of the MAIN sample. By contrast, since the IRAC/ch2 band can suffer from dust attenuation in high-redshift universe, the IRAC selection of the SECONDARY sample might cause a selection bias against heavily attenuated sources.

\begin{figure*}[h]
 \epsscale{0.5}
 \plotone{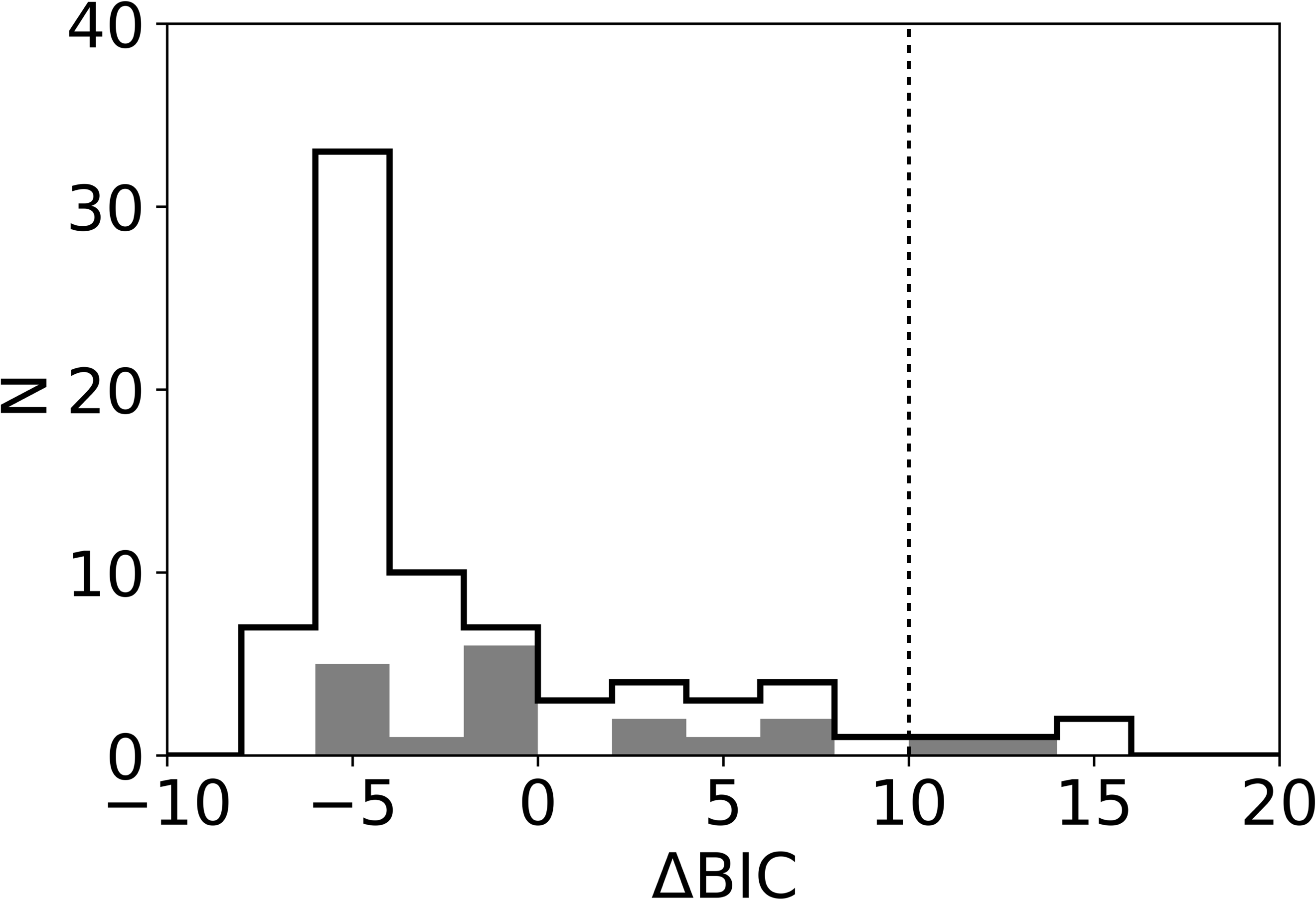}
 \caption{Distribution of $\Delta \mathrm{BIC}$ in the ALCS tier-1 samples (ALCS-XAGNs are excluded). One source belonging to the SECONDARY sample lies off the plot with $\Delta\mathrm{BIC}=118$. The vertical dotted line shows $\Delta \mathrm{BIC}=10$. The multiply-imaged sources are counted individually. \label{figure:BIC}}
\end{figure*}

Adhering to the methodology outlined in Section~\ref{subsubsection:AGNemission}, we identify six AGN candidates within the tier-1 sample (excluding ALCS-XAGNs); hereafter, we refer to these AGN candidates as ``ALCS-nonXAGNs". Figure~\ref{figure:BIC} shows the histogram of $\Delta \mathrm{BIC}$. Here we note that R0032-ID57 is also classified as an AGN with the BIC selection ($\Delta\mathrm{BIC}=13.3$). However, the other multiple images (R0032-ID32/53/55/58) are not classified as AGNs, and no obvious signs of AGN are found in the SED of R0032-ID57. Therefore, we do not consider R0032-ID57 as a host of an AGN.

\subsection{SFR versus stellar mass}\label{subsection:Ms-SFR}

\begin{figure*}[t]
 \epsscale{1.17}
 \plotone{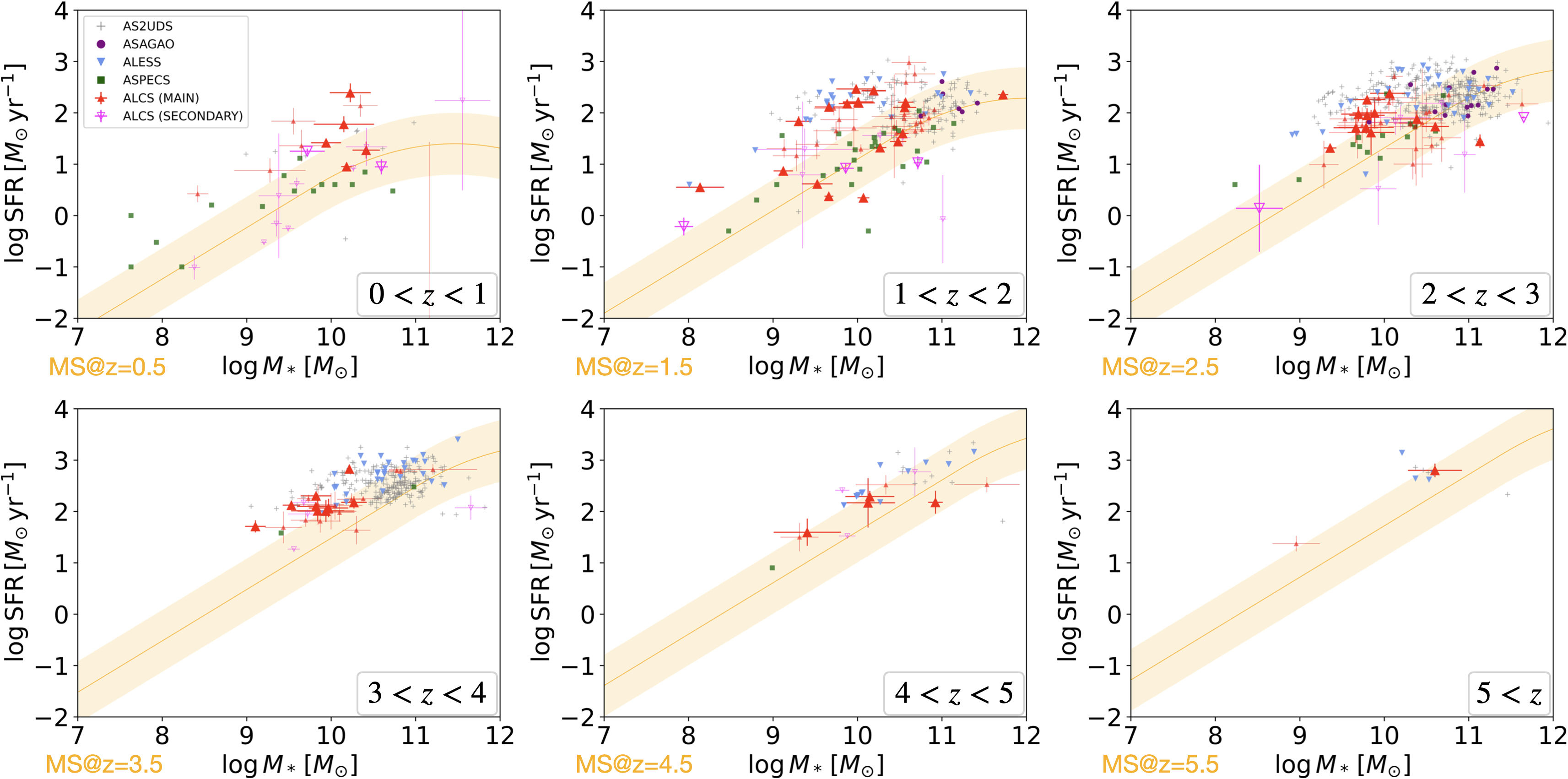}
 \caption{Relation between stellar mass ($M_*$) and SFR for galaxies detected in major ALMA surveys. The orange solid lines represent the star-formation main sequence at each redshift \citep{2015A&A...575A..74S}. The light orange areas show the ranges four times above or below the main sequence. The large symbols show the galaxies with spectroscopic redshifts, while the small symbols show those with photometric redshifts. The multiply-imaged sources are plotted individually. The stellar masses and the SFRs of the ALCS sources are corrected for lensing magnification. \label{figure:Ms_SFR}}
\end{figure*}

Figure~\ref{figure:Ms_SFR} plots the relation between stellar mass and SFR. For comparison, we plot the star-forming ``main sequence'' given by \citet{2015A&A...575A..74S}. This figure shows that the majority of the ALCS sources are in the star-forming main sequence (82/162 [51\%] sources are in the ranges four times above or below the star-forming main sequence), while some of them show intense starburst activities (69/162 [43\%]) and the others (11/162 [7\%]) are below the main sequence. This trend is similar to what is seen for the ALESS, AS2UDS, and ASAGAO samples, although the ALCS sample contains sources with even lower SFRs and stellar masses. On the other hand, the ALCS sample shows higher SFRs compared to the ASPECS sample. These differences simply reflect the different sample selections. Since the effective depth of ALCS is deeper than those of ALESS, AS2UDS, and ASAGAO due to the lensing effect, ALCS was able to detect galaxies with lower SFRs ($M_*$ is typically lower, according to the star-forming main sequence). At the same time, because the survey area with strong lensing effects is limited, the ALCS sample generally has higher SFRs compared with ASPECS, which is a homogeneously deep survey with ALMA.

\subsection{IRX versus stellar mass}\label{subsubsection:IRX_Ms}

\begin{figure*}[htb]
 \epsscale{1.17}
 \plotone{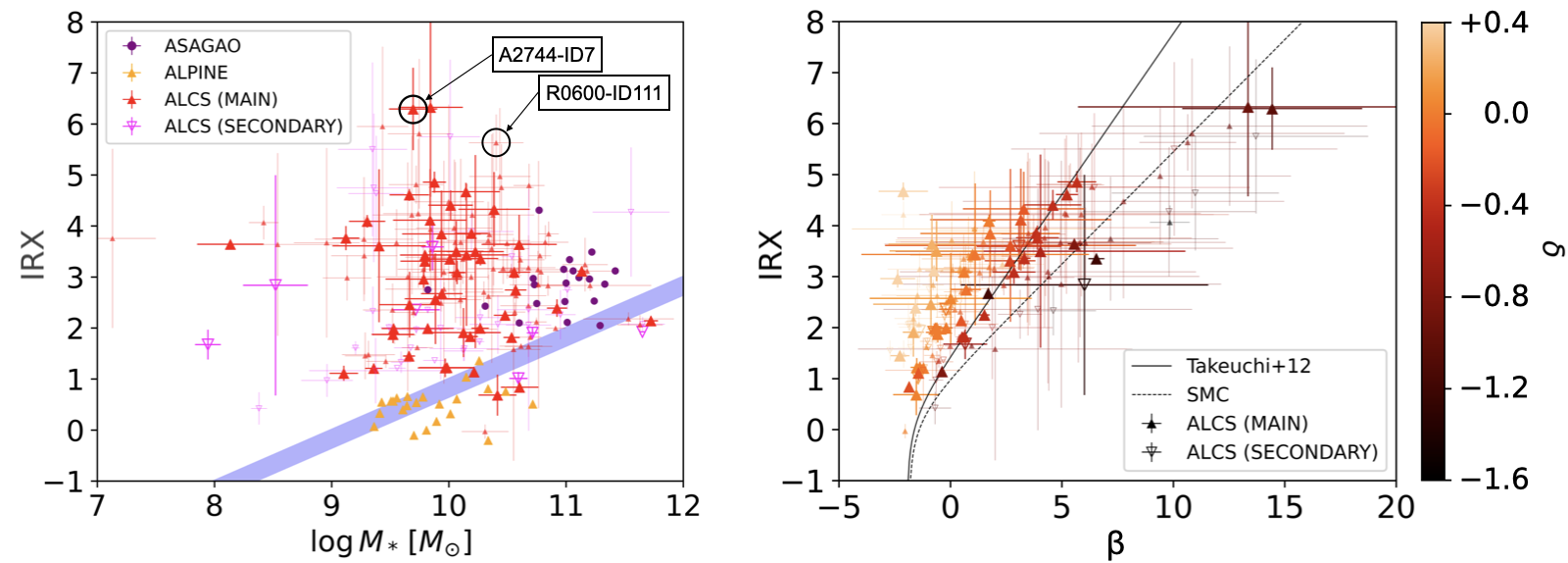}
 \caption{(Left) Relation of IRX versus stellar mass in major ALMA surveys. The blue line shows the conventional relation of optical or UV-selected galaxies at $z\sim$2--3 compiled by \citet{2016ApJ...833...72B}. (Right) Relation of IRX versus observed UV slope ($\beta$) in the ALCS sample. The color map shows the power-law index to modify the original Calzetti law ($\delta$). The black solid and dashed lines show the local relationship derived by \citet{2012ApJ...755..144T} and the so-called SMC IRX--$\beta$ relationship, respectively. In these figures, the ALCS tier-1/2/3/4/5/6 samples are plotted. The large symbols show the galaxies with spectroscopic redshifts, while the small symbols show those with photometric redshifts. The multiply-imaged sources are plotted individually. The stellar masses of the ALCS sources are corrected for lensing magnification. \label{figure:beta_IRX}}
\end{figure*}

\begin{figure*}[htb]
 \epsscale{1.17}
 \plotone{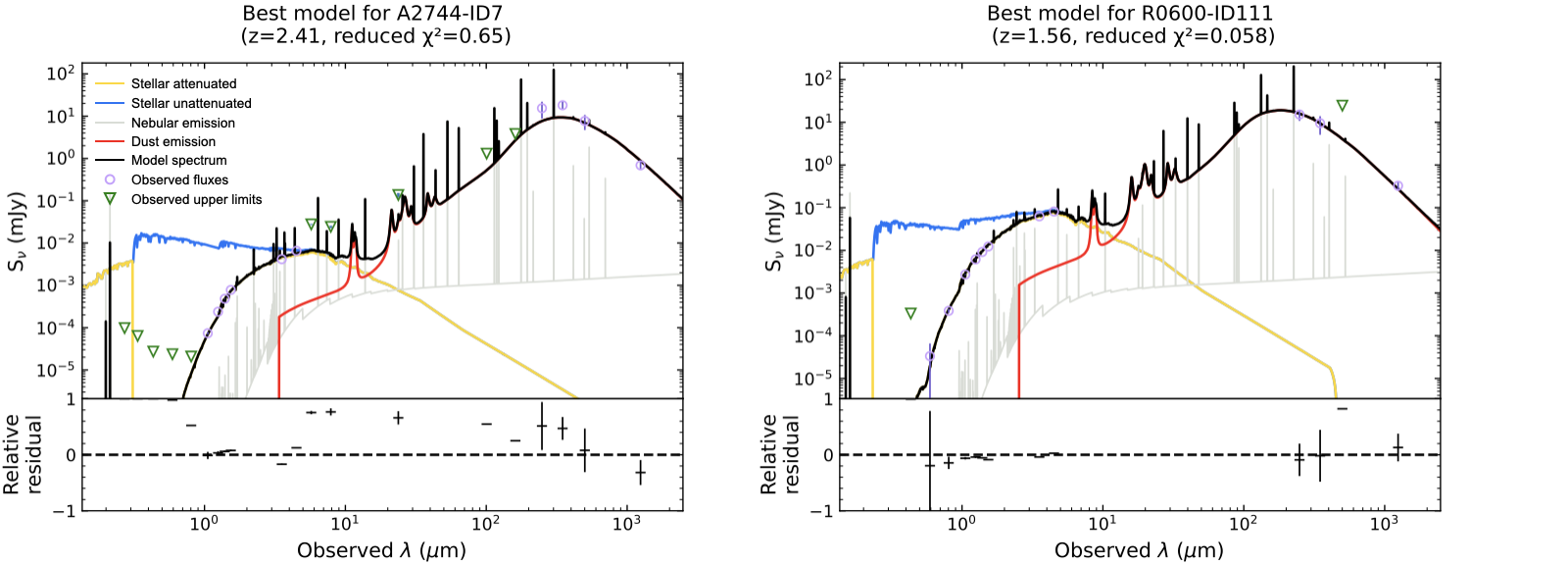}
 \caption{The SEDs of the galaxies that have especially high IRX. The SEDs are NOT corrected for lensing magnification. The colors and the symbols are the same as those in Figure~\ref{figure:example}. \label{figure:highIRX}}
\end{figure*}

The left panel of Figure~\ref{figure:beta_IRX} shows the relation between stellar mass and IRX. For comparison, we plot the conventional relationship of optical or UV selected galaxies at $z\sim 2$--3 studied by \citet{2010ApJ...712.1070R}, \citet{2014ApJ...795..104W}, and \citet{2016A&A...587A.122A}, as compiled by \citet{2016ApJ...833...72B}. We see that the ALCS sample has higher IRXs than the conventional relationship, implying that these galaxies are more deeply enshrouded by dust than optical- or UV-selected galaxies. We also find two extremely dusty-obscured galaxies that have especially high IRX (A2744-ID7 [$\mathrm{IRX}=6.3\pm0.8$] and R0600-ID111 [$\mathrm{IRX}=5.6\pm0.6$]). Figure~\ref{figure:highIRX} shows the SEDs of the two sources. These galaxies have steeper attenuation curves than the average (the steepness of the attenuation slopes are $-1.0\pm0.2$ and $-0.8\pm0.1$, respectively), which may result in more UV absorption and thus larger IRX. Recently, \citet{2023ApJ...945L..25K} found a potential major merger at $z=2.58$ that is deeply enshrouded by dust utilizing James Webb Space Telescope (JWST) observations. This might suggest that high-IRX sources in SMG samples are important targets for JWST to reveal the evolutional stages of distant galaxies and SMBHs. We note that there still remain some sources that potentially have high IRXs (R0032-ID281 [$6.3\pm1.8$], ACT0102-ID22 [$6.0\pm1.6$], A3192-ID131 [$5.8\pm1.5$], R0032-ID245 [$5.7\pm1.5$], and R0949-ID14 [$5.5\pm1.7$]), which belong to the tier-3/4 samples. Moreover, the tier-7/8 samples may also contain such a population, although they are not analyzed in this study. Deeper observations in optical and near-infrared bands are needed to confirm the nature of these sources.

\subsection{IRX versus observed UV slope}\label{subsubsection:IRX_beta}

The right panel of Figure~\ref{figure:beta_IRX} shows the relation between IRX and observed UV slope. For comparison, we plot the local relation derived by \citet{2012ApJ...755..144T} and the so-called SMC IRX--$\beta$ relation, which is based on the obervational results of \citet{1982A&A...113L..15L}, \citet{1984A&A...132..389P}, and \citet{1985A&A...149..330B} (see also \citealt{1992ApJ...395..130P,1998ApJ...508..539P,2016ApJ...833..254S}). The color map shows the steepness of the attenuation slopes. We see that the ALCS sample is distributed above the local relation at $\beta\lesssim3$, while it is close to the SMC IRX--$\beta$ relation at $\beta\gtrsim3$. This can be explained by the sample selection. Since the ALCS sample is selected by millimeter observations, it is more sensitive to high-IRX sources than UV/optical selected samples. On the other hand, since the tier-1/2/3/4/5/6 samples are also selected by HST or IRAC observations, those samples miss sources that have both high IRX and small $\beta$, which are faint even in the near-infrared band.

We also confirm a large scatter in the steepness of attenuation slopes, which strongly affects the IRX--$\beta$ relation, i.e., galaxies with flatter attenuation slopes (high $\delta$) tend to show bluer UV spectrum (low $\beta$). As will be discussed in Section~\ref{subsection:attenuation}, the steepness of the attenuation slope is weakly correlated with the stellar mass. This might account for the differences in the IRX--$\beta$ relation for different stellar masses reported in recent studies (e.g., \citealt{2020ApJ...902..112B}). However, the IRX--$\beta$ relation can also be strongly affected by the geometry of dust and stellar components (e.g., \citealt{2017MNRAS.472.2315P,2018MNRAS.474.1718N,2020MNRAS.491.4724F}). Spatially-resolved SED analysis with high-resolution observation will be helpful to reveal the dust properties in high-redshift star-forming galaxies.

\section{Discussion} \label{section:discussion}

\subsection{Dust properties}

In this subsection, we discuss the dust properties of the ALCS sample. We focus on the dust attenuation law, dust temperature, and dust mass.

\subsubsection{Dust Attenuation Law}\label{subsection:attenuation}

\begin{figure*}[t]
 \epsscale{1.17}
 \plotone{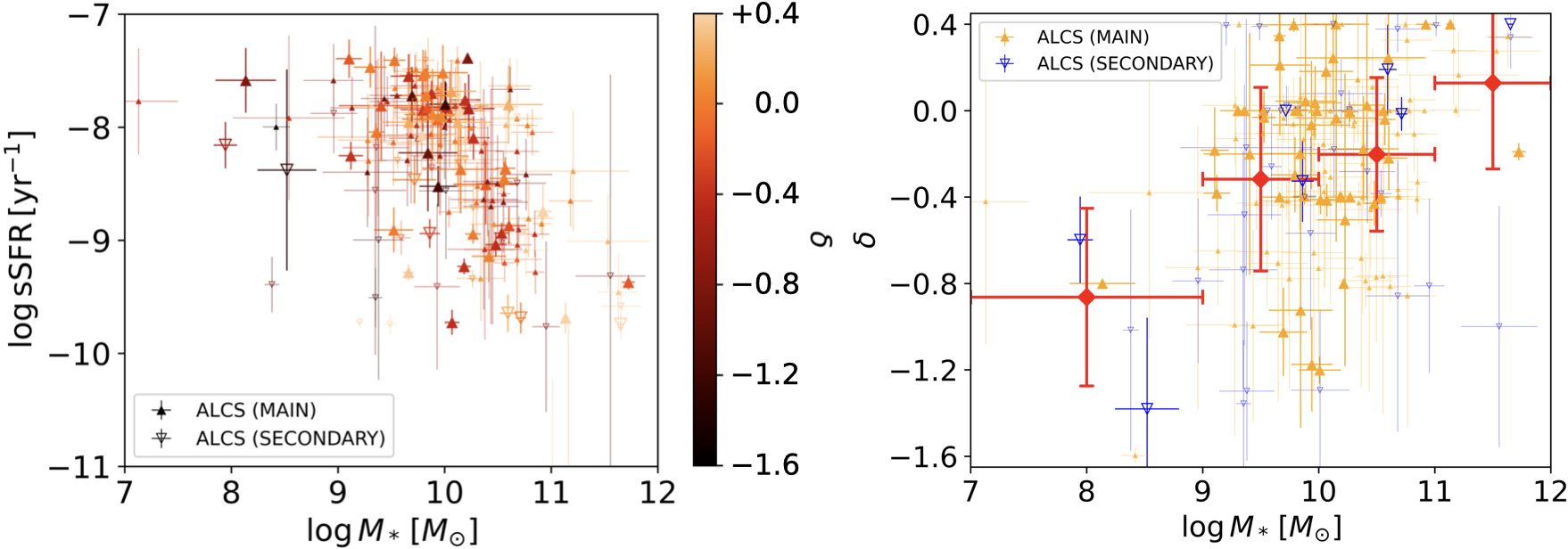}
 \caption{(Left) The color-coded plot of $\delta$ (power-law index to modify the original Calzetti law, representing the steepness of the attenuation curve) on the sSFR versus stellar mass plane. (Right) Relation between $\delta$ and stellar mass. The red points show the mean values and standard deviations of $\delta$ in different stellar-mass bins. In these figures, the ALCS tier-1/2/3/4/5/6 samples are plotted. The large symbols show the galaxies that have spectroscopic redshifts, while the small symbols show those with photometric redshifts. The multiply-imaged sources are plotted individually, but they are counted only once in deriving the average values. The stellar masses are corrected for lensing magnification. \label{figure:delta}}
\end{figure*}

The dust attenuation law is critical to study the stellar populations and thus the evolutionary scenario of galaxies. Over time, numerical studies have revealed the extensive variation in attenuation curves among different galaxies and even within a single galaxy depending on the line of sight (e.g., \citealt{1975A&A....44..195N,1981A&A....99L...5R,1980Natur.283..725N}). The main features to characterize an attenuation curve are the steepness of the curve and the depth of the 2175 \AA\ bump (UV bump). Unfortunately, it is challenging to measure the depth of UV bumps with the poor photometric data of high-redshift galaxies. Therefore, in this paper, we focus only on the steepness of the attenuation curves of the ALCS sample.

In the modified Calzetti starburst attenuation law, the steepness of the attenuation curve is quantified by $\delta$, the power-law index to modify the original Calzetti law \citep{2000ApJ...533..682C}. The left panel of Figure~\ref{figure:delta} gives a color-coded plot of $\delta$ on the sSFR versus stellar mass plane. We confirm that no strong correlation exists between $\delta$ and sSFR. The right panel of Figure~\ref{figure:delta} plots the relation between stellar mass and $\delta$. A weak correlation is noticed between the steepness of attenuation curve and stellar mass, i.e., galaxies with high stellar masses tend to have flatter attenuation curves. This is consistent with the trend observed in local galaxies \citep{2018ApJ...859...11S}.

\subsubsection{Dust temperature}

\begin{figure*}[htb]
 \epsscale{1.17}
 \plotone{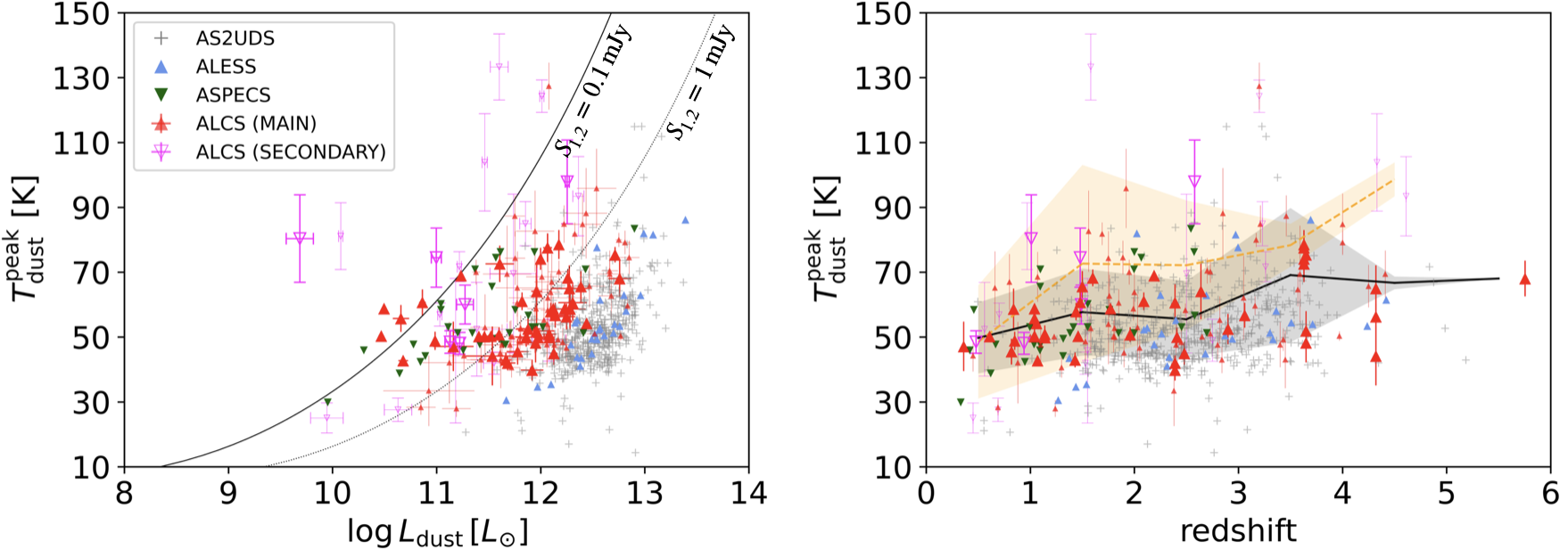}
 \caption{(Left) Characteristic dust temperature versus dust luminosity. The black solid and dotted lines represent the flux limit at 1.2 mm, where we assume an optically-thin graybody with an emissivity index of 1.8 at $z=3$. (Right) Characteristic dust temperature versus redshift. The black solid and orange dashed lines show the mean characteristic dust temperature of the MAIN and the SECONDARY samples in each redshift bin, respectively. The semi-transparent regions show the standard deviations (i.e., $\pm 1\sigma$) from the mean values. In these plots, the ALCS, ASPECS, and AS2UDS samples are limited to the sources that are detected with Herschel over 2$\sigma$. The ALESS sample is also limited to the well-sampled subset reported in \citet{2021ApJ...919...30D}. The large symbols show the galaxies that have spectroscopic redshifts, while the small symbols show those with photometric redshifts. The multiply-imaged sources are plotted individually, but they are counted only once in deriving the average values. The dust luminosities are corrected for lensing magnification. \label{figure:temperature}}
\end{figure*}

The cosmological evolution of dust temperature is a key issue for understanding star-formation activity in the high-redshift universe, as assumptions regarding dust temperature can significantly impact dust luminosity estimations and, consequently, star-formation rates. Some studies have claimed a global increase in dust temperature with redshift across $z=0\text{--}9$ (c.f., \citealt{2018A&A...609A..30S,2022MNRAS.516L..30V}). On the other hand, other studies have pointed out that the increase of dust temperature only arises from the sample selection bias, and no cosmological evolution in dust temperature at a given far-infrared luminosity is required (e.g., \citealt{2020MNRAS.494.3828D,2022ApJ...930..142D}). For the ALCS sample, this issue has already been discussed in \citet{2022ApJ...932...77S}. They conclude that no evolution has been found in the dust temperatures of Luminous Infrared Galaxies (LIRGs) ($L_{\mathrm{IR}}<10^{12}L_{\odot}$) at $z\leq2$ and ULIRGs ($L_{\mathrm{IR}}>10^{12}L_{\odot}$) at $z\simeq 1\text{--}4$. In this paper, we just discuss the robust properties of dust temperature in the ALCS sample and compare them with other SMG samples.

Figure~\ref{figure:temperature} compares the dust temperatures obtained from major ALMA surveys. For a conservative discussion, we only plot the sources that are detected with Herschel over 2$\sigma$ for the ALCS, ASPECS, and AS2UDS samples, where the dust temperatures are more reliably constrained than the other sources. For the same reason, the ALESS sample is limited to the well-sampled subset reported in \citet{2021ApJ...919...30D}. One should be aware that these selections can cause biases and the subsamples might not reflect the averaged properties of the whole samples. The left panel plots characteristic dust temperature versus dust luminosity. We confirm that the galaxies that have higher dust luminosities tend to have higher dust temperatures. This can be explained by the selection bias, that galaxies with colder dust are brighter in the sub/millimeter band in the same luminosity range. We also confirm that, thanks to the lensing effect, the ALCS sample probes a broader range in lower dust luminosities than the conventional SMG samples (ALESS and AS2UDS) in the same temperature range. The right panel plots characteristic dust temperature versus redshift. The black solid and orange dashed lines show the average characteristic dust temperatures of the ALCS MAIN and the SECONDARY samples in each redshift bin, respectively. We confirm good agreement between the ALCS and the ASPECS samples, whereas a small systematic offset is found between the ALCS and the ALESS/AS2UDS samples. This discrepancy can also be attributed to the selection, showing that the ALCS sample includes sources with higher dust temperatures than the conventional SMG samples in the same redshift range. 

We note that M0416-ID156 and A2537-ID24/66 (a multiply-imaged source) exhibit especially high dust temperatures ($T_{\mathrm{dust}} > 70\,\mathrm{K}$, corresponding to $T_{\mathrm{dust}}^{\mathrm{pead}} > 119\,\mathrm{K}$). These results might suggest an intense star-formation activity in a compact morphology \citep{2021ApJ...910...89B} or a ``dubbed starburst" as shown in \citet{2022A&A...659A.196G}. However, we should be cautious about the potential uncertainty of the SED fitting. For instance, the far-infrared SED of M0416-ID156 is not well-fitted by the EThemis model. This may be attributed to the oversimplification of the model, and can cause large uncertainty in dust temperature and dust luminosity. Conversely, the SEDs of A2537-ID24/66 are fairly well-fitted. Nevertheless, there remains a possibility that their high dust temperatures are actually caused by the remaining blending effect of low-redshift sources. Future multi-band observation with ALMA will be helpful to confirm the nature of these sources.

\subsubsection{Dust mass}

\begin{figure*}[htb]
 \epsscale{1.17}
 \plotone{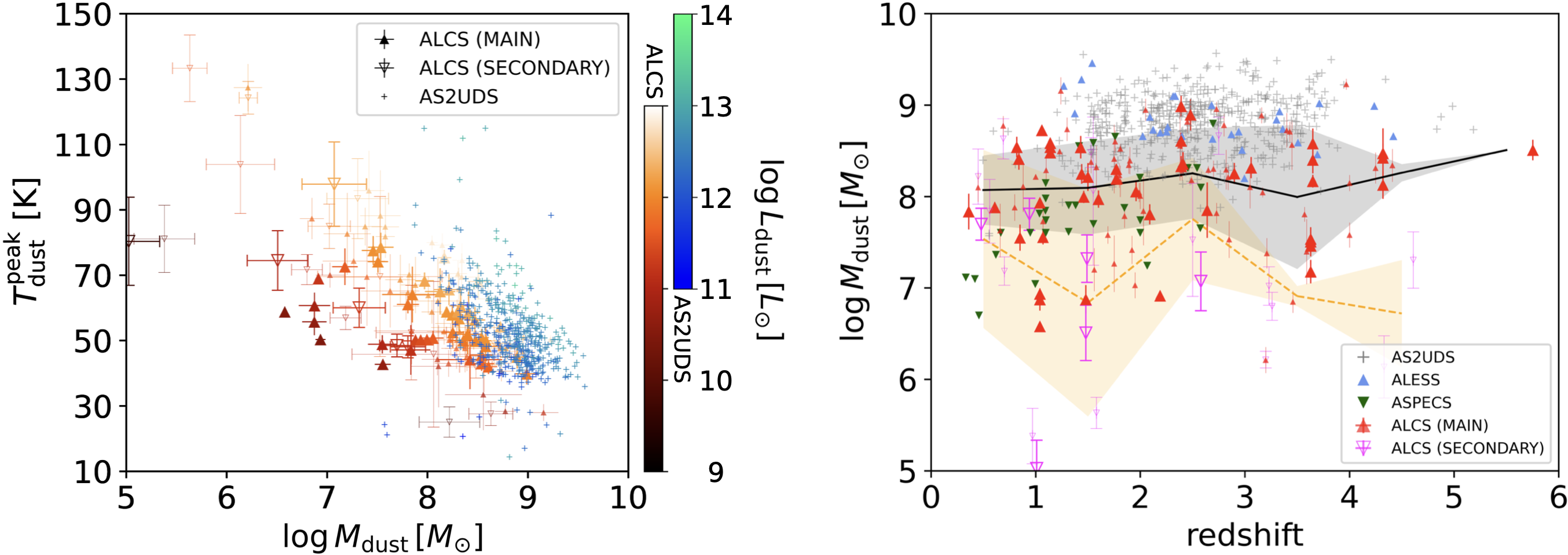}
 \caption{(Left) Characteristic dust temperature versus dust mass. The color bars show the dust luminosities. (Right) Dust mass versus redshift. The black solid and orange dashed lines show the mean dust mass for the MAIN and the SECONDARY samples in each redshift bin, respectively. The standard deviations are shown by the semi-transparent regions. In these plots, the ALCS, ASPECS, and AS2UDS samples are limited to the sources that are detected with Herschel over 2$\sigma$. The ALESS sample is also limited to the well-sampled subset reported in \citet{2021ApJ...919...30D}. The large symbols show the galaxies that have spectroscopic redshifts, while the small symbols show those with photometric redshifts. The multiply-imaged sources are plotted individually. The dust masses and dust luminosities are corrected for lensing magnification. \label{figure:dustmass}}
\end{figure*} 

The dust mass of a galaxy is also an essential parameter because dust is an efficient catalyst of molecular hydrogen formation in the interstellar medium (e.g., \citealt{1963ApJ...138..393G,2004ApJ...604..222C,2009A&A...496..365C}) and plays as a possible enhancer of star-formation activities in high-redshift universe \citep{2002MNRAS.337..921H}. Since the dust mass of a galaxy is dominated by cold interstellar dust, submillimeter observations are suited for establishing a dust-mass-selected sample. Recently, \citet{2020MNRAS.494.3828D} have suggested that the dust mass of AS2UDS sample is tightly correlated with the flux density at 870 $\mu$m. Moreover, \citet{2021MNRAS.500..942D} have shown that 870 $\mu$m selection (AS2UDS) provides a more uniformly selected dust mass sample across $z=1\text{--}6$ than 450 $\mu$m selection (STUDIES; \citealt{2017ApJ...850...37W}). Since the ALCS sample is selected by longer wavelengths than those of AS2UDS, it is expected to provide a uniformly selected dust mass sample like AS2UDS. However, the estimation of dust mass is sensitive to dust temperature and dust opacity \citet{2021ApJ...919...30D}. Deep multi-wavelength observations in far-infrared bands are needed to solve these degeneracies.

The left panel of Figure~\ref{figure:dustmass} compares the dust temperature, dust mass, and dust luminosity in the ALCS and the AS2UDS samples. For conservative discussion, we again impose the same far-infrared selections as those in Figure~\ref{figure:temperature}. We confirm that the ALCS sample contains galaxies with lower dust masses than the AS2UDS sample in the same dust temperature range. This can be attributed to the selection effect, showing that the ALCS sample is able to probe galaxies that have lower dust mass than the conventional SMG samples. The right panel plots dust mass versus redshift. The black solid and orange dashed lines show the average dust mass of the ALCS MAIN and the SECONDARY samples in each redshift bin, respectively. This figure shows that there is no obvious evolution in the mean dust mass of the ALCS sample but the ALCS sample spreads widely in a lower dust mass range than the conventional SMG samples. Actually, as shown in the middle panel of Figure~\ref{figure:distribution}, ALCS MAIN sample behaves as a dust-mass-selected sample ranging $\log M_*/M_{\odot}=8\text{--}9$ across $z$=1--6 before lensing correction. Therefore, the large scatter found in the dust mass of the ALCS sample just arises from the variation in the magnification factors.

\subsection{AGN}

In this subsection, we discuss the properties of the ALCS-nonXAGNs. We also discuss the cosmological evolution of the AGN luminosity density inferred from our sample.

\subsubsection{Comparison with X-ray luminosity upperbounds}

The left panel of Figure~\ref{figure:AGN} plots the X-ray luminosity upperbounds with various assumptions on line-of-sight obscuration, against bolometric AGN luminosity derived by \texttt{CIGALE}. For comparison, we plot the empirical bolometric-to-X-ray luminosity relation of local AGNs ($L_{\mathrm{bol}}/L_{\text{2--10 keV}}=20$; \citealt{2007MNRAS.381.1235V})\footnote{Note that this relationship has a large scatter ($L_{\mathrm{bol}}/L_{\mathrm{2\text{--}10\,keV}}=6\text{--}400$) and shows potential strong dependence on the Eddington ratio (see also \citealt{2012MNRAS.425..623L} and \citealt{2017ApJ...844...10B} for other calibrations).}. We find that most of the ALCS-nonXAGNs are consistent with their X-ray upperbands only in the heavily-obscured cases, where the line-of-sight hydrogen column density ($N_{\mathrm{H}}^{\mathrm{LOS}}$) is over $10^{24.5}\,\mathrm{cm}^{-2}$. This suggests that these AGNs are heavily obscured or Compton-thick AGNs. Recently, \citet{2017ApJ...839...58S} have reported that the gas column density inferred from the dust surface density in the AS2UDS sample is $N_{\mathrm{H}}=9.8^{+1.4}_{-0.7}\times 10^{23}\,\mathrm{cm}^{-2}$, which is equivalent to the heavy-obscuration case in Figure~\ref{figure:AGN}. Hence, the heavy obscuration of our AGN sample might be attributed to the host galaxies. Moreover, it has been suggested that some AGNs in local ULIRGs show ``X-ray weak'' features ($\kappa_{\text{2--10 keV}}>100$; \citealt{2015ApJ...814...56T,2021ApJS..257...61Y}). If this is applicable in our sample, the ALCS-nonXAGNs can be consistent with their X-ray luminosity upperbounds even in the less obscured cases ($N_{\mathrm{H}}^{\mathrm{LOS}} \leq 10^{23}\,\mathrm{cm}^{-2}$).

\subsubsection{Cosmological evolution of AGN luminosity density}

\begin{figure*}[htb]
  \epsscale{1.17}
  \plotone{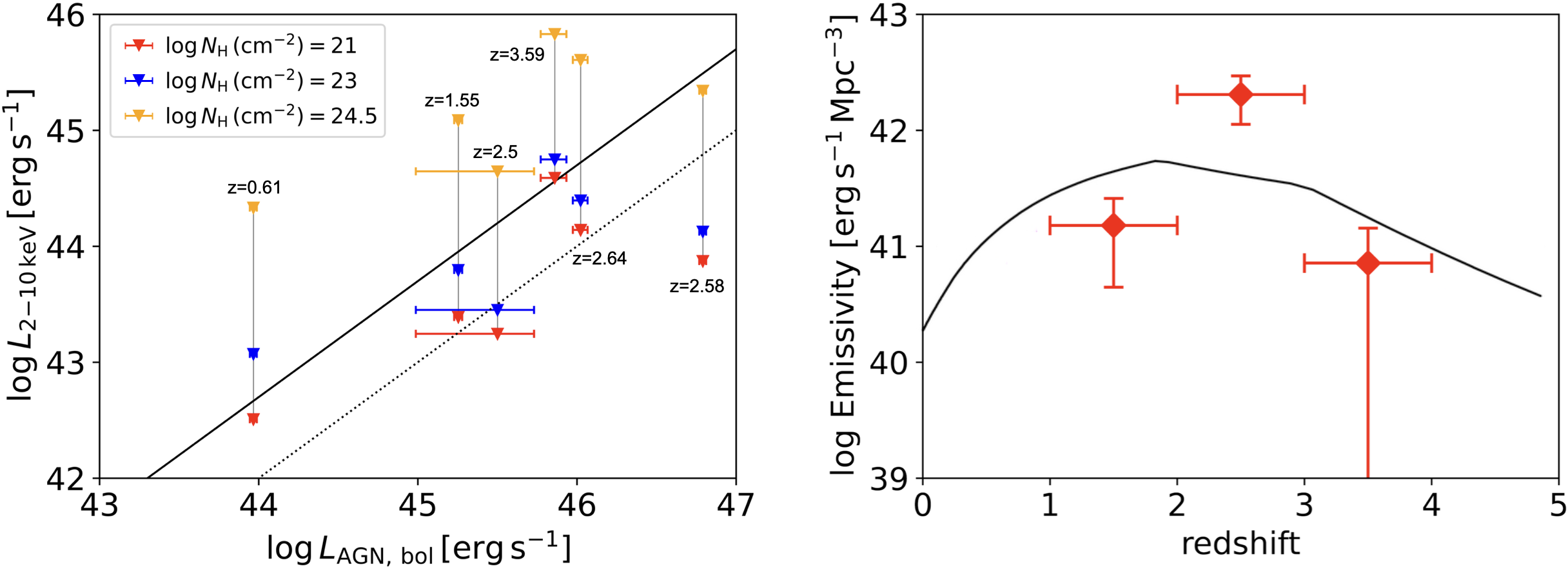}
  \caption{(Left) X-ray luminosity upperbound with various assumptions of line-of-sight obscuration, plotted against bolometric AGN luminosity derived by SED analysis. The colors correspond to the line-of-sight hydrogen column density assumed in calculating the X-ray luminosity upperbound. The black solid line shows the empirical bolometric-to-X-ray luminosity relation of local AGNs ($L_{\mathrm{bol}}/L_{\text{2--10 keV}}=20$; \citealt{2007MNRAS.381.1235V}), and the black dotted line shows the ``X-ray weak'' case ($L_{\mathrm{bol}}/L_{\text{2--10 keV}}=100$). In this plot, the AGN luminosities and X-ray luminosity upperbounds are corrected for lensing magnification. (Right) Cosmological evolution of AGN luminosity density. The error bars only show the statistical errors that arise from the number of sources in each redshift bin. The black solid line shows that determined from X-ray surveys by \citet{2014ApJ...786..104U}. \label{figure:AGN}}
\end{figure*}
 
We calculate the cosmological evolution of the AGN luminosity density (growth rate of SMBHs) in high-redshift U/LIRGs ($10^{11}L_{\odot} < L_{\mathrm{IR}} < 10^{13}L_{\odot}$). First, we estimate the average $L_{\mathrm{AGN}}$-to-$L_{\mathrm{IR}}$ ratio in redshift bins of $z=1\text{--}2$, $z=2\text{--}3$, and $z=3\text{--}4$. These values are calculated by comparing the summed best-fit AGN luminosities and infrared luminosities of the U/LIRGs in the tier-1 sample in each redshift bin.\footnote{For the ALCS-XAGNs, we employ the values derived by the SED analysis performed by \citet{2023ApJ...945..121U}.} Then, we apply the luminosity ratio to the infrared luminosity density of high-redshift U/LIRGs and obtain the AGN luminosity density. The infrared luminosity density is calculated by integrating the infrared luminosity function derived by \citet{2023arXiv230301658F} over $10^{11}L_{\odot}<L_{\mathrm{IR}}<10^{13}L_{\odot}$. We note that the infrared luminosity mentioned in this subsection refers to the one that does not contain the AGN component (same as the dust luminosity from star-formation activity).

The right panel of Figure~\ref{figure:AGN} shows the cosmological evolution of AGN luminosity density inferred from the ALCS tier-1 sample (including ALCS-XAGNs). We also plot the AGN luminosity density estimated by X-ray surveys below 10 keV \citep{2014ApJ...786..104U}. We confirm a possible excess at $z=2$--3. This might show that a significant fraction of AGNs in this epoch are heavily obscured and might be missed in previous X-ray surveys. However, we must bear in mind that the AGN detection by SED analysis should have large uncertainties, and the results should highly depend on the model assumptions. Future sensitive hard X-ray (${>}10$ keV) observations will be useful to reveal the nature of heavily obscured or Compton-thick AGNs in the high redshift universe.

\section{Summary} \label{section:summary}

We report the multi-wavelength properties of millimeter galaxies detected in the ALMA Lensing Cluster Survey (ALCS). The main conclusions are summarized as follows:
\begin{enumerate}
 \item We performed UV to millimeter SED modeling of the whole ALCS sample (except for the cluster members and the ALCS-XAGNs), utilizing the photometric data obtained with HST, Spitzer, Herschel, and ALMA. 
 \item We confirm that the majority of the ALCS sources lie on the star-forming main sequence, while a smaller fraction shows intense starburst activities. This trend is the same as for other ALMA-detected SMG samples, e.g., ALESS, AS2UDS, and ASAGAO, although the ALCS sample contains galaxies with even lower SFRs and stellar masses thanks to the lensing effect.
 \item We find two extremely dust-obscured galaxies (A2744-ID7 [$z=2.41$] and R0600-ID111 [z=1.56]). These galaxies have steeper attenuation slopes than the average, which may result in more UV absorption and hence higher IRX.
 \item We confirm that the ALCS sample exhibits a wider range in dust temperatures than the conventional SMG samples (ALESS, and AS2UDS) in the same redshift range. We also confirm that the ALCS sample shows a large scatter in a lower dust mass range than the conventional SMG samples after lensing correction.
 \item We identify six AGN candidates that are not detected in the archival Chandra data. The X-ray upperbounds indicate that these AGNs are heavily obscured or Compton-thick AGNs. We also find that the AGN luminosity density inferred from the ALCS tier-1 sample (including ALCS-XAGNs) shows a possible excess at $z=2\text{--}3$ compared with that determined from X-ray surveys below 10 keV \citep{2014ApJ...786..104U}. This suggests that a significant fraction of AGNs in this epoch are heavily obscured by gas and dust and might be missed in the previous X-ray observations.
\end{enumerate}

\begin{acknowledgments}

This publication uses data from the ALMA programs: ADS/JAO. ALMA \#2018.1.00035.L, \#2013.1.00999.S, and \#2015.1.01425.S. ALMA is a partnership of ESO (representing its member states), NSF (USA) and NINS (Japan), together with NRC (Canada), MOST and ASIAA (Taiwan), and KASI (Republic of Korea), in cooperation with the Republic of Chile. The Joint ALMA Observatory is operated by ESO, AUI/NRAO and NAOJ. This work has been financially supported by JSPS KAKENHI Grant Numbers 22J22795 (R.U.), 20H01946 (Y.U.), 17H06130 (K.K., Y.U.), 19K14759 (Y.T.), 22K20391, 23K13154 (S.Y.), and 22H01266 (Y.T.). This work has also been supported by the NAOJ ALMA Scientific Research Grant Number 2017-06B (K.K.). I.S. acknowledges support from STFC (ST/X001075/1). S.Y. is grateful for support from RIKEN Special Postdoctoral Researcher Program. We additionally acknowledge support from ANID - Millennium Science Initiative Program - ICN12\_009 (FEB), CATA-BASAL - ACE210002 (FEB) and FB210003 (FEB), and FONDECYT Regular - 1200495 (FEB). Numerical computations were carried out using the SuMIRe cluster operated by the Extragalactic OIR group at ASIAA. DE acknowledges support from a Beatriz Galindo senior fellowship (BG20/00224) from the Spanish Ministry of Science and Innovation, projects PID2020-114414GB-100 and PID2020-113689GB-I00 financed by MCIN/AEI/10.13039/501100011033, project P20\_00334 financed by the Junta de Andalucía, project A-FQM-510-UGR20 of the FEDER/Junta de Andalucía-Consejería de Transformación Económica, Industria, Conocimientoy Universidades.

\end{acknowledgments}

\vspace{5mm}

\facilities{ALMA, Chandra, HST, Spitzer, Herschel}

\software{
CIGALE v2022.0 \citep{2019A&A...622A.103B,2022ApJ...927..192Y},
MAGPHYS high-z extension (v2) \citep{2015ApJ...806..110D,2020ApJ...888..108B},
MAGPHYS photo-z extension \citep{2019ApJ...882...61B},
HEAsoft v6.27 \citep{2014ascl.soft08004N},  
CIAO v4.12 \citep{2006SPIE.6270E..1VF}, 
CASA v5.4.0 \citep{2007ASPC..376..127M},
SExtractor v2.5.0 \citep{1996A&AS..117..393B},
DustEM v4.2 \citep{2011A&A...525A.103C}
}

\appendix

\section{Detailed Description of X-ray Analysis and X-ray Luminosity Upperbounds}\label{appendix:xray}

The X-ray spectrum of an AGN mainly consists of three components: (1)~a direct component from the nucleus, (2)~reflection components from the torus and/or the accretion disk, and (3)~a scattered component and emission from photoionized plasma. Since the emission from photoionized plasma is not dominant above rest-frame 2 keV, we exclude this component from our model. We disregard the reflection component from the accretion disk, whose interpretation is under debate (see e.g., \citealt{2019ApJ...875..115O}). Furthermore, we do not consider the X-ray emission from star-formation activity, which is much fainter than the AGN component above rest-frame 2 keV. Hence, in this study, we only consider a direct component, a reflection component from a torus, and a scattered component. This model is described as follows in the XSPEC terminology:
\begin{eqnarray}
 \mathrm{model\,1} &=& \texttt{phabs} * (\texttt{zphabs}*\texttt{cabs}*\texttt{zcutoffpl} + \texttt{const}*\texttt{zcutoffpl} + \texttt{atable\{xclumpy\_v01\_RC.fits\}}\\
 &+& \texttt{atable\{xclumpy\_v01\_RL.fits\}})\nonumber
\end{eqnarray}
\begin{enumerate}
 \item The first term (\texttt{phabs}) represents the photo-electric absorption by our Galaxy, which is fixed at the value estimated from the source position using the method of \citet{2013MNRAS.431..394W}.
 \item The first term in the parentheses represents the direct component. The \texttt{zphabs} and \texttt{cabs} represent photo-electric absorption and Compton scattering by torus matter. The line-of-sight hydrogen column density is tested at $\log N_{\mathrm{H}}^{\mathrm{LOS}}=21,\,23,\,24.5$. The photon index and cutoff energy are fixed at $\Gamma=1.9$ and $E_{\mathrm{cut}}=370\,\mathrm{keV}$, respectively, as typical values of local AGNs \citep{2017ApJS..233...17R}.
 \item The second term in the parentheses represents the unabsorbed scattered component. The parameters of \texttt{zcutoffpl} are linked to those of the direct component. We fix the scattered fraction (\texttt{const}) to 0.01 as a typical value \citep{2017ApJS..233...17R,2021MNRAS.504..428G}.
 \item The third and fourth terms in the parentheses represent the reflection continuum and fluorescence lines from the torus. We model these components with the XCLUMPY model \citep{2019ApJ...877...95T}. The photon index, cutoff energy, and normalization are linked to those of the direct component. We fixed the torus angular width at $\sigma=20\degr$. The inclination angle is fixed at $i=30\degr$ when $\log N_{\mathrm{H}}^{\mathrm{LOS}}/\mathrm{cm}^{-2}=21$, and $i=70\degr$ when $\log N_{\mathrm{H}}^{\mathrm{LOS}}/\mathrm{cm}^{-2}=23$ and 24.5. The equatorial hydrogen column density ($N_{\mathrm{H}}^{\mathrm{Equ}}$) is set to be consistent with the line-of-sight hydrogen column density and the torus geometry (see Equation 3 in \citealt{2019ApJ...877...95T}). 
\end{enumerate}
X-ray spectra of distant AGNs are often analyzed with simpler models, such as an absorbed power law model. Accordingly, we also estimate the X-ray luminosity upperbounds with a simple model described as follows in the XSPEC terminology:
\begin{equation}
 \mathrm{model\,2} = \texttt{phabs} * \texttt{zphabs}*\texttt{cabs}*\texttt{zcutoffpl}
\end{equation}
The meanings of each symbol in model 2 are the same as those in the direct component of model 1. The line-of-sight hydrogen column density, photon index, and cutoff energy are fixed to $\log N_{\mathrm{H}}^{\mathrm{LOS}}/\mathrm{cm}^{-2}=22$, $\Gamma=1.9$, and $E_{\mathrm{cut}}=370\,\mathrm{keV}$, respectively.

\section{Detailed description of SED modeling}\label{appendix:freeparameter}

\begin{figure}[h]
 \epsscale{1.16}
 \plotone{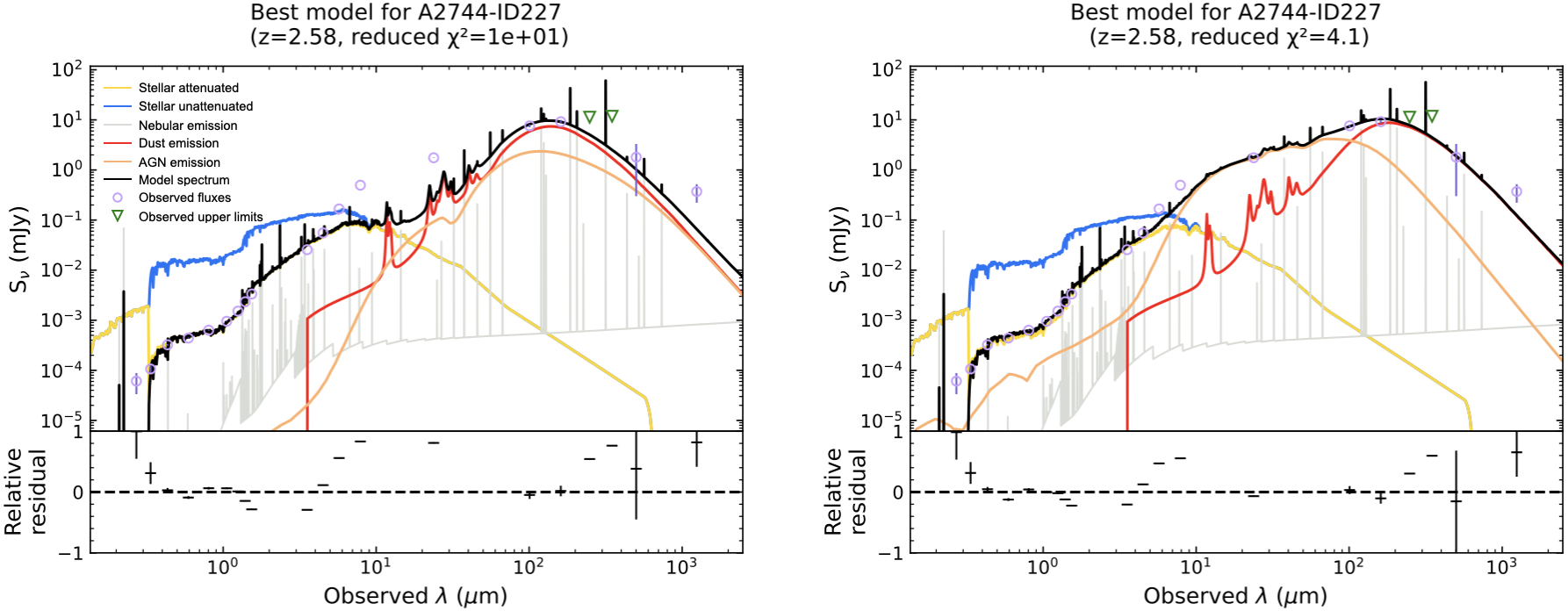}
 \caption{The SED and the best-fit model of A2744-ID227 analyzed with the basic parameter set (left panel) and those with the extended parameter set (right panel). The SEDs are NOT corrected for lensing magnification. The colors and the symbols are the same as those in Figure~\ref{figure:example}. \label{figure:failure}}
\end{figure}

First, the ALCS tier-1/2/5/6 samples were analyzed with a basic parameter set, which is summarized in Table~\ref{table:modelbasic}. This adequately reproduces all the SEDs ($\chi^2/\mathrm{dof}<5$) except for A2744-ID227. The left panel of Figure~\ref{figure:failure} shows the SED and the best-fit models of A2744-ID227 analyzed with the basic parameter set. We noticed a significant excess in the mid-infrared band, suggesting the existence of a type-2 AGN. Hence, we reanalyzed the SED with an extended parameter set of the AGN module, which is given in Table~\ref{table:modelA2744-ID227}. The right panel of Figure~\ref{figure:failure} shows the SED and the best-fit models of A2744-ID227 analyzed with the extended parameter set. This operation greatly improved the goodness of the fit for this object ($\chi^2/\mathrm{dof}=10.6 \to 4.1$). Second, the ALCS tier-3/4 sources were analyzed with a limited parameter set for the dust emission model to avoid unreasonable fitting results. This parameter set is summarized in Table~\ref{table:model57}. This adequately reproduces all the SEDs in the tier-3/4 samples ($\chi^2/\mathrm{dof}<5$).

\begin{figure}
 \epsscale{0.6}
 \plotone{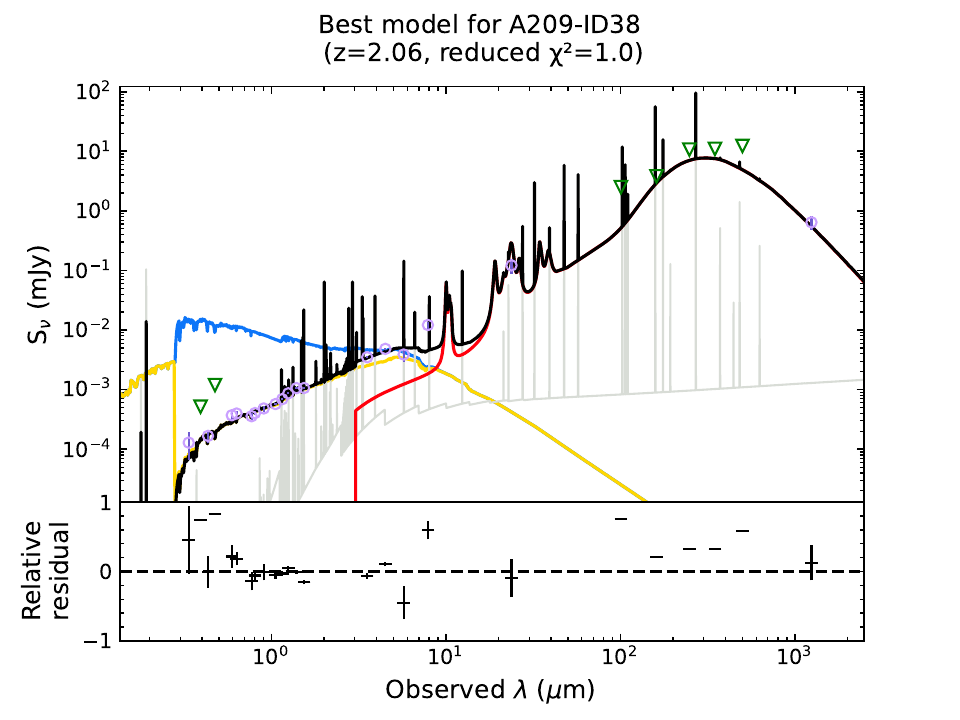}
 \caption{Example of the SED and the best-fit model of the ALCS sample. The complete figure set (160 images) is available in the online journal. \label{figure:SED}}
\end{figure}

The observed SEDs and best-fit models of all the objects are shown in Figure~\ref{figure:SED}. We note that the best-fit SEDs of A383-ID50 and M0416-ID156 look inconsistent with the observed ALMA fluxes ($\log F_{\mathrm{obs\,[1.2\,mm]}}/F_{\mathrm{model}}>1$). This may be caused by the oversimplification of the dust SED model. We also note that the best-fit SED of ACT0102-ID50 looks inconsistent with the photometric data of Herschel ($\log F_{\mathrm{obs\,[250\,\mu m]}}/F_{\mathrm{model}}>1$). This might be caused by the blending effect from nearby sources or additional star-formation activities that are deeply embedded in dust and not detected in optical to near-infrared bands.

\begin{table}
 \caption{Basic parameter set used for the SED modeling}
 \begin{tabularx}{\linewidth}{@{}lll}
   \hline\hline
   Parameter &\hspace{0pt}Symbol &\hspace{0pt}Value\\
   \hline
   \multicolumn3c{SFH (sfhdelayed)}\\
   \hline
   e-folding time of the main stellar population &\hspace{0pt}$\tau_{\mathrm{main}}$ [Myr] &\hspace{0pt}100, 316, 1000, 3162, 10000\\
   Age of the main stellar population &\hspace{0pt}$\mathrm{age_{main}}$ [Myr] &\hspace{0pt}100, 158, 251, 398, 631, 1000, 1585, 2512, 3981, 6310\\
   e-folding time of the late burst population &\hspace{0pt}$\tau_{\mathrm{burst}}$ [Myr] &\hspace{0pt}100\\
   Age of the late burst population &\hspace{0pt}$\mathrm{age_{burst}}$ [Myr] &\hspace{0pt}5, 10, 50, 100\\
   Mass fraction of the late burst population &\hspace{0pt}$f_{\mathrm{burst}}$ &\hspace{0pt}0, 0.05, 0.1, 0.3\\
   \hline
   \multicolumn3c{SSP (bc03; \citealt{2003MNRAS.344.1000B})}\\
   \hline
   IMF of the stellar model &\hspace{0pt} &\hspace{0pt} \cite{2003PASP..115..763C}\\
   Metalicity of the stellar model &\hspace{0pt} &\hspace{0pt} 0.02\\
   \hline
   \multicolumn3c{Dust Attenuation (dustatt\_modified\_starburst; \citealt{2000ApJ...533..682C})}\\
   \hline
   The color excess of the nebular lines. &\hspace{0pt}$E(B-V)_{\mathrm{lines}}$ &\hspace{0pt}0.1, 0.2, 0.4, 0.6, 0.8, 1.0, 1.2, 1.4, 1.6, 1.8, 2.0, 2.2,\\
   &\hspace{0pt} &\hspace{0pt}2.4\\
   Reduction factor to apply $E(B-V)_{\mathrm{lines}}$ to calculate &\hspace{0pt}$E(B-V)_{\mathrm{factor}}$ &\hspace{0pt}0.44\\
   the stellar continuum attenuation & & \\
   UV bump ampliturde &\hspace{0pt} &\hspace{0pt} 0, 3 (MW)\\
   Power-law index to modify the attenuation curve &\hspace{0pt}$\delta$ &\hspace{0pt}--1.6, --1.2, --0.8, --0.4, 0.0, 0.4\\
   \hline
   \multicolumn3c{Dust emission (EThemis)}\\
   \hline
   Mass fraction of the small hydrocarbon solids &\hspace{0pt}$q_{\mathrm{hac}}$ &\hspace{0pt}0.01, 0.02, 0.06\\
   Minimum radiation field &\hspace{0pt}$\log U_{\mathrm{min}}$ &\hspace{0pt}--1.0, --0.2, 0.4, 0.8, 1.2, 1.6, 2.0, 2.2, 2.4, 2.6, 2.8, 3.0,\\
   &\hspace{0pt} &\hspace{0pt}3.2, 3.4, 3.6, 3.8\\
   Power-law index of the starlight intensity distribution &\hspace{0pt}$\alpha$ &\hspace{0pt}2.5\\
   Mass fraction of dust illuminated with $U=U_{\mathrm{min}}$ &\hspace{0pt}$1-\gamma$ &\hspace{0pt}0.95, 0.99\\
   \hline
   \multicolumn3c{AGN emission (skirtor2016; \citealt{2012MNRAS.420.2756S,2016MNRAS.458.2288S})}\\
   \hline
   Average edge-on optical depth at 9.7 micron &\hspace{0pt}$\tau_{\rm 9.7}$ &\hspace{0pt}11\\
   Radial gradient of dust density &\hspace{0pt}$p$ &\hspace{0pt}1.0\\
   Dust density gradient with polar angle &\hspace{0pt}$q$ &\hspace{0pt}1.0\\
   Half-opening angle of the dust-free cone &\hspace{0pt}$\Delta$\,[\degr] &\hspace{0pt}40\\
   Ratio of outer to inner radius &\hspace{0pt}$R$ &\hspace{0pt}20\\
   Inclination &\hspace{0pt}$\theta$\,[\degr] &\hspace{0pt}30, 70\\
   Fraction of AGN IR luminosity to total IR luminosity &\hspace{0pt}$f_{\rm AGN}$ &\hspace{0pt}0.0, 0.1, 0.3, 0.5, 0.7, 0.9 (0.0 for the SECONDARY sample)\\
   Extinction in polar direction &\hspace{0pt}$E(B-V)$ &\hspace{0pt}0.3\\
   Temperature of the polar dust &\hspace{0pt}$T_{\mathrm{pol}}$\,[K] &\hspace{0pt}150\\
   \hline
 \end{tabularx}\label{table:modelbasic}
\end{table}

\begin{table}
 \caption{Parameter set of AGN component used for A2744-ID227}
 \begin{tabularx}{\linewidth}{@{}lll}
   \hline\hline
   Parameter &\hspace{0pt}Symbol &\hspace{19pt}Value\\
   \hline
   \multicolumn3c{\hspace{131pt}AGN emission (skirtor2016; \citealt{2012MNRAS.420.2756S,2016MNRAS.458.2288S})}\\
   \hline
   Average edge-on optical depth at 9.7 micron &\hspace{0pt}$\tau_{\rm 9.7}$ &\hspace{19pt}3, 11\\
   Radial gradient of dust density &\hspace{0pt}$p$ &\hspace{19pt}1.0\\
   Dust density gradient with polar angle &\hspace{0pt}$q$ &\hspace{19pt}1.0\\
   Half-opening angle of the dust-free cone &\hspace{0pt}$\Delta$\,[\degr] &\hspace{19pt}40\\
   Ratio of outer to inner radius &\hspace{0pt}$R$ &\hspace{19pt}20\\
   Inclination &\hspace{0pt}$\theta$\,[\degr] &\hspace{19pt}70\\
   Fraction of AGN IR luminosity to total IR luminosity &\hspace{0pt}$f_{\rm AGN}$ &\hspace{19pt}0.0, 0.1, 0.3, 0.5, 0.7, 0.9\\
   Extinction in polar direction &\hspace{0pt}$E(B-V)$ &\hspace{19pt}0.0, 0.3\\
   Temperature of the polar dust &\hspace{0pt}$T_{\mathrm{pol}}$\,[K] &\hspace{19pt}150\\
   \hline
 \end{tabularx}\label{table:modelA2744-ID227}
\end{table}

\begin{table}
 \caption{Parameter set of dust emission component used for tier-3/4 samples}
 \begin{tabularx}{\linewidth}{@{}lll}
   \hline\hline
   Parameter &\hspace{0pt}Symbol &\hspace{25pt}Value\\
   \hline
   \multicolumn3c{Dust emission (EThemis)}\\
   \hline
   Mass fraction of the small hydrocarbon solids &\hspace{0pt}$q_{\mathrm{hac}}$ &\hspace{25pt}0.01\\
   Minimum radiation field &\hspace{0pt}$\log U_{\mathrm{min}}$ &\hspace{25pt}--1.0, --0.2, 0.4, 0.8, 1.2, 1.6, 2.0, 2.2, 2.4, 2.6, 2.8, 3.0,\\
   &\hspace{0pt} &\hspace{25pt}3.2, 3.4, 3.6, 3.8\\
   Power-law index of the starlight intensity distribution &\hspace{0pt}$\alpha$ &\hspace{25pt}2.5\\
   Mass fraction of dust illuminated with $U=U_{\mathrm{min}}$ &\hspace{0pt}$1-\gamma$ &\hspace{25pt}0.99\\
   \hline
 \end{tabularx}\label{table:model57}
\end{table}

\figsetstart
\figsetnum{16}
\figsettitle{The SEDs and the best-fit models of the ALCS tier-1/2/3/4/5/6 samples.}

\figsetgrpstart
\figsetgrpnum{16.1}
\figsetgrptitle{A209-ID38}
\figsetplot{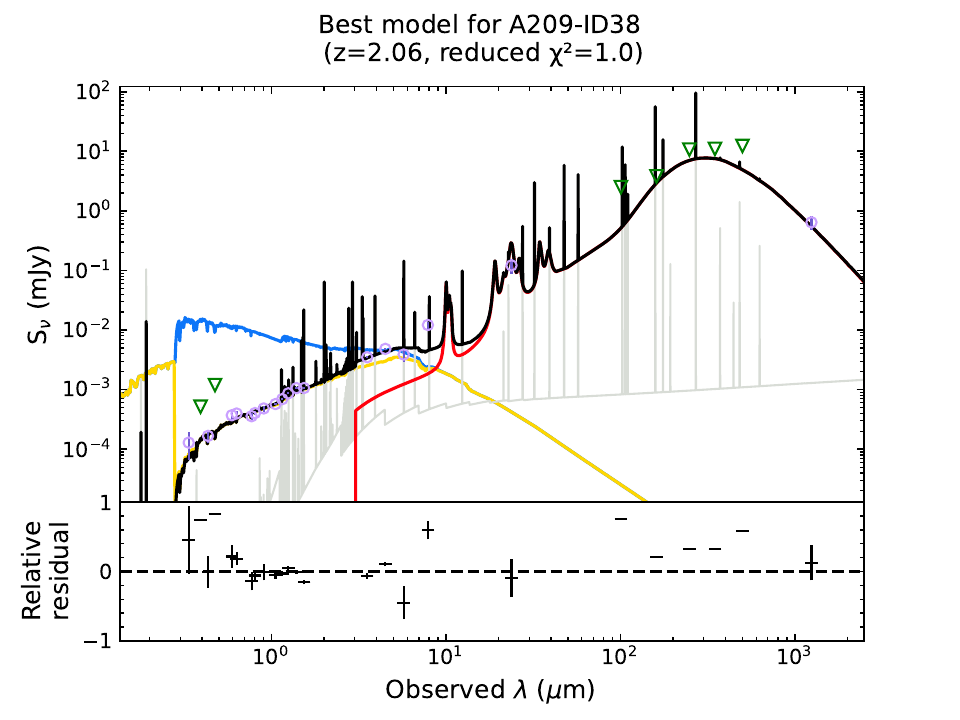}
\figsetgrpnote{The SED and the best-fit model of A209-ID38. Lensing magnification is NOT corrected in this figure. The black solid line represents the composite spectrum of the galaxy. The yellow line illustrates the stellar emission attenuated by interstellar dust. The blue line depicts the unattenuated stellar emission for reference purpose. The orange line corresponds to the emission from an AGN. The red line shows the infrared emission from interstellar dust. The gray line denotes the nebulae emission. The observed data points are represented by purple circles, accompanied by 1$\sigma$ error bars. The bottom panel displays the relative residuals.}
\figsetgrpend

\figsetgrpstart
\figsetgrpnum{16.2}
\figsetgrptitle{A2163-ID11}
\figsetplot{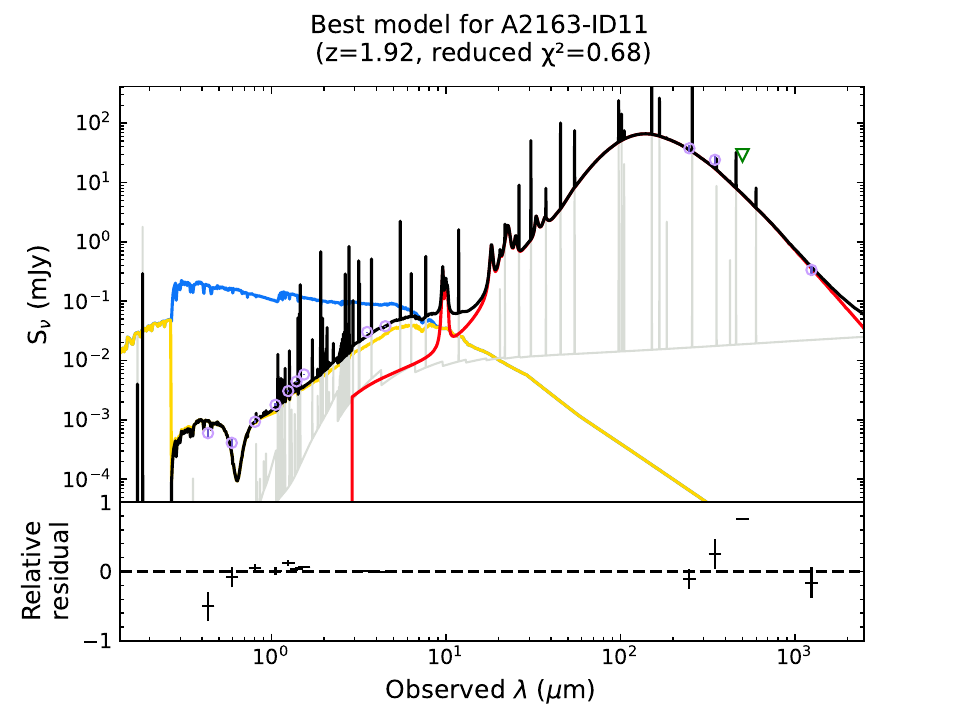}
\figsetgrpnote{The SED and the best-fit model of A2163-ID11. Lensing magnification is NOT corrected in this figure. The black solid line represents the composite spectrum of the galaxy. The yellow line illustrates the stellar emission attenuated by interstellar dust. The blue line depicts the unattenuated stellar emission for reference purpose. The orange line corresponds to the emission from an AGN. The red line shows the infrared emission from interstellar dust. The gray line denotes the nebulae emission. The observed data points are represented by purple circles, accompanied by 1$\sigma$ error bars. The bottom panel displays the relative residuals.}
\figsetgrpend

\figsetgrpstart
\figsetgrpnum{16.3}
\figsetgrptitle{A2537-ID24}
\figsetplot{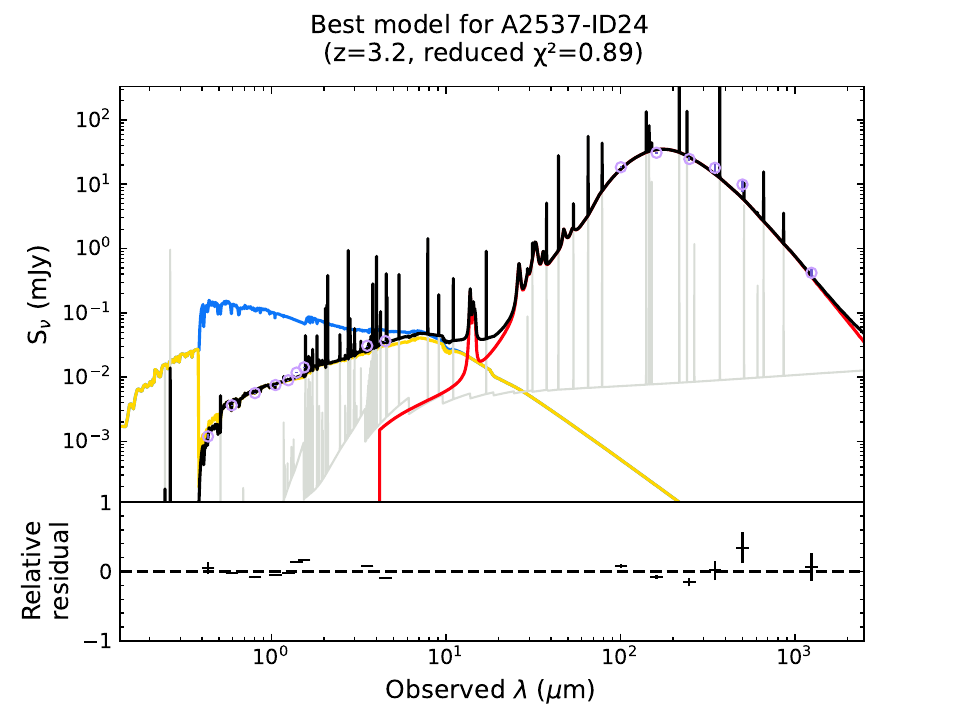}
\figsetgrpnote{The SED and the best-fit model of A2537-ID24. Lensing magnification is NOT corrected in this figure. The black solid line represents the composite spectrum of the galaxy. The yellow line illustrates the stellar emission attenuated by interstellar dust. The blue line depicts the unattenuated stellar emission for reference purpose. The orange line corresponds to the emission from an AGN. The red line shows the infrared emission from interstellar dust. The gray line denotes the nebulae emission. The observed data points are represented by purple circles, accompanied by 1$\sigma$ error bars. The bottom panel displays the relative residuals.}
\figsetgrpend

\figsetgrpstart
\figsetgrpnum{16.4}
\figsetgrptitle{A2537-ID42}
\figsetplot{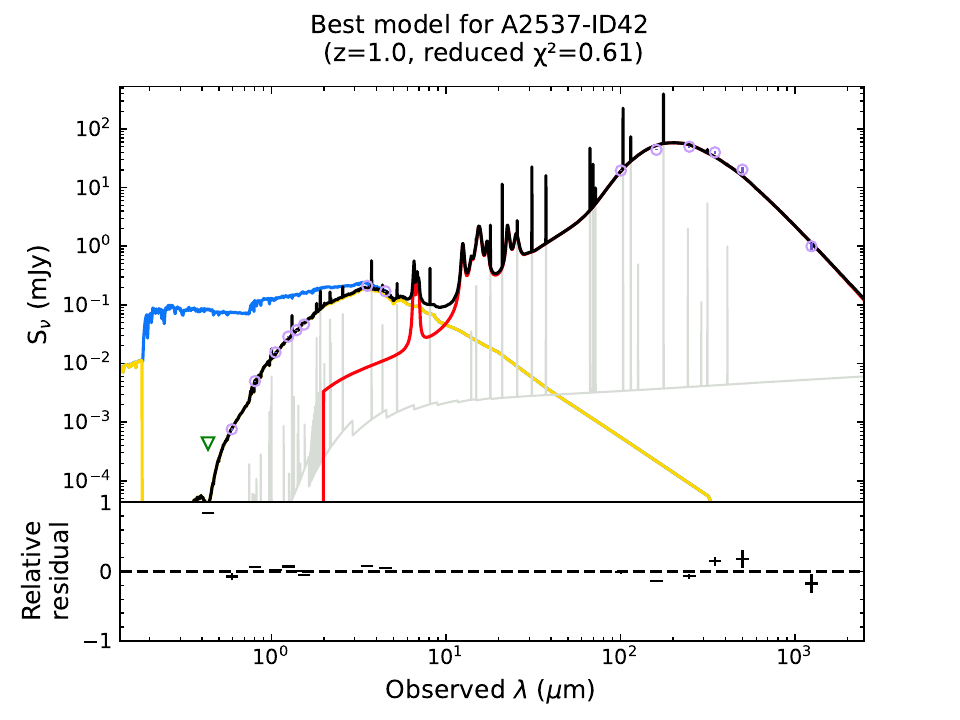}
\figsetgrpnote{The SED and the best-fit model of A2537-ID42. Lensing magnification is NOT corrected in this figure. The black solid line represents the composite spectrum of the galaxy. The yellow line illustrates the stellar emission attenuated by interstellar dust. The blue line depicts the unattenuated stellar emission for reference purpose. The orange line corresponds to the emission from an AGN. The red line shows the infrared emission from interstellar dust. The gray line denotes the nebulae emission. The observed data points are represented by purple circles, accompanied by 1$\sigma$ error bars. The bottom panel displays the relative residuals.}
\figsetgrpend

\figsetgrpstart
\figsetgrpnum{16.5}
\figsetgrptitle{A2537-ID49}
\figsetplot{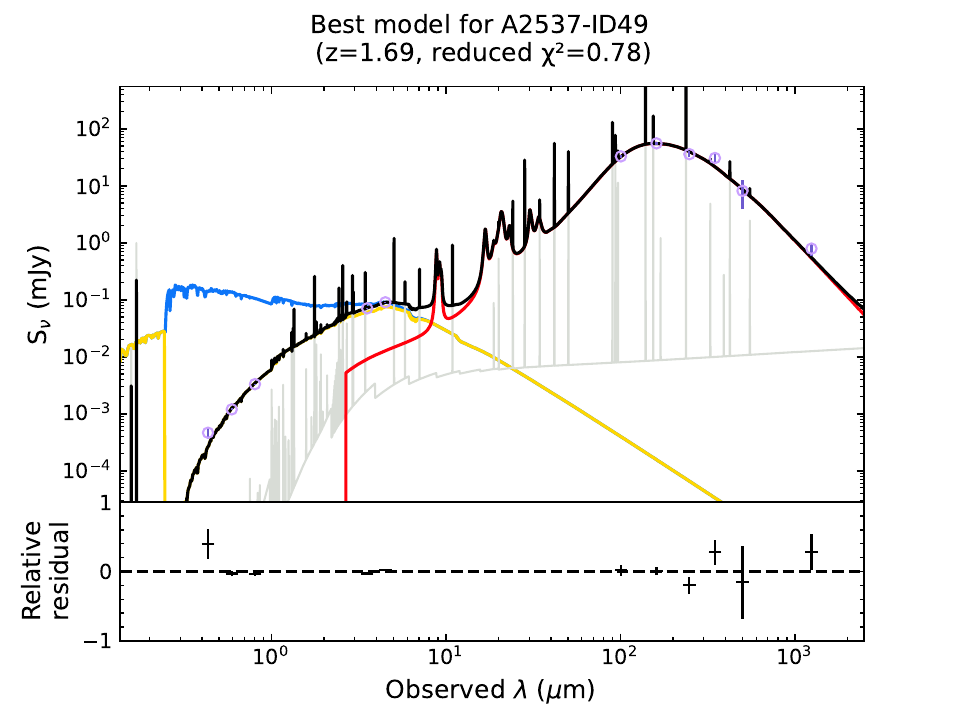}
\figsetgrpnote{The SED and the best-fit model of A2537-ID49. Lensing magnification is NOT corrected in this figure. The black solid line represents the composite spectrum of the galaxy. The yellow line illustrates the stellar emission attenuated by interstellar dust. The blue line depicts the unattenuated stellar emission for reference purpose. The orange line corresponds to the emission from an AGN. The red line shows the infrared emission from interstellar dust. The gray line denotes the nebulae emission. The observed data points are represented by purple circles, accompanied by 1$\sigma$ error bars. The bottom panel displays the relative residuals.}
\figsetgrpend

\figsetgrpstart
\figsetgrpnum{16.6}
\figsetgrptitle{A2537-ID6}
\figsetplot{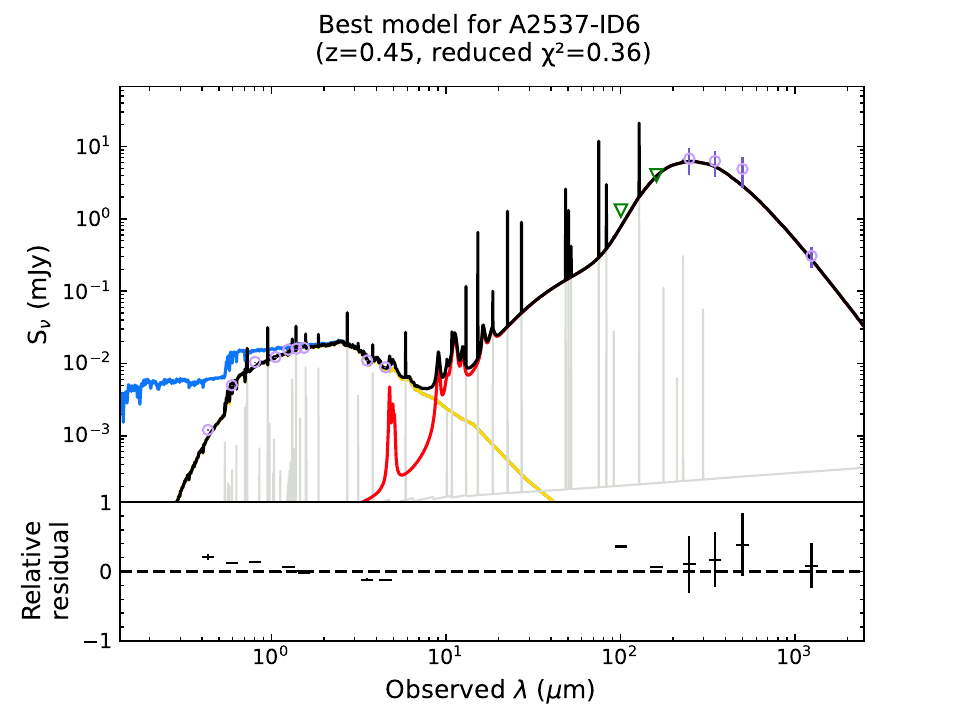}
\figsetgrpnote{The SED and the best-fit model of A2537-ID6. Lensing magnification is NOT corrected in this figure. The black solid line represents the composite spectrum of the galaxy. The yellow line illustrates the stellar emission attenuated by interstellar dust. The blue line depicts the unattenuated stellar emission for reference purpose. The orange line corresponds to the emission from an AGN. The red line shows the infrared emission from interstellar dust. The gray line denotes the nebulae emission. The observed data points are represented by purple circles, accompanied by 1$\sigma$ error bars. The bottom panel displays the relative residuals.}
\figsetgrpend

\figsetgrpstart
\figsetgrpnum{16.7}
\figsetgrptitle{A2537-ID66}
\figsetplot{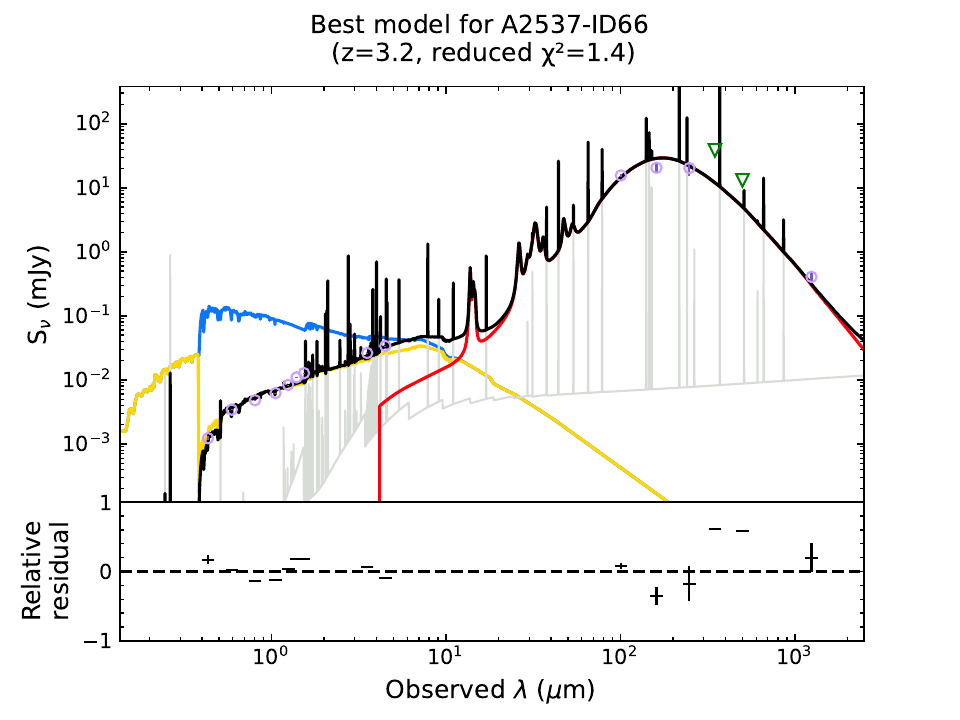}
\figsetgrpnote{The SED and the best-fit model of A2537-ID66. Lensing magnification is NOT corrected in this figure. The black solid line represents the composite spectrum of the galaxy. The yellow line illustrates the stellar emission attenuated by interstellar dust. The blue line depicts the unattenuated stellar emission for reference purpose. The orange line corresponds to the emission from an AGN. The red line shows the infrared emission from interstellar dust. The gray line denotes the nebulae emission. The observed data points are represented by purple circles, accompanied by 1$\sigma$ error bars. The bottom panel displays the relative residuals.}
\figsetgrpend

\figsetgrpstart
\figsetgrpnum{16.8}
\figsetgrptitle{A2744-ID17}
\figsetplot{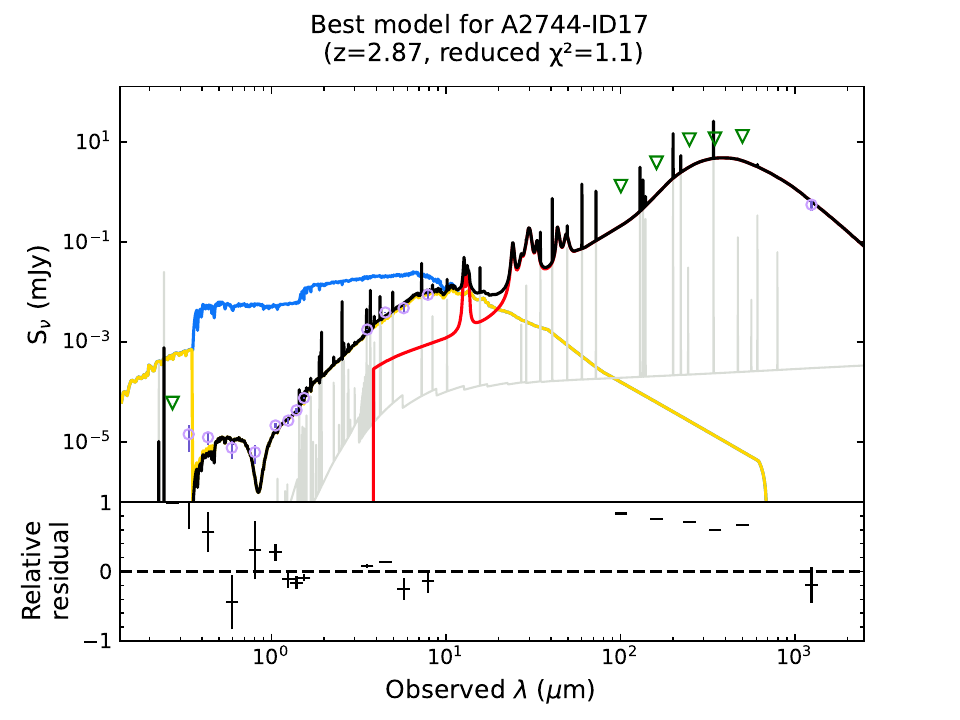}
\figsetgrpnote{The SED and the best-fit model of A2744-ID17. Lensing magnification is NOT corrected in this figure. The black solid line represents the composite spectrum of the galaxy. The yellow line illustrates the stellar emission attenuated by interstellar dust. The blue line depicts the unattenuated stellar emission for reference purpose. The orange line corresponds to the emission from an AGN. The red line shows the infrared emission from interstellar dust. The gray line denotes the nebulae emission. The observed data points are represented by purple circles, accompanied by 1$\sigma$ error bars. The bottom panel displays the relative residuals.}
\figsetgrpend

\figsetgrpstart
\figsetgrpnum{16.9}
\figsetgrptitle{A2744-ID176}
\figsetplot{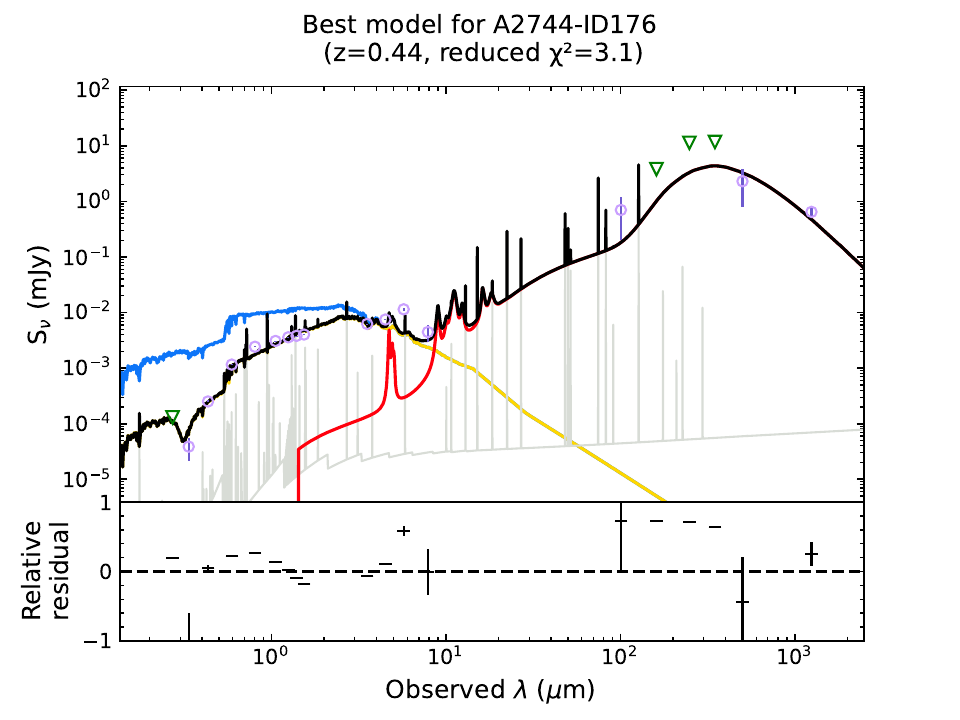}
\figsetgrpnote{The SED and the best-fit model of A2744-ID176. Lensing magnification is NOT corrected in this figure. The black solid line represents the composite spectrum of the galaxy. The yellow line illustrates the stellar emission attenuated by interstellar dust. The blue line depicts the unattenuated stellar emission for reference purpose. The orange line corresponds to the emission from an AGN. The red line shows the infrared emission from interstellar dust. The gray line denotes the nebulae emission. The observed data points are represented by purple circles, accompanied by 1$\sigma$ error bars. The bottom panel displays the relative residuals.}
\figsetgrpend

\figsetgrpstart
\figsetgrpnum{16.10}
\figsetgrptitle{A2744-ID178}
\figsetplot{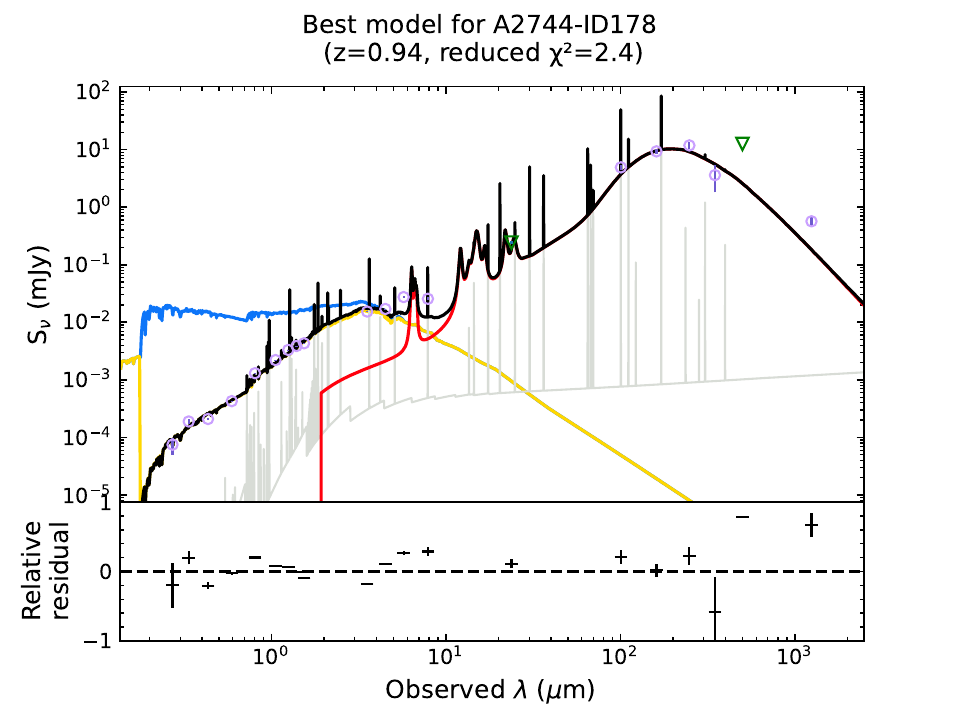}
\figsetgrpnote{The SED and the best-fit model of A2744-ID178. Lensing magnification is NOT corrected in this figure. The black solid line represents the composite spectrum of the galaxy. The yellow line illustrates the stellar emission attenuated by interstellar dust. The blue line depicts the unattenuated stellar emission for reference purpose. The orange line corresponds to the emission from an AGN. The red line shows the infrared emission from interstellar dust. The gray line denotes the nebulae emission. The observed data points are represented by purple circles, accompanied by 1$\sigma$ error bars. The bottom panel displays the relative residuals.}
\figsetgrpend

\figsetgrpstart
\figsetgrpnum{16.11}
\figsetgrptitle{A2744-ID21}
\figsetplot{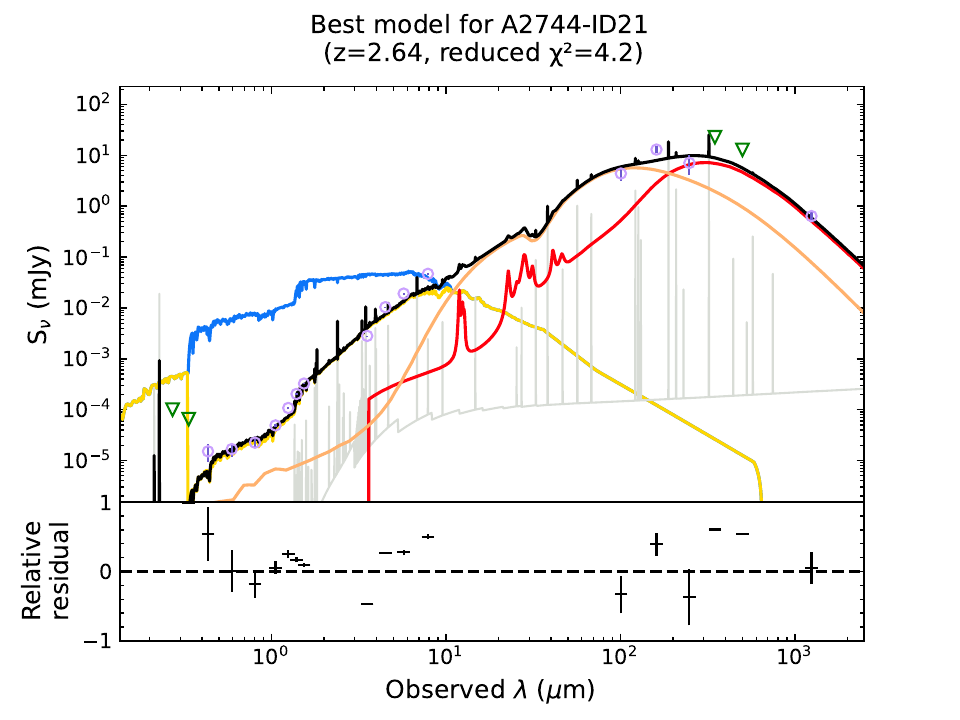}
\figsetgrpnote{The SED and the best-fit model of A2744-ID21. Lensing magnification is NOT corrected in this figure. The black solid line represents the composite spectrum of the galaxy. The yellow line illustrates the stellar emission attenuated by interstellar dust. The blue line depicts the unattenuated stellar emission for reference purpose. The orange line corresponds to the emission from an AGN. The red line shows the infrared emission from interstellar dust. The gray line denotes the nebulae emission. The observed data points are represented by purple circles, accompanied by 1$\sigma$ error bars. The bottom panel displays the relative residuals.}
\figsetgrpend

\figsetgrpstart
\figsetgrpnum{16.12}
\figsetgrptitle{A2744-ID227}
\figsetplot{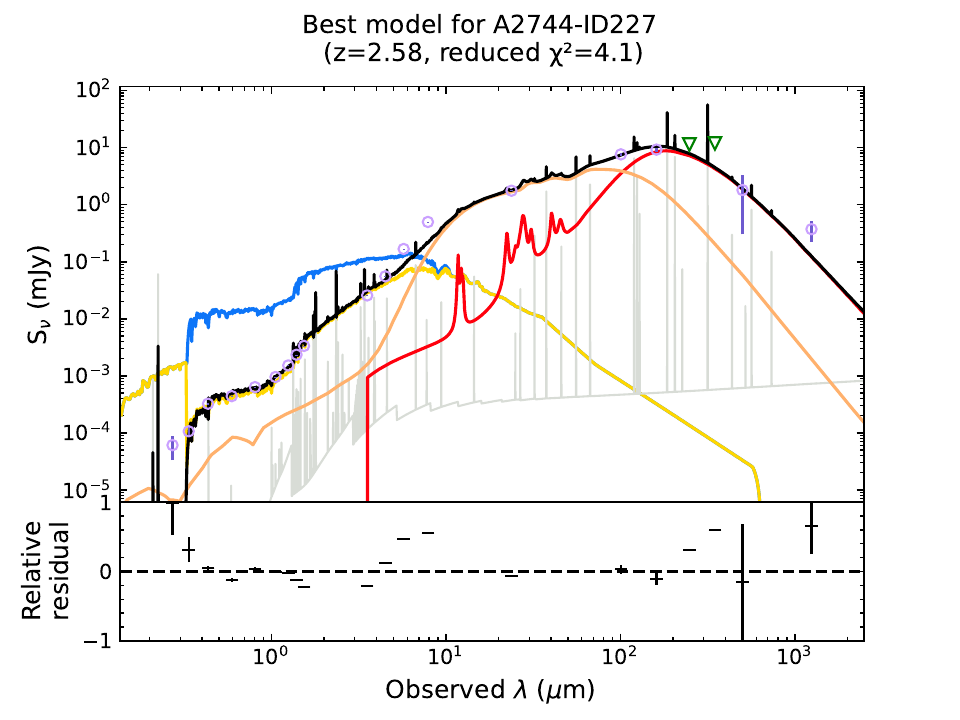}
\figsetgrpnote{The SED and the best-fit model of A2744-ID227. Lensing magnification is NOT corrected in this figure. The black solid line represents the composite spectrum of the galaxy. The yellow line illustrates the stellar emission attenuated by interstellar dust. The blue line depicts the unattenuated stellar emission for reference purpose. The orange line corresponds to the emission from an AGN. The red line shows the infrared emission from interstellar dust. The gray line denotes the nebulae emission. The observed data points are represented by purple circles, accompanied by 1$\sigma$ error bars. The bottom panel displays the relative residuals.}
\figsetgrpend

\figsetgrpstart
\figsetgrpnum{16.13}
\figsetgrptitle{A2744-ID319}
\figsetplot{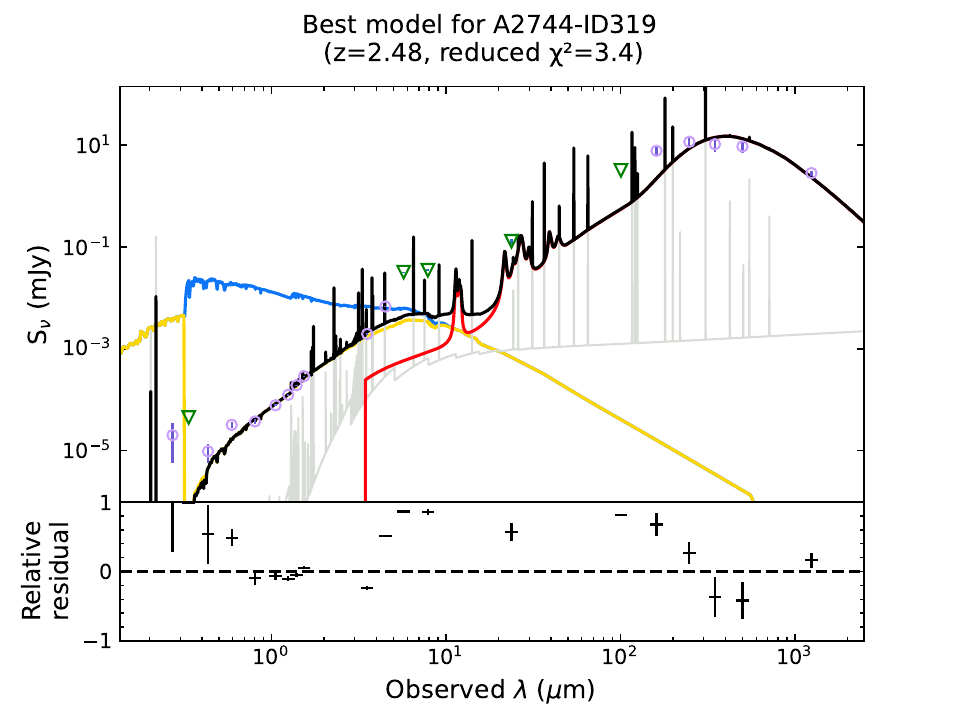}
\figsetgrpnote{The SED and the best-fit model of A2744-ID319. Lensing magnification is NOT corrected in this figure. The black solid line represents the composite spectrum of the galaxy. The yellow line illustrates the stellar emission attenuated by interstellar dust. The blue line depicts the unattenuated stellar emission for reference purpose. The orange line corresponds to the emission from an AGN. The red line shows the infrared emission from interstellar dust. The gray line denotes the nebulae emission. The observed data points are represented by purple circles, accompanied by 1$\sigma$ error bars. The bottom panel displays the relative residuals.}
\figsetgrpend

\figsetgrpstart
\figsetgrpnum{16.14}
\figsetgrptitle{A2744-ID33}
\figsetplot{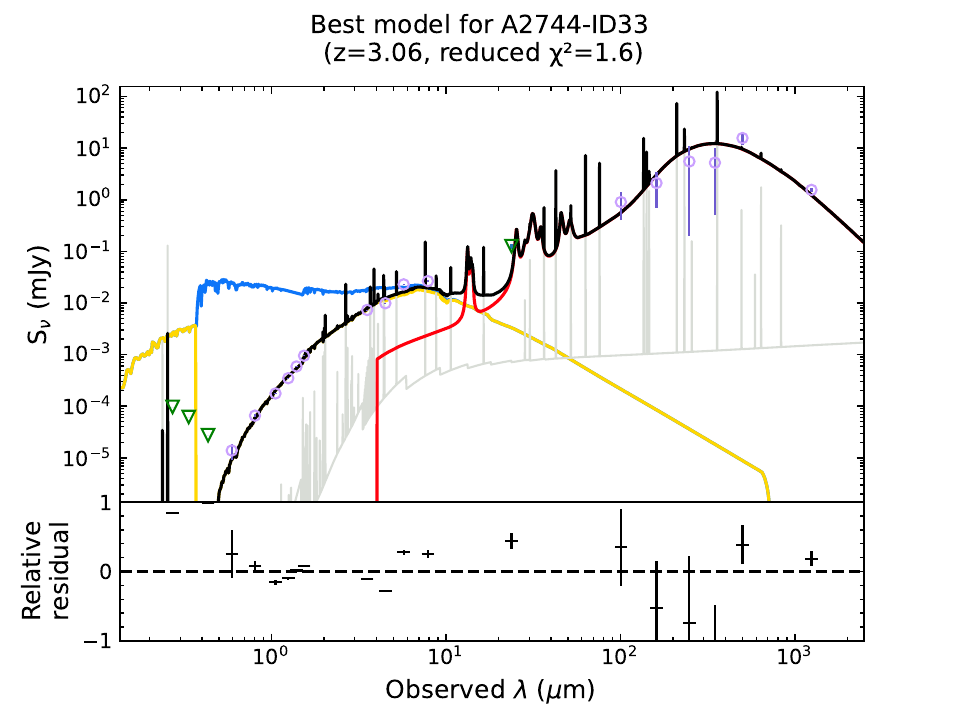}
\figsetgrpnote{The SED and the best-fit model of A2744-ID33. Lensing magnification is NOT corrected in this figure. The black solid line represents the composite spectrum of the galaxy. The yellow line illustrates the stellar emission attenuated by interstellar dust. The blue line depicts the unattenuated stellar emission for reference purpose. The orange line corresponds to the emission from an AGN. The red line shows the infrared emission from interstellar dust. The gray line denotes the nebulae emission. The observed data points are represented by purple circles, accompanied by 1$\sigma$ error bars. The bottom panel displays the relative residuals.}
\figsetgrpend

\figsetgrpstart
\figsetgrpnum{16.15}
\figsetgrptitle{A2744-ID47}
\figsetplot{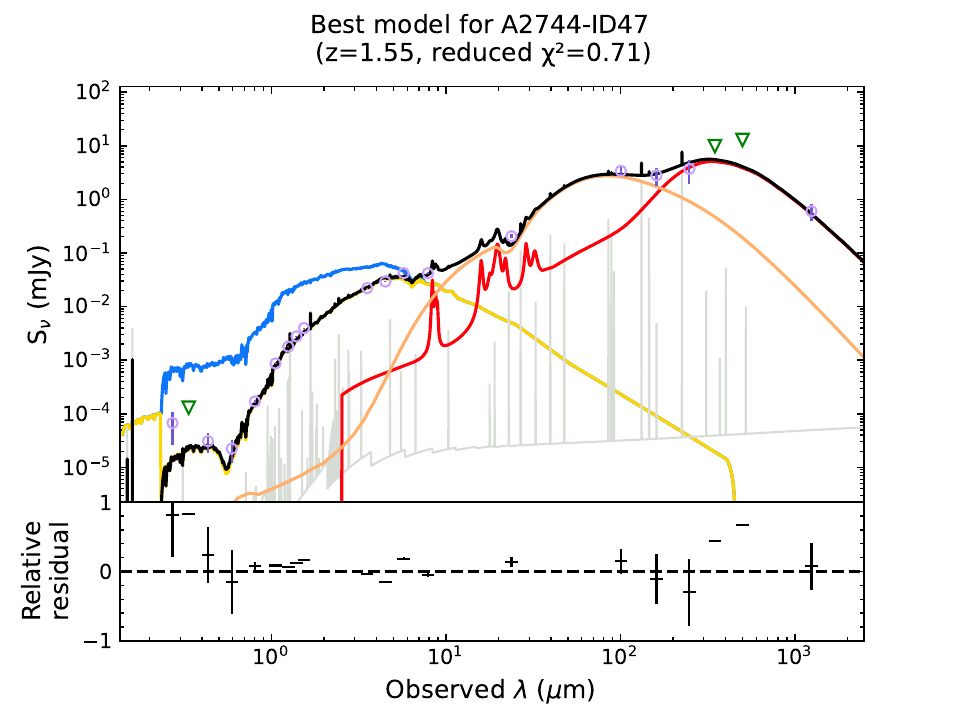}
\figsetgrpnote{The SED and the best-fit model of A2744-ID47. Lensing magnification is NOT corrected in this figure. The black solid line represents the composite spectrum of the galaxy. The yellow line illustrates the stellar emission attenuated by interstellar dust. The blue line depicts the unattenuated stellar emission for reference purpose. The orange line corresponds to the emission from an AGN. The red line shows the infrared emission from interstellar dust. The gray line denotes the nebulae emission. The observed data points are represented by purple circles, accompanied by 1$\sigma$ error bars. The bottom panel displays the relative residuals.}
\figsetgrpend

\figsetgrpstart
\figsetgrpnum{16.16}
\figsetgrptitle{A2744-ID56}
\figsetplot{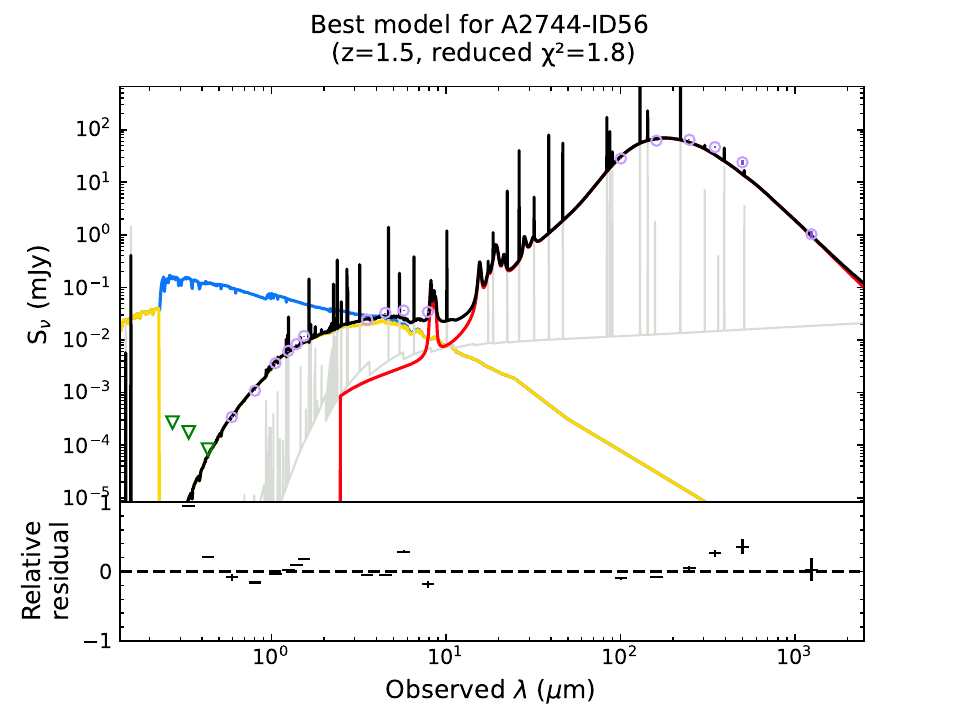}
\figsetgrpnote{The SED and the best-fit model of A2744-ID56. Lensing magnification is NOT corrected in this figure. The black solid line represents the composite spectrum of the galaxy. The yellow line illustrates the stellar emission attenuated by interstellar dust. The blue line depicts the unattenuated stellar emission for reference purpose. The orange line corresponds to the emission from an AGN. The red line shows the infrared emission from interstellar dust. The gray line denotes the nebulae emission. The observed data points are represented by purple circles, accompanied by 1$\sigma$ error bars. The bottom panel displays the relative residuals.}
\figsetgrpend

\figsetgrpstart
\figsetgrpnum{16.17}
\figsetgrptitle{A2744-ID7}
\figsetplot{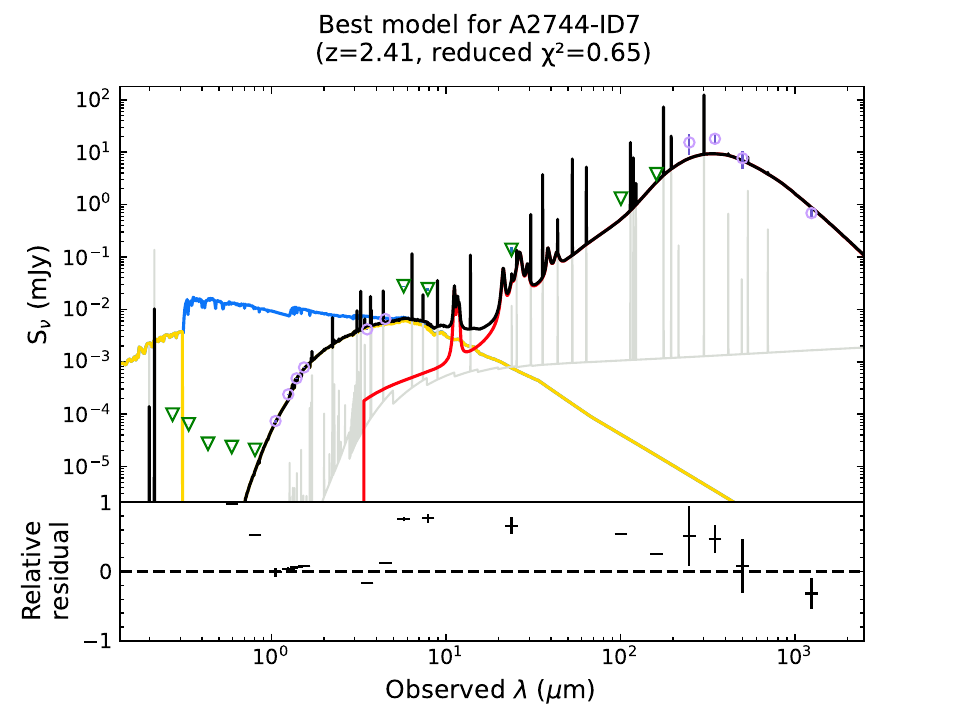}
\figsetgrpnote{The SED and the best-fit model of A2744-ID7. Lensing magnification is NOT corrected in this figure. The black solid line represents the composite spectrum of the galaxy. The yellow line illustrates the stellar emission attenuated by interstellar dust. The blue line depicts the unattenuated stellar emission for reference purpose. The orange line corresponds to the emission from an AGN. The red line shows the infrared emission from interstellar dust. The gray line denotes the nebulae emission. The observed data points are represented by purple circles, accompanied by 1$\sigma$ error bars. The bottom panel displays the relative residuals.}
\figsetgrpend

\figsetgrpstart
\figsetgrpnum{16.18}
\figsetgrptitle{A2744-ID81}
\figsetplot{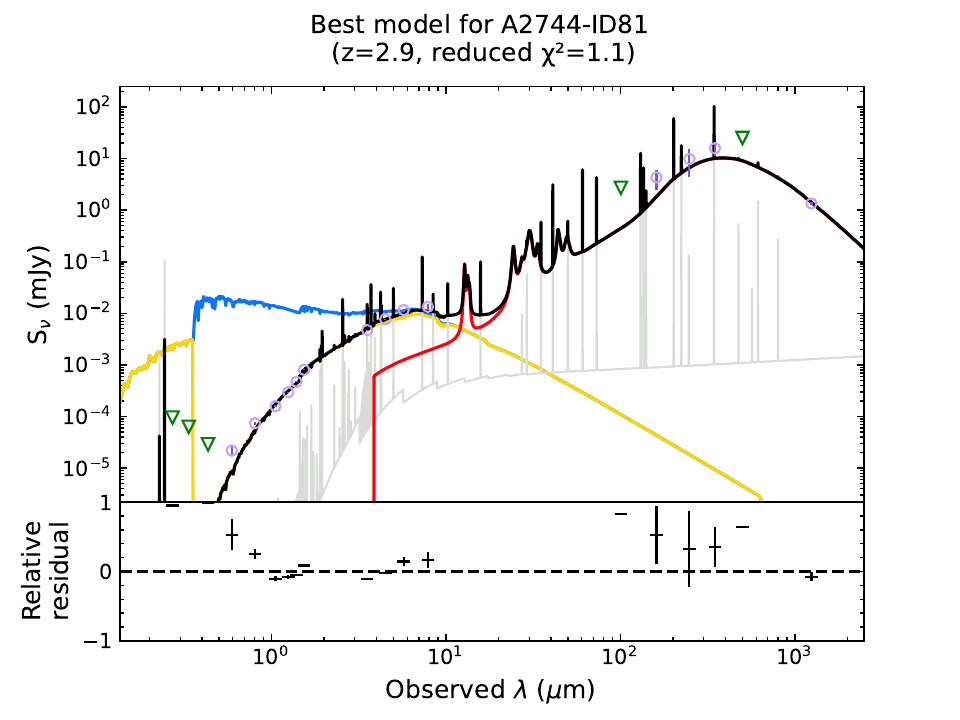}
\figsetgrpnote{The SED and the best-fit model of A2744-ID81. Lensing magnification is NOT corrected in this figure. The black solid line represents the composite spectrum of the galaxy. The yellow line illustrates the stellar emission attenuated by interstellar dust. The blue line depicts the unattenuated stellar emission for reference purpose. The orange line corresponds to the emission from an AGN. The red line shows the infrared emission from interstellar dust. The gray line denotes the nebulae emission. The observed data points are represented by purple circles, accompanied by 1$\sigma$ error bars. The bottom panel displays the relative residuals.}
\figsetgrpend

\figsetgrpstart
\figsetgrpnum{16.19}
\figsetgrptitle{A3192-ID131}
\figsetplot{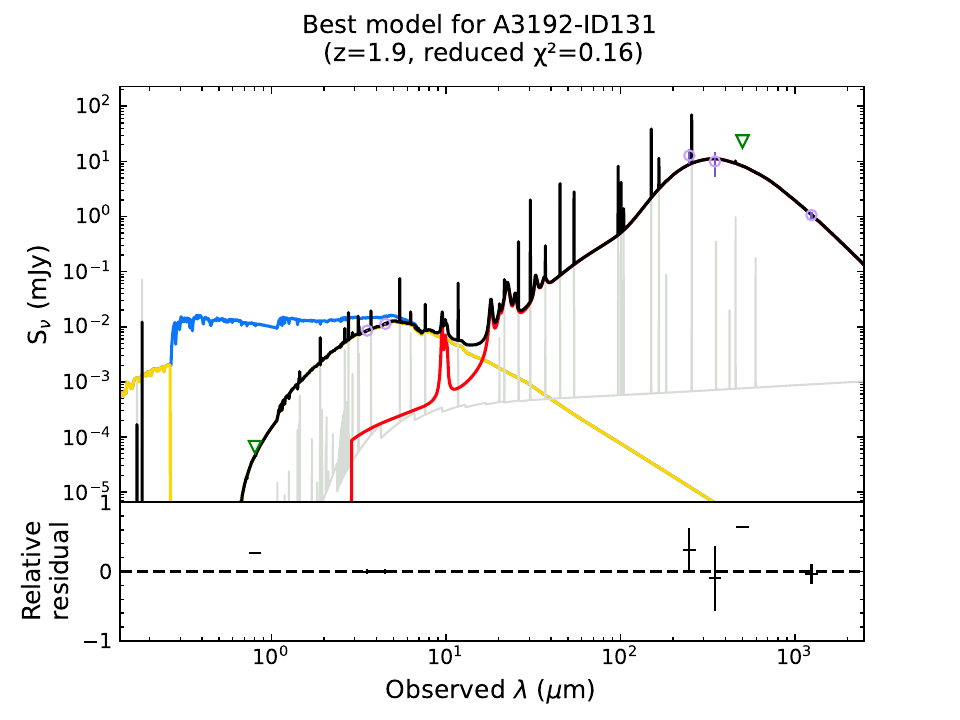}
\figsetgrpnote{The SED and the best-fit model of A3192-ID131. Lensing magnification is NOT corrected in this figure. The black solid line represents the composite spectrum of the galaxy. The yellow line illustrates the stellar emission attenuated by interstellar dust. The blue line depicts the unattenuated stellar emission for reference purpose. The orange line corresponds to the emission from an AGN. The red line shows the infrared emission from interstellar dust. The gray line denotes the nebulae emission. The observed data points are represented by purple circles, accompanied by 1$\sigma$ error bars. The bottom panel displays the relative residuals.}
\figsetgrpend

\figsetgrpstart
\figsetgrpnum{16.20}
\figsetgrptitle{A3192-ID138}
\figsetplot{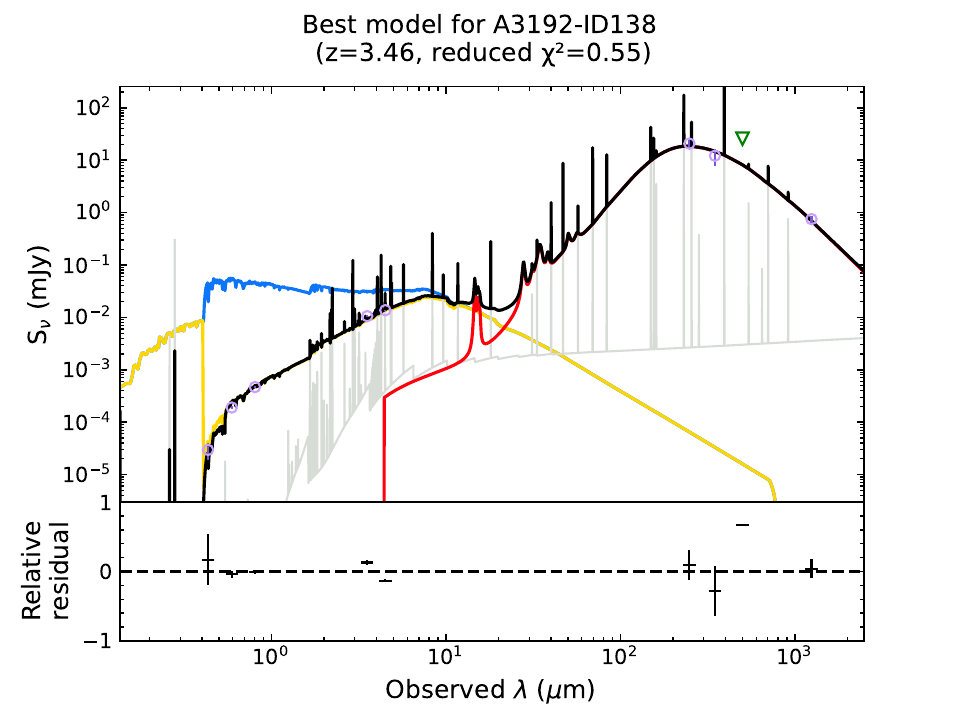}
\figsetgrpnote{The SED and the best-fit model of A3192-ID138. Lensing magnification is NOT corrected in this figure. The black solid line represents the composite spectrum of the galaxy. The yellow line illustrates the stellar emission attenuated by interstellar dust. The blue line depicts the unattenuated stellar emission for reference purpose. The orange line corresponds to the emission from an AGN. The red line shows the infrared emission from interstellar dust. The gray line denotes the nebulae emission. The observed data points are represented by purple circles, accompanied by 1$\sigma$ error bars. The bottom panel displays the relative residuals.}
\figsetgrpend

\figsetgrpstart
\figsetgrpnum{16.21}
\figsetgrptitle{A3192-ID154}
\figsetplot{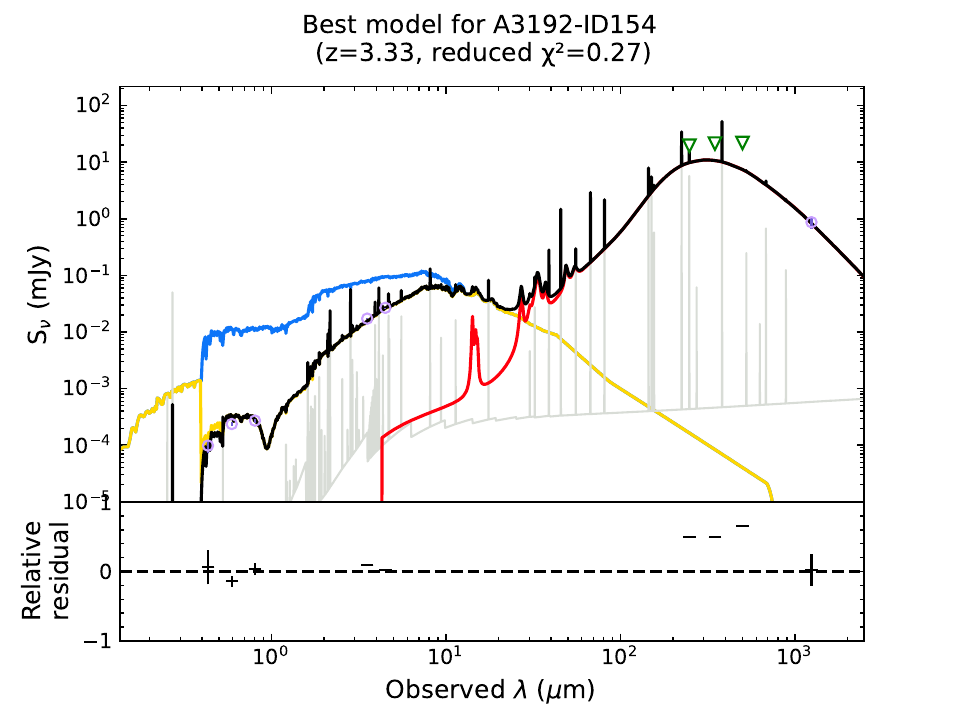}
\figsetgrpnote{The SED and the best-fit model of A3192-ID154. Lensing magnification is NOT corrected in this figure. The black solid line represents the composite spectrum of the galaxy. The yellow line illustrates the stellar emission attenuated by interstellar dust. The blue line depicts the unattenuated stellar emission for reference purpose. The orange line corresponds to the emission from an AGN. The red line shows the infrared emission from interstellar dust. The gray line denotes the nebulae emission. The observed data points are represented by purple circles, accompanied by 1$\sigma$ error bars. The bottom panel displays the relative residuals.}
\figsetgrpend

\figsetgrpstart
\figsetgrpnum{16.22}
\figsetgrptitle{A3192-ID31}
\figsetplot{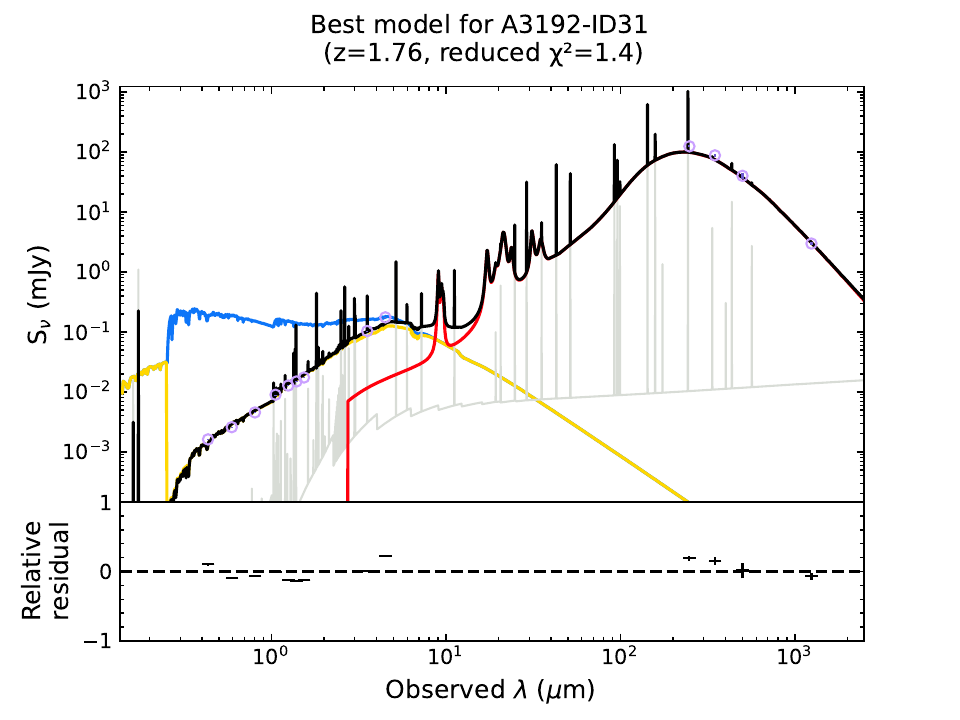}
\figsetgrpnote{The SED and the best-fit model of A3192-ID31. Lensing magnification is NOT corrected in this figure. The black solid line represents the composite spectrum of the galaxy. The yellow line illustrates the stellar emission attenuated by interstellar dust. The blue line depicts the unattenuated stellar emission for reference purpose. The orange line corresponds to the emission from an AGN. The red line shows the infrared emission from interstellar dust. The gray line denotes the nebulae emission. The observed data points are represented by purple circles, accompanied by 1$\sigma$ error bars. The bottom panel displays the relative residuals.}
\figsetgrpend

\figsetgrpstart
\figsetgrpnum{16.23}
\figsetgrptitle{A3192-ID40}
\figsetplot{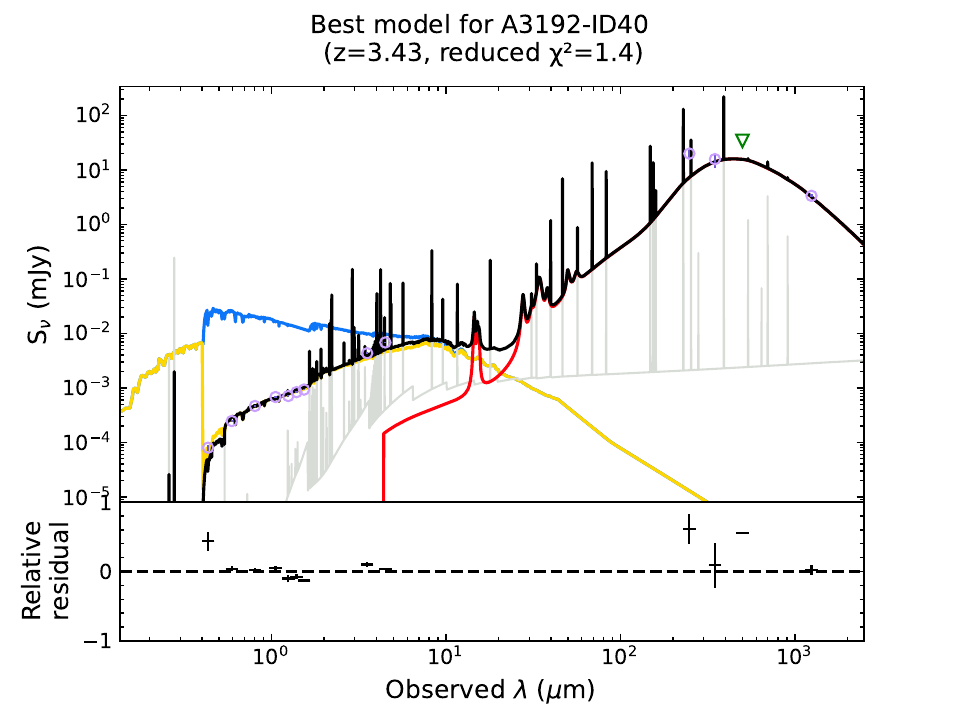}
\figsetgrpnote{The SED and the best-fit model of A3192-ID40. Lensing magnification is NOT corrected in this figure. The black solid line represents the composite spectrum of the galaxy. The yellow line illustrates the stellar emission attenuated by interstellar dust. The blue line depicts the unattenuated stellar emission for reference purpose. The orange line corresponds to the emission from an AGN. The red line shows the infrared emission from interstellar dust. The gray line denotes the nebulae emission. The observed data points are represented by purple circles, accompanied by 1$\sigma$ error bars. The bottom panel displays the relative residuals.}
\figsetgrpend

\figsetgrpstart
\figsetgrpnum{16.24}
\figsetgrptitle{A3192-ID83}
\figsetplot{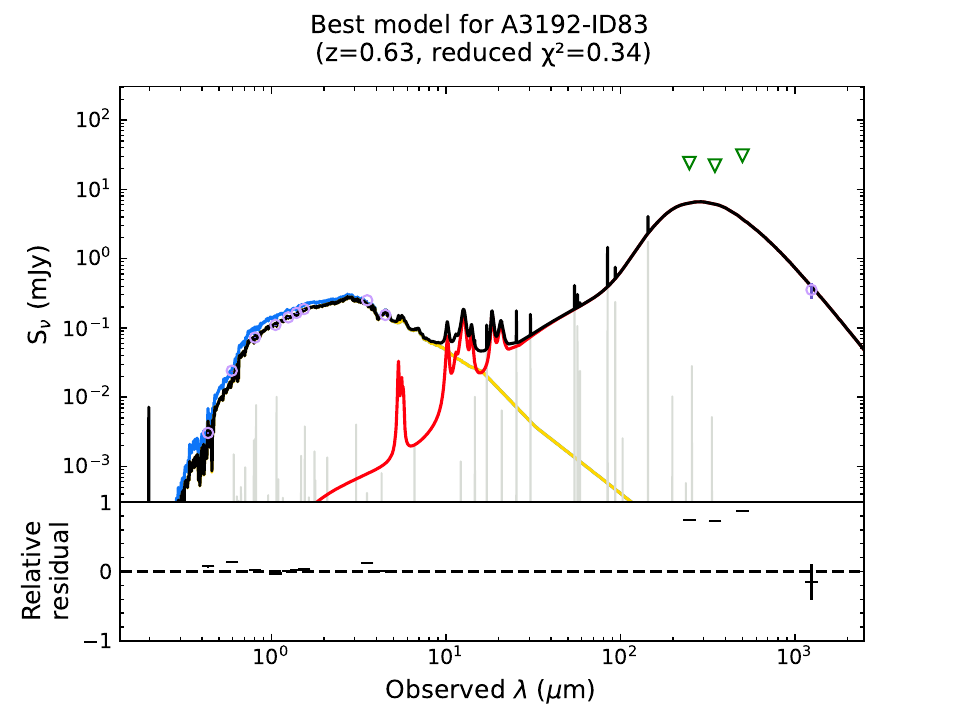}
\figsetgrpnote{The SED and the best-fit model of A3192-ID83. Lensing magnification is NOT corrected in this figure. The black solid line represents the composite spectrum of the galaxy. The yellow line illustrates the stellar emission attenuated by interstellar dust. The blue line depicts the unattenuated stellar emission for reference purpose. The orange line corresponds to the emission from an AGN. The red line shows the infrared emission from interstellar dust. The gray line denotes the nebulae emission. The observed data points are represented by purple circles, accompanied by 1$\sigma$ error bars. The bottom panel displays the relative residuals.}
\figsetgrpend

\figsetgrpstart
\figsetgrpnum{16.25}
\figsetgrptitle{A370-ID103}
\figsetplot{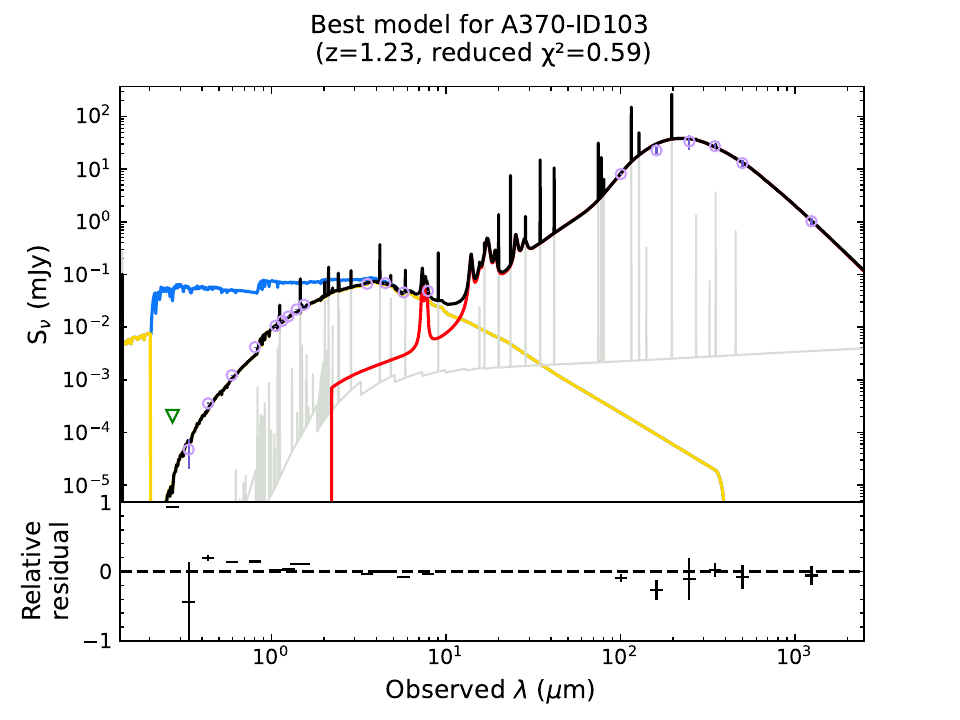}
\figsetgrpnote{The SED and the best-fit model of A370-ID103. Lensing magnification is NOT corrected in this figure. The black solid line represents the composite spectrum of the galaxy. The yellow line illustrates the stellar emission attenuated by interstellar dust. The blue line depicts the unattenuated stellar emission for reference purpose. The orange line corresponds to the emission from an AGN. The red line shows the infrared emission from interstellar dust. The gray line denotes the nebulae emission. The observed data points are represented by purple circles, accompanied by 1$\sigma$ error bars. The bottom panel displays the relative residuals.}
\figsetgrpend

\figsetgrpstart
\figsetgrpnum{16.26}
\figsetgrptitle{A370-ID146}
\figsetplot{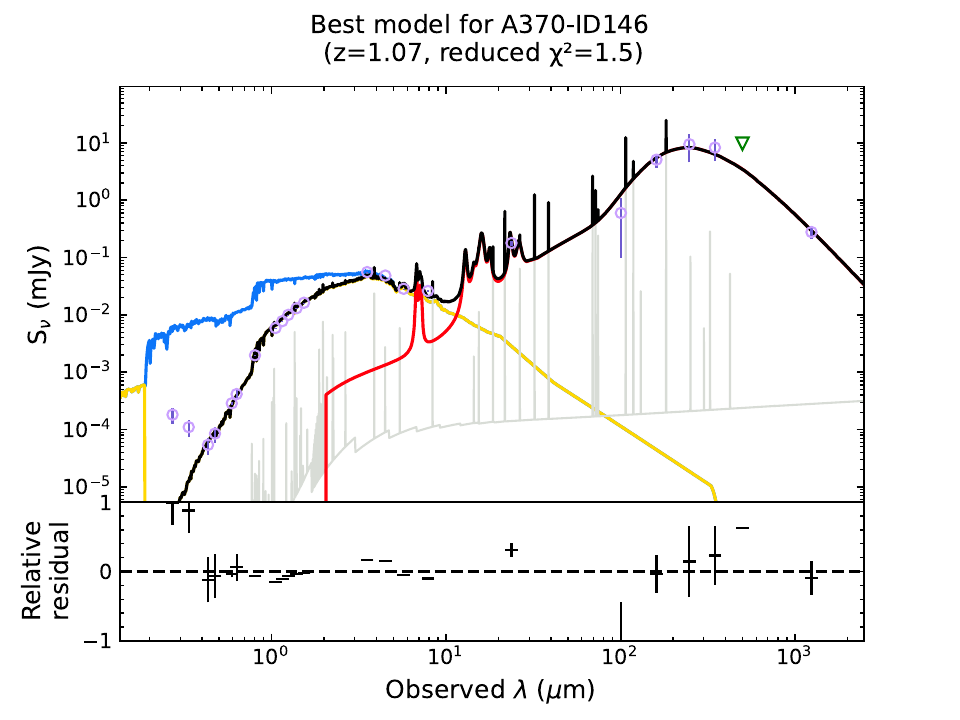}
\figsetgrpnote{The SED and the best-fit model of A370-ID146. Lensing magnification is NOT corrected in this figure. The black solid line represents the composite spectrum of the galaxy. The yellow line illustrates the stellar emission attenuated by interstellar dust. The blue line depicts the unattenuated stellar emission for reference purpose. The orange line corresponds to the emission from an AGN. The red line shows the infrared emission from interstellar dust. The gray line denotes the nebulae emission. The observed data points are represented by purple circles, accompanied by 1$\sigma$ error bars. The bottom panel displays the relative residuals.}
\figsetgrpend

\figsetgrpstart
\figsetgrpnum{16.27}
\figsetgrptitle{A370-ID172}
\figsetplot{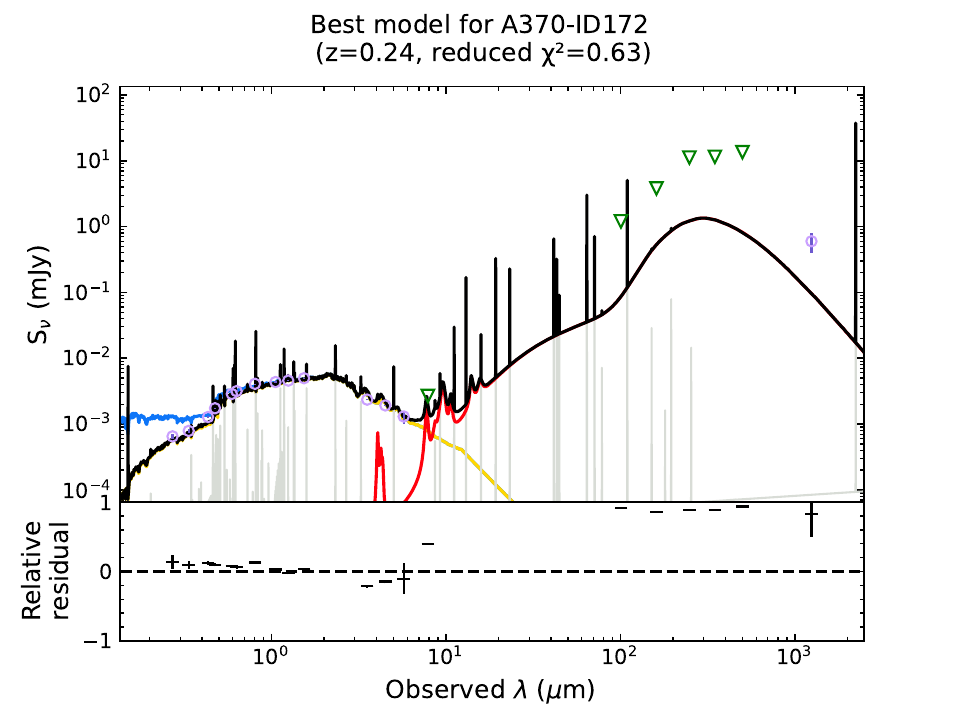}
\figsetgrpnote{The SED and the best-fit model of A370-ID172. Lensing magnification is NOT corrected in this figure. The black solid line represents the composite spectrum of the galaxy. The yellow line illustrates the stellar emission attenuated by interstellar dust. The blue line depicts the unattenuated stellar emission for reference purpose. The orange line corresponds to the emission from an AGN. The red line shows the infrared emission from interstellar dust. The gray line denotes the nebulae emission. The observed data points are represented by purple circles, accompanied by 1$\sigma$ error bars. The bottom panel displays the relative residuals.}
\figsetgrpend

\figsetgrpstart
\figsetgrpnum{16.28}
\figsetgrptitle{A370-ID18}
\figsetplot{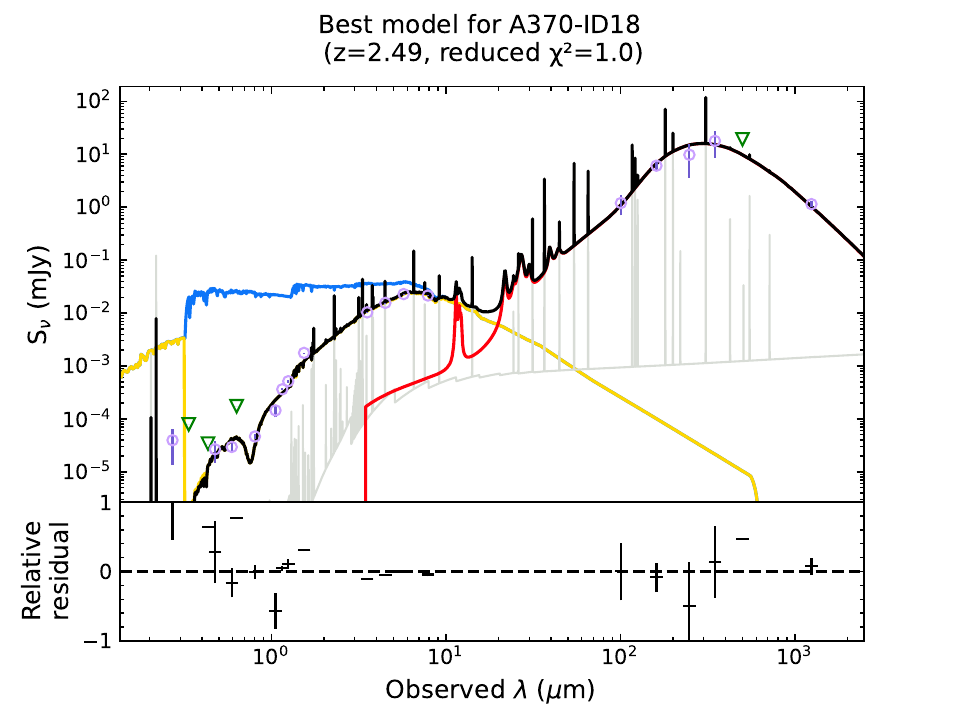}
\figsetgrpnote{The SED and the best-fit model of A370-ID18. Lensing magnification is NOT corrected in this figure. The black solid line represents the composite spectrum of the galaxy. The yellow line illustrates the stellar emission attenuated by interstellar dust. The blue line depicts the unattenuated stellar emission for reference purpose. The orange line corresponds to the emission from an AGN. The red line shows the infrared emission from interstellar dust. The gray line denotes the nebulae emission. The observed data points are represented by purple circles, accompanied by 1$\sigma$ error bars. The bottom panel displays the relative residuals.}
\figsetgrpend

\figsetgrpstart
\figsetgrpnum{16.29}
\figsetgrptitle{A370-ID27}
\figsetplot{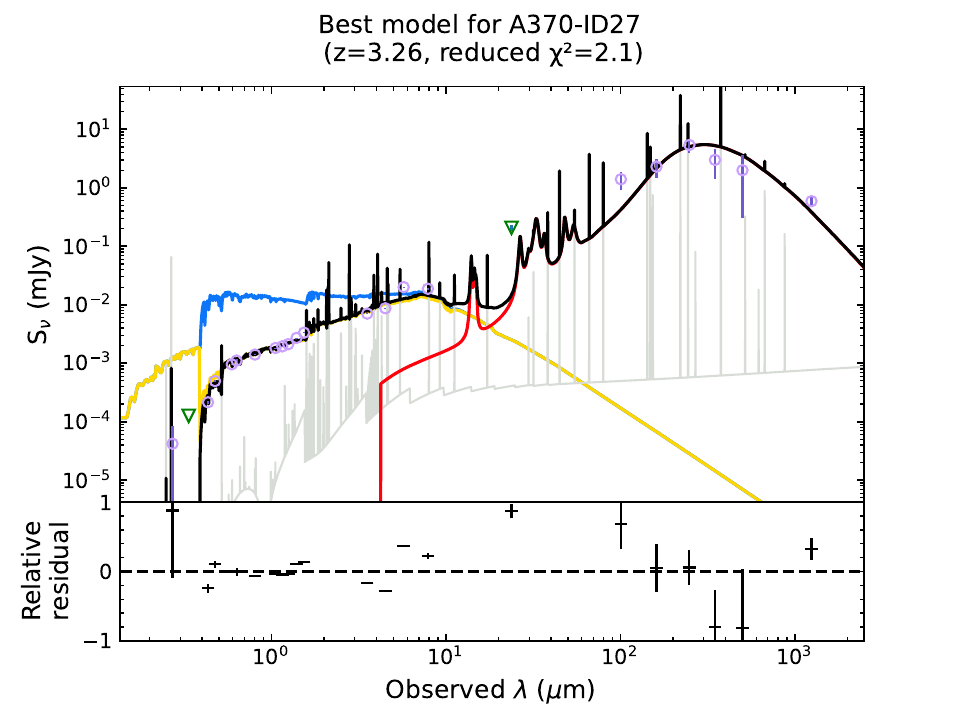}
\figsetgrpnote{The SED and the best-fit model of A370-ID27. Lensing magnification is NOT corrected in this figure. The black solid line represents the composite spectrum of the galaxy. The yellow line illustrates the stellar emission attenuated by interstellar dust. The blue line depicts the unattenuated stellar emission for reference purpose. The orange line corresponds to the emission from an AGN. The red line shows the infrared emission from interstellar dust. The gray line denotes the nebulae emission. The observed data points are represented by purple circles, accompanied by 1$\sigma$ error bars. The bottom panel displays the relative residuals.}
\figsetgrpend

\figsetgrpstart
\figsetgrpnum{16.30}
\figsetgrptitle{A370-ID31}
\figsetplot{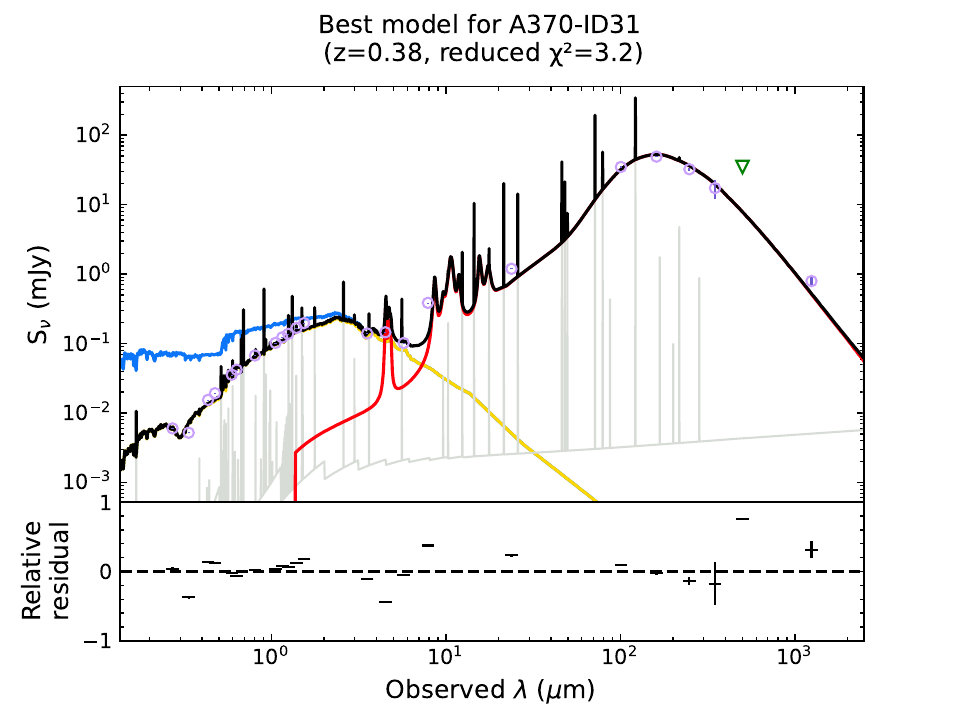}
\figsetgrpnote{The SED and the best-fit model of A370-ID31. Lensing magnification is NOT corrected in this figure. The black solid line represents the composite spectrum of the galaxy. The yellow line illustrates the stellar emission attenuated by interstellar dust. The blue line depicts the unattenuated stellar emission for reference purpose. The orange line corresponds to the emission from an AGN. The red line shows the infrared emission from interstellar dust. The gray line denotes the nebulae emission. The observed data points are represented by purple circles, accompanied by 1$\sigma$ error bars. The bottom panel displays the relative residuals.}
\figsetgrpend

\figsetgrpstart
\figsetgrpnum{16.31}
\figsetgrptitle{A383-ID24}
\figsetplot{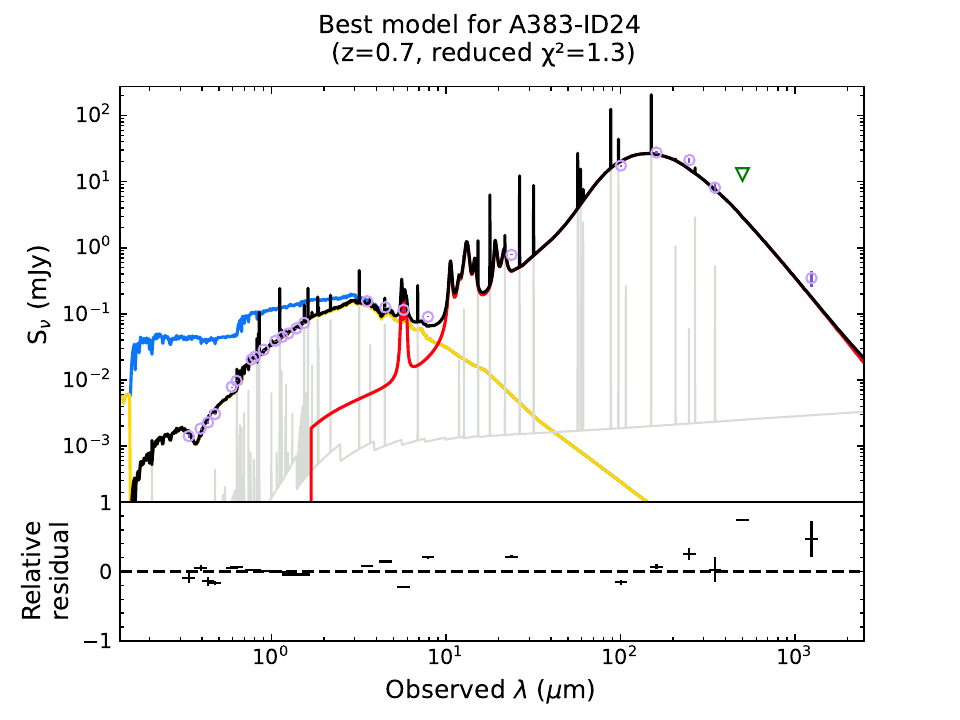}
\figsetgrpnote{The SED and the best-fit model of A383-ID24. Lensing magnification is NOT corrected in this figure. The black solid line represents the composite spectrum of the galaxy. The yellow line illustrates the stellar emission attenuated by interstellar dust. The blue line depicts the unattenuated stellar emission for reference purpose. The orange line corresponds to the emission from an AGN. The red line shows the infrared emission from interstellar dust. The gray line denotes the nebulae emission. The observed data points are represented by purple circles, accompanied by 1$\sigma$ error bars. The bottom panel displays the relative residuals.}
\figsetgrpend

\figsetgrpstart
\figsetgrpnum{16.32}
\figsetgrptitle{A383-ID50}
\figsetplot{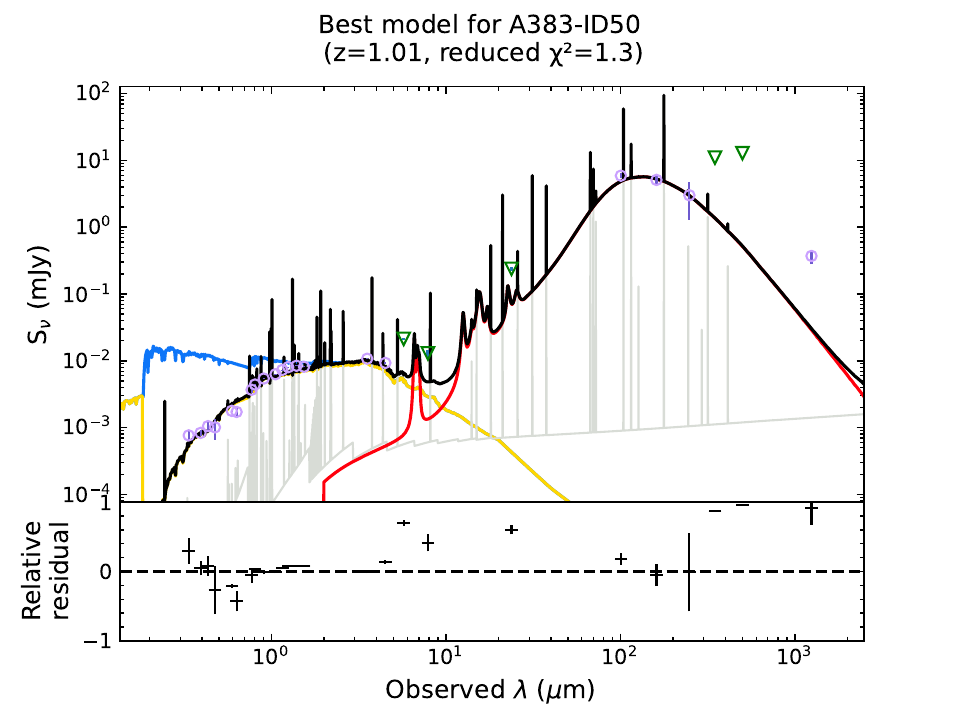}
\figsetgrpnote{The SED and the best-fit model of A383-ID50. Lensing magnification is NOT corrected in this figure. The black solid line represents the composite spectrum of the galaxy. The yellow line illustrates the stellar emission attenuated by interstellar dust. The blue line depicts the unattenuated stellar emission for reference purpose. The orange line corresponds to the emission from an AGN. The red line shows the infrared emission from interstellar dust. The gray line denotes the nebulae emission. The observed data points are represented by purple circles, accompanied by 1$\sigma$ error bars. The bottom panel displays the relative residuals.}
\figsetgrpend

\figsetgrpstart
\figsetgrpnum{16.33}
\figsetgrptitle{ACT0102-ID11}
\figsetplot{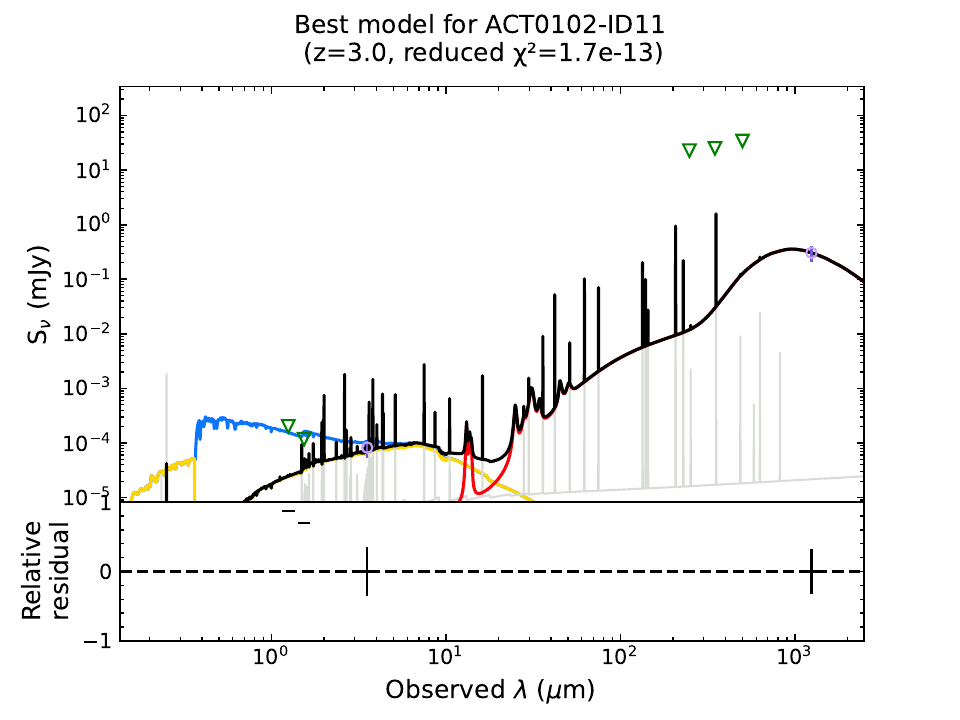}
\figsetgrpnote{The SED and the best-fit model of ACT0102-ID11. Lensing magnification is NOT corrected in this figure. The black solid line represents the composite spectrum of the galaxy. The yellow line illustrates the stellar emission attenuated by interstellar dust. The blue line depicts the unattenuated stellar emission for reference purpose. The orange line corresponds to the emission from an AGN. The red line shows the infrared emission from interstellar dust. The gray line denotes the nebulae emission. The observed data points are represented by purple circles, accompanied by 1$\sigma$ error bars. The bottom panel displays the relative residuals.}
\figsetgrpend

\figsetgrpstart
\figsetgrpnum{16.34}
\figsetgrptitle{ACT0102-ID118}
\figsetplot{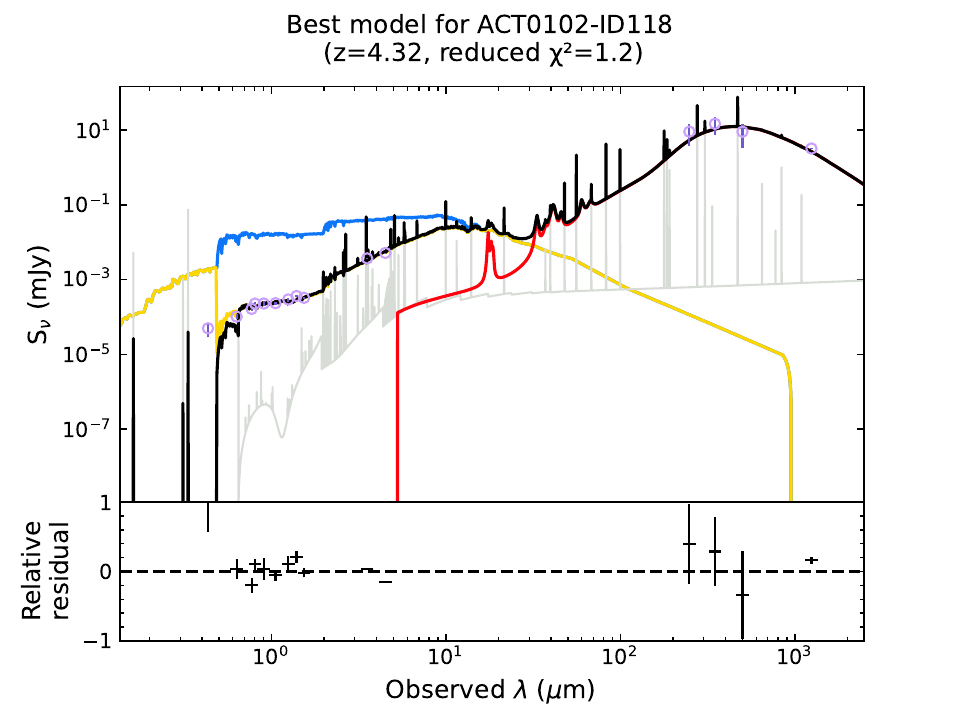}
\figsetgrpnote{The SED and the best-fit model of ACT0102-ID118. Lensing magnification is NOT corrected in this figure. The black solid line represents the composite spectrum of the galaxy. The yellow line illustrates the stellar emission attenuated by interstellar dust. The blue line depicts the unattenuated stellar emission for reference purpose. The orange line corresponds to the emission from an AGN. The red line shows the infrared emission from interstellar dust. The gray line denotes the nebulae emission. The observed data points are represented by purple circles, accompanied by 1$\sigma$ error bars. The bottom panel displays the relative residuals.}
\figsetgrpend

\figsetgrpstart
\figsetgrpnum{16.35}
\figsetgrptitle{ACT0102-ID128}
\figsetplot{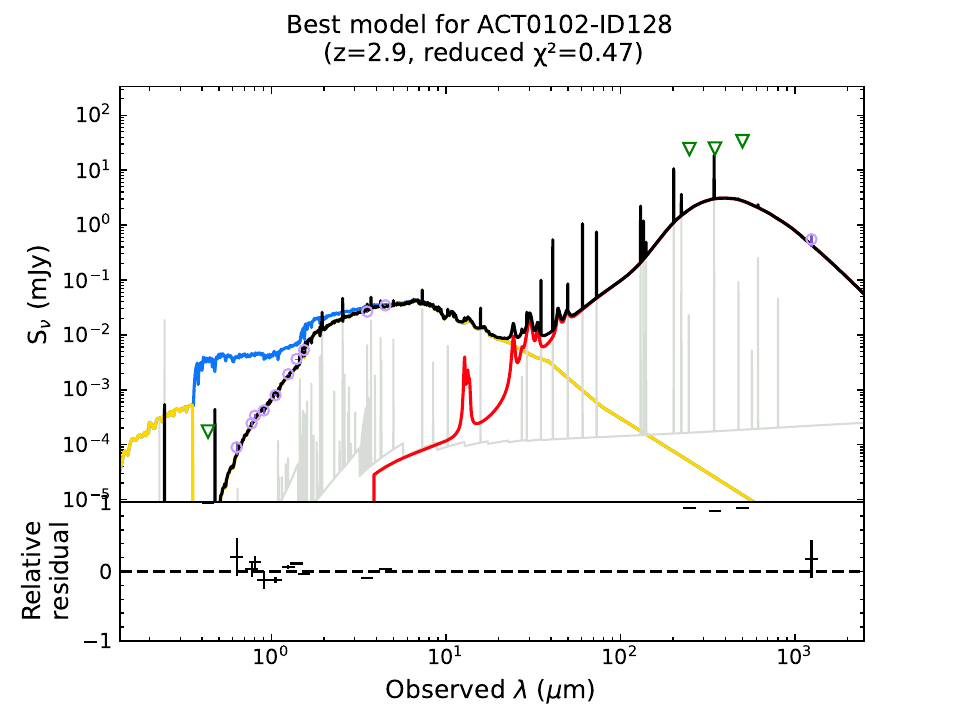}
\figsetgrpnote{The SED and the best-fit model of ACT0102-ID128. Lensing magnification is NOT corrected in this figure. The black solid line represents the composite spectrum of the galaxy. The yellow line illustrates the stellar emission attenuated by interstellar dust. The blue line depicts the unattenuated stellar emission for reference purpose. The orange line corresponds to the emission from an AGN. The red line shows the infrared emission from interstellar dust. The gray line denotes the nebulae emission. The observed data points are represented by purple circles, accompanied by 1$\sigma$ error bars. The bottom panel displays the relative residuals.}
\figsetgrpend

\figsetgrpstart
\figsetgrpnum{16.36}
\figsetgrptitle{ACT0102-ID160}
\figsetplot{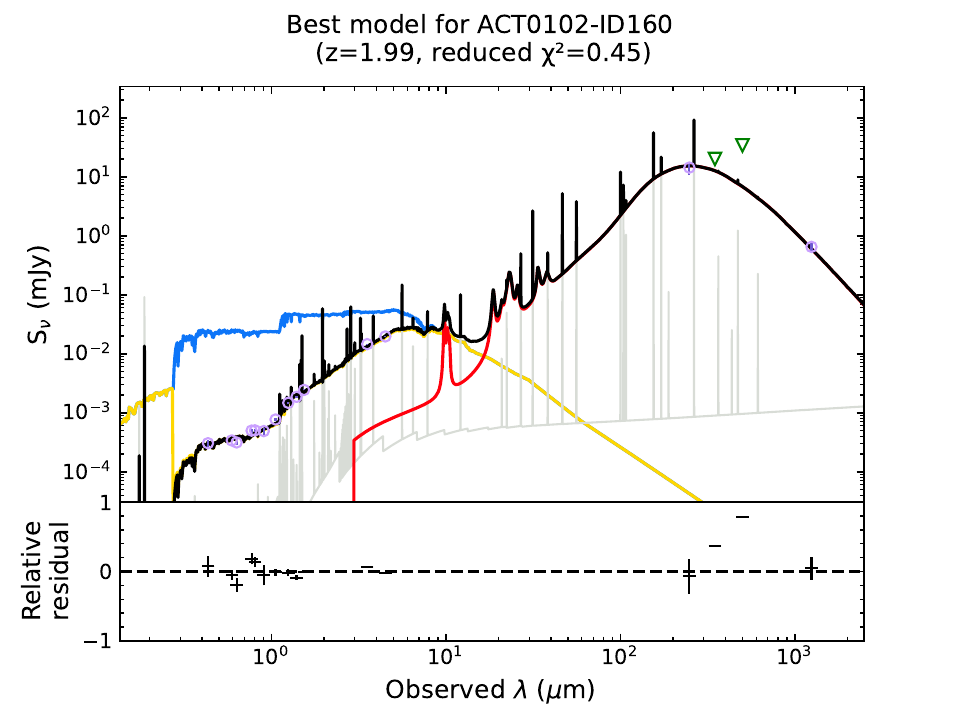}
\figsetgrpnote{The SED and the best-fit model of ACT0102-ID160. Lensing magnification is NOT corrected in this figure. The black solid line represents the composite spectrum of the galaxy. The yellow line illustrates the stellar emission attenuated by interstellar dust. The blue line depicts the unattenuated stellar emission for reference purpose. The orange line corresponds to the emission from an AGN. The red line shows the infrared emission from interstellar dust. The gray line denotes the nebulae emission. The observed data points are represented by purple circles, accompanied by 1$\sigma$ error bars. The bottom panel displays the relative residuals.}
\figsetgrpend

\figsetgrpstart
\figsetgrpnum{16.37}
\figsetgrptitle{ACT0102-ID215}
\figsetplot{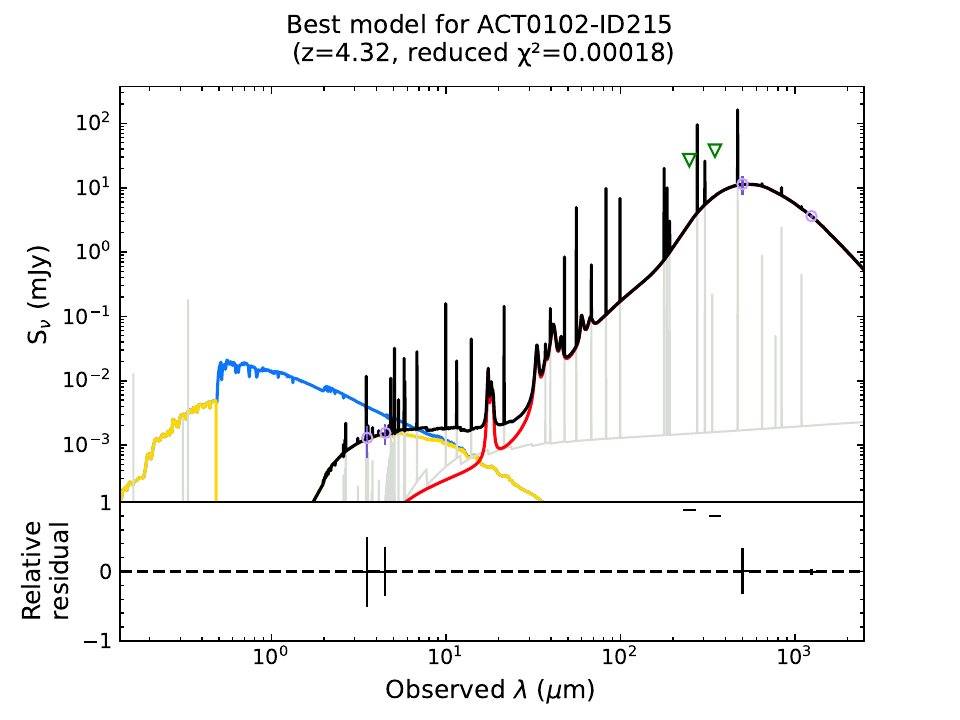}
\figsetgrpnote{The SED and the best-fit model of ACT0102-ID215. Lensing magnification is NOT corrected in this figure. The black solid line represents the composite spectrum of the galaxy. The yellow line illustrates the stellar emission attenuated by interstellar dust. The blue line depicts the unattenuated stellar emission for reference purpose. The orange line corresponds to the emission from an AGN. The red line shows the infrared emission from interstellar dust. The gray line denotes the nebulae emission. The observed data points are represented by purple circles, accompanied by 1$\sigma$ error bars. The bottom panel displays the relative residuals.}
\figsetgrpend

\figsetgrpstart
\figsetgrpnum{16.38}
\figsetgrptitle{ACT0102-ID22}
\figsetplot{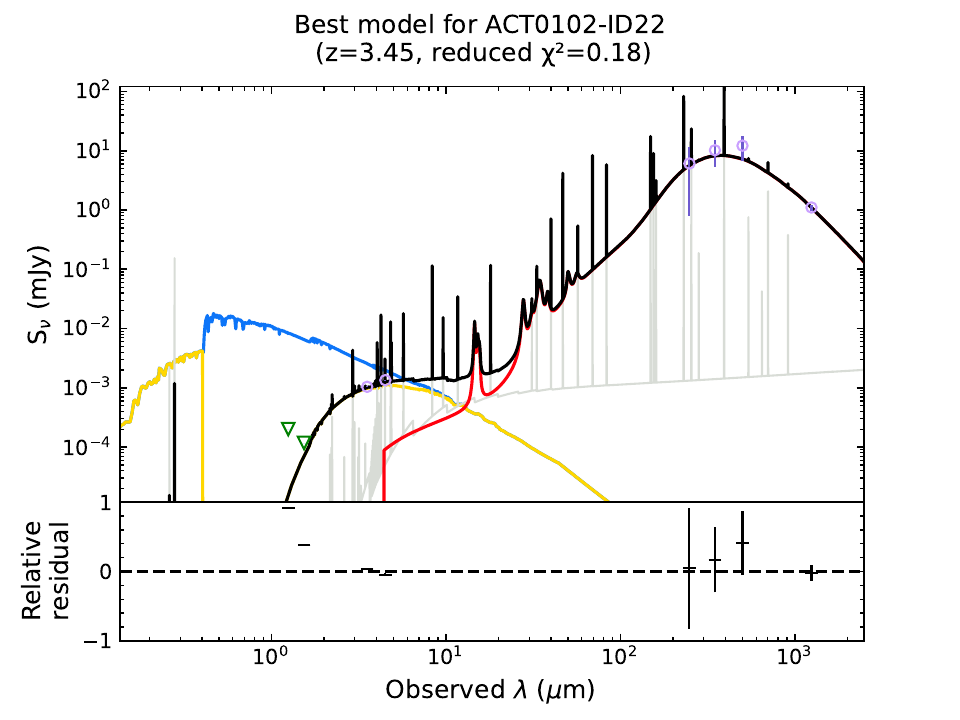}
\figsetgrpnote{The SED and the best-fit model of ACT0102-ID22. Lensing magnification is NOT corrected in this figure. The black solid line represents the composite spectrum of the galaxy. The yellow line illustrates the stellar emission attenuated by interstellar dust. The blue line depicts the unattenuated stellar emission for reference purpose. The orange line corresponds to the emission from an AGN. The red line shows the infrared emission from interstellar dust. The gray line denotes the nebulae emission. The observed data points are represented by purple circles, accompanied by 1$\sigma$ error bars. The bottom panel displays the relative residuals.}
\figsetgrpend

\figsetgrpstart
\figsetgrpnum{16.39}
\figsetgrptitle{ACT0102-ID223}
\figsetplot{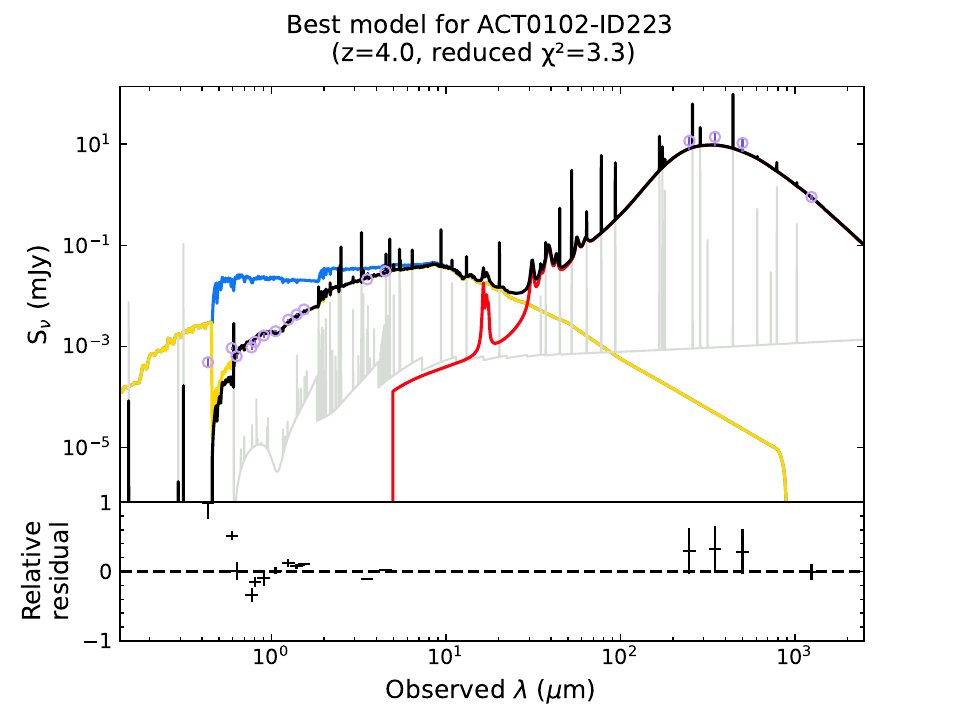}
\figsetgrpnote{The SED and the best-fit model of ACT0102-ID223. Lensing magnification is NOT corrected in this figure. The black solid line represents the composite spectrum of the galaxy. The yellow line illustrates the stellar emission attenuated by interstellar dust. The blue line depicts the unattenuated stellar emission for reference purpose. The orange line corresponds to the emission from an AGN. The red line shows the infrared emission from interstellar dust. The gray line denotes the nebulae emission. The observed data points are represented by purple circles, accompanied by 1$\sigma$ error bars. The bottom panel displays the relative residuals.}
\figsetgrpend

\figsetgrpstart
\figsetgrpnum{16.40}
\figsetgrptitle{ACT0102-ID224}
\figsetplot{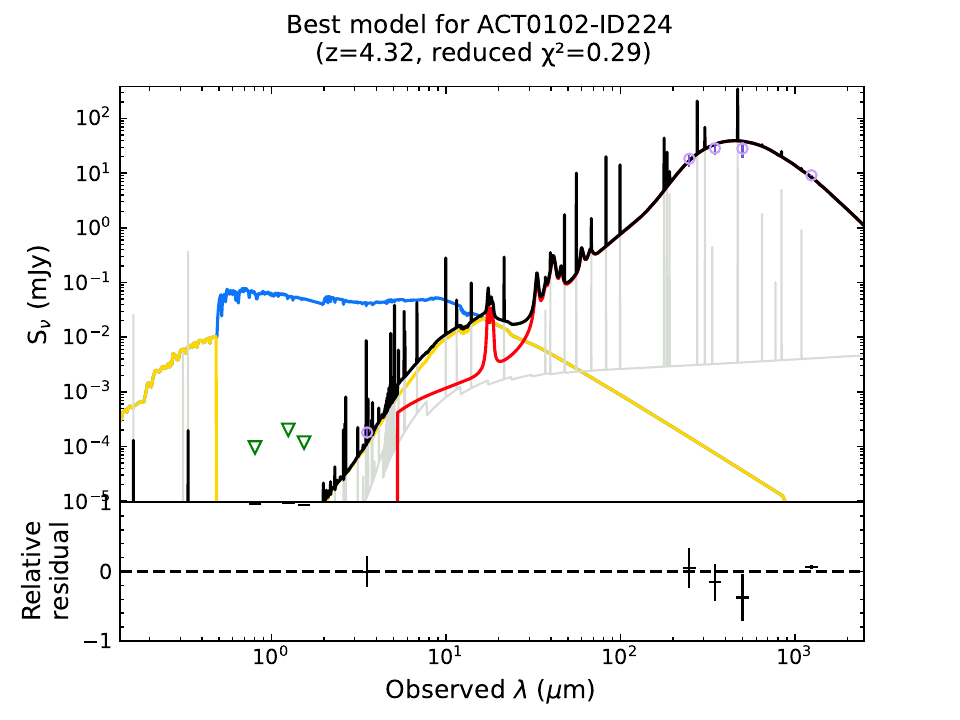}
\figsetgrpnote{The SED and the best-fit model of ACT0102-ID224. Lensing magnification is NOT corrected in this figure. The black solid line represents the composite spectrum of the galaxy. The yellow line illustrates the stellar emission attenuated by interstellar dust. The blue line depicts the unattenuated stellar emission for reference purpose. The orange line corresponds to the emission from an AGN. The red line shows the infrared emission from interstellar dust. The gray line denotes the nebulae emission. The observed data points are represented by purple circles, accompanied by 1$\sigma$ error bars. The bottom panel displays the relative residuals.}
\figsetgrpend

\figsetgrpstart
\figsetgrpnum{16.41}
\figsetgrptitle{ACT0102-ID241}
\figsetplot{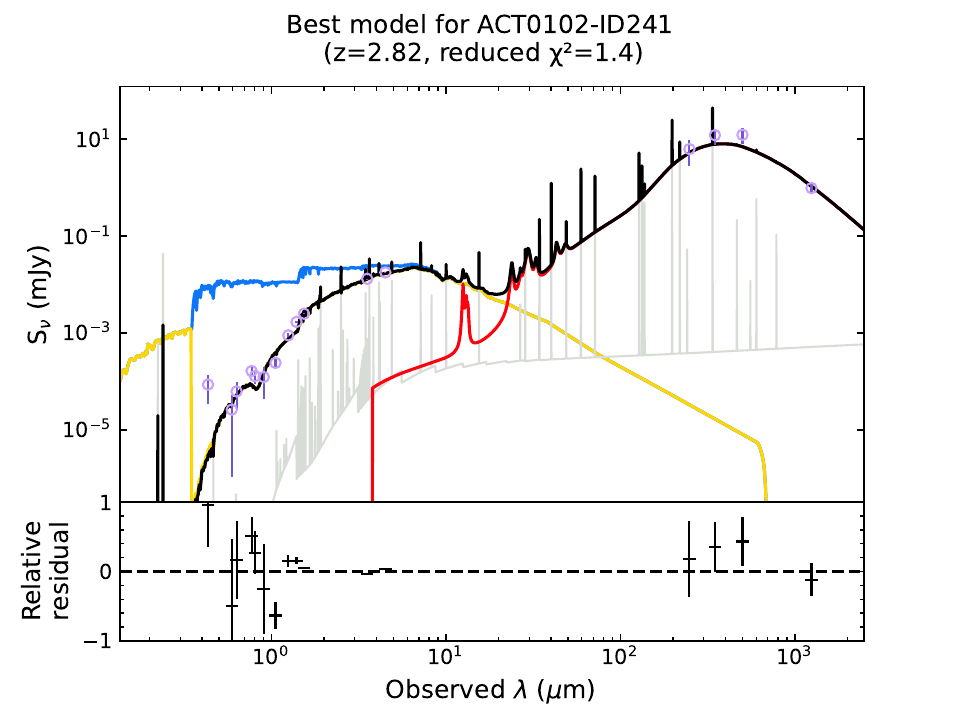}
\figsetgrpnote{The SED and the best-fit model of ACT0102-ID241. Lensing magnification is NOT corrected in this figure. The black solid line represents the composite spectrum of the galaxy. The yellow line illustrates the stellar emission attenuated by interstellar dust. The blue line depicts the unattenuated stellar emission for reference purpose. The orange line corresponds to the emission from an AGN. The red line shows the infrared emission from interstellar dust. The gray line denotes the nebulae emission. The observed data points are represented by purple circles, accompanied by 1$\sigma$ error bars. The bottom panel displays the relative residuals.}
\figsetgrpend

\figsetgrpstart
\figsetgrpnum{16.42}
\figsetgrptitle{ACT0102-ID276}
\figsetplot{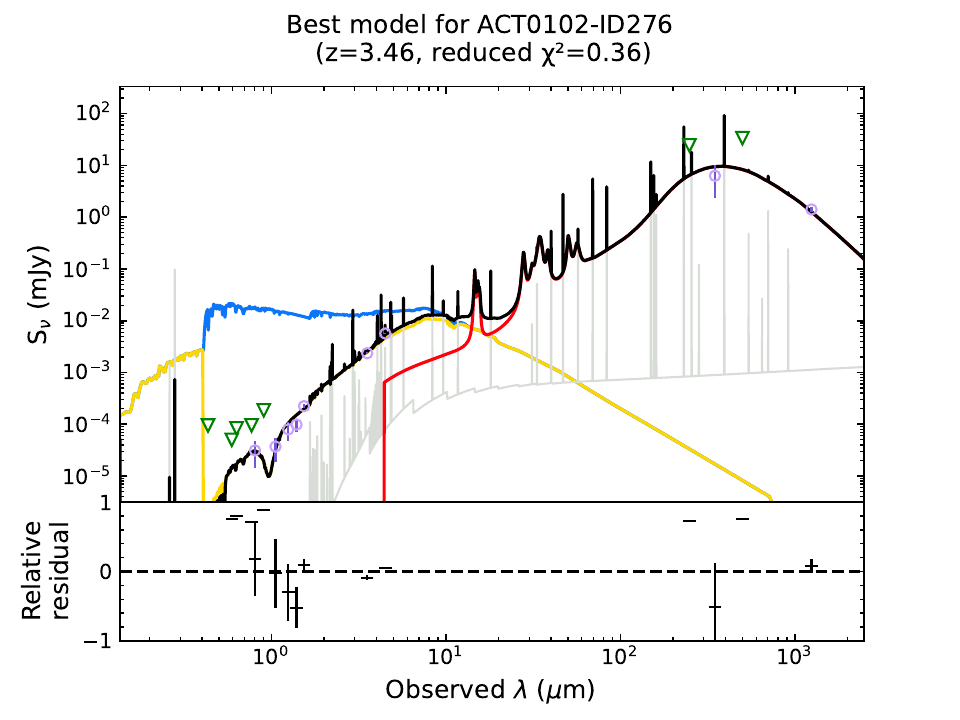}
\figsetgrpnote{The SED and the best-fit model of ACT0102-ID276. Lensing magnification is NOT corrected in this figure. The black solid line represents the composite spectrum of the galaxy. The yellow line illustrates the stellar emission attenuated by interstellar dust. The blue line depicts the unattenuated stellar emission for reference purpose. The orange line corresponds to the emission from an AGN. The red line shows the infrared emission from interstellar dust. The gray line denotes the nebulae emission. The observed data points are represented by purple circles, accompanied by 1$\sigma$ error bars. The bottom panel displays the relative residuals.}
\figsetgrpend

\figsetgrpstart
\figsetgrpnum{16.43}
\figsetgrptitle{ACT0102-ID294}
\figsetplot{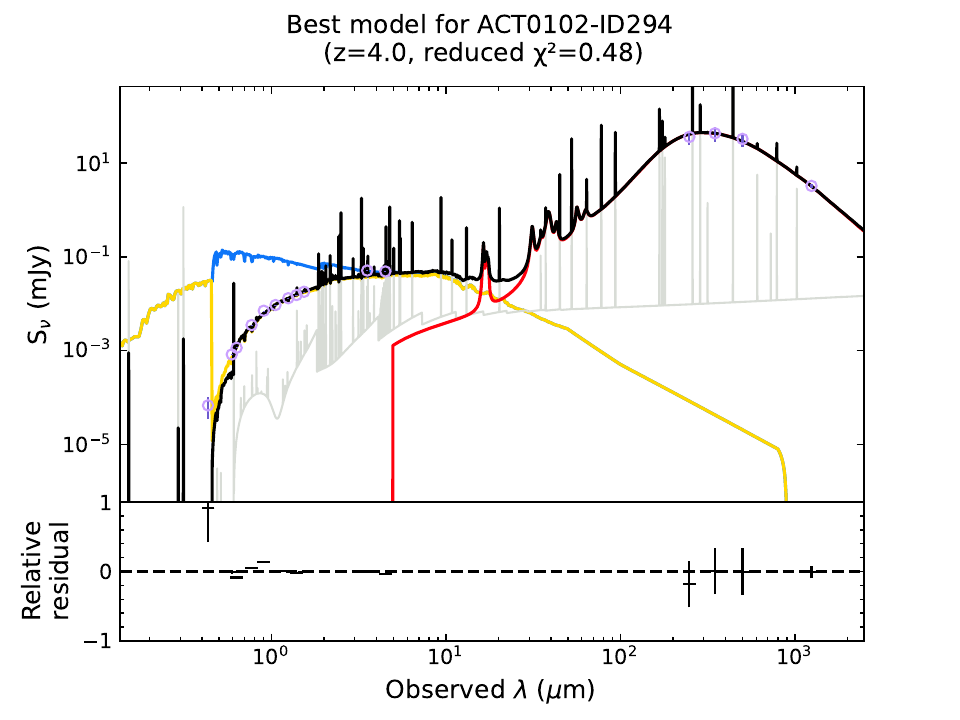}
\figsetgrpnote{The SED and the best-fit model of ACT0102-ID294. Lensing magnification is NOT corrected in this figure. The black solid line represents the composite spectrum of the galaxy. The yellow line illustrates the stellar emission attenuated by interstellar dust. The blue line depicts the unattenuated stellar emission for reference purpose. The orange line corresponds to the emission from an AGN. The red line shows the infrared emission from interstellar dust. The gray line denotes the nebulae emission. The observed data points are represented by purple circles, accompanied by 1$\sigma$ error bars. The bottom panel displays the relative residuals.}
\figsetgrpend

\figsetgrpstart
\figsetgrpnum{16.44}
\figsetgrptitle{ACT0102-ID50}
\figsetplot{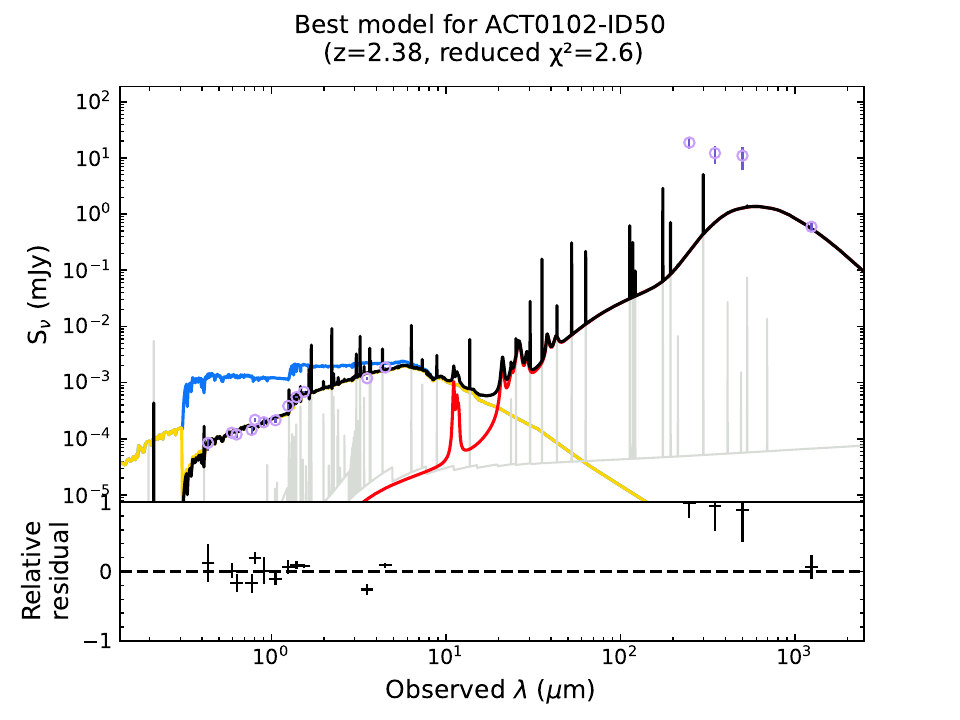}
\figsetgrpnote{The SED and the best-fit model of ACT0102-ID50. Lensing magnification is NOT corrected in this figure. The black solid line represents the composite spectrum of the galaxy. The yellow line illustrates the stellar emission attenuated by interstellar dust. The blue line depicts the unattenuated stellar emission for reference purpose. The orange line corresponds to the emission from an AGN. The red line shows the infrared emission from interstellar dust. The gray line denotes the nebulae emission. The observed data points are represented by purple circles, accompanied by 1$\sigma$ error bars. The bottom panel displays the relative residuals.}
\figsetgrpend

\figsetgrpstart
\figsetgrpnum{16.45}
\figsetgrptitle{ACT0102-ID52}
\figsetplot{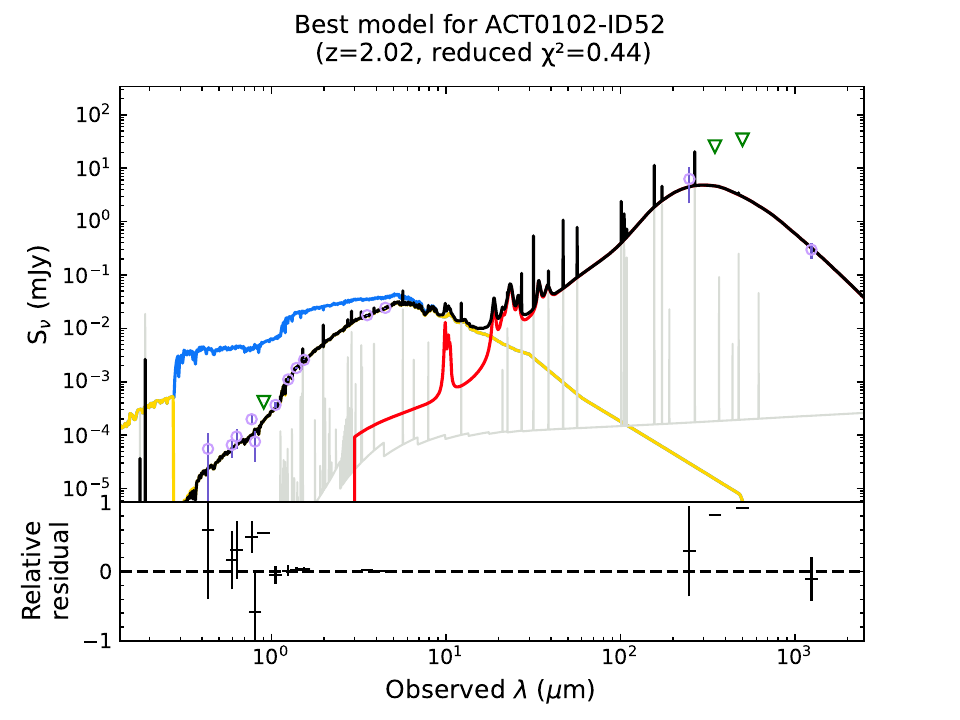}
\figsetgrpnote{The SED and the best-fit model of ACT0102-ID52. Lensing magnification is NOT corrected in this figure. The black solid line represents the composite spectrum of the galaxy. The yellow line illustrates the stellar emission attenuated by interstellar dust. The blue line depicts the unattenuated stellar emission for reference purpose. The orange line corresponds to the emission from an AGN. The red line shows the infrared emission from interstellar dust. The gray line denotes the nebulae emission. The observed data points are represented by purple circles, accompanied by 1$\sigma$ error bars. The bottom panel displays the relative residuals.}
\figsetgrpend

\figsetgrpstart
\figsetgrpnum{16.46}
\figsetgrptitle{AS1063-ID147}
\figsetplot{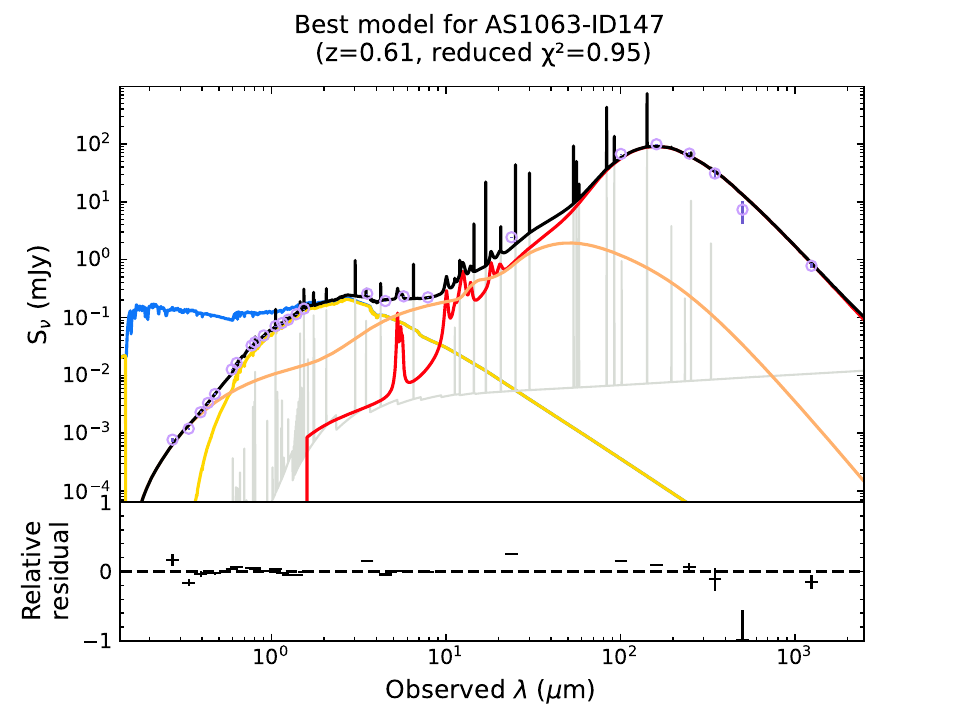}
\figsetgrpnote{The SED and the best-fit model of AS1063-ID147. Lensing magnification is NOT corrected in this figure. The black solid line represents the composite spectrum of the galaxy. The yellow line illustrates the stellar emission attenuated by interstellar dust. The blue line depicts the unattenuated stellar emission for reference purpose. The orange line corresponds to the emission from an AGN. The red line shows the infrared emission from interstellar dust. The gray line denotes the nebulae emission. The observed data points are represented by purple circles, accompanied by 1$\sigma$ error bars. The bottom panel displays the relative residuals.}
\figsetgrpend

\figsetgrpstart
\figsetgrpnum{16.47}
\figsetgrptitle{AS1063-ID15}
\figsetplot{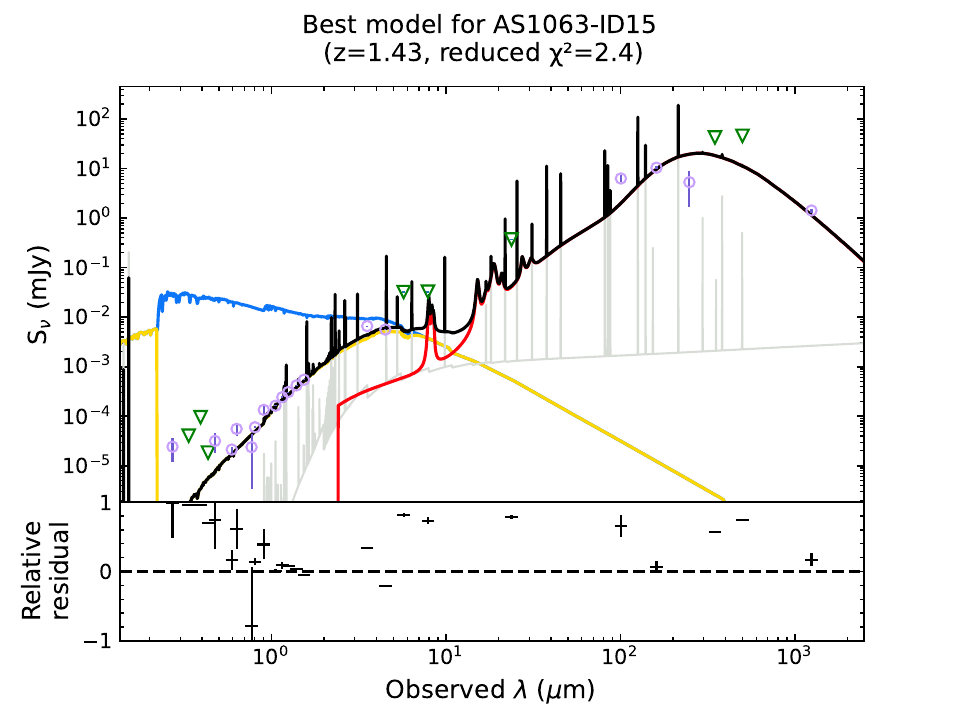}
\figsetgrpnote{The SED and the best-fit model of AS1063-ID15. Lensing magnification is NOT corrected in this figure. The black solid line represents the composite spectrum of the galaxy. The yellow line illustrates the stellar emission attenuated by interstellar dust. The blue line depicts the unattenuated stellar emission for reference purpose. The orange line corresponds to the emission from an AGN. The red line shows the infrared emission from interstellar dust. The gray line denotes the nebulae emission. The observed data points are represented by purple circles, accompanied by 1$\sigma$ error bars. The bottom panel displays the relative residuals.}
\figsetgrpend

\figsetgrpstart
\figsetgrpnum{16.48}
\figsetgrptitle{AS1063-ID17}
\figsetplot{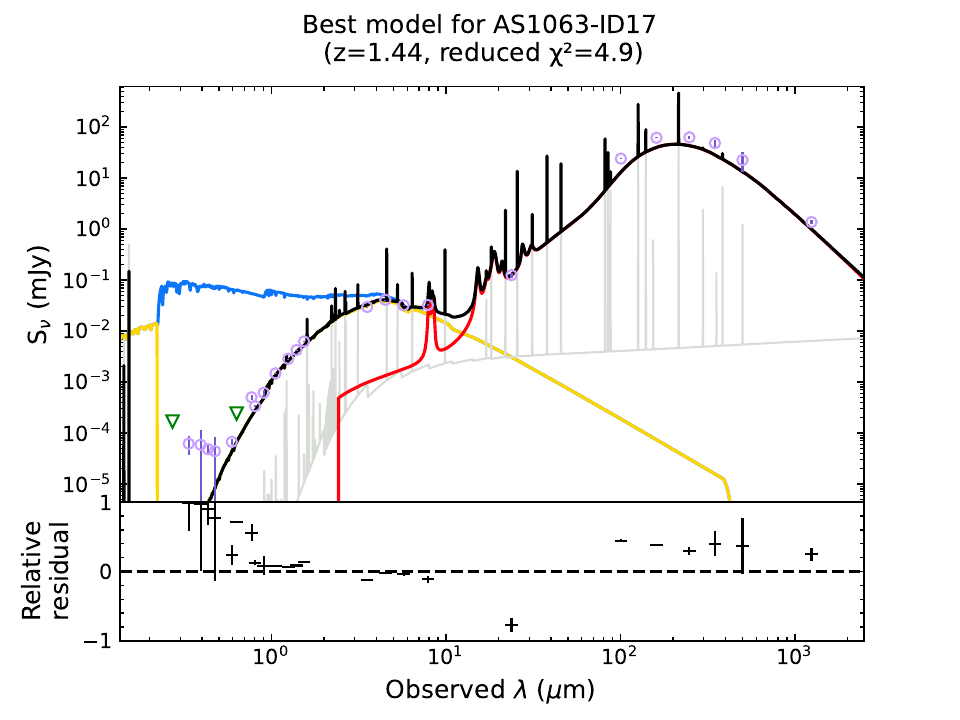}
\figsetgrpnote{The SED and the best-fit model of AS1063-ID17. Lensing magnification is NOT corrected in this figure. The black solid line represents the composite spectrum of the galaxy. The yellow line illustrates the stellar emission attenuated by interstellar dust. The blue line depicts the unattenuated stellar emission for reference purpose. The orange line corresponds to the emission from an AGN. The red line shows the infrared emission from interstellar dust. The gray line denotes the nebulae emission. The observed data points are represented by purple circles, accompanied by 1$\sigma$ error bars. The bottom panel displays the relative residuals.}
\figsetgrpend

\figsetgrpstart
\figsetgrpnum{16.49}
\figsetgrptitle{AS1063-ID222}
\figsetplot{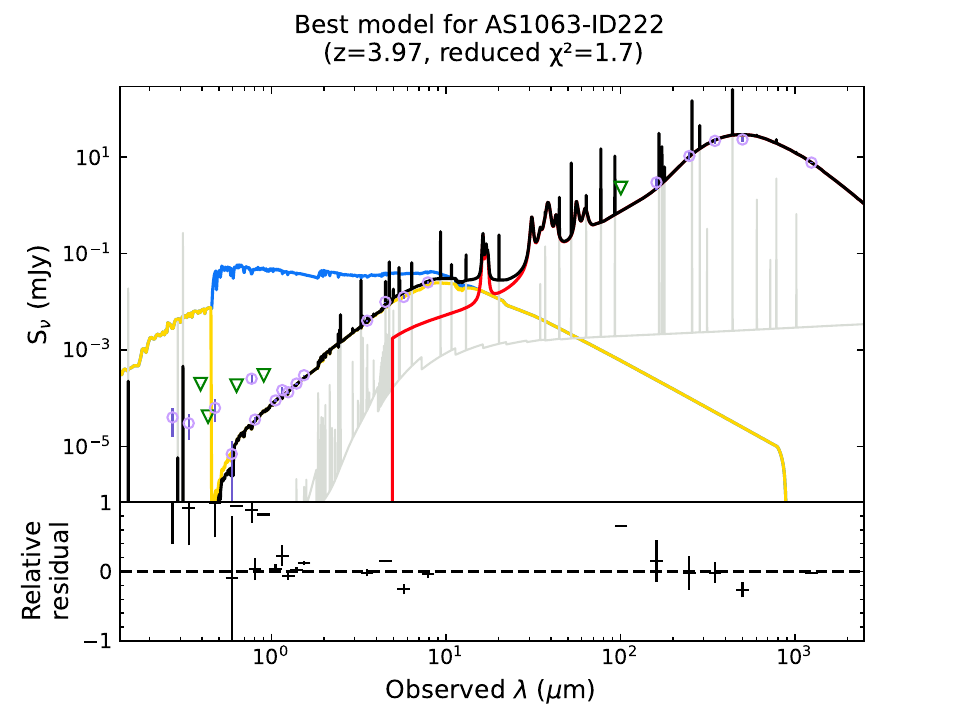}
\figsetgrpnote{The SED and the best-fit model of AS1063-ID222. Lensing magnification is NOT corrected in this figure. The black solid line represents the composite spectrum of the galaxy. The yellow line illustrates the stellar emission attenuated by interstellar dust. The blue line depicts the unattenuated stellar emission for reference purpose. The orange line corresponds to the emission from an AGN. The red line shows the infrared emission from interstellar dust. The gray line denotes the nebulae emission. The observed data points are represented by purple circles, accompanied by 1$\sigma$ error bars. The bottom panel displays the relative residuals.}
\figsetgrpend

\figsetgrpstart
\figsetgrpnum{16.50}
\figsetgrptitle{AS295-ID269}
\figsetplot{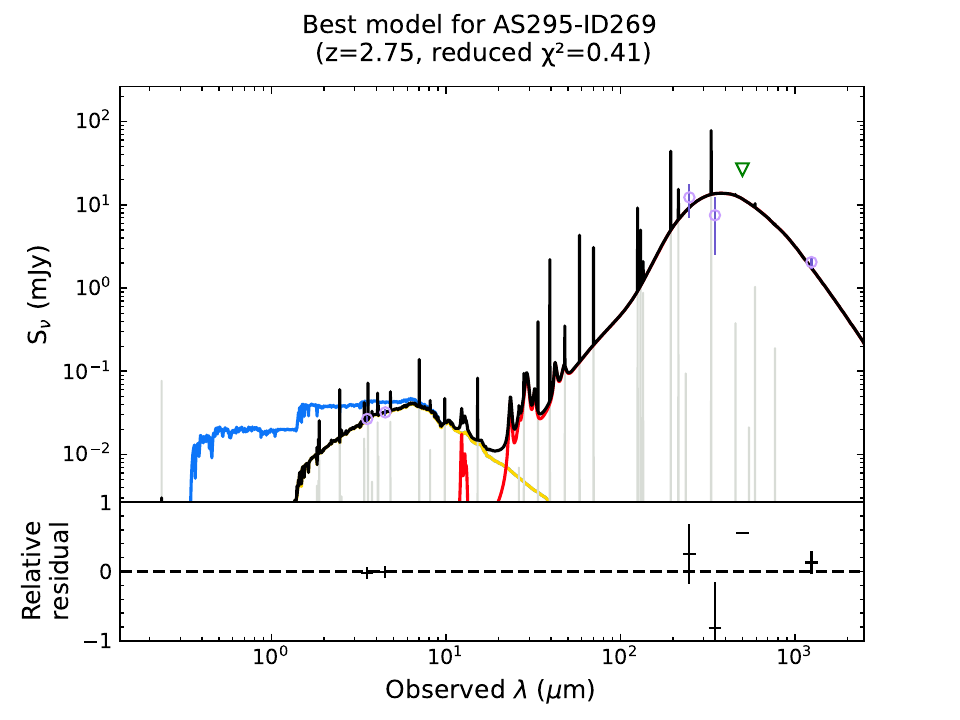}
\figsetgrpnote{The SED and the best-fit model of AS295-ID269. Lensing magnification is NOT corrected in this figure. The black solid line represents the composite spectrum of the galaxy. The yellow line illustrates the stellar emission attenuated by interstellar dust. The blue line depicts the unattenuated stellar emission for reference purpose. The orange line corresponds to the emission from an AGN. The red line shows the infrared emission from interstellar dust. The gray line denotes the nebulae emission. The observed data points are represented by purple circles, accompanied by 1$\sigma$ error bars. The bottom panel displays the relative residuals.}
\figsetgrpend

\figsetgrpstart
\figsetgrpnum{16.51}
\figsetgrptitle{AS295-ID9}
\figsetplot{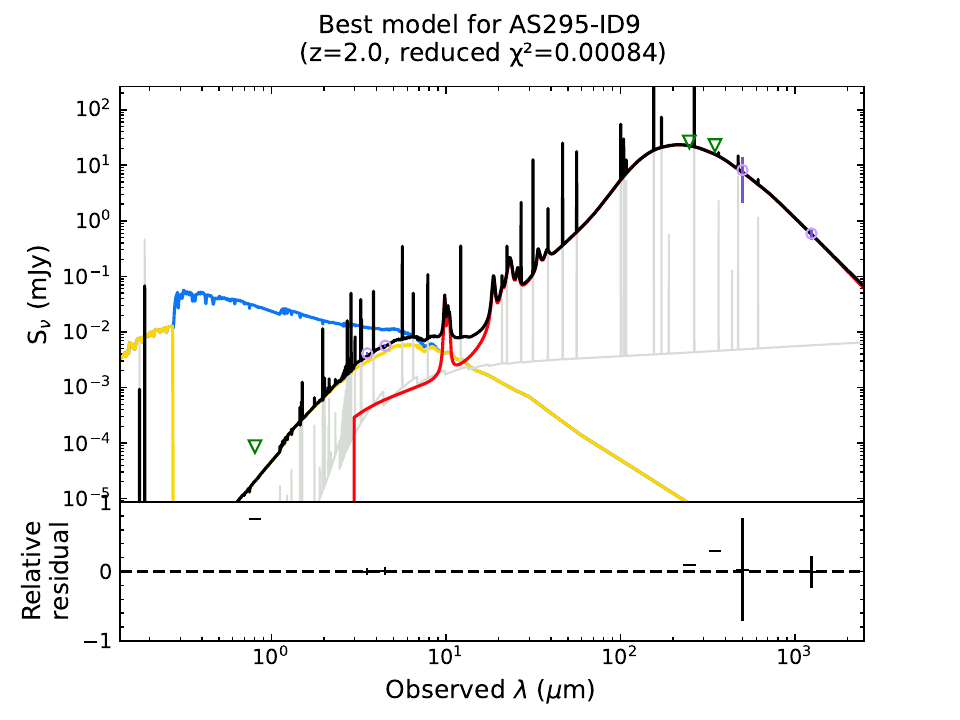}
\figsetgrpnote{The SED and the best-fit model of AS295-ID9. Lensing magnification is NOT corrected in this figure. The black solid line represents the composite spectrum of the galaxy. The yellow line illustrates the stellar emission attenuated by interstellar dust. The blue line depicts the unattenuated stellar emission for reference purpose. The orange line corresponds to the emission from an AGN. The red line shows the infrared emission from interstellar dust. The gray line denotes the nebulae emission. The observed data points are represented by purple circles, accompanied by 1$\sigma$ error bars. The bottom panel displays the relative residuals.}
\figsetgrpend

\figsetgrpstart
\figsetgrpnum{16.52}
\figsetgrptitle{M0035-ID33}
\figsetplot{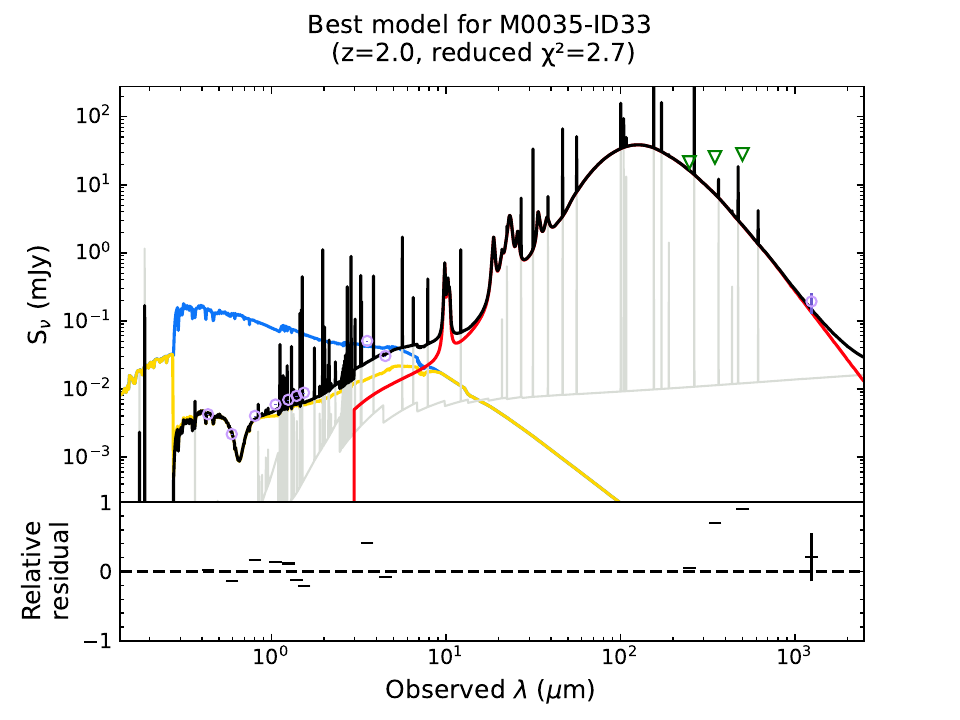}
\figsetgrpnote{The SED and the best-fit model of M0035-ID33. Lensing magnification is NOT corrected in this figure. The black solid line represents the composite spectrum of the galaxy. The yellow line illustrates the stellar emission attenuated by interstellar dust. The blue line depicts the unattenuated stellar emission for reference purpose. The orange line corresponds to the emission from an AGN. The red line shows the infrared emission from interstellar dust. The gray line denotes the nebulae emission. The observed data points are represented by purple circles, accompanied by 1$\sigma$ error bars. The bottom panel displays the relative residuals.}
\figsetgrpend

\figsetgrpstart
\figsetgrpnum{16.53}
\figsetgrptitle{M0035-ID41}
\figsetplot{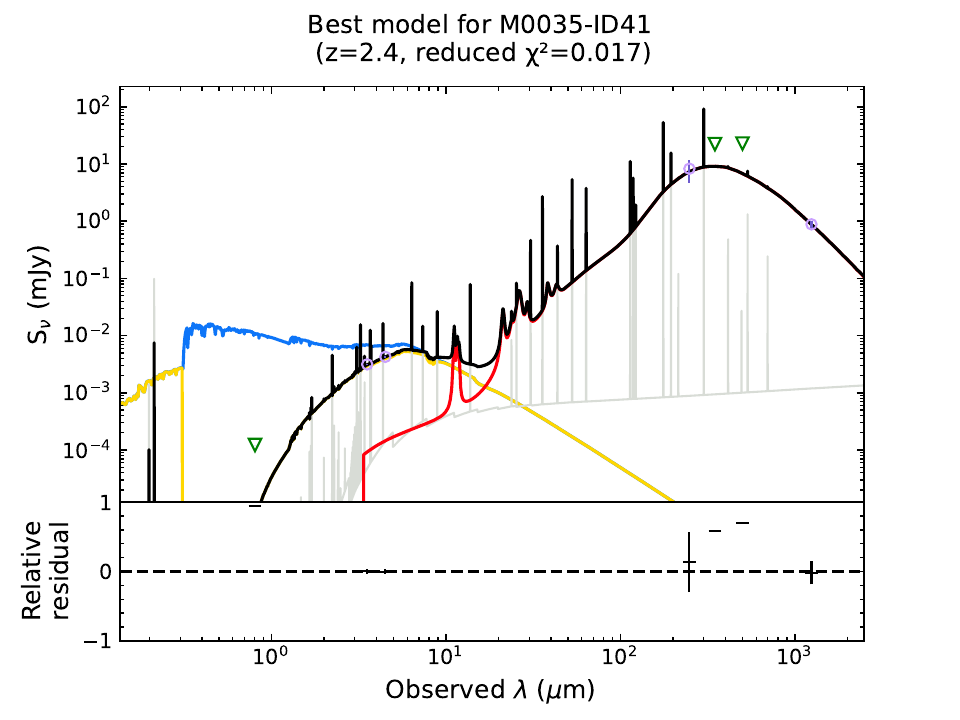}
\figsetgrpnote{The SED and the best-fit model of M0035-ID41. Lensing magnification is NOT corrected in this figure. The black solid line represents the composite spectrum of the galaxy. The yellow line illustrates the stellar emission attenuated by interstellar dust. The blue line depicts the unattenuated stellar emission for reference purpose. The orange line corresponds to the emission from an AGN. The red line shows the infrared emission from interstellar dust. The gray line denotes the nebulae emission. The observed data points are represented by purple circles, accompanied by 1$\sigma$ error bars. The bottom panel displays the relative residuals.}
\figsetgrpend

\figsetgrpstart
\figsetgrpnum{16.54}
\figsetgrptitle{M0035-ID68}
\figsetplot{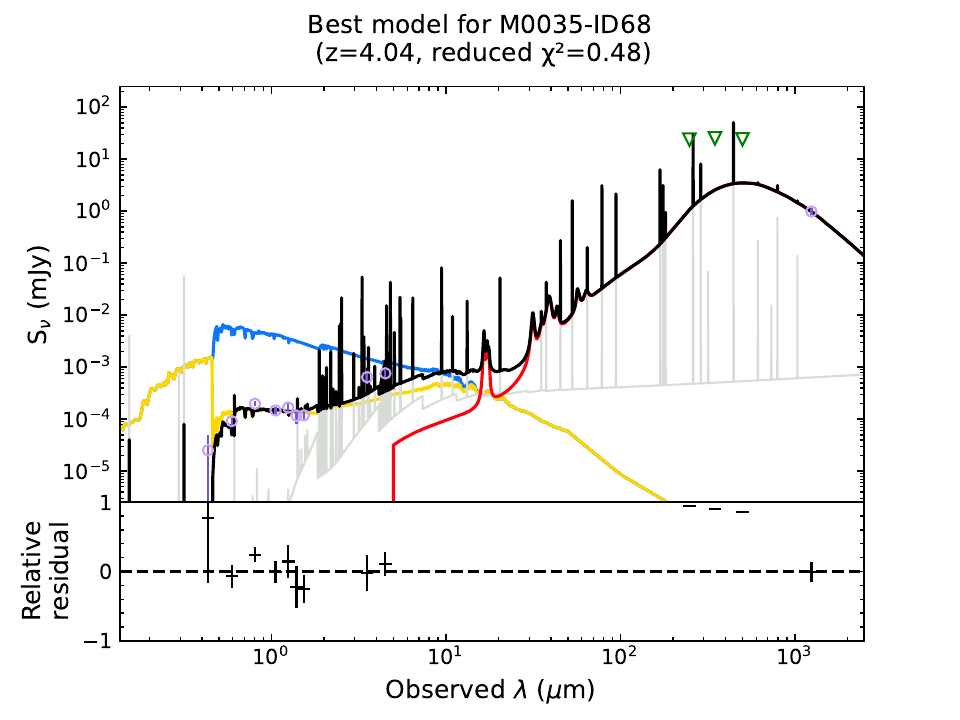}
\figsetgrpnote{The SED and the best-fit model of M0035-ID68. Lensing magnification is NOT corrected in this figure. The black solid line represents the composite spectrum of the galaxy. The yellow line illustrates the stellar emission attenuated by interstellar dust. The blue line depicts the unattenuated stellar emission for reference purpose. The orange line corresponds to the emission from an AGN. The red line shows the infrared emission from interstellar dust. The gray line denotes the nebulae emission. The observed data points are represented by purple circles, accompanied by 1$\sigma$ error bars. The bottom panel displays the relative residuals.}
\figsetgrpend

\figsetgrpstart
\figsetgrpnum{16.55}
\figsetgrptitle{M0035-ID94}
\figsetplot{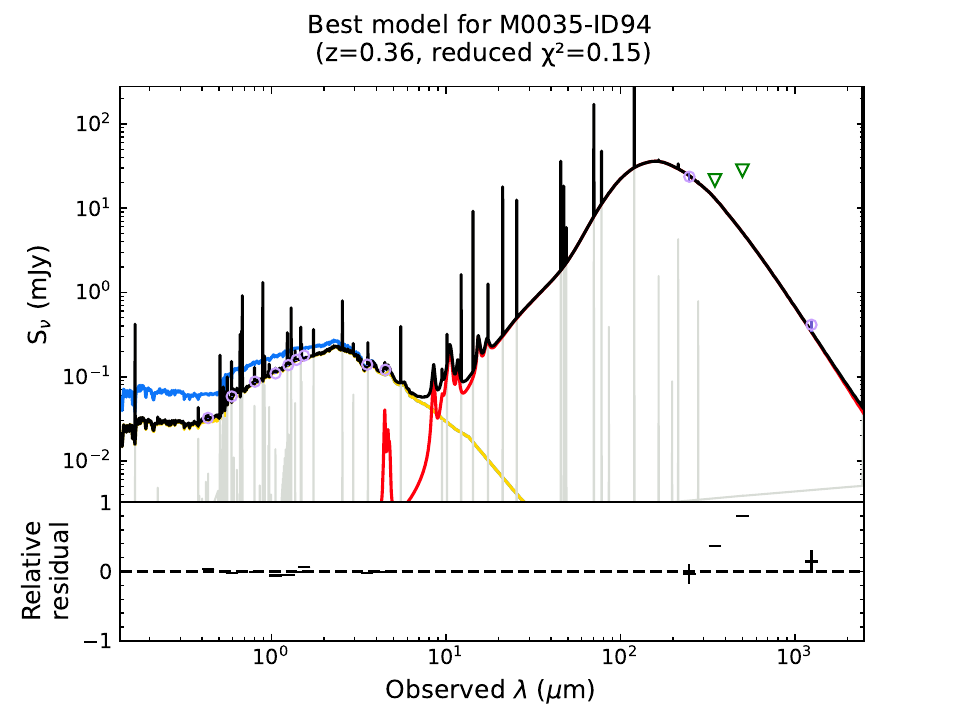}
\figsetgrpnote{The SED and the best-fit model of M0035-ID94. Lensing magnification is NOT corrected in this figure. The black solid line represents the composite spectrum of the galaxy. The yellow line illustrates the stellar emission attenuated by interstellar dust. The blue line depicts the unattenuated stellar emission for reference purpose. The orange line corresponds to the emission from an AGN. The red line shows the infrared emission from interstellar dust. The gray line denotes the nebulae emission. The observed data points are represented by purple circles, accompanied by 1$\sigma$ error bars. The bottom panel displays the relative residuals.}
\figsetgrpend

\figsetgrpstart
\figsetgrpnum{16.56}
\figsetgrptitle{M0159-ID24}
\figsetplot{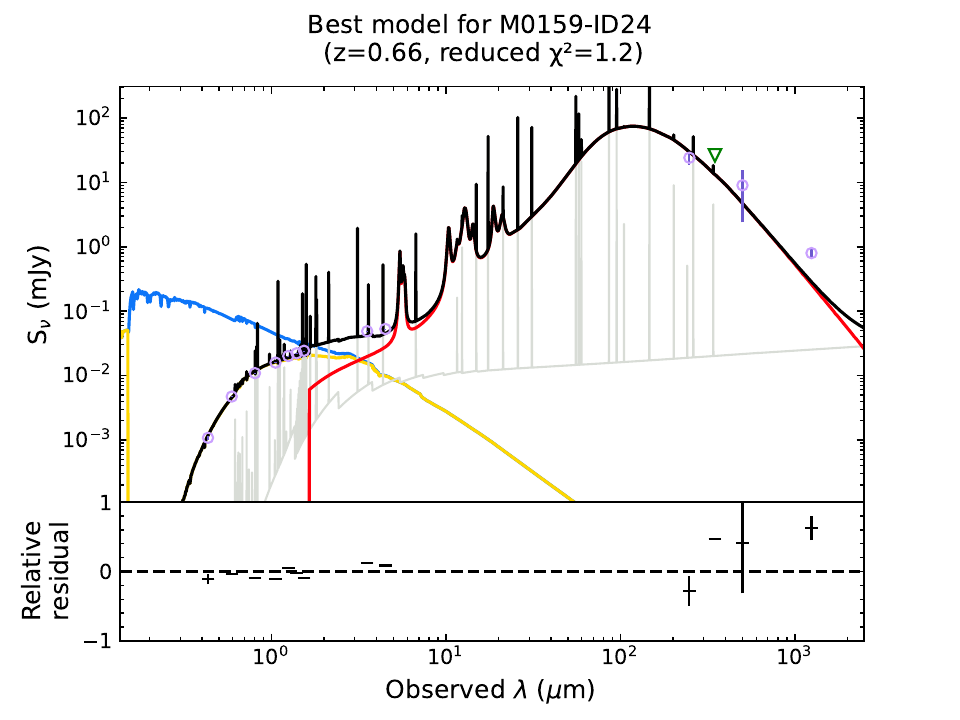}
\figsetgrpnote{The SED and the best-fit model of M0159-ID24. Lensing magnification is NOT corrected in this figure. The black solid line represents the composite spectrum of the galaxy. The yellow line illustrates the stellar emission attenuated by interstellar dust. The blue line depicts the unattenuated stellar emission for reference purpose. The orange line corresponds to the emission from an AGN. The red line shows the infrared emission from interstellar dust. The gray line denotes the nebulae emission. The observed data points are represented by purple circles, accompanied by 1$\sigma$ error bars. The bottom panel displays the relative residuals.}
\figsetgrpend

\figsetgrpstart
\figsetgrpnum{16.57}
\figsetgrptitle{M0159-ID5}
\figsetplot{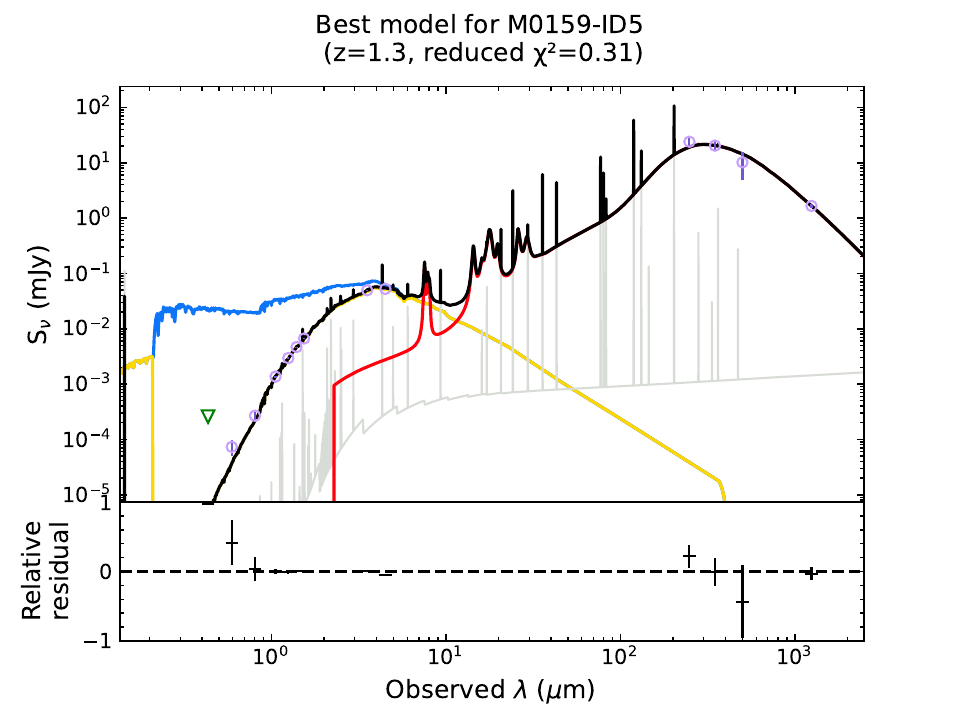}
\figsetgrpnote{The SED and the best-fit model of M0159-ID5. Lensing magnification is NOT corrected in this figure. The black solid line represents the composite spectrum of the galaxy. The yellow line illustrates the stellar emission attenuated by interstellar dust. The blue line depicts the unattenuated stellar emission for reference purpose. The orange line corresponds to the emission from an AGN. The red line shows the infrared emission from interstellar dust. The gray line denotes the nebulae emission. The observed data points are represented by purple circles, accompanied by 1$\sigma$ error bars. The bottom panel displays the relative residuals.}
\figsetgrpend

\figsetgrpstart
\figsetgrpnum{16.58}
\figsetgrptitle{M0159-ID61}
\figsetplot{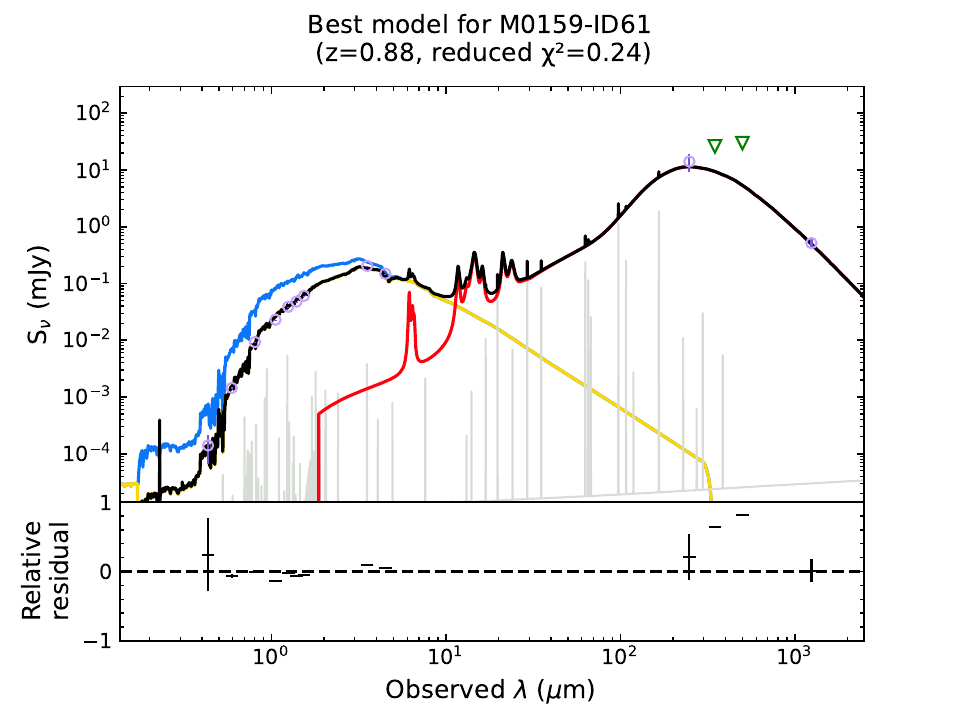}
\figsetgrpnote{The SED and the best-fit model of M0159-ID61. Lensing magnification is NOT corrected in this figure. The black solid line represents the composite spectrum of the galaxy. The yellow line illustrates the stellar emission attenuated by interstellar dust. The blue line depicts the unattenuated stellar emission for reference purpose. The orange line corresponds to the emission from an AGN. The red line shows the infrared emission from interstellar dust. The gray line denotes the nebulae emission. The observed data points are represented by purple circles, accompanied by 1$\sigma$ error bars. The bottom panel displays the relative residuals.}
\figsetgrpend

\figsetgrpstart
\figsetgrpnum{16.59}
\figsetgrptitle{M0257-ID13}
\figsetplot{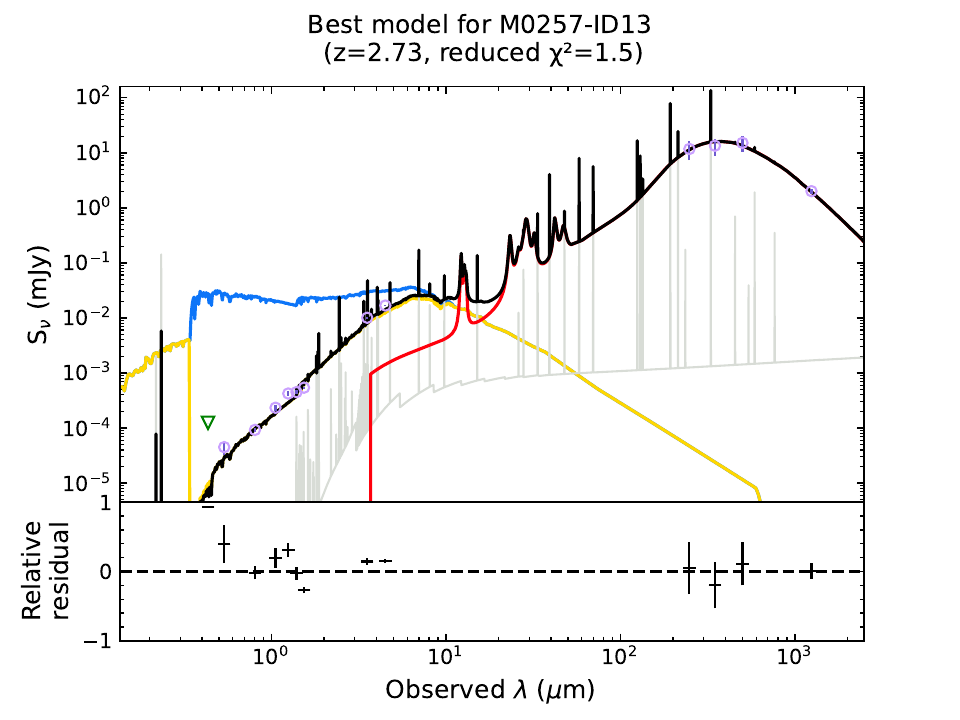}
\figsetgrpnote{The SED and the best-fit model of M0257-ID13. Lensing magnification is NOT corrected in this figure. The black solid line represents the composite spectrum of the galaxy. The yellow line illustrates the stellar emission attenuated by interstellar dust. The blue line depicts the unattenuated stellar emission for reference purpose. The orange line corresponds to the emission from an AGN. The red line shows the infrared emission from interstellar dust. The gray line denotes the nebulae emission. The observed data points are represented by purple circles, accompanied by 1$\sigma$ error bars. The bottom panel displays the relative residuals.}
\figsetgrpend

\figsetgrpstart
\figsetgrpnum{16.60}
\figsetgrptitle{M0416-ID120}
\figsetplot{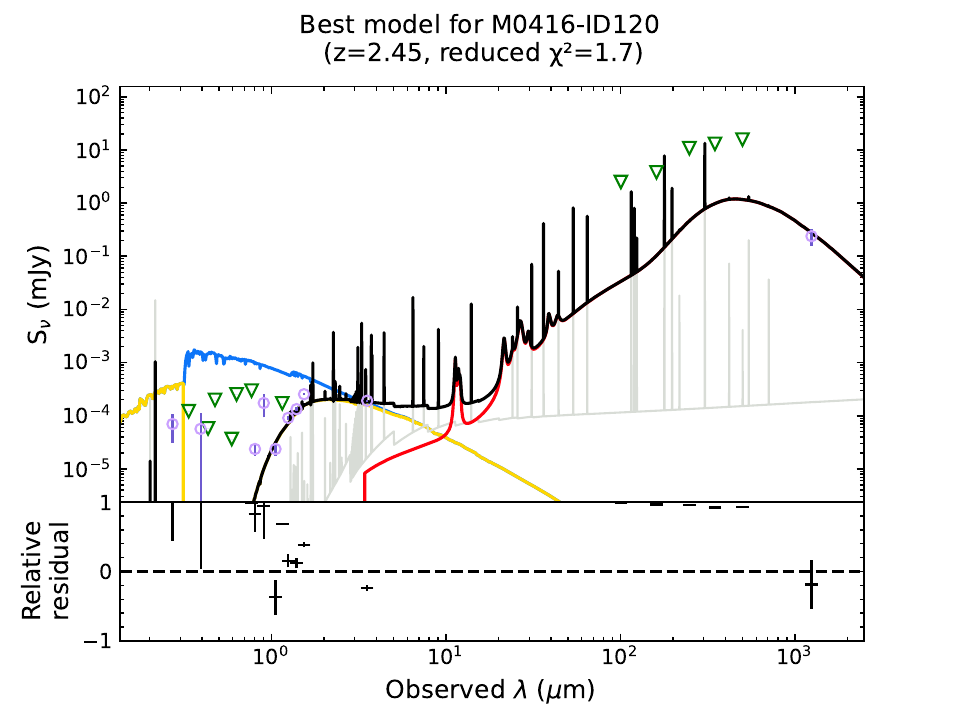}
\figsetgrpnote{The SED and the best-fit model of M0416-ID120. Lensing magnification is NOT corrected in this figure. The black solid line represents the composite spectrum of the galaxy. The yellow line illustrates the stellar emission attenuated by interstellar dust. The blue line depicts the unattenuated stellar emission for reference purpose. The orange line corresponds to the emission from an AGN. The red line shows the infrared emission from interstellar dust. The gray line denotes the nebulae emission. The observed data points are represented by purple circles, accompanied by 1$\sigma$ error bars. The bottom panel displays the relative residuals.}
\figsetgrpend

\figsetgrpstart
\figsetgrpnum{16.61}
\figsetgrptitle{M0416-ID138}
\figsetplot{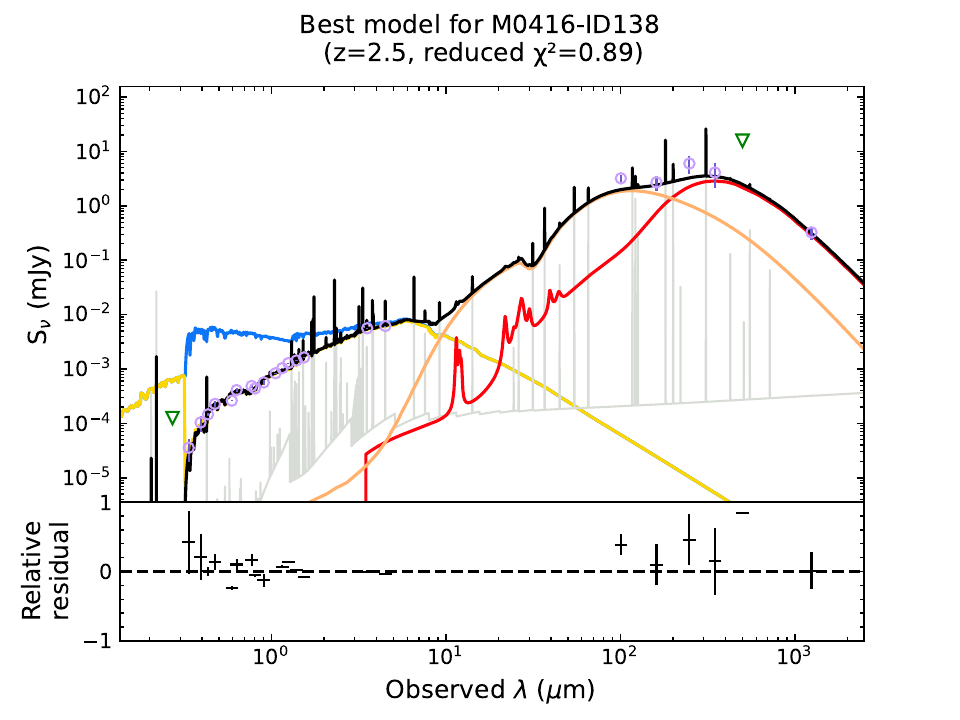}
\figsetgrpnote{The SED and the best-fit model of M0416-ID138. Lensing magnification is NOT corrected in this figure. The black solid line represents the composite spectrum of the galaxy. The yellow line illustrates the stellar emission attenuated by interstellar dust. The blue line depicts the unattenuated stellar emission for reference purpose. The orange line corresponds to the emission from an AGN. The red line shows the infrared emission from interstellar dust. The gray line denotes the nebulae emission. The observed data points are represented by purple circles, accompanied by 1$\sigma$ error bars. The bottom panel displays the relative residuals.}
\figsetgrpend

\figsetgrpstart
\figsetgrpnum{16.62}
\figsetgrptitle{M0416-ID156}
\figsetplot{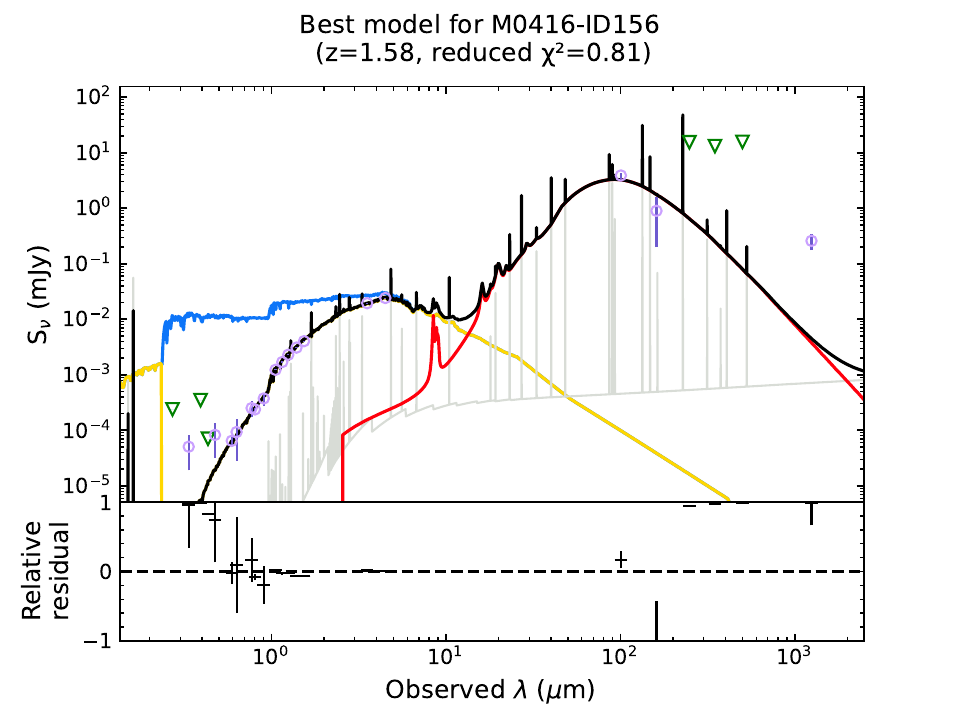}
\figsetgrpnote{The SED and the best-fit model of M0416-ID156. Lensing magnification is NOT corrected in this figure. The black solid line represents the composite spectrum of the galaxy. The yellow line illustrates the stellar emission attenuated by interstellar dust. The blue line depicts the unattenuated stellar emission for reference purpose. The orange line corresponds to the emission from an AGN. The red line shows the infrared emission from interstellar dust. The gray line denotes the nebulae emission. The observed data points are represented by purple circles, accompanied by 1$\sigma$ error bars. The bottom panel displays the relative residuals.}
\figsetgrpend

\figsetgrpstart
\figsetgrpnum{16.63}
\figsetgrptitle{M0416-ID160}
\figsetplot{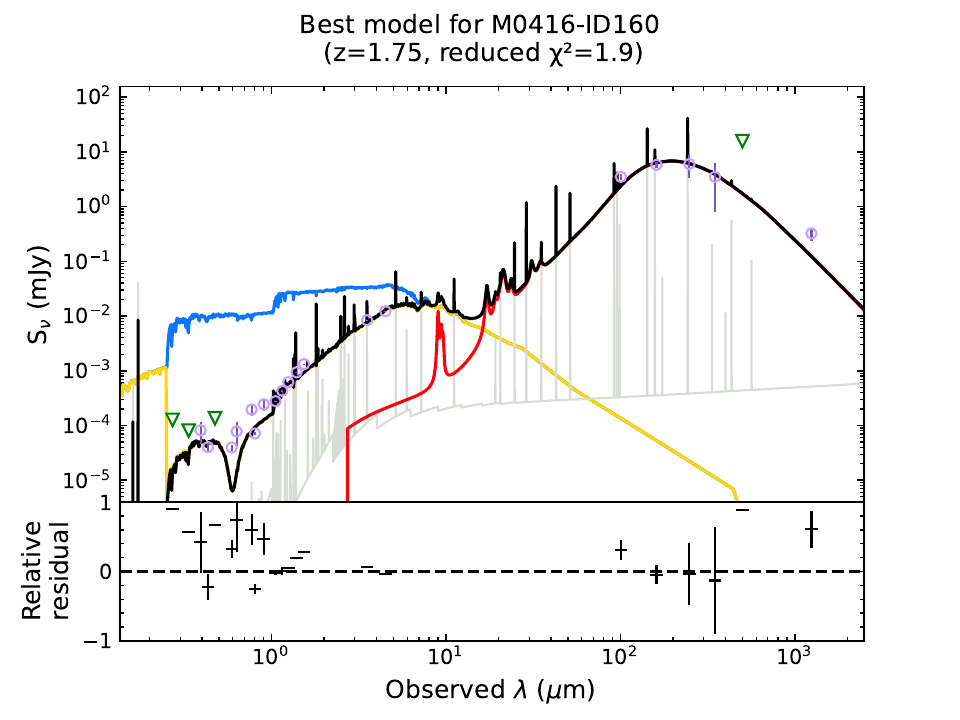}
\figsetgrpnote{The SED and the best-fit model of M0416-ID160. Lensing magnification is NOT corrected in this figure. The black solid line represents the composite spectrum of the galaxy. The yellow line illustrates the stellar emission attenuated by interstellar dust. The blue line depicts the unattenuated stellar emission for reference purpose. The orange line corresponds to the emission from an AGN. The red line shows the infrared emission from interstellar dust. The gray line denotes the nebulae emission. The observed data points are represented by purple circles, accompanied by 1$\sigma$ error bars. The bottom panel displays the relative residuals.}
\figsetgrpend

\figsetgrpstart
\figsetgrpnum{16.64}
\figsetgrptitle{M0416-ID51}
\figsetplot{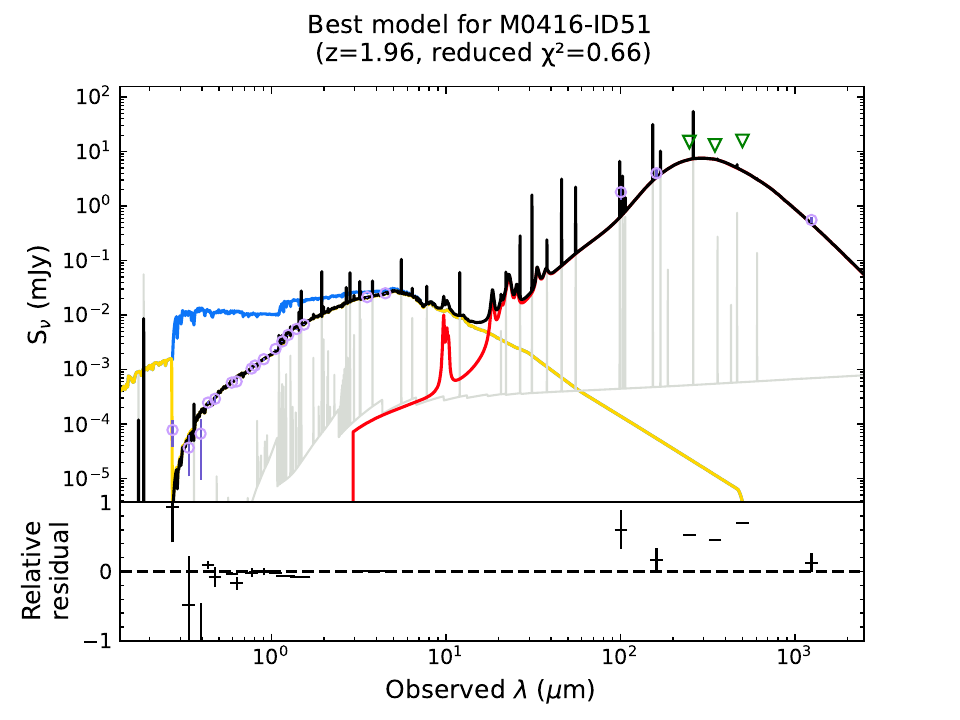}
\figsetgrpnote{The SED and the best-fit model of M0416-ID51. Lensing magnification is NOT corrected in this figure. The black solid line represents the composite spectrum of the galaxy. The yellow line illustrates the stellar emission attenuated by interstellar dust. The blue line depicts the unattenuated stellar emission for reference purpose. The orange line corresponds to the emission from an AGN. The red line shows the infrared emission from interstellar dust. The gray line denotes the nebulae emission. The observed data points are represented by purple circles, accompanied by 1$\sigma$ error bars. The bottom panel displays the relative residuals.}
\figsetgrpend

\figsetgrpstart
\figsetgrpnum{16.65}
\figsetgrptitle{M0416-ID79}
\figsetplot{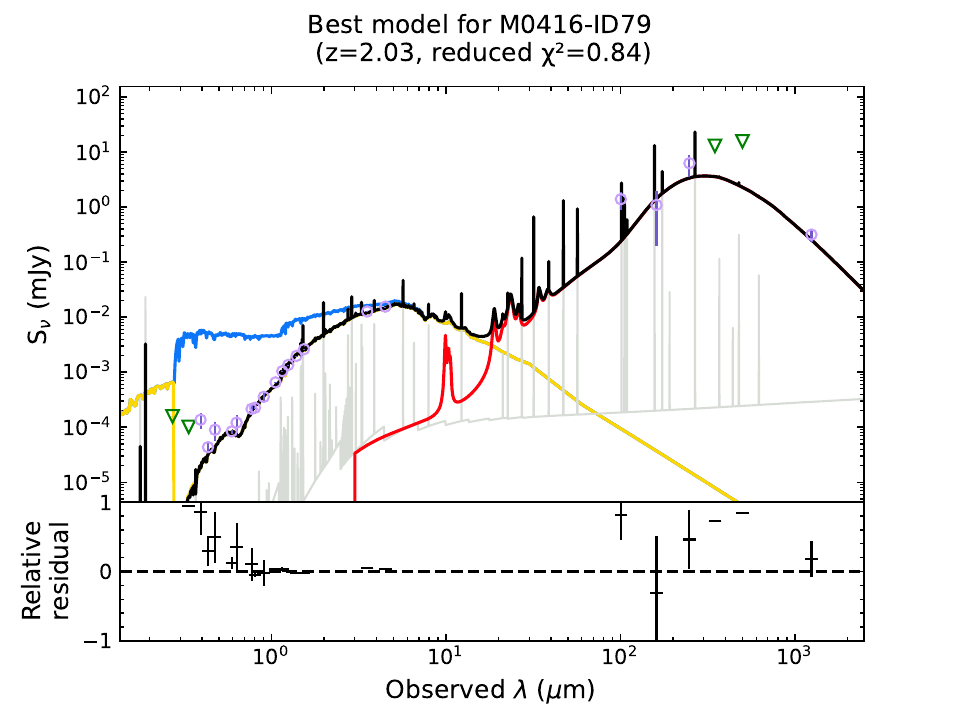}
\figsetgrpnote{The SED and the best-fit model of M0416-ID79. Lensing magnification is NOT corrected in this figure. The black solid line represents the composite spectrum of the galaxy. The yellow line illustrates the stellar emission attenuated by interstellar dust. The blue line depicts the unattenuated stellar emission for reference purpose. The orange line corresponds to the emission from an AGN. The red line shows the infrared emission from interstellar dust. The gray line denotes the nebulae emission. The observed data points are represented by purple circles, accompanied by 1$\sigma$ error bars. The bottom panel displays the relative residuals.}
\figsetgrpend

\figsetgrpstart
\figsetgrpnum{16.66}
\figsetgrptitle{M0417-ID121}
\figsetplot{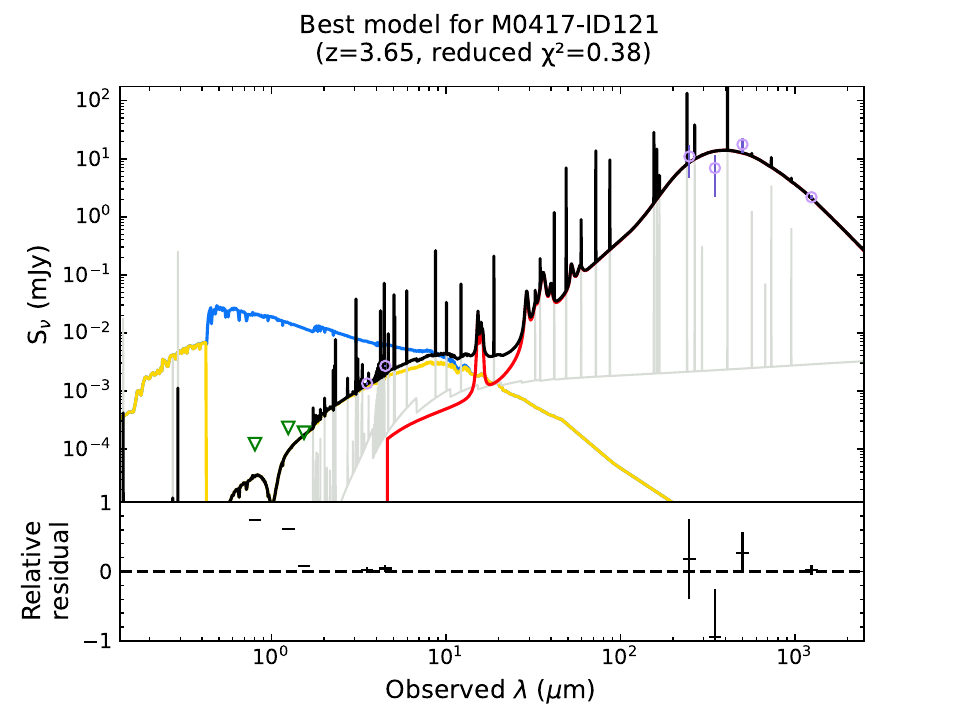}
\figsetgrpnote{The SED and the best-fit model of M0417-ID121. Lensing magnification is NOT corrected in this figure. The black solid line represents the composite spectrum of the galaxy. The yellow line illustrates the stellar emission attenuated by interstellar dust. The blue line depicts the unattenuated stellar emission for reference purpose. The orange line corresponds to the emission from an AGN. The red line shows the infrared emission from interstellar dust. The gray line denotes the nebulae emission. The observed data points are represented by purple circles, accompanied by 1$\sigma$ error bars. The bottom panel displays the relative residuals.}
\figsetgrpend

\figsetgrpstart
\figsetgrpnum{16.67}
\figsetgrptitle{M0417-ID204}
\figsetplot{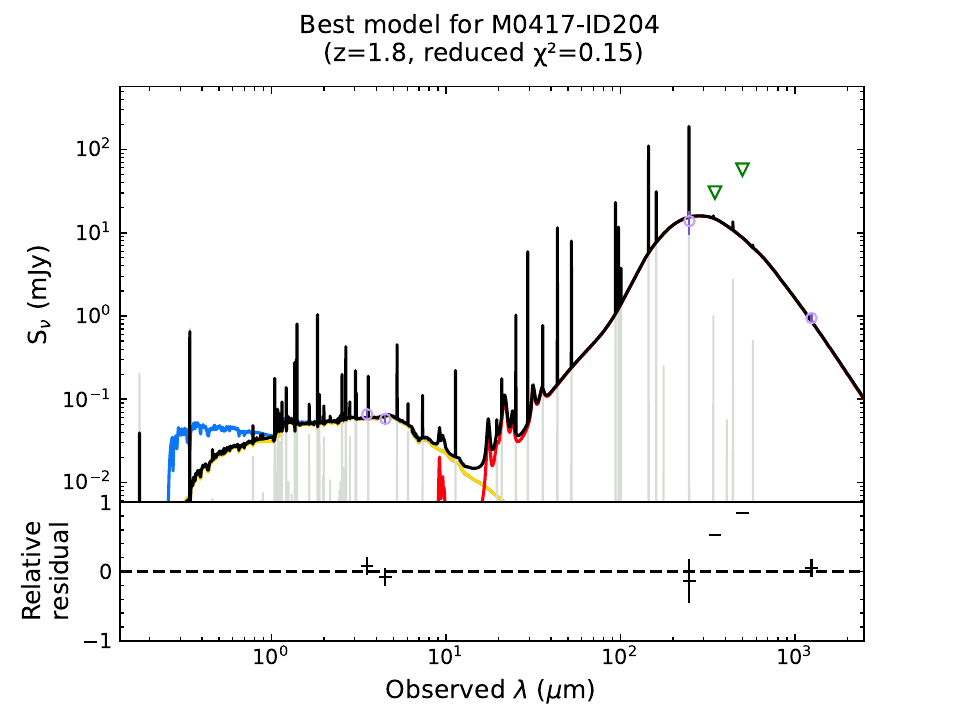}
\figsetgrpnote{The SED and the best-fit model of M0417-ID204. Lensing magnification is NOT corrected in this figure. The black solid line represents the composite spectrum of the galaxy. The yellow line illustrates the stellar emission attenuated by interstellar dust. The blue line depicts the unattenuated stellar emission for reference purpose. The orange line corresponds to the emission from an AGN. The red line shows the infrared emission from interstellar dust. The gray line denotes the nebulae emission. The observed data points are represented by purple circles, accompanied by 1$\sigma$ error bars. The bottom panel displays the relative residuals.}
\figsetgrpend

\figsetgrpstart
\figsetgrpnum{16.68}
\figsetgrptitle{M0417-ID218}
\figsetplot{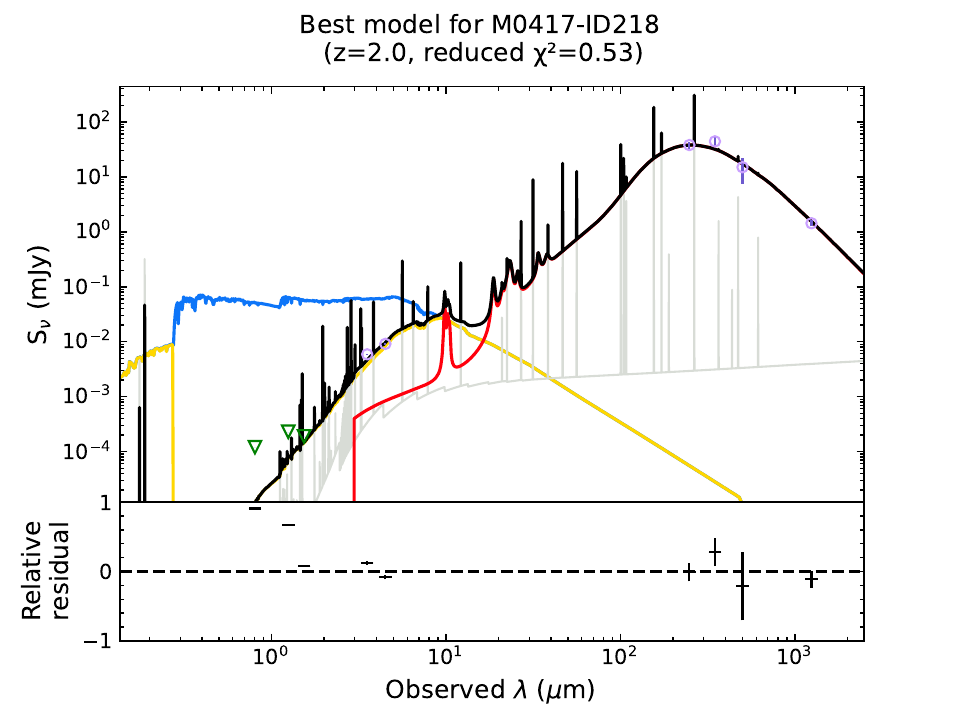}
\figsetgrpnote{The SED and the best-fit model of M0417-ID218. Lensing magnification is NOT corrected in this figure. The black solid line represents the composite spectrum of the galaxy. The yellow line illustrates the stellar emission attenuated by interstellar dust. The blue line depicts the unattenuated stellar emission for reference purpose. The orange line corresponds to the emission from an AGN. The red line shows the infrared emission from interstellar dust. The gray line denotes the nebulae emission. The observed data points are represented by purple circles, accompanied by 1$\sigma$ error bars. The bottom panel displays the relative residuals.}
\figsetgrpend

\figsetgrpstart
\figsetgrpnum{16.69}
\figsetgrptitle{M0417-ID221}
\figsetplot{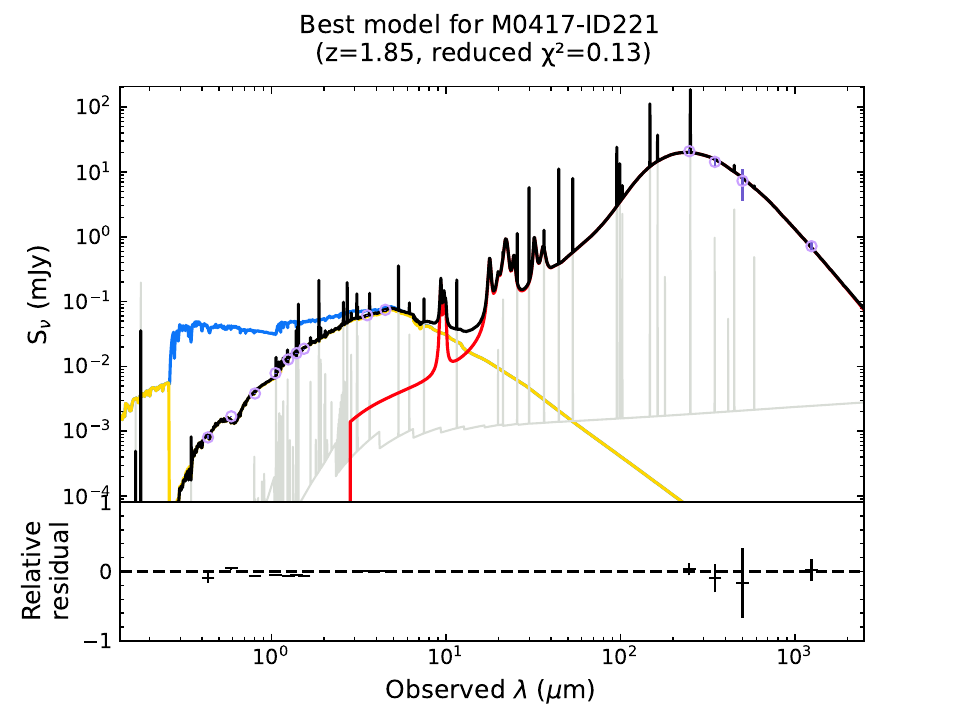}
\figsetgrpnote{The SED and the best-fit model of M0417-ID221. Lensing magnification is NOT corrected in this figure. The black solid line represents the composite spectrum of the galaxy. The yellow line illustrates the stellar emission attenuated by interstellar dust. The blue line depicts the unattenuated stellar emission for reference purpose. The orange line corresponds to the emission from an AGN. The red line shows the infrared emission from interstellar dust. The gray line denotes the nebulae emission. The observed data points are represented by purple circles, accompanied by 1$\sigma$ error bars. The bottom panel displays the relative residuals.}
\figsetgrpend

\figsetgrpstart
\figsetgrpnum{16.70}
\figsetgrptitle{M0417-ID223}
\figsetplot{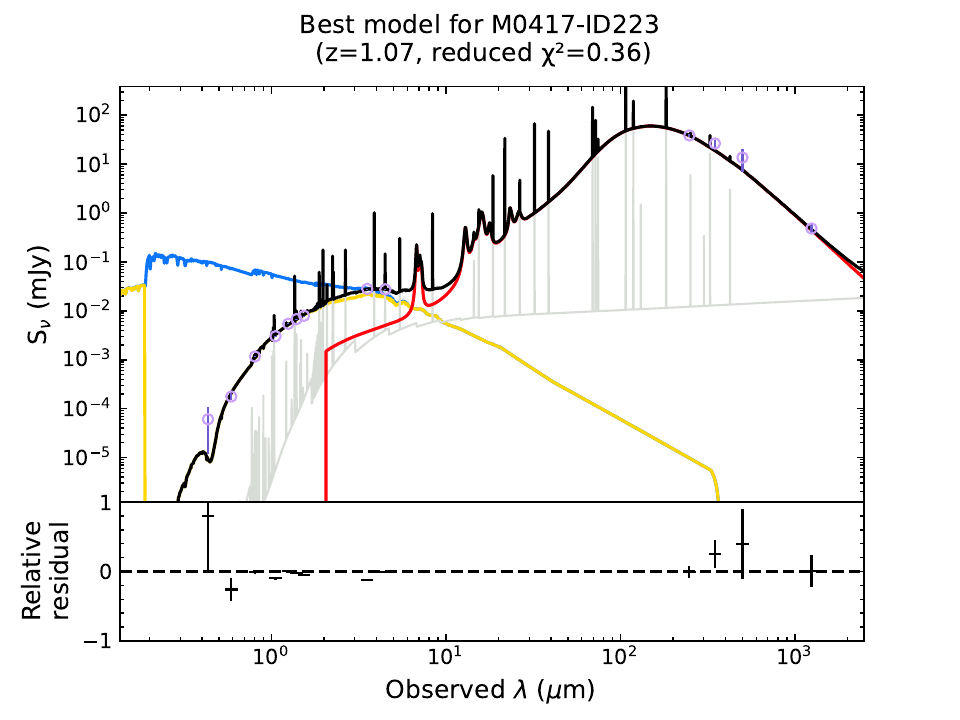}
\figsetgrpnote{The SED and the best-fit model of M0417-ID223. Lensing magnification is NOT corrected in this figure. The black solid line represents the composite spectrum of the galaxy. The yellow line illustrates the stellar emission attenuated by interstellar dust. The blue line depicts the unattenuated stellar emission for reference purpose. The orange line corresponds to the emission from an AGN. The red line shows the infrared emission from interstellar dust. The gray line denotes the nebulae emission. The observed data points are represented by purple circles, accompanied by 1$\sigma$ error bars. The bottom panel displays the relative residuals.}
\figsetgrpend

\figsetgrpstart
\figsetgrpnum{16.71}
\figsetgrptitle{M0417-ID46}
\figsetplot{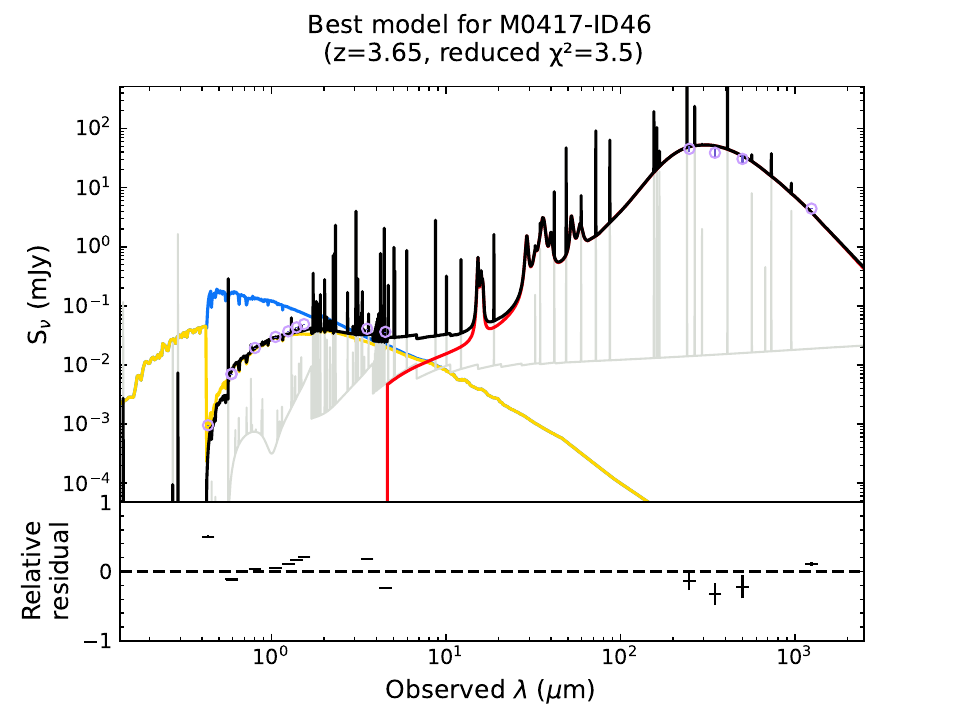}
\figsetgrpnote{The SED and the best-fit model of M0417-ID46. Lensing magnification is NOT corrected in this figure. The black solid line represents the composite spectrum of the galaxy. The yellow line illustrates the stellar emission attenuated by interstellar dust. The blue line depicts the unattenuated stellar emission for reference purpose. The orange line corresponds to the emission from an AGN. The red line shows the infrared emission from interstellar dust. The gray line denotes the nebulae emission. The observed data points are represented by purple circles, accompanied by 1$\sigma$ error bars. The bottom panel displays the relative residuals.}
\figsetgrpend

\figsetgrpstart
\figsetgrpnum{16.72}
\figsetgrptitle{M0417-ID49}
\figsetplot{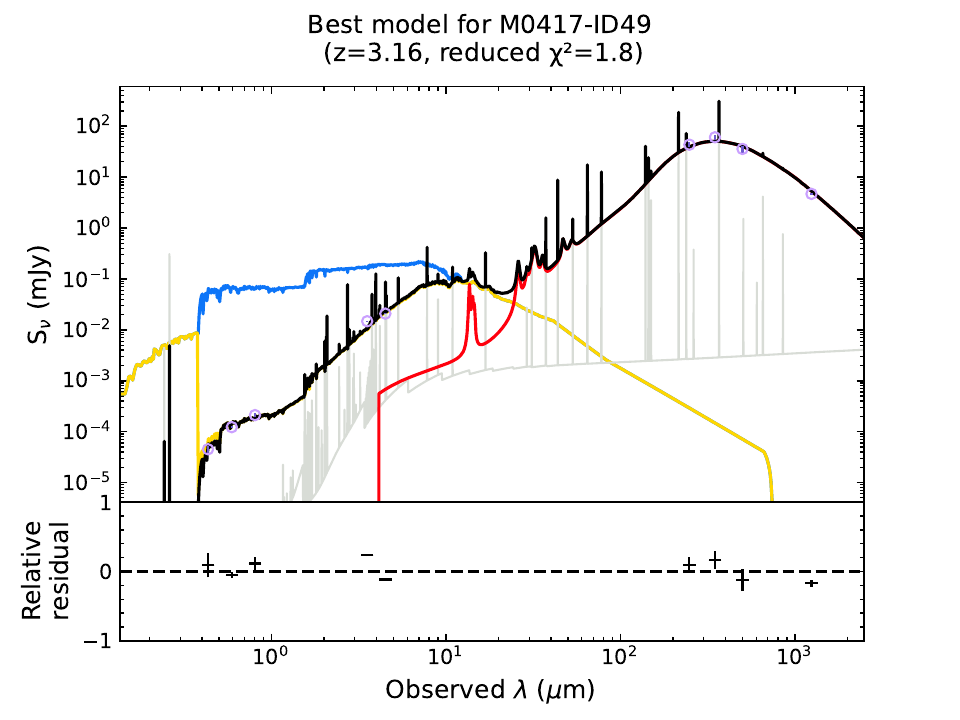}
\figsetgrpnote{The SED and the best-fit model of M0417-ID49. Lensing magnification is NOT corrected in this figure. The black solid line represents the composite spectrum of the galaxy. The yellow line illustrates the stellar emission attenuated by interstellar dust. The blue line depicts the unattenuated stellar emission for reference purpose. The orange line corresponds to the emission from an AGN. The red line shows the infrared emission from interstellar dust. The gray line denotes the nebulae emission. The observed data points are represented by purple circles, accompanied by 1$\sigma$ error bars. The bottom panel displays the relative residuals.}
\figsetgrpend

\figsetgrpstart
\figsetgrpnum{16.73}
\figsetgrptitle{M0417-ID58}
\figsetplot{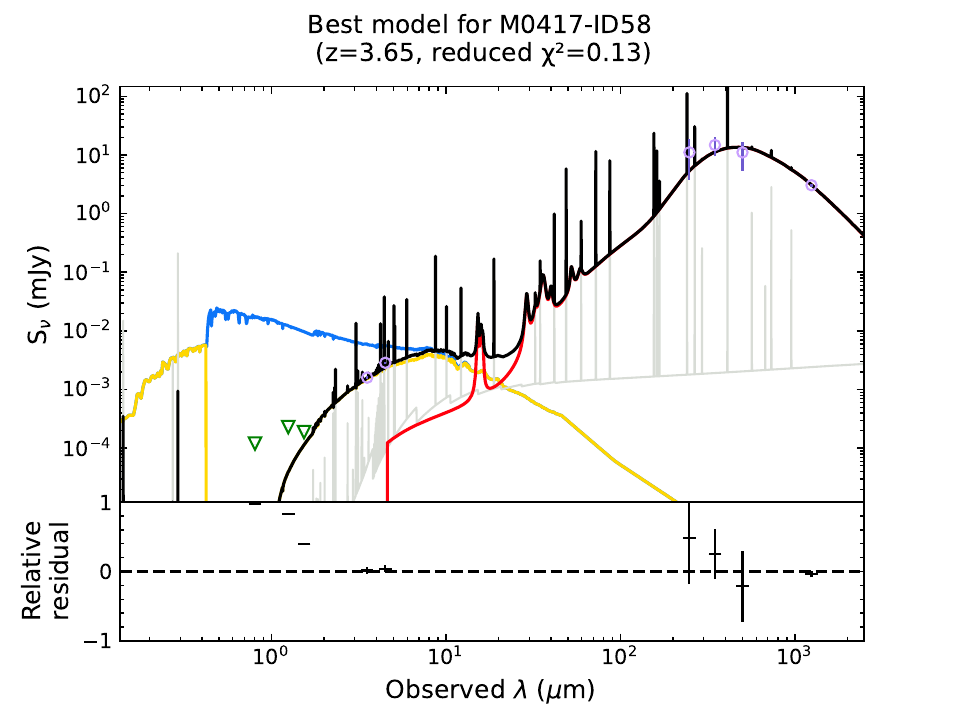}
\figsetgrpnote{The SED and the best-fit model of M0417-ID58. Lensing magnification is NOT corrected in this figure. The black solid line represents the composite spectrum of the galaxy. The yellow line illustrates the stellar emission attenuated by interstellar dust. The blue line depicts the unattenuated stellar emission for reference purpose. The orange line corresponds to the emission from an AGN. The red line shows the infrared emission from interstellar dust. The gray line denotes the nebulae emission. The observed data points are represented by purple circles, accompanied by 1$\sigma$ error bars. The bottom panel displays the relative residuals.}
\figsetgrpend

\figsetgrpstart
\figsetgrpnum{16.74}
\figsetgrptitle{M0429-ID27}
\figsetplot{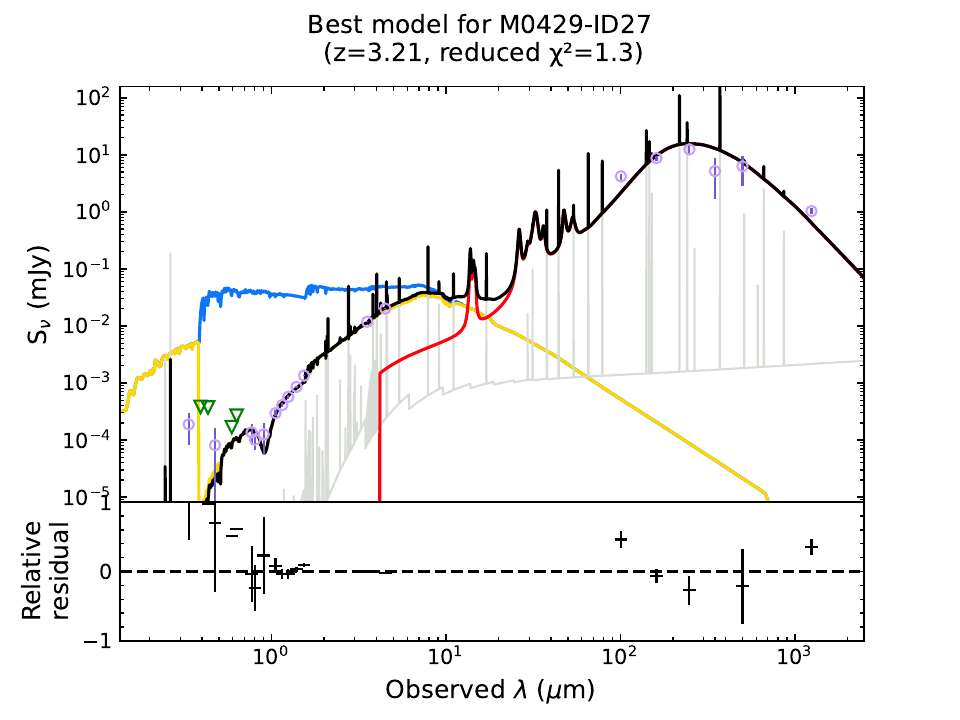}
\figsetgrpnote{The SED and the best-fit model of M0429-ID27. Lensing magnification is NOT corrected in this figure. The black solid line represents the composite spectrum of the galaxy. The yellow line illustrates the stellar emission attenuated by interstellar dust. The blue line depicts the unattenuated stellar emission for reference purpose. The orange line corresponds to the emission from an AGN. The red line shows the infrared emission from interstellar dust. The gray line denotes the nebulae emission. The observed data points are represented by purple circles, accompanied by 1$\sigma$ error bars. The bottom panel displays the relative residuals.}
\figsetgrpend

\figsetgrpstart
\figsetgrpnum{16.75}
\figsetgrptitle{M0429-ID4}
\figsetplot{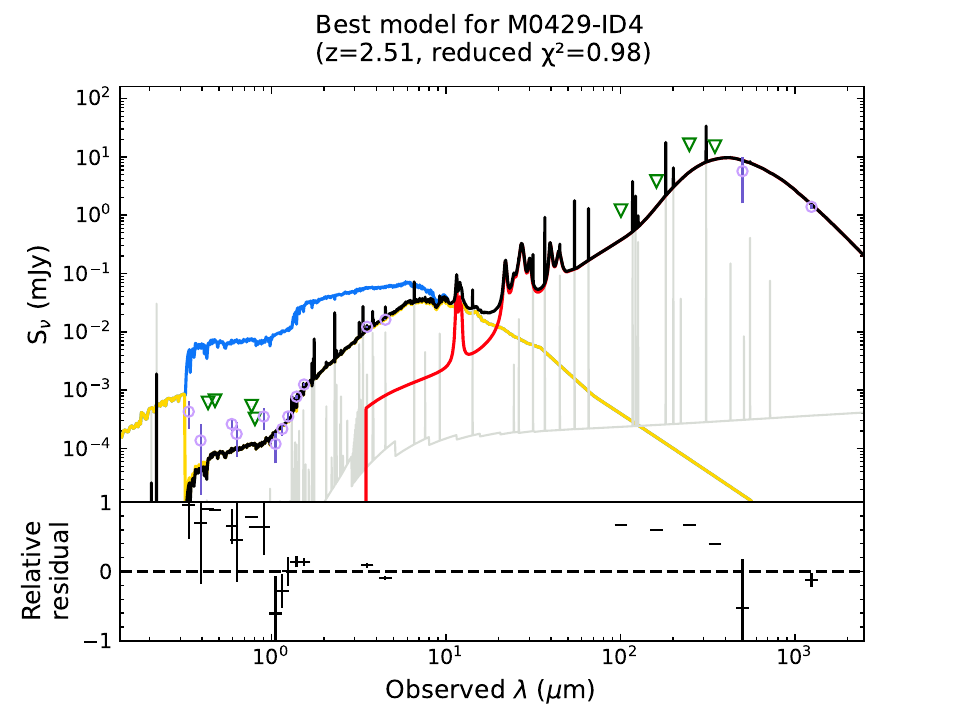}
\figsetgrpnote{The SED and the best-fit model of M0429-ID4. Lensing magnification is NOT corrected in this figure. The black solid line represents the composite spectrum of the galaxy. The yellow line illustrates the stellar emission attenuated by interstellar dust. The blue line depicts the unattenuated stellar emission for reference purpose. The orange line corresponds to the emission from an AGN. The red line shows the infrared emission from interstellar dust. The gray line denotes the nebulae emission. The observed data points are represented by purple circles, accompanied by 1$\sigma$ error bars. The bottom panel displays the relative residuals.}
\figsetgrpend

\figsetgrpstart
\figsetgrpnum{16.76}
\figsetgrptitle{M0553-ID133}
\figsetplot{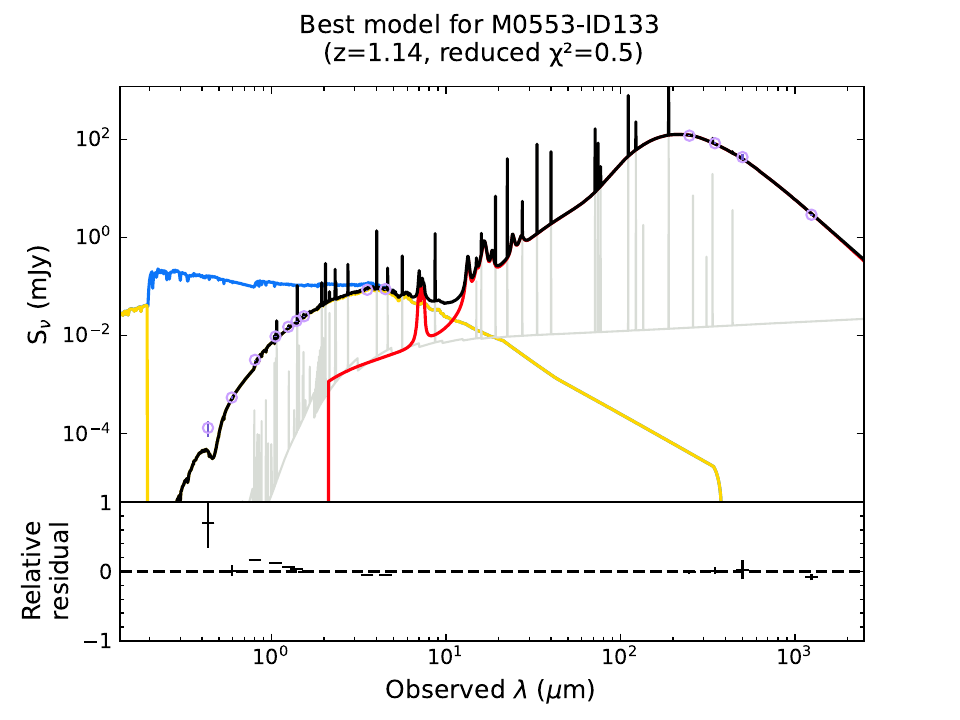}
\figsetgrpnote{The SED and the best-fit model of M0553-ID133. Lensing magnification is NOT corrected in this figure. The black solid line represents the composite spectrum of the galaxy. The yellow line illustrates the stellar emission attenuated by interstellar dust. The blue line depicts the unattenuated stellar emission for reference purpose. The orange line corresponds to the emission from an AGN. The red line shows the infrared emission from interstellar dust. The gray line denotes the nebulae emission. The observed data points are represented by purple circles, accompanied by 1$\sigma$ error bars. The bottom panel displays the relative residuals.}
\figsetgrpend

\figsetgrpstart
\figsetgrpnum{16.77}
\figsetgrptitle{M0553-ID17}
\figsetplot{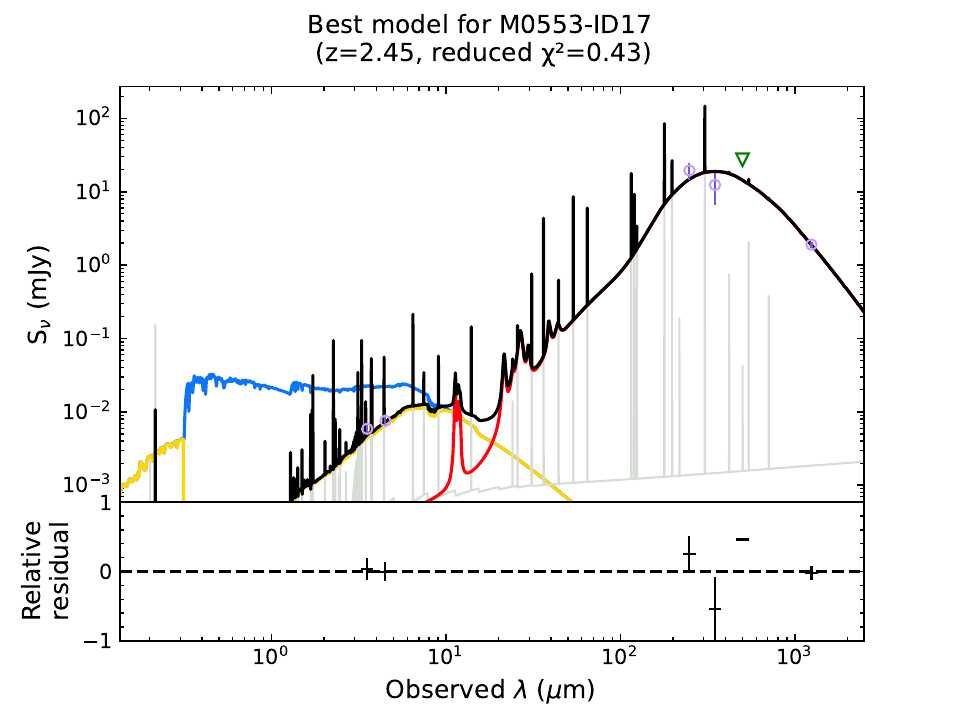}
\figsetgrpnote{The SED and the best-fit model of M0553-ID17. Lensing magnification is NOT corrected in this figure. The black solid line represents the composite spectrum of the galaxy. The yellow line illustrates the stellar emission attenuated by interstellar dust. The blue line depicts the unattenuated stellar emission for reference purpose. The orange line corresponds to the emission from an AGN. The red line shows the infrared emission from interstellar dust. The gray line denotes the nebulae emission. The observed data points are represented by purple circles, accompanied by 1$\sigma$ error bars. The bottom panel displays the relative residuals.}
\figsetgrpend

\figsetgrpstart
\figsetgrpnum{16.78}
\figsetgrptitle{M0553-ID18}
\figsetplot{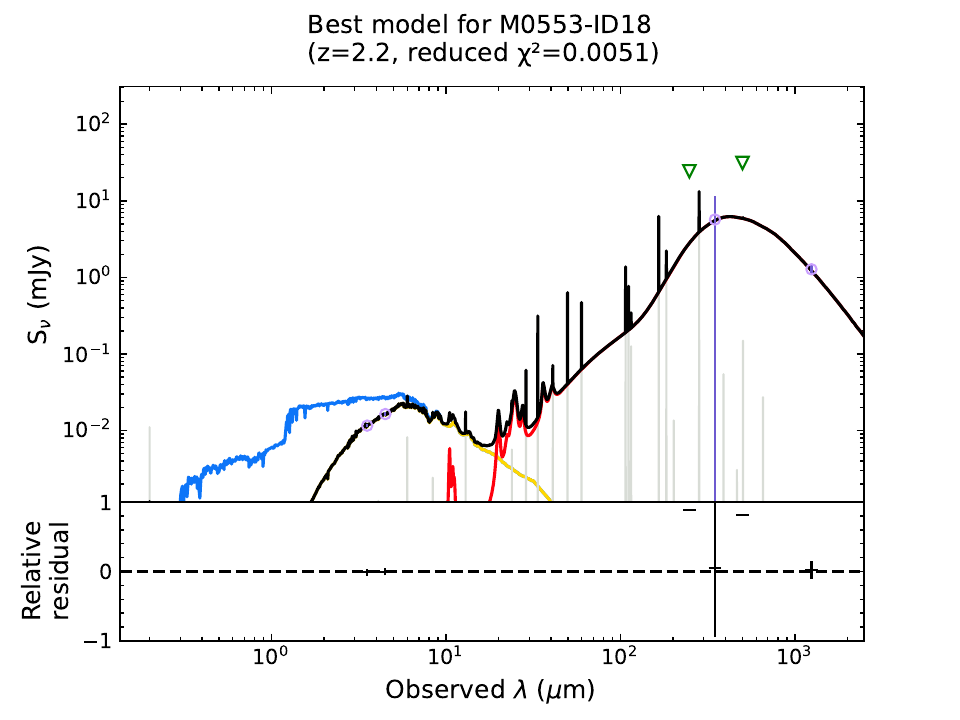}
\figsetgrpnote{The SED and the best-fit model of M0553-ID18. Lensing magnification is NOT corrected in this figure. The black solid line represents the composite spectrum of the galaxy. The yellow line illustrates the stellar emission attenuated by interstellar dust. The blue line depicts the unattenuated stellar emission for reference purpose. The orange line corresponds to the emission from an AGN. The red line shows the infrared emission from interstellar dust. The gray line denotes the nebulae emission. The observed data points are represented by purple circles, accompanied by 1$\sigma$ error bars. The bottom panel displays the relative residuals.}
\figsetgrpend

\figsetgrpstart
\figsetgrpnum{16.79}
\figsetgrptitle{M0553-ID190}
\figsetplot{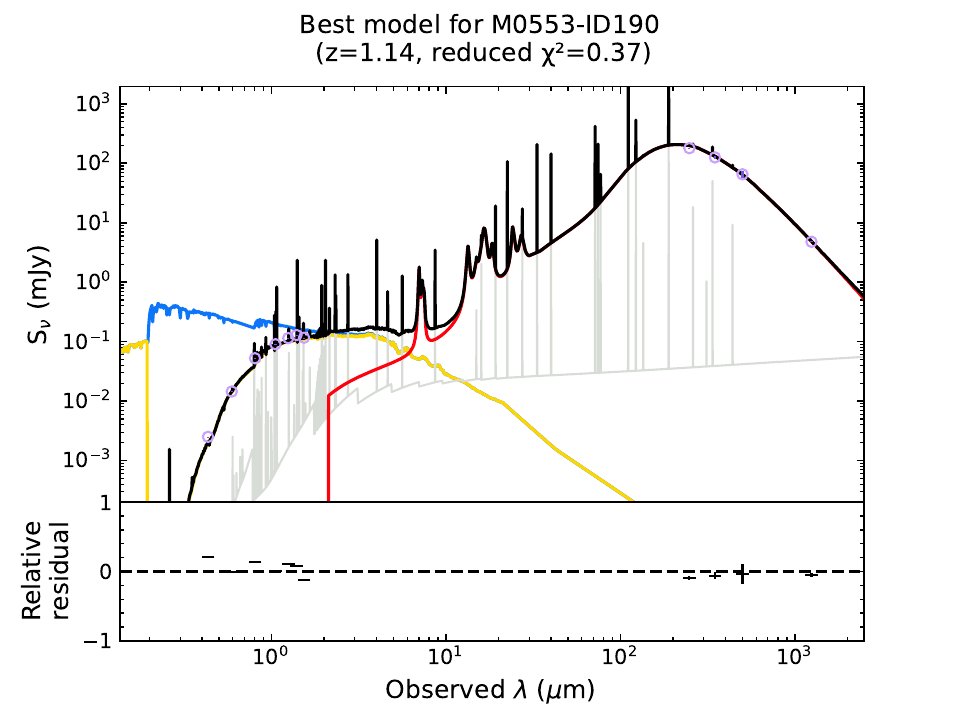}
\figsetgrpnote{The SED and the best-fit model of M0553-ID190. Lensing magnification is NOT corrected in this figure. The black solid line represents the composite spectrum of the galaxy. The yellow line illustrates the stellar emission attenuated by interstellar dust. The blue line depicts the unattenuated stellar emission for reference purpose. The orange line corresponds to the emission from an AGN. The red line shows the infrared emission from interstellar dust. The gray line denotes the nebulae emission. The observed data points are represented by purple circles, accompanied by 1$\sigma$ error bars. The bottom panel displays the relative residuals.}
\figsetgrpend

\figsetgrpstart
\figsetgrpnum{16.80}
\figsetgrptitle{M0553-ID200}
\figsetplot{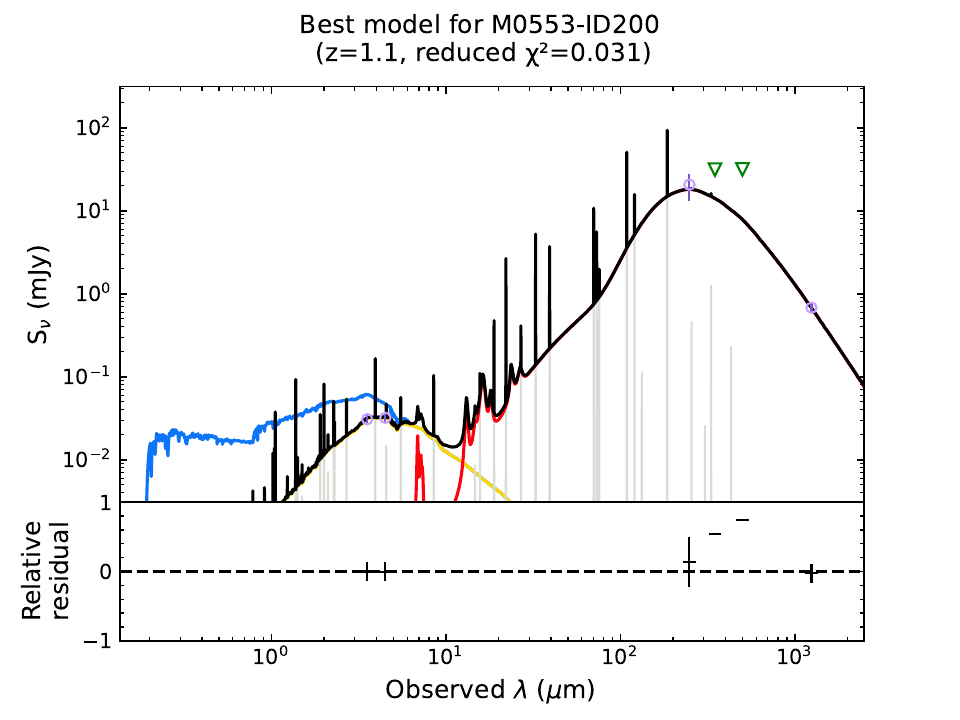}
\figsetgrpnote{The SED and the best-fit model of M0553-ID200. Lensing magnification is NOT corrected in this figure. The black solid line represents the composite spectrum of the galaxy. The yellow line illustrates the stellar emission attenuated by interstellar dust. The blue line depicts the unattenuated stellar emission for reference purpose. The orange line corresponds to the emission from an AGN. The red line shows the infrared emission from interstellar dust. The gray line denotes the nebulae emission. The observed data points are represented by purple circles, accompanied by 1$\sigma$ error bars. The bottom panel displays the relative residuals.}
\figsetgrpend

\figsetgrpstart
\figsetgrpnum{16.81}
\figsetgrptitle{M0553-ID249}
\figsetplot{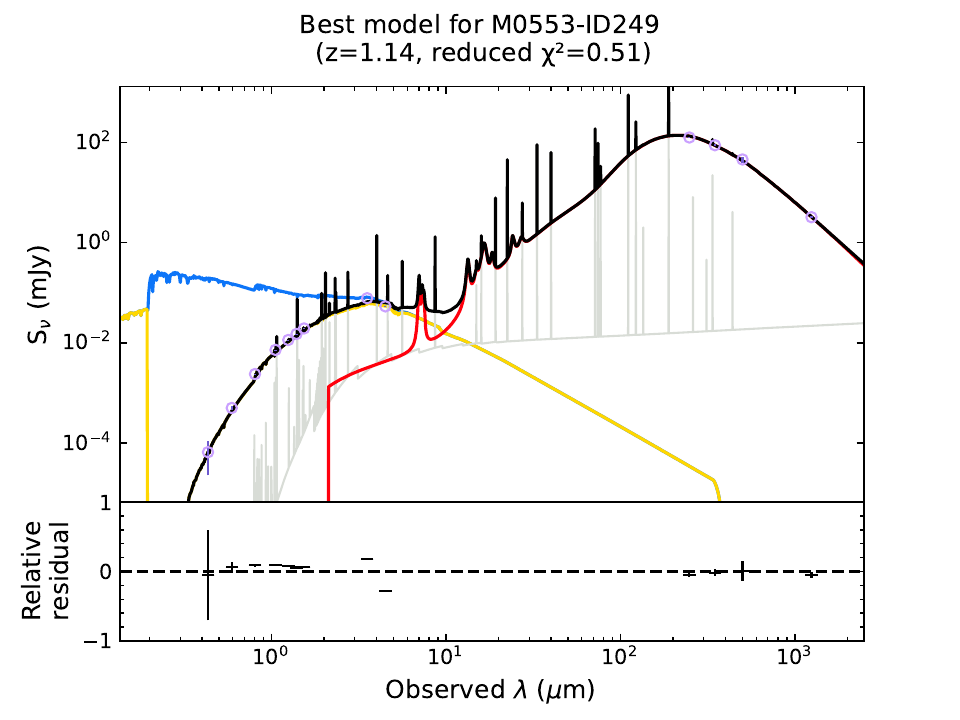}
\figsetgrpnote{The SED and the best-fit model of M0553-ID249. Lensing magnification is NOT corrected in this figure. The black solid line represents the composite spectrum of the galaxy. The yellow line illustrates the stellar emission attenuated by interstellar dust. The blue line depicts the unattenuated stellar emission for reference purpose. The orange line corresponds to the emission from an AGN. The red line shows the infrared emission from interstellar dust. The gray line denotes the nebulae emission. The observed data points are represented by purple circles, accompanied by 1$\sigma$ error bars. The bottom panel displays the relative residuals.}
\figsetgrpend

\figsetgrpstart
\figsetgrpnum{16.82}
\figsetgrptitle{M0553-ID275}
\figsetplot{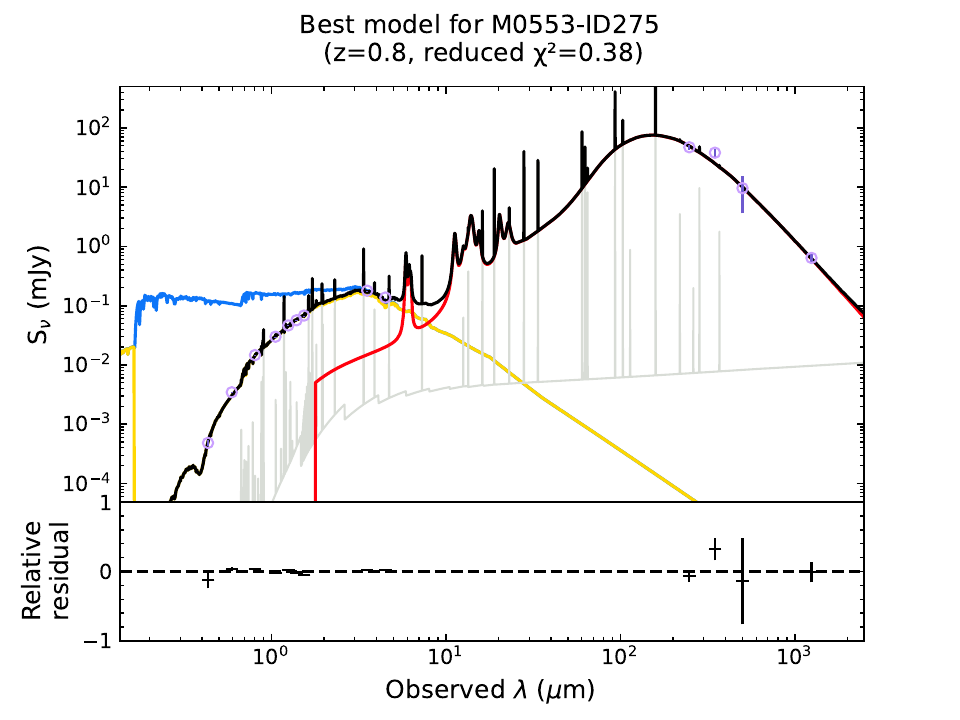}
\figsetgrpnote{The SED and the best-fit model of M0553-ID275. Lensing magnification is NOT corrected in this figure. The black solid line represents the composite spectrum of the galaxy. The yellow line illustrates the stellar emission attenuated by interstellar dust. The blue line depicts the unattenuated stellar emission for reference purpose. The orange line corresponds to the emission from an AGN. The red line shows the infrared emission from interstellar dust. The gray line denotes the nebulae emission. The observed data points are represented by purple circles, accompanied by 1$\sigma$ error bars. The bottom panel displays the relative residuals.}
\figsetgrpend

\figsetgrpstart
\figsetgrpnum{16.83}
\figsetgrptitle{M0553-ID303}
\figsetplot{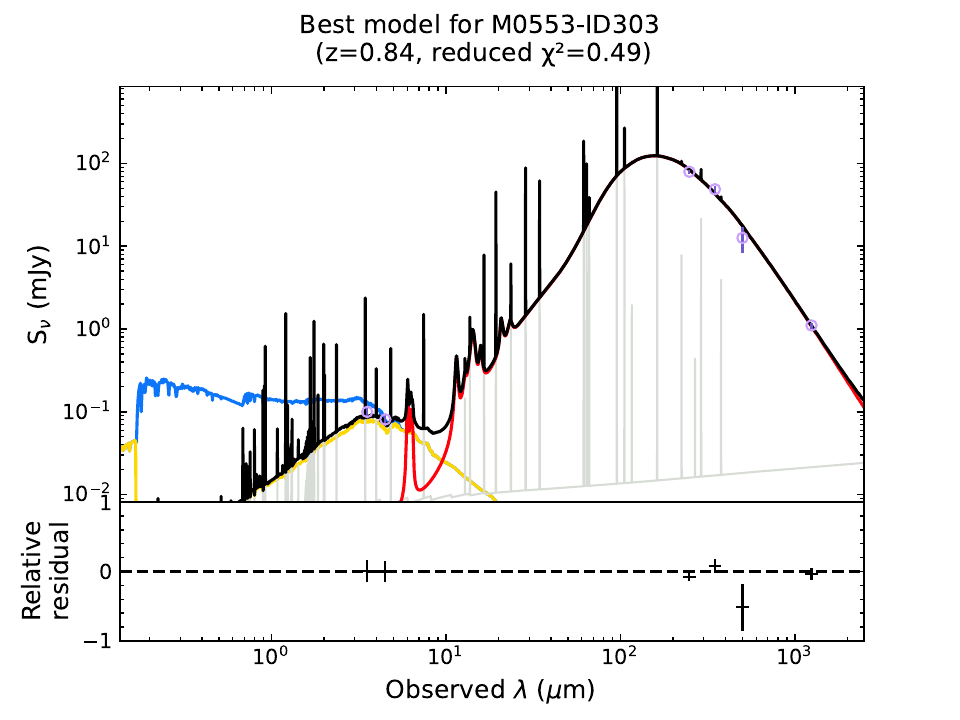}
\figsetgrpnote{The SED and the best-fit model of M0553-ID303. Lensing magnification is NOT corrected in this figure. The black solid line represents the composite spectrum of the galaxy. The yellow line illustrates the stellar emission attenuated by interstellar dust. The blue line depicts the unattenuated stellar emission for reference purpose. The orange line corresponds to the emission from an AGN. The red line shows the infrared emission from interstellar dust. The gray line denotes the nebulae emission. The observed data points are represented by purple circles, accompanied by 1$\sigma$ error bars. The bottom panel displays the relative residuals.}
\figsetgrpend

\figsetgrpstart
\figsetgrpnum{16.84}
\figsetgrptitle{M0553-ID355}
\figsetplot{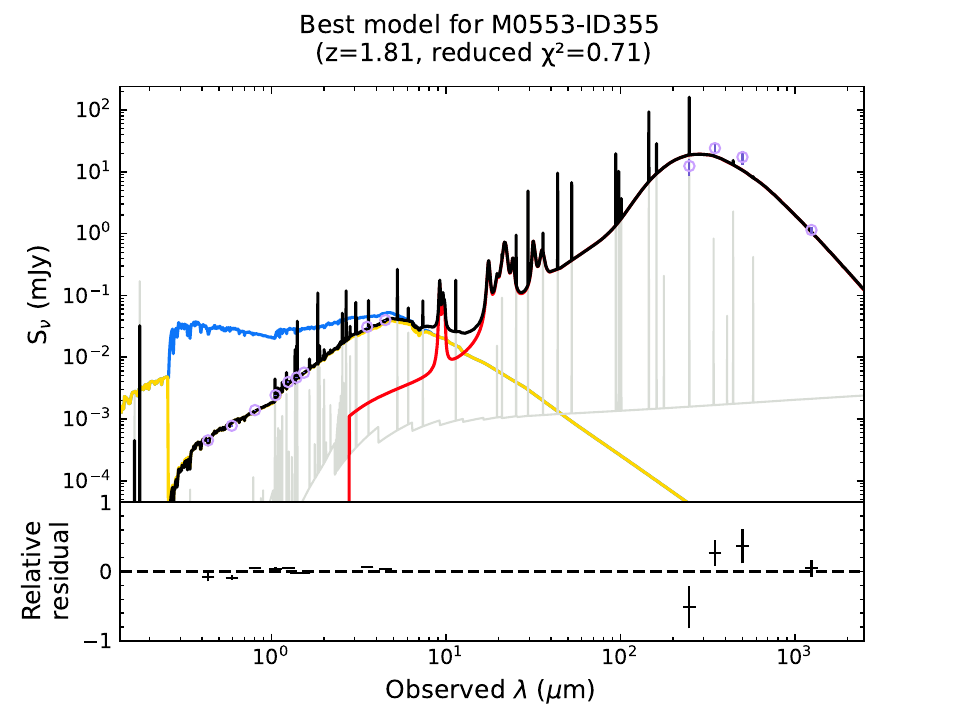}
\figsetgrpnote{The SED and the best-fit model of M0553-ID355. Lensing magnification is NOT corrected in this figure. The black solid line represents the composite spectrum of the galaxy. The yellow line illustrates the stellar emission attenuated by interstellar dust. The blue line depicts the unattenuated stellar emission for reference purpose. The orange line corresponds to the emission from an AGN. The red line shows the infrared emission from interstellar dust. The gray line denotes the nebulae emission. The observed data points are represented by purple circles, accompanied by 1$\sigma$ error bars. The bottom panel displays the relative residuals.}
\figsetgrpend

\figsetgrpstart
\figsetgrpnum{16.85}
\figsetgrptitle{M0553-ID375}
\figsetplot{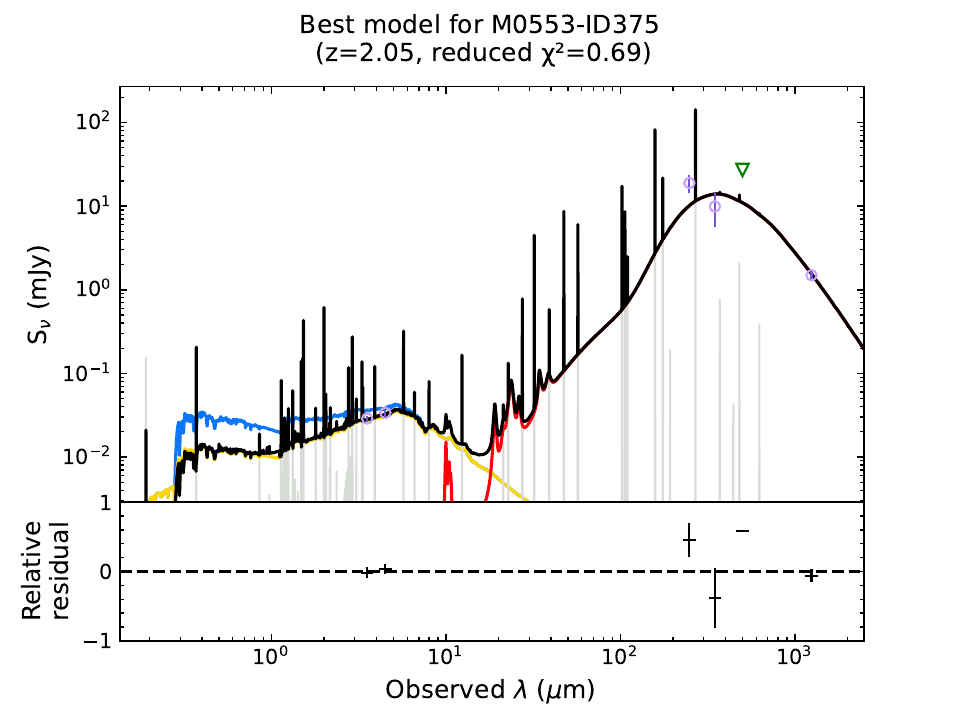}
\figsetgrpnote{The SED and the best-fit model of M0553-ID375. Lensing magnification is NOT corrected in this figure. The black solid line represents the composite spectrum of the galaxy. The yellow line illustrates the stellar emission attenuated by interstellar dust. The blue line depicts the unattenuated stellar emission for reference purpose. The orange line corresponds to the emission from an AGN. The red line shows the infrared emission from interstellar dust. The gray line denotes the nebulae emission. The observed data points are represented by purple circles, accompanied by 1$\sigma$ error bars. The bottom panel displays the relative residuals.}
\figsetgrpend

\figsetgrpstart
\figsetgrpnum{16.86}
\figsetgrptitle{M0553-ID398}
\figsetplot{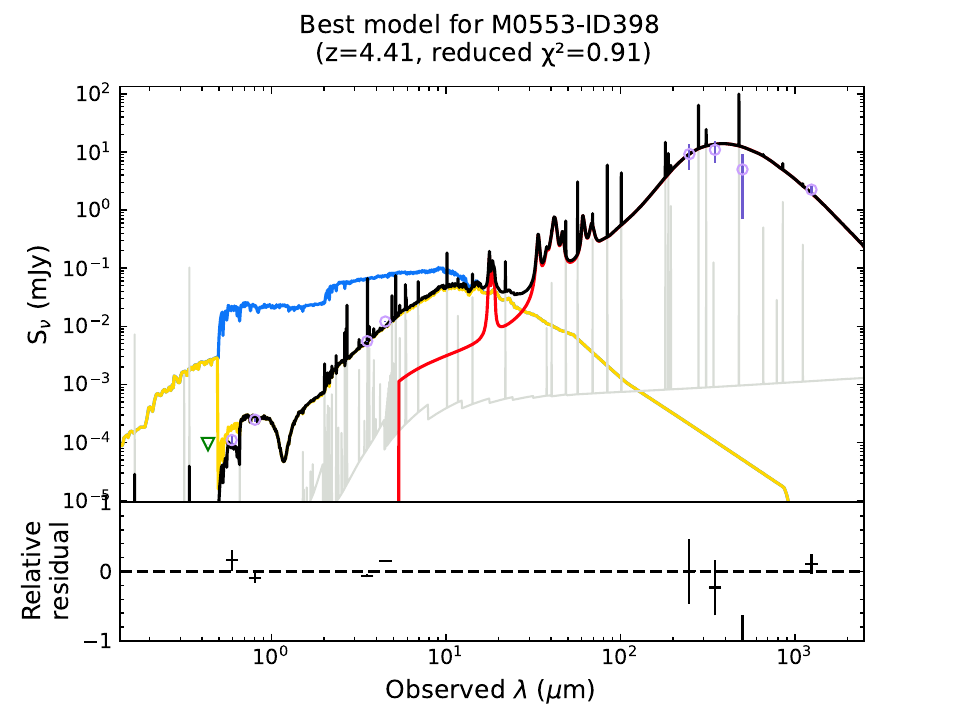}
\figsetgrpnote{The SED and the best-fit model of M0553-ID398. Lensing magnification is NOT corrected in this figure. The black solid line represents the composite spectrum of the galaxy. The yellow line illustrates the stellar emission attenuated by interstellar dust. The blue line depicts the unattenuated stellar emission for reference purpose. The orange line corresponds to the emission from an AGN. The red line shows the infrared emission from interstellar dust. The gray line denotes the nebulae emission. The observed data points are represented by purple circles, accompanied by 1$\sigma$ error bars. The bottom panel displays the relative residuals.}
\figsetgrpend

\figsetgrpstart
\figsetgrpnum{16.87}
\figsetgrptitle{M0553-ID58}
\figsetplot{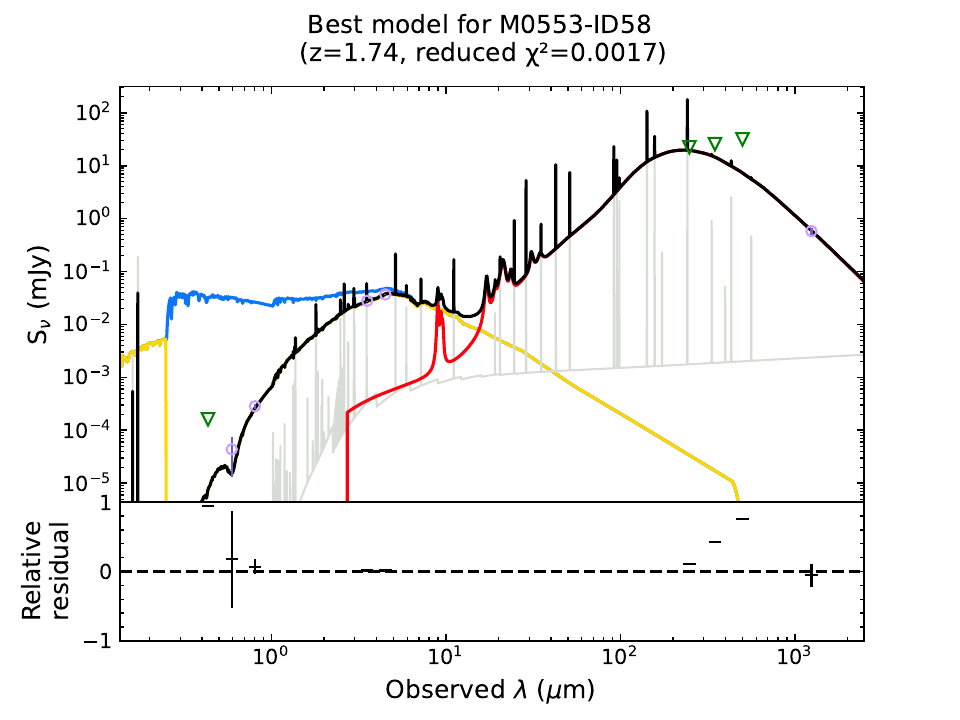}
\figsetgrpnote{The SED and the best-fit model of M0553-ID58. Lensing magnification is NOT corrected in this figure. The black solid line represents the composite spectrum of the galaxy. The yellow line illustrates the stellar emission attenuated by interstellar dust. The blue line depicts the unattenuated stellar emission for reference purpose. The orange line corresponds to the emission from an AGN. The red line shows the infrared emission from interstellar dust. The gray line denotes the nebulae emission. The observed data points are represented by purple circles, accompanied by 1$\sigma$ error bars. The bottom panel displays the relative residuals.}
\figsetgrpend

\figsetgrpstart
\figsetgrpnum{16.88}
\figsetgrptitle{M0553-ID61}
\figsetplot{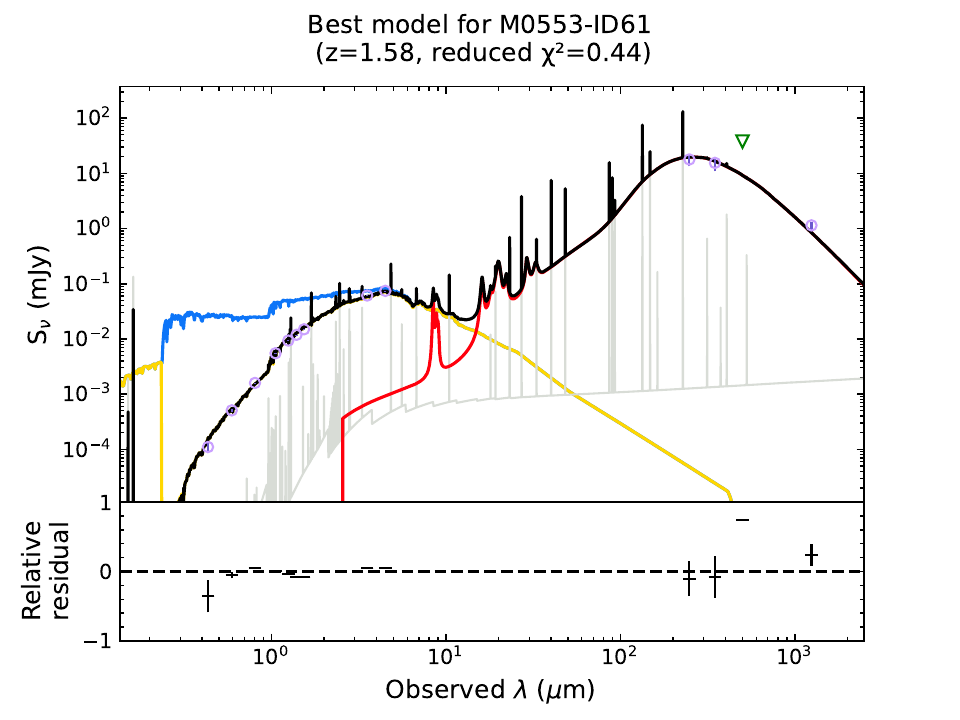}
\figsetgrpnote{The SED and the best-fit model of M0553-ID61. Lensing magnification is NOT corrected in this figure. The black solid line represents the composite spectrum of the galaxy. The yellow line illustrates the stellar emission attenuated by interstellar dust. The blue line depicts the unattenuated stellar emission for reference purpose. The orange line corresponds to the emission from an AGN. The red line shows the infrared emission from interstellar dust. The gray line denotes the nebulae emission. The observed data points are represented by purple circles, accompanied by 1$\sigma$ error bars. The bottom panel displays the relative residuals.}
\figsetgrpend

\figsetgrpstart
\figsetgrpnum{16.89}
\figsetgrptitle{M0553-ID67}
\figsetplot{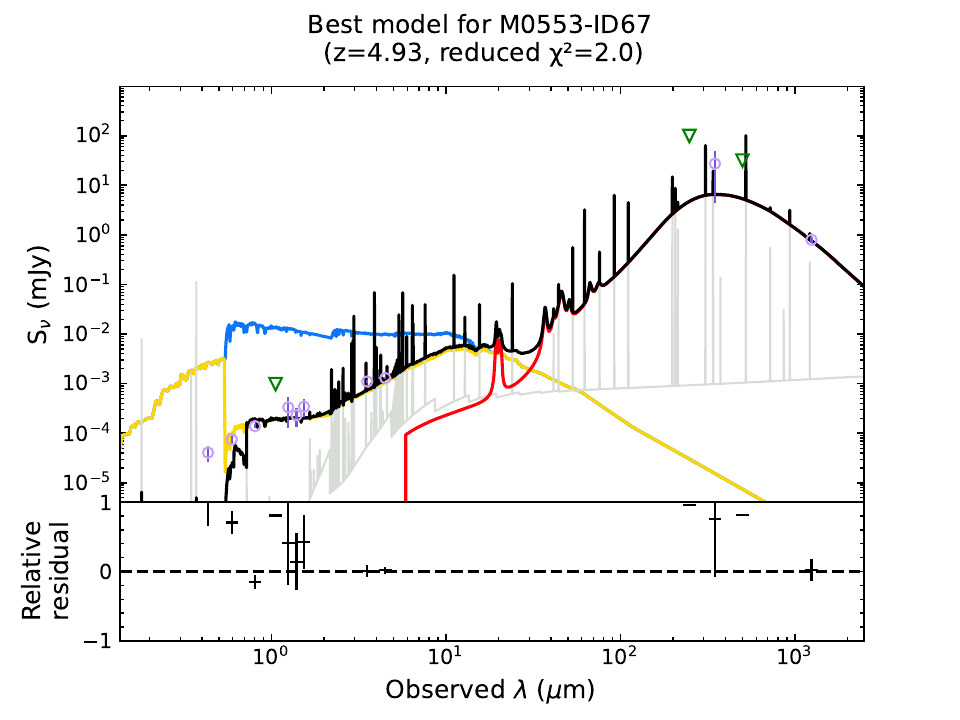}
\figsetgrpnote{The SED and the best-fit model of M0553-ID67. Lensing magnification is NOT corrected in this figure. The black solid line represents the composite spectrum of the galaxy. The yellow line illustrates the stellar emission attenuated by interstellar dust. The blue line depicts the unattenuated stellar emission for reference purpose. The orange line corresponds to the emission from an AGN. The red line shows the infrared emission from interstellar dust. The gray line denotes the nebulae emission. The observed data points are represented by purple circles, accompanied by 1$\sigma$ error bars. The bottom panel displays the relative residuals.}
\figsetgrpend

\figsetgrpstart
\figsetgrpnum{16.90}
\figsetgrptitle{M1115-ID2}
\figsetplot{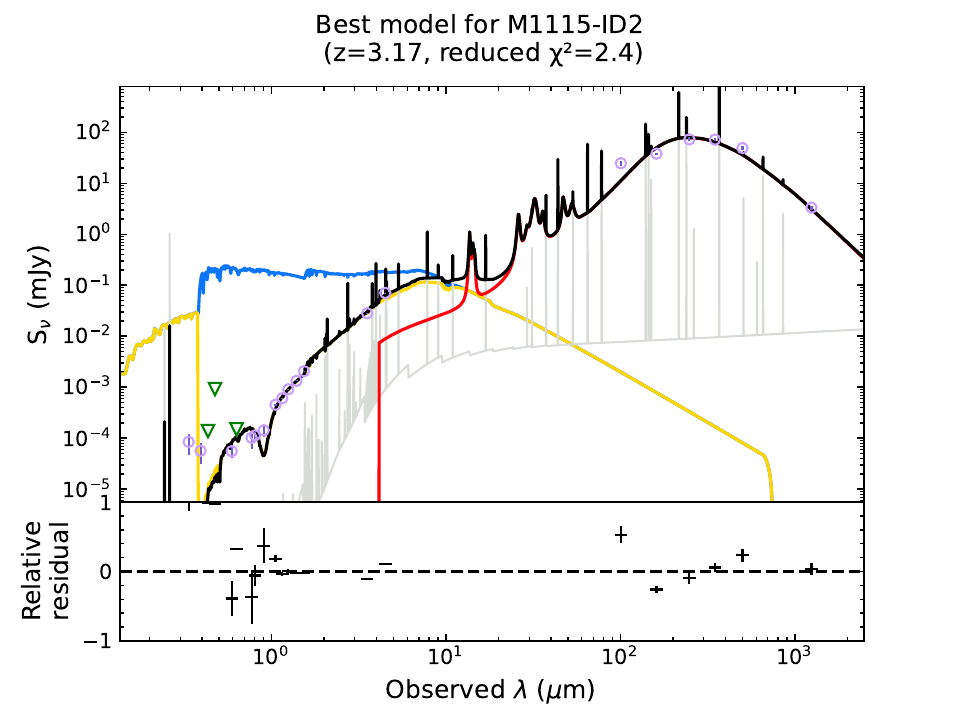}
\figsetgrpnote{The SED and the best-fit model of M1115-ID2. Lensing magnification is NOT corrected in this figure. The black solid line represents the composite spectrum of the galaxy. The yellow line illustrates the stellar emission attenuated by interstellar dust. The blue line depicts the unattenuated stellar emission for reference purpose. The orange line corresponds to the emission from an AGN. The red line shows the infrared emission from interstellar dust. The gray line denotes the nebulae emission. The observed data points are represented by purple circles, accompanied by 1$\sigma$ error bars. The bottom panel displays the relative residuals.}
\figsetgrpend

\figsetgrpstart
\figsetgrpnum{16.91}
\figsetgrptitle{M1115-ID33}
\figsetplot{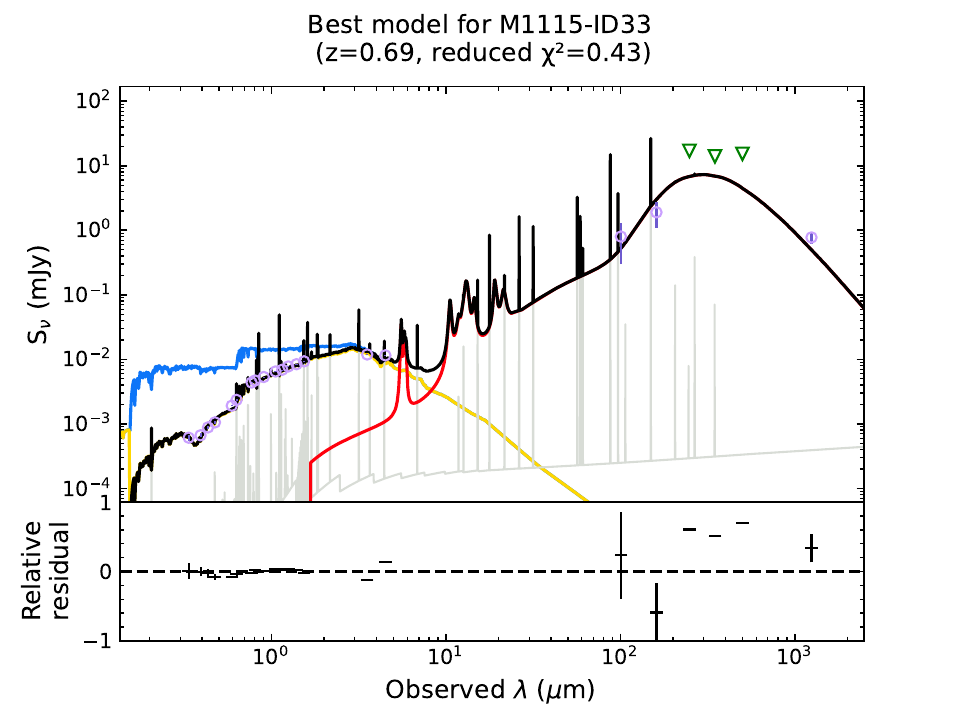}
\figsetgrpnote{The SED and the best-fit model of M1115-ID33. Lensing magnification is NOT corrected in this figure. The black solid line represents the composite spectrum of the galaxy. The yellow line illustrates the stellar emission attenuated by interstellar dust. The blue line depicts the unattenuated stellar emission for reference purpose. The orange line corresponds to the emission from an AGN. The red line shows the infrared emission from interstellar dust. The gray line denotes the nebulae emission. The observed data points are represented by purple circles, accompanied by 1$\sigma$ error bars. The bottom panel displays the relative residuals.}
\figsetgrpend

\figsetgrpstart
\figsetgrpnum{16.92}
\figsetgrptitle{M1115-ID34}
\figsetplot{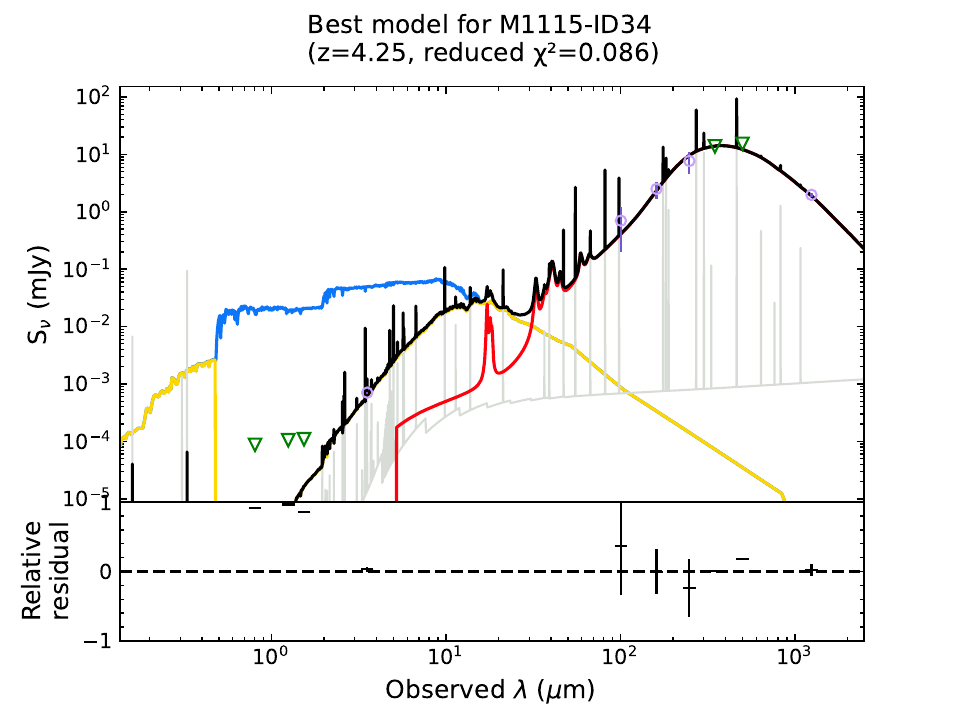}
\figsetgrpnote{The SED and the best-fit model of M1115-ID34. Lensing magnification is NOT corrected in this figure. The black solid line represents the composite spectrum of the galaxy. The yellow line illustrates the stellar emission attenuated by interstellar dust. The blue line depicts the unattenuated stellar emission for reference purpose. The orange line corresponds to the emission from an AGN. The red line shows the infrared emission from interstellar dust. The gray line denotes the nebulae emission. The observed data points are represented by purple circles, accompanied by 1$\sigma$ error bars. The bottom panel displays the relative residuals.}
\figsetgrpend

\figsetgrpstart
\figsetgrpnum{16.93}
\figsetgrptitle{M1115-ID4}
\figsetplot{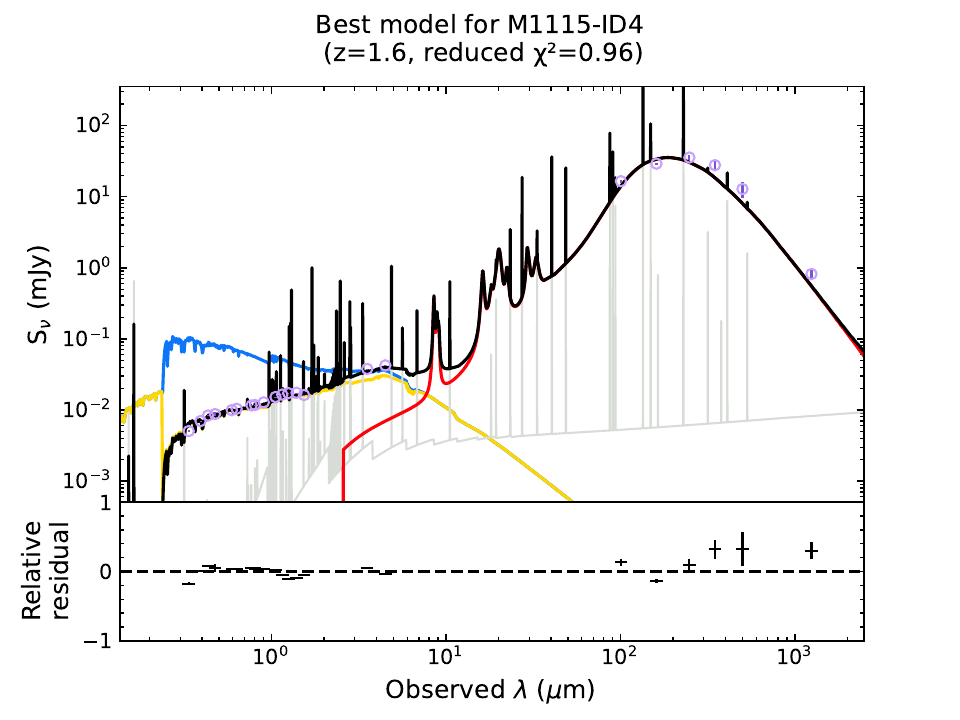}
\figsetgrpnote{The SED and the best-fit model of M1115-ID4. Lensing magnification is NOT corrected in this figure. The black solid line represents the composite spectrum of the galaxy. The yellow line illustrates the stellar emission attenuated by interstellar dust. The blue line depicts the unattenuated stellar emission for reference purpose. The orange line corresponds to the emission from an AGN. The red line shows the infrared emission from interstellar dust. The gray line denotes the nebulae emission. The observed data points are represented by purple circles, accompanied by 1$\sigma$ error bars. The bottom panel displays the relative residuals.}
\figsetgrpend

\figsetgrpstart
\figsetgrpnum{16.94}
\figsetgrptitle{M1149-ID27}
\figsetplot{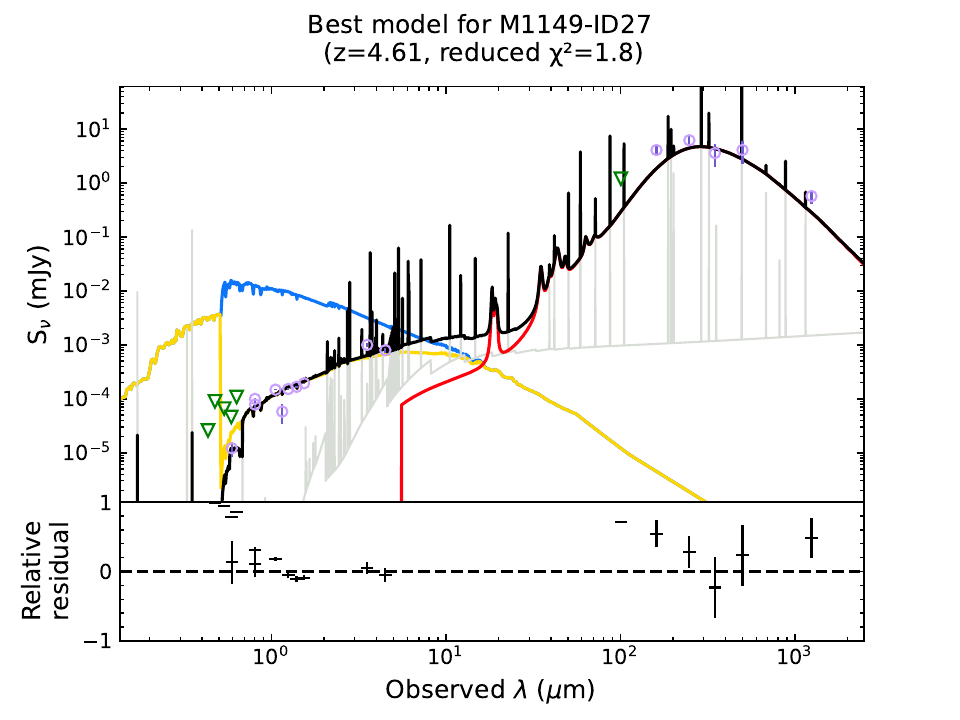}
\figsetgrpnote{The SED and the best-fit model of M1149-ID27. Lensing magnification is NOT corrected in this figure. The black solid line represents the composite spectrum of the galaxy. The yellow line illustrates the stellar emission attenuated by interstellar dust. The blue line depicts the unattenuated stellar emission for reference purpose. The orange line corresponds to the emission from an AGN. The red line shows the infrared emission from interstellar dust. The gray line denotes the nebulae emission. The observed data points are represented by purple circles, accompanied by 1$\sigma$ error bars. The bottom panel displays the relative residuals.}
\figsetgrpend

\figsetgrpstart
\figsetgrpnum{16.95}
\figsetgrptitle{M1149-ID77}
\figsetplot{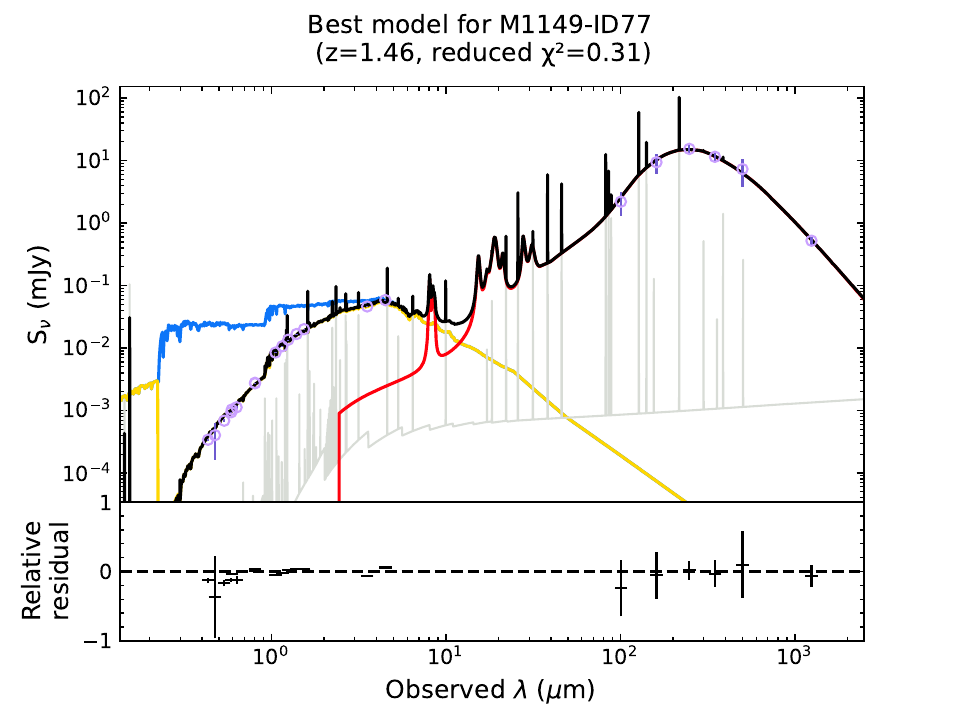}
\figsetgrpnote{The SED and the best-fit model of M1149-ID77. Lensing magnification is NOT corrected in this figure. The black solid line represents the composite spectrum of the galaxy. The yellow line illustrates the stellar emission attenuated by interstellar dust. The blue line depicts the unattenuated stellar emission for reference purpose. The orange line corresponds to the emission from an AGN. The red line shows the infrared emission from interstellar dust. The gray line denotes the nebulae emission. The observed data points are represented by purple circles, accompanied by 1$\sigma$ error bars. The bottom panel displays the relative residuals.}
\figsetgrpend

\figsetgrpstart
\figsetgrpnum{16.96}
\figsetgrptitle{M1149-ID95}
\figsetplot{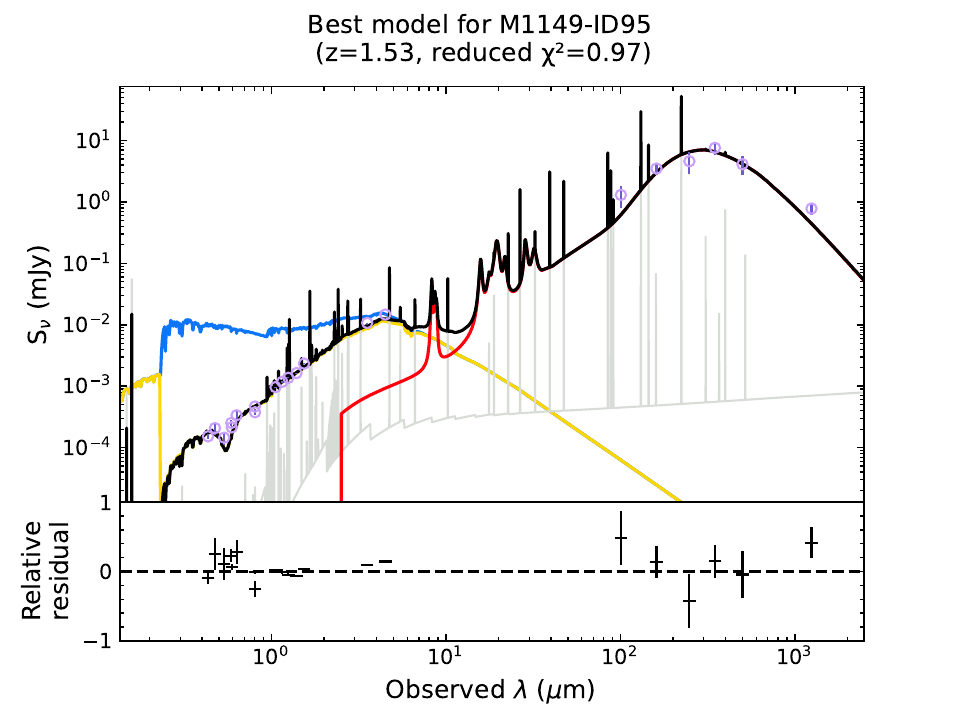}
\figsetgrpnote{The SED and the best-fit model of M1149-ID95. Lensing magnification is NOT corrected in this figure. The black solid line represents the composite spectrum of the galaxy. The yellow line illustrates the stellar emission attenuated by interstellar dust. The blue line depicts the unattenuated stellar emission for reference purpose. The orange line corresponds to the emission from an AGN. The red line shows the infrared emission from interstellar dust. The gray line denotes the nebulae emission. The observed data points are represented by purple circles, accompanied by 1$\sigma$ error bars. The bottom panel displays the relative residuals.}
\figsetgrpend

\figsetgrpstart
\figsetgrpnum{16.97}
\figsetgrptitle{M1206-ID27}
\figsetplot{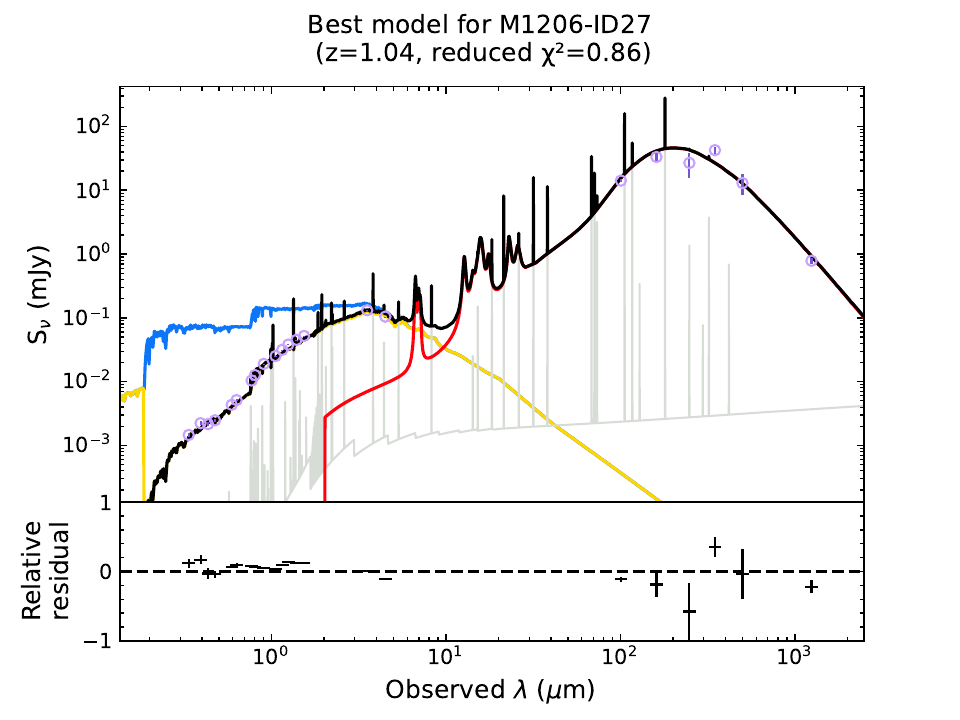}
\figsetgrpnote{The SED and the best-fit model of M1206-ID27. Lensing magnification is NOT corrected in this figure. The black solid line represents the composite spectrum of the galaxy. The yellow line illustrates the stellar emission attenuated by interstellar dust. The blue line depicts the unattenuated stellar emission for reference purpose. The orange line corresponds to the emission from an AGN. The red line shows the infrared emission from interstellar dust. The gray line denotes the nebulae emission. The observed data points are represented by purple circles, accompanied by 1$\sigma$ error bars. The bottom panel displays the relative residuals.}
\figsetgrpend

\figsetgrpstart
\figsetgrpnum{16.98}
\figsetgrptitle{M1206-ID38}
\figsetplot{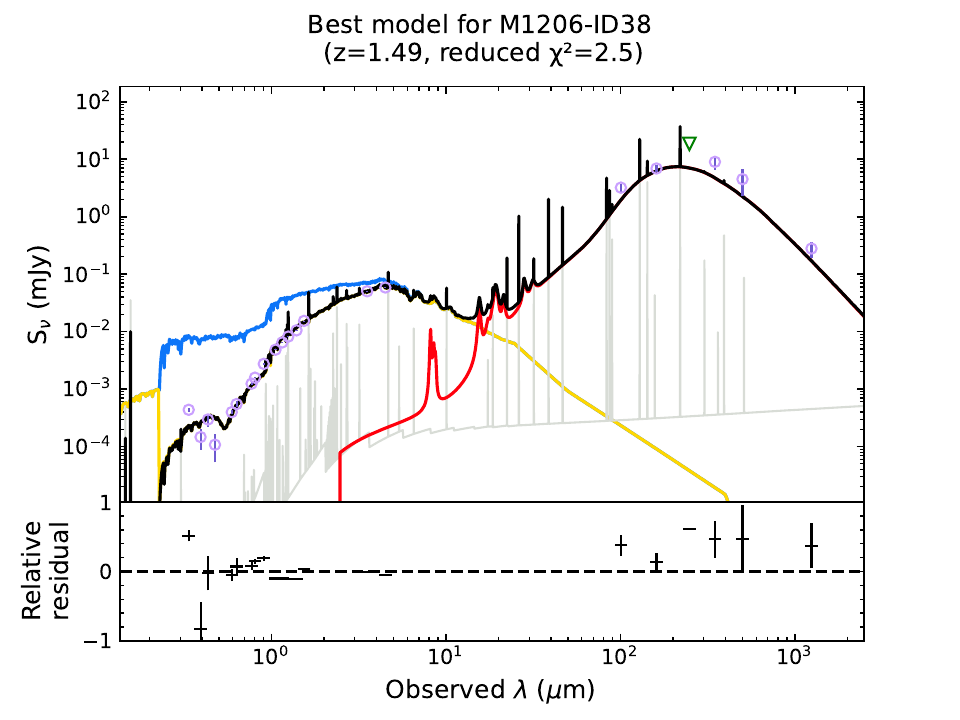}
\figsetgrpnote{The SED and the best-fit model of M1206-ID38. Lensing magnification is NOT corrected in this figure. The black solid line represents the composite spectrum of the galaxy. The yellow line illustrates the stellar emission attenuated by interstellar dust. The blue line depicts the unattenuated stellar emission for reference purpose. The orange line corresponds to the emission from an AGN. The red line shows the infrared emission from interstellar dust. The gray line denotes the nebulae emission. The observed data points are represented by purple circles, accompanied by 1$\sigma$ error bars. The bottom panel displays the relative residuals.}
\figsetgrpend

\figsetgrpstart
\figsetgrpnum{16.99}
\figsetgrptitle{M1206-ID54}
\figsetplot{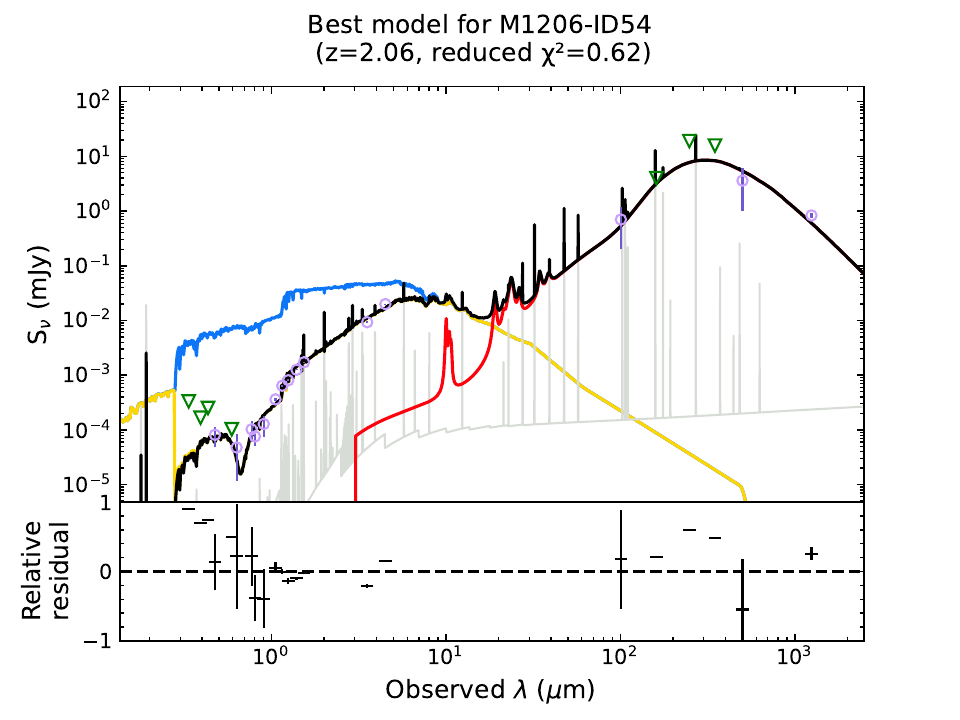}
\figsetgrpnote{The SED and the best-fit model of M1206-ID54. Lensing magnification is NOT corrected in this figure. The black solid line represents the composite spectrum of the galaxy. The yellow line illustrates the stellar emission attenuated by interstellar dust. The blue line depicts the unattenuated stellar emission for reference purpose. The orange line corresponds to the emission from an AGN. The red line shows the infrared emission from interstellar dust. The gray line denotes the nebulae emission. The observed data points are represented by purple circles, accompanied by 1$\sigma$ error bars. The bottom panel displays the relative residuals.}
\figsetgrpend

\figsetgrpstart
\figsetgrpnum{16.100}
\figsetgrptitle{M1206-ID55}
\figsetplot{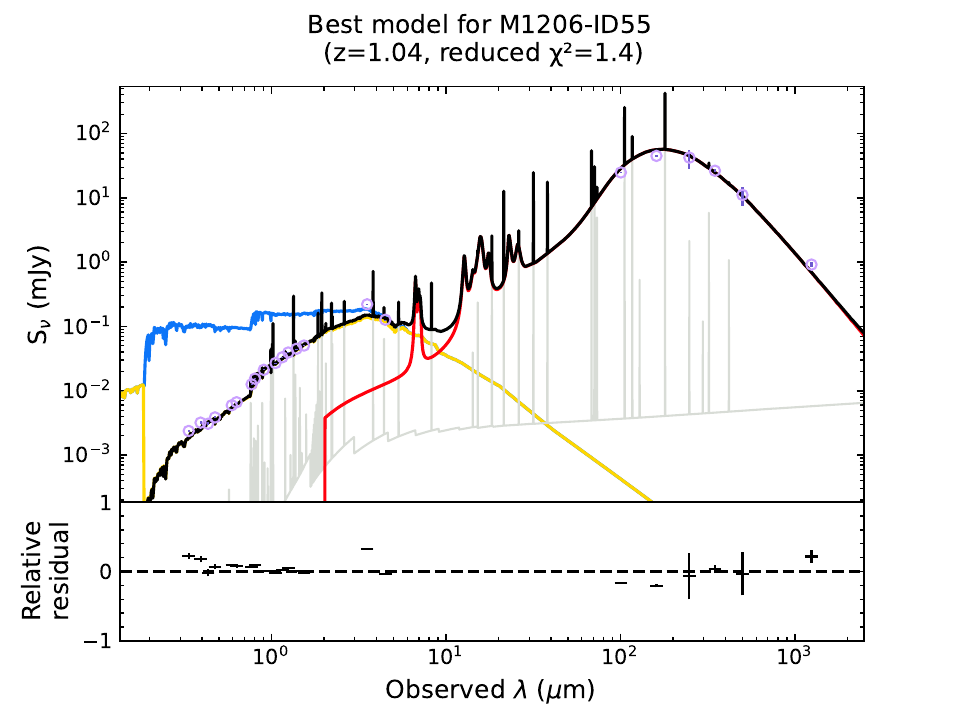}
\figsetgrpnote{The SED and the best-fit model of M1206-ID55. Lensing magnification is NOT corrected in this figure. The black solid line represents the composite spectrum of the galaxy. The yellow line illustrates the stellar emission attenuated by interstellar dust. The blue line depicts the unattenuated stellar emission for reference purpose. The orange line corresponds to the emission from an AGN. The red line shows the infrared emission from interstellar dust. The gray line denotes the nebulae emission. The observed data points are represented by purple circles, accompanied by 1$\sigma$ error bars. The bottom panel displays the relative residuals.}
\figsetgrpend

\figsetgrpstart
\figsetgrpnum{16.101}
\figsetgrptitle{M1206-ID60}
\figsetplot{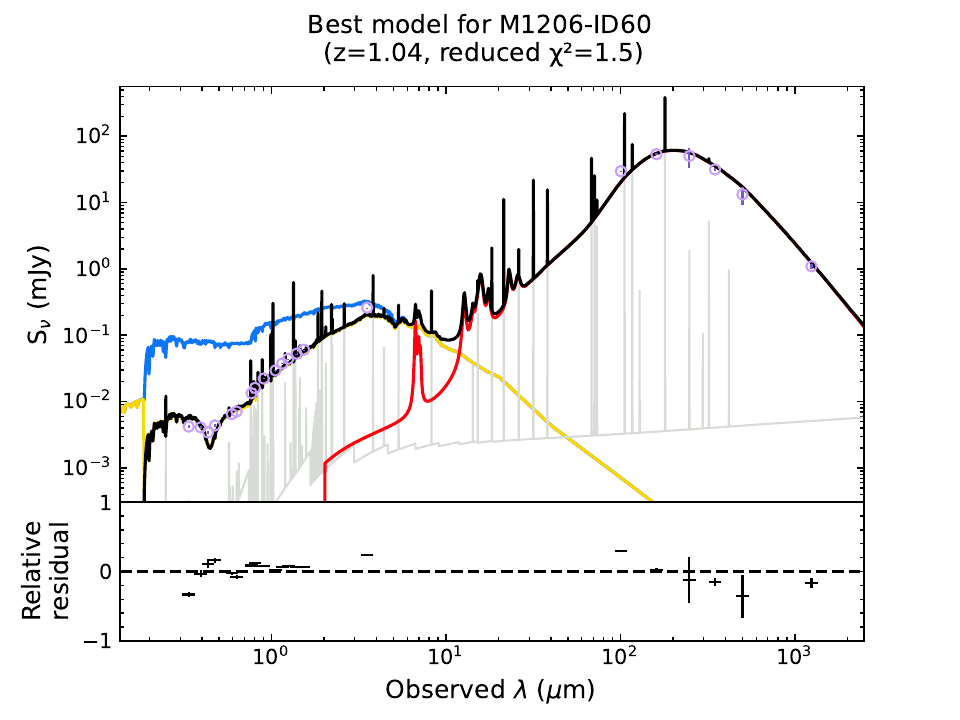}
\figsetgrpnote{The SED and the best-fit model of M1206-ID60. Lensing magnification is NOT corrected in this figure. The black solid line represents the composite spectrum of the galaxy. The yellow line illustrates the stellar emission attenuated by interstellar dust. The blue line depicts the unattenuated stellar emission for reference purpose. The orange line corresponds to the emission from an AGN. The red line shows the infrared emission from interstellar dust. The gray line denotes the nebulae emission. The observed data points are represented by purple circles, accompanied by 1$\sigma$ error bars. The bottom panel displays the relative residuals.}
\figsetgrpend

\figsetgrpstart
\figsetgrpnum{16.102}
\figsetgrptitle{M1206-ID61}
\figsetplot{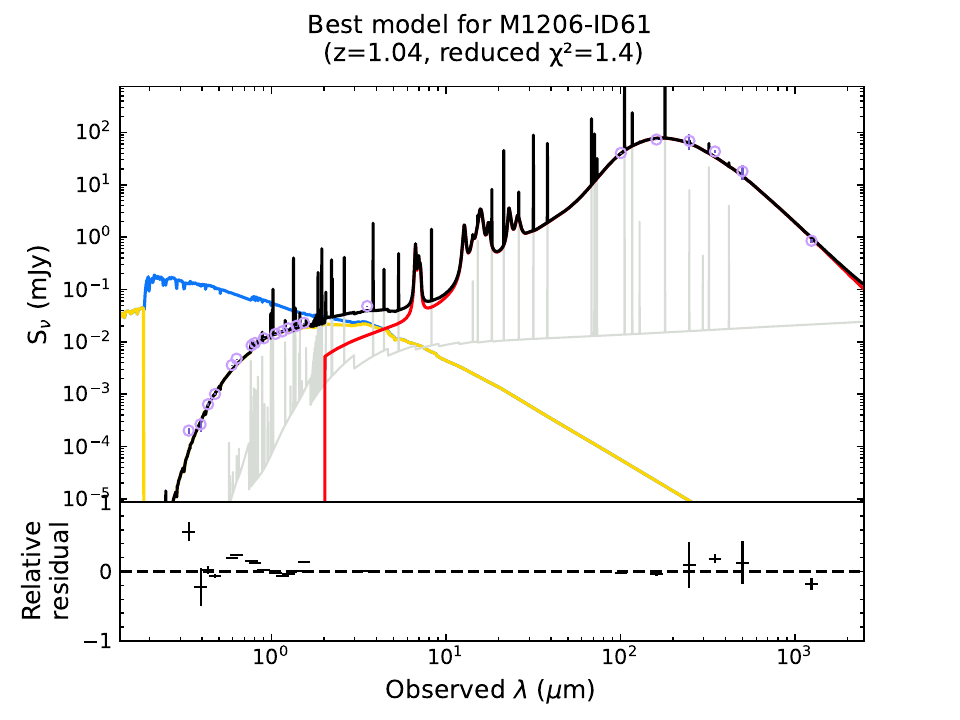}
\figsetgrpnote{The SED and the best-fit model of M1206-ID61. Lensing magnification is NOT corrected in this figure. The black solid line represents the composite spectrum of the galaxy. The yellow line illustrates the stellar emission attenuated by interstellar dust. The blue line depicts the unattenuated stellar emission for reference purpose. The orange line corresponds to the emission from an AGN. The red line shows the infrared emission from interstellar dust. The gray line denotes the nebulae emission. The observed data points are represented by purple circles, accompanied by 1$\sigma$ error bars. The bottom panel displays the relative residuals.}
\figsetgrpend

\figsetgrpstart
\figsetgrpnum{16.103}
\figsetgrptitle{M1206-ID84}
\figsetplot{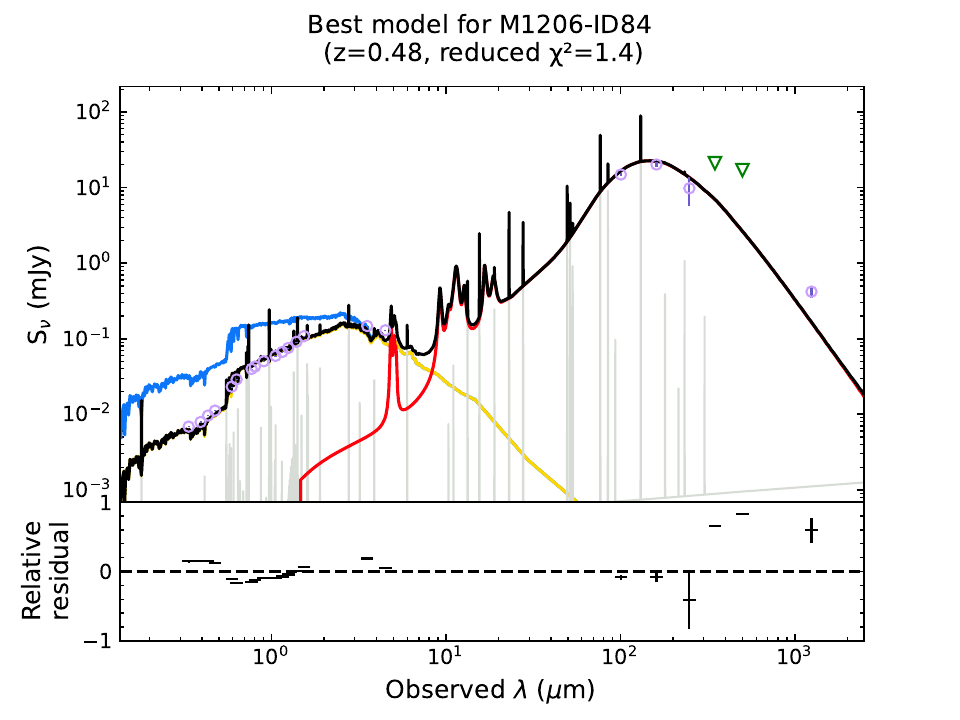}
\figsetgrpnote{The SED and the best-fit model of M1206-ID84. Lensing magnification is NOT corrected in this figure. The black solid line represents the composite spectrum of the galaxy. The yellow line illustrates the stellar emission attenuated by interstellar dust. The blue line depicts the unattenuated stellar emission for reference purpose. The orange line corresponds to the emission from an AGN. The red line shows the infrared emission from interstellar dust. The gray line denotes the nebulae emission. The observed data points are represented by purple circles, accompanied by 1$\sigma$ error bars. The bottom panel displays the relative residuals.}
\figsetgrpend

\figsetgrpstart
\figsetgrpnum{16.104}
\figsetgrptitle{M1311-ID27}
\figsetplot{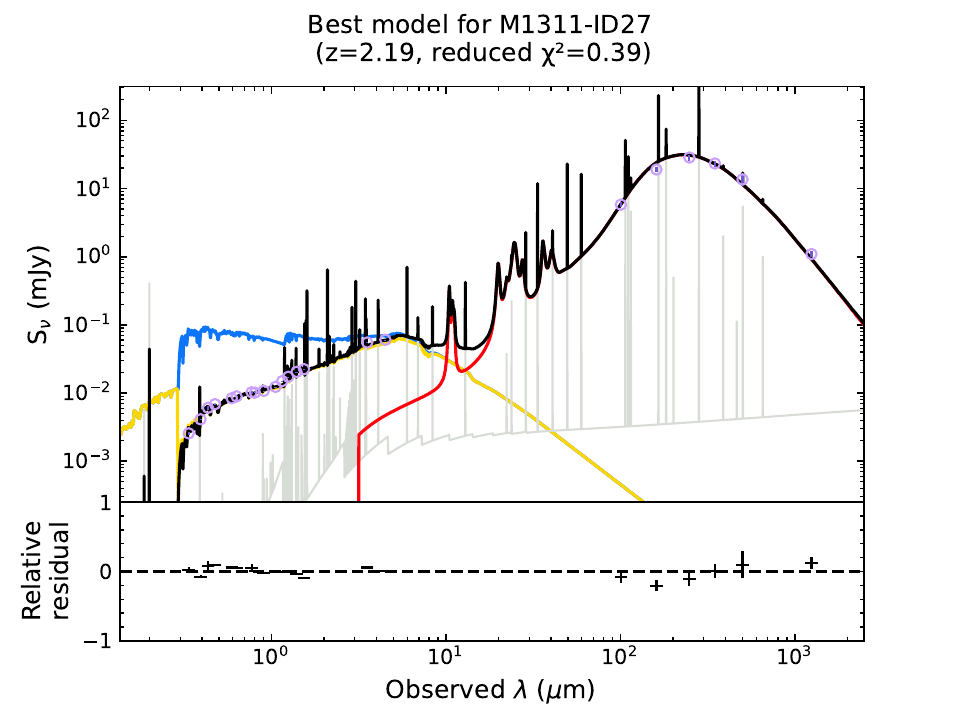}
\figsetgrpnote{The SED and the best-fit model of M1311-ID27. Lensing magnification is NOT corrected in this figure. The black solid line represents the composite spectrum of the galaxy. The yellow line illustrates the stellar emission attenuated by interstellar dust. The blue line depicts the unattenuated stellar emission for reference purpose. The orange line corresponds to the emission from an AGN. The red line shows the infrared emission from interstellar dust. The gray line denotes the nebulae emission. The observed data points are represented by purple circles, accompanied by 1$\sigma$ error bars. The bottom panel displays the relative residuals.}
\figsetgrpend

\figsetgrpstart
\figsetgrpnum{16.105}
\figsetgrptitle{M1311-ID33}
\figsetplot{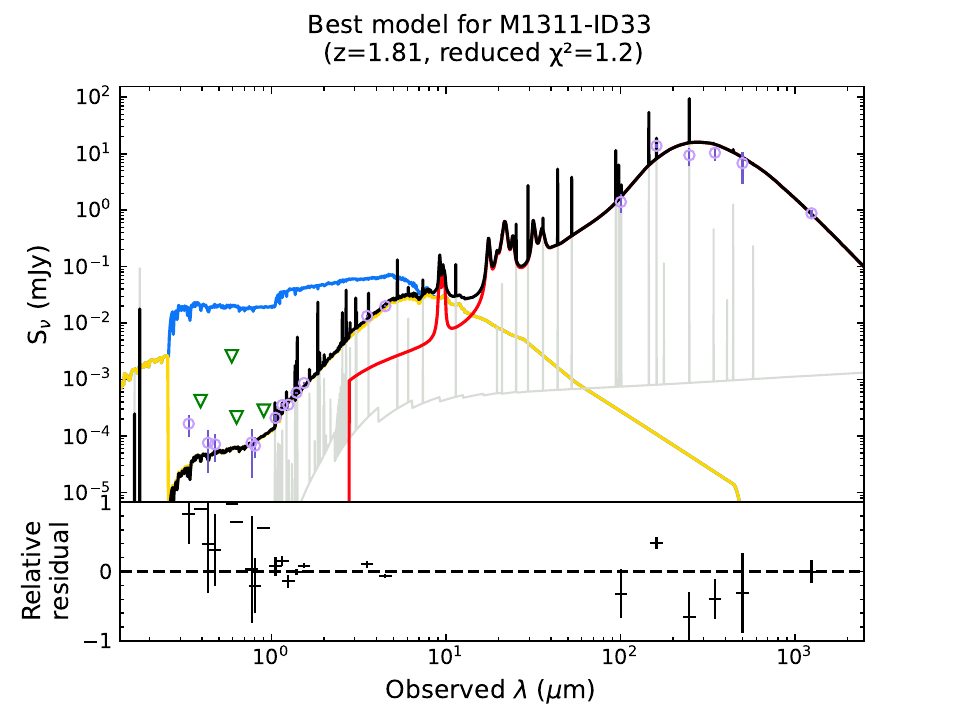}
\figsetgrpnote{The SED and the best-fit model of M1311-ID33. Lensing magnification is NOT corrected in this figure. The black solid line represents the composite spectrum of the galaxy. The yellow line illustrates the stellar emission attenuated by interstellar dust. The blue line depicts the unattenuated stellar emission for reference purpose. The orange line corresponds to the emission from an AGN. The red line shows the infrared emission from interstellar dust. The gray line denotes the nebulae emission. The observed data points are represented by purple circles, accompanied by 1$\sigma$ error bars. The bottom panel displays the relative residuals.}
\figsetgrpend

\figsetgrpstart
\figsetgrpnum{16.106}
\figsetgrptitle{M1423-ID38}
\figsetplot{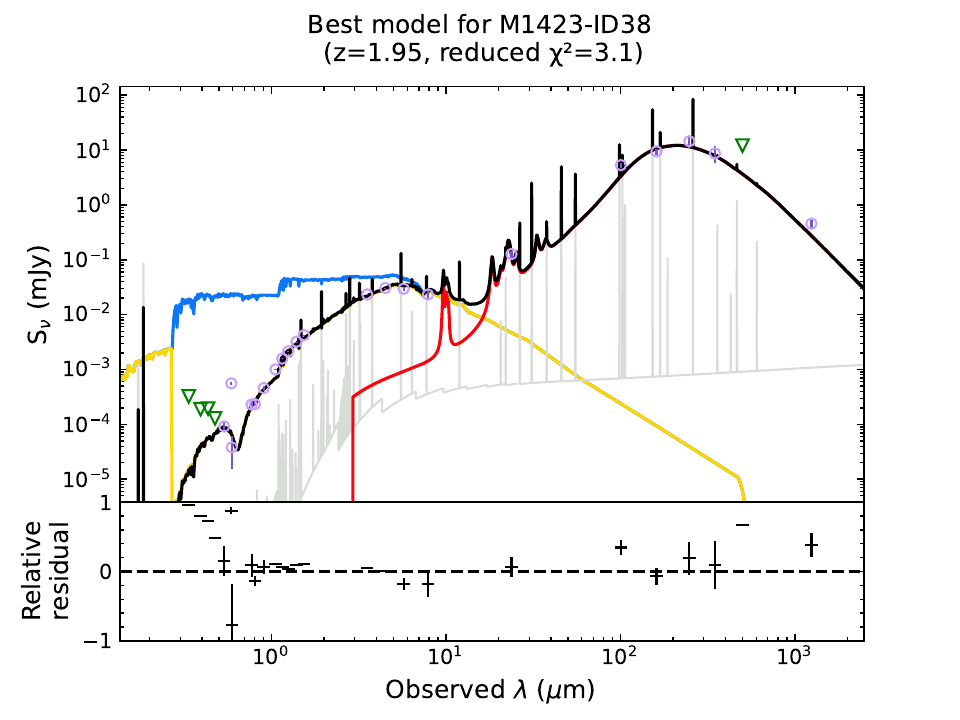}
\figsetgrpnote{The SED and the best-fit model of M1423-ID38. Lensing magnification is NOT corrected in this figure. The black solid line represents the composite spectrum of the galaxy. The yellow line illustrates the stellar emission attenuated by interstellar dust. The blue line depicts the unattenuated stellar emission for reference purpose. The orange line corresponds to the emission from an AGN. The red line shows the infrared emission from interstellar dust. The gray line denotes the nebulae emission. The observed data points are represented by purple circles, accompanied by 1$\sigma$ error bars. The bottom panel displays the relative residuals.}
\figsetgrpend

\figsetgrpstart
\figsetgrpnum{16.107}
\figsetgrptitle{M1423-ID52}
\figsetplot{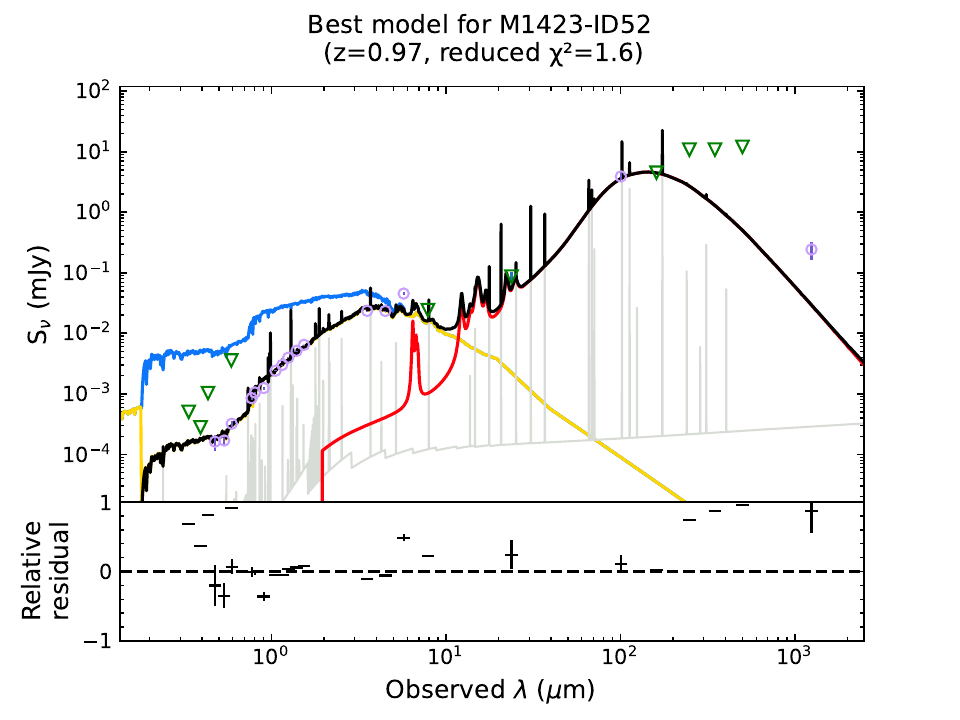}
\figsetgrpnote{The SED and the best-fit model of M1423-ID52. Lensing magnification is NOT corrected in this figure. The black solid line represents the composite spectrum of the galaxy. The yellow line illustrates the stellar emission attenuated by interstellar dust. The blue line depicts the unattenuated stellar emission for reference purpose. The orange line corresponds to the emission from an AGN. The red line shows the infrared emission from interstellar dust. The gray line denotes the nebulae emission. The observed data points are represented by purple circles, accompanied by 1$\sigma$ error bars. The bottom panel displays the relative residuals.}
\figsetgrpend

\figsetgrpstart
\figsetgrpnum{16.108}
\figsetgrptitle{M1423-ID76}
\figsetplot{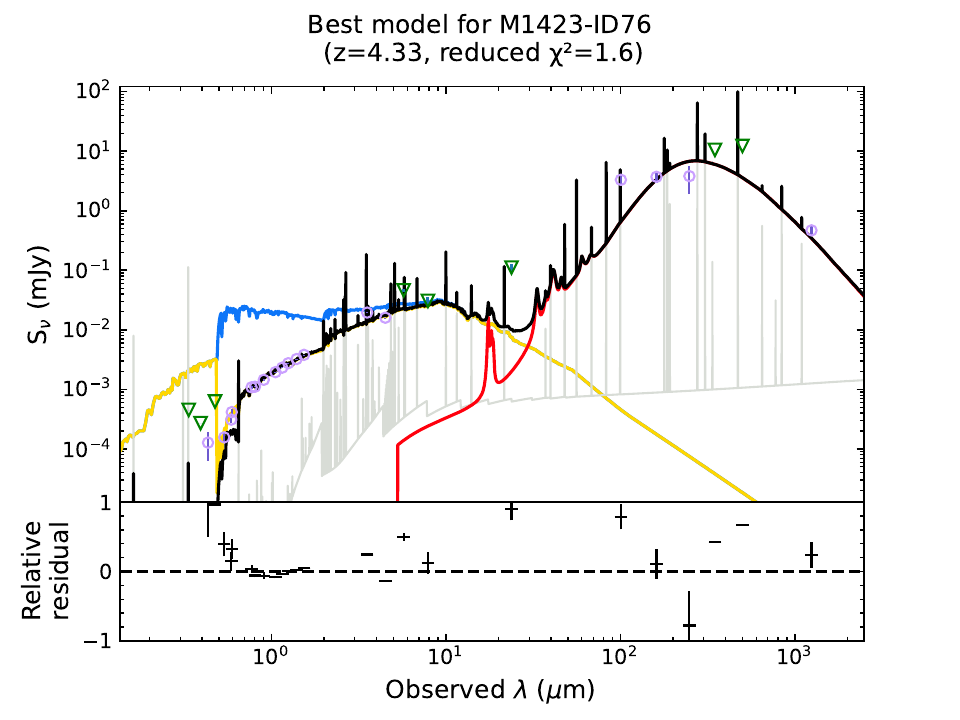}
\figsetgrpnote{The SED and the best-fit model of M1423-ID76. Lensing magnification is NOT corrected in this figure. The black solid line represents the composite spectrum of the galaxy. The yellow line illustrates the stellar emission attenuated by interstellar dust. The blue line depicts the unattenuated stellar emission for reference purpose. The orange line corresponds to the emission from an AGN. The red line shows the infrared emission from interstellar dust. The gray line denotes the nebulae emission. The observed data points are represented by purple circles, accompanied by 1$\sigma$ error bars. The bottom panel displays the relative residuals.}
\figsetgrpend

\figsetgrpstart
\figsetgrpnum{16.109}
\figsetgrptitle{M1931-ID47}
\figsetplot{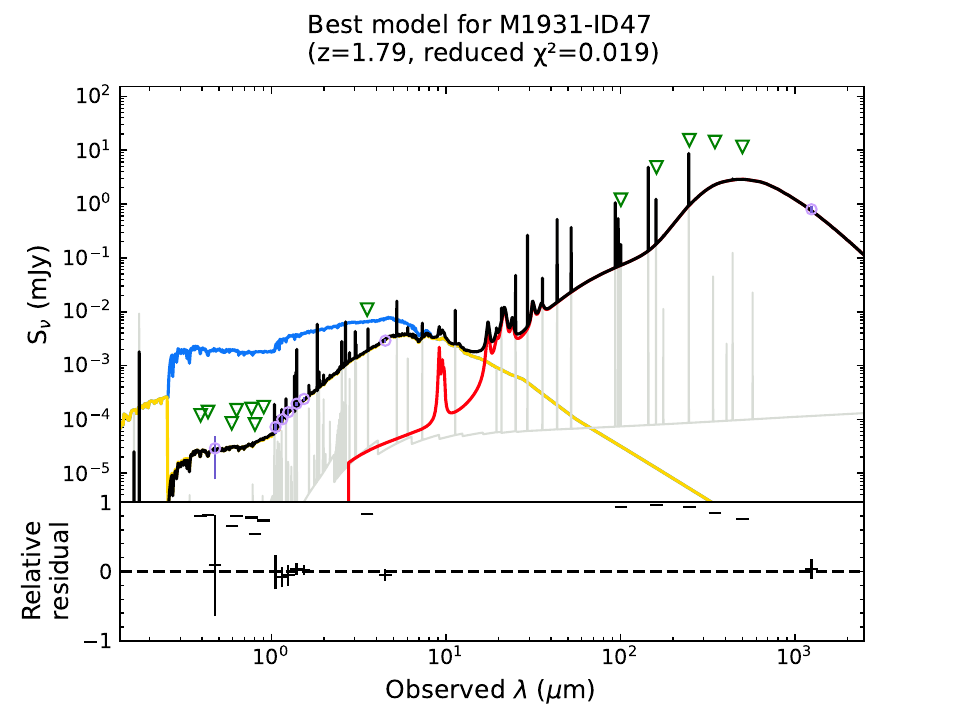}
\figsetgrpnote{The SED and the best-fit model of M1931-ID47. Lensing magnification is NOT corrected in this figure. The black solid line represents the composite spectrum of the galaxy. The yellow line illustrates the stellar emission attenuated by interstellar dust. The blue line depicts the unattenuated stellar emission for reference purpose. The orange line corresponds to the emission from an AGN. The red line shows the infrared emission from interstellar dust. The gray line denotes the nebulae emission. The observed data points are represented by purple circles, accompanied by 1$\sigma$ error bars. The bottom panel displays the relative residuals.}
\figsetgrpend

\figsetgrpstart
\figsetgrpnum{16.110}
\figsetgrptitle{M1931-ID55}
\figsetplot{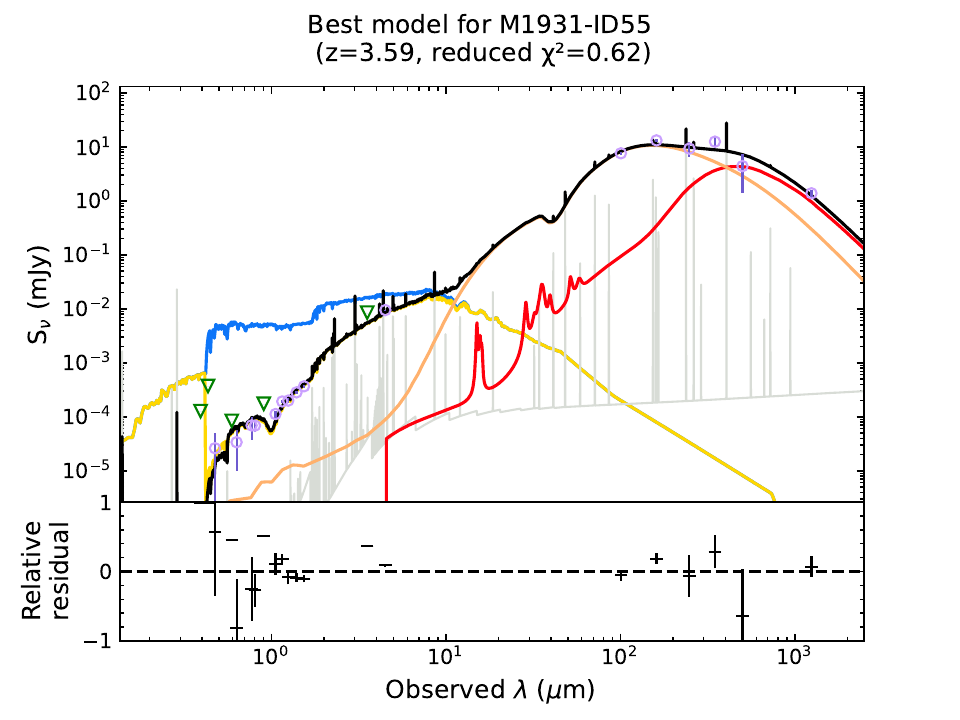}
\figsetgrpnote{The SED and the best-fit model of M1931-ID55. Lensing magnification is NOT corrected in this figure. The black solid line represents the composite spectrum of the galaxy. The yellow line illustrates the stellar emission attenuated by interstellar dust. The blue line depicts the unattenuated stellar emission for reference purpose. The orange line corresponds to the emission from an AGN. The red line shows the infrared emission from interstellar dust. The gray line denotes the nebulae emission. The observed data points are represented by purple circles, accompanied by 1$\sigma$ error bars. The bottom panel displays the relative residuals.}
\figsetgrpend

\figsetgrpstart
\figsetgrpnum{16.111}
\figsetgrptitle{M1931-ID61}
\figsetplot{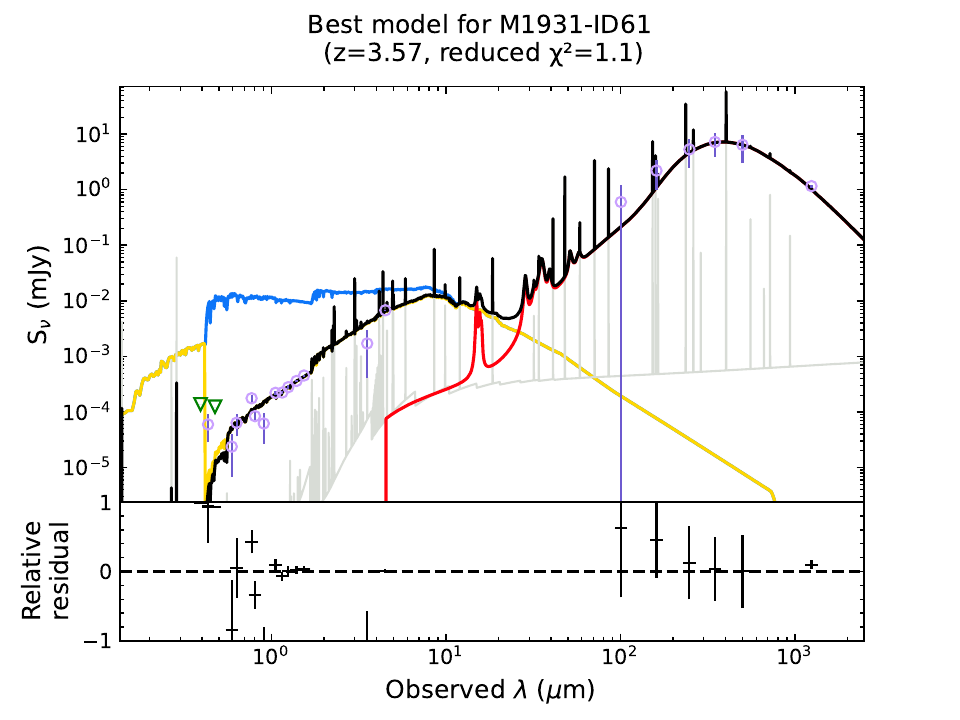}
\figsetgrpnote{The SED and the best-fit model of M1931-ID61. Lensing magnification is NOT corrected in this figure. The black solid line represents the composite spectrum of the galaxy. The yellow line illustrates the stellar emission attenuated by interstellar dust. The blue line depicts the unattenuated stellar emission for reference purpose. The orange line corresponds to the emission from an AGN. The red line shows the infrared emission from interstellar dust. The gray line denotes the nebulae emission. The observed data points are represented by purple circles, accompanied by 1$\sigma$ error bars. The bottom panel displays the relative residuals.}
\figsetgrpend

\figsetgrpstart
\figsetgrpnum{16.112}
\figsetgrptitle{M1931-ID69}
\figsetplot{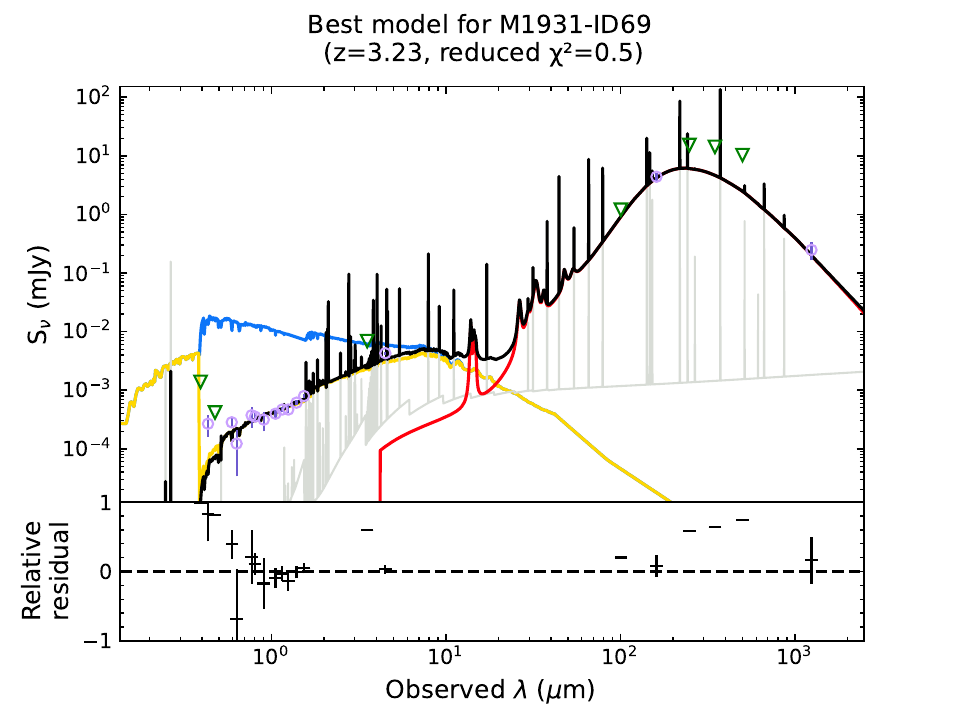}
\figsetgrpnote{The SED and the best-fit model of M1931-ID69. Lensing magnification is NOT corrected in this figure. The black solid line represents the composite spectrum of the galaxy. The yellow line illustrates the stellar emission attenuated by interstellar dust. The blue line depicts the unattenuated stellar emission for reference purpose. The orange line corresponds to the emission from an AGN. The red line shows the infrared emission from interstellar dust. The gray line denotes the nebulae emission. The observed data points are represented by purple circles, accompanied by 1$\sigma$ error bars. The bottom panel displays the relative residuals.}
\figsetgrpend

\figsetgrpstart
\figsetgrpnum{16.113}
\figsetgrptitle{M2129-ID24}
\figsetplot{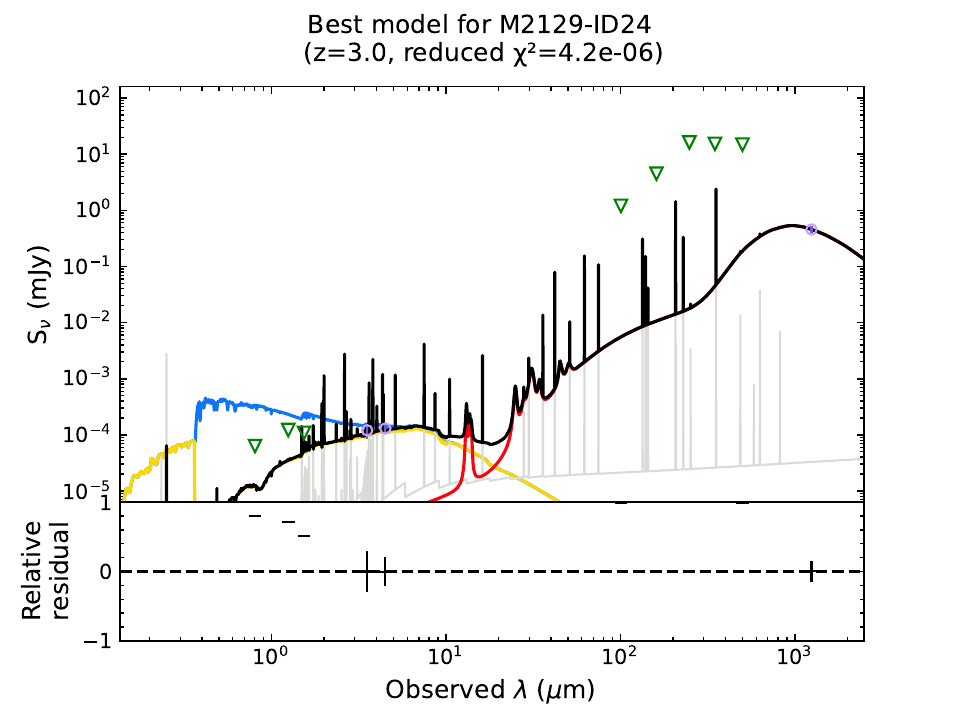}
\figsetgrpnote{The SED and the best-fit model of M2129-ID24. Lensing magnification is NOT corrected in this figure. The black solid line represents the composite spectrum of the galaxy. The yellow line illustrates the stellar emission attenuated by interstellar dust. The blue line depicts the unattenuated stellar emission for reference purpose. The orange line corresponds to the emission from an AGN. The red line shows the infrared emission from interstellar dust. The gray line denotes the nebulae emission. The observed data points are represented by purple circles, accompanied by 1$\sigma$ error bars. The bottom panel displays the relative residuals.}
\figsetgrpend

\figsetgrpstart
\figsetgrpnum{16.114}
\figsetgrptitle{M2129-ID46}
\figsetplot{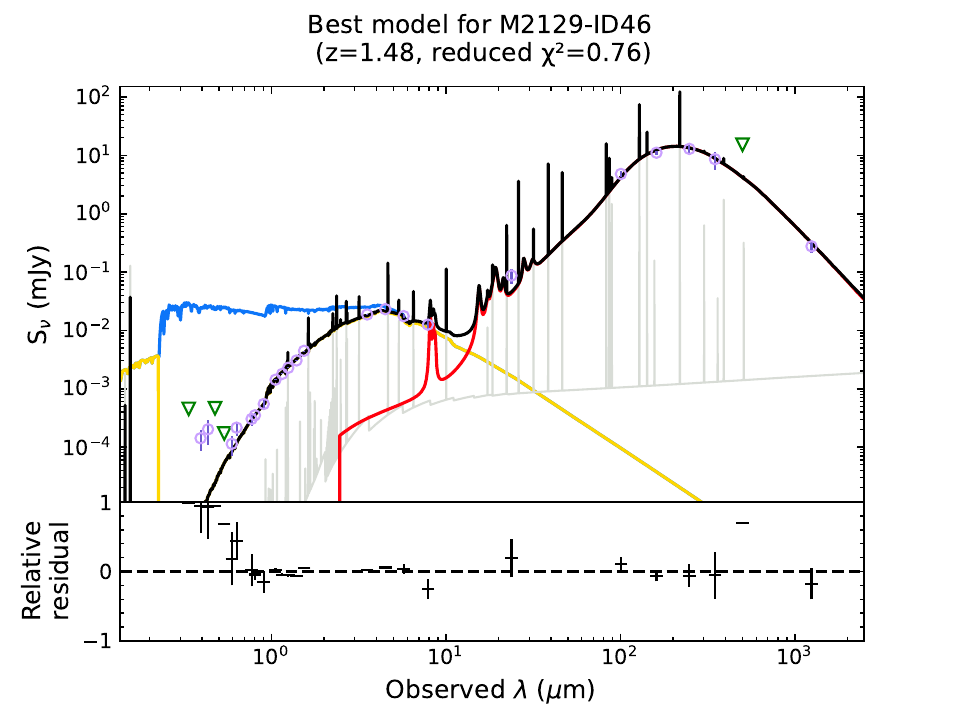}
\figsetgrpnote{The SED and the best-fit model of M2129-ID46. Lensing magnification is NOT corrected in this figure. The black solid line represents the composite spectrum of the galaxy. The yellow line illustrates the stellar emission attenuated by interstellar dust. The blue line depicts the unattenuated stellar emission for reference purpose. The orange line corresponds to the emission from an AGN. The red line shows the infrared emission from interstellar dust. The gray line denotes the nebulae emission. The observed data points are represented by purple circles, accompanied by 1$\sigma$ error bars. The bottom panel displays the relative residuals.}
\figsetgrpend

\figsetgrpstart
\figsetgrpnum{16.115}
\figsetgrptitle{M2129-ID62}
\figsetplot{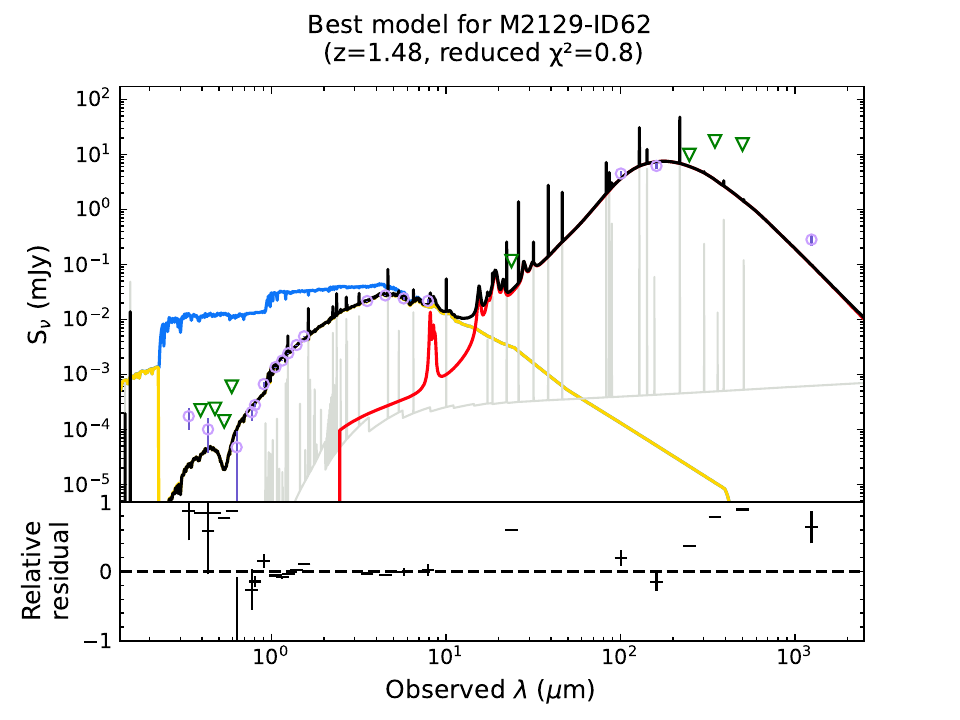}
\figsetgrpnote{The SED and the best-fit model of M2129-ID62. Lensing magnification is NOT corrected in this figure. The black solid line represents the composite spectrum of the galaxy. The yellow line illustrates the stellar emission attenuated by interstellar dust. The blue line depicts the unattenuated stellar emission for reference purpose. The orange line corresponds to the emission from an AGN. The red line shows the infrared emission from interstellar dust. The gray line denotes the nebulae emission. The observed data points are represented by purple circles, accompanied by 1$\sigma$ error bars. The bottom panel displays the relative residuals.}
\figsetgrpend

\figsetgrpstart
\figsetgrpnum{16.116}
\figsetgrptitle{P171-ID161}
\figsetplot{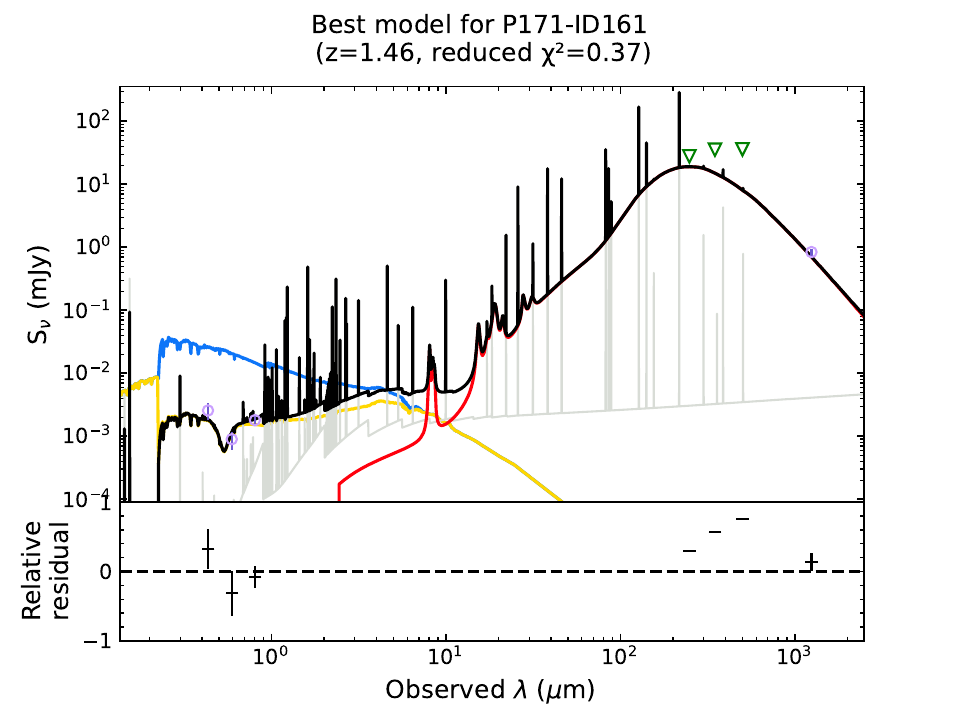}
\figsetgrpnote{The SED and the best-fit model of P171-ID161. Lensing magnification is NOT corrected in this figure. The black solid line represents the composite spectrum of the galaxy. The yellow line illustrates the stellar emission attenuated by interstellar dust. The blue line depicts the unattenuated stellar emission for reference purpose. The orange line corresponds to the emission from an AGN. The red line shows the infrared emission from interstellar dust. The gray line denotes the nebulae emission. The observed data points are represented by purple circles, accompanied by 1$\sigma$ error bars. The bottom panel displays the relative residuals.}
\figsetgrpend

\figsetgrpstart
\figsetgrpnum{16.117}
\figsetgrptitle{P171-ID162}
\figsetplot{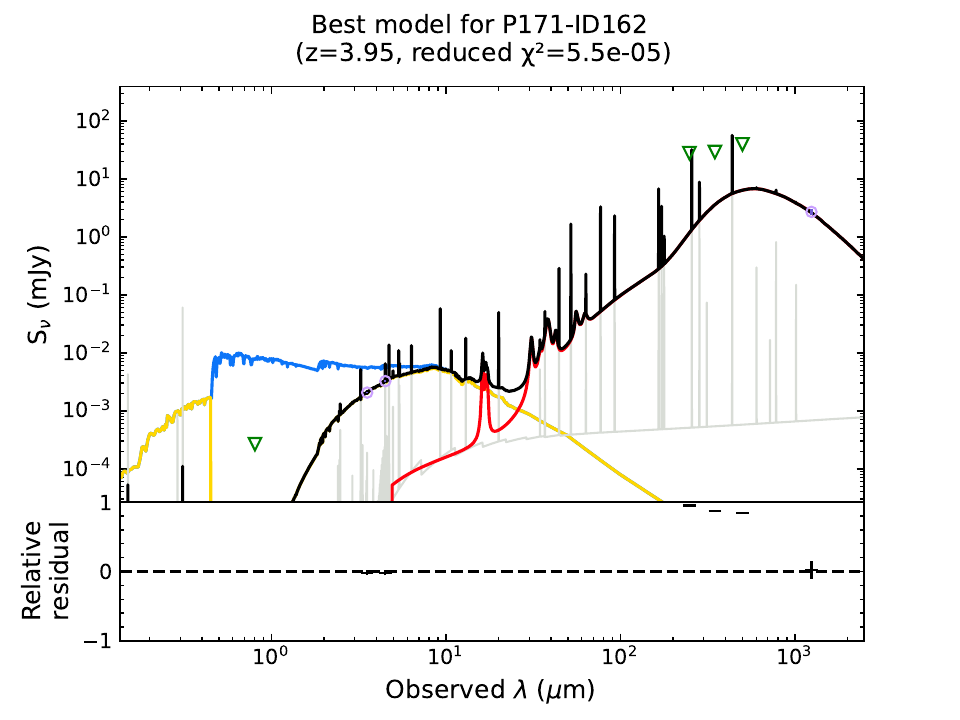}
\figsetgrpnote{The SED and the best-fit model of P171-ID162. Lensing magnification is NOT corrected in this figure. The black solid line represents the composite spectrum of the galaxy. The yellow line illustrates the stellar emission attenuated by interstellar dust. The blue line depicts the unattenuated stellar emission for reference purpose. The orange line corresponds to the emission from an AGN. The red line shows the infrared emission from interstellar dust. The gray line denotes the nebulae emission. The observed data points are represented by purple circles, accompanied by 1$\sigma$ error bars. The bottom panel displays the relative residuals.}
\figsetgrpend

\figsetgrpstart
\figsetgrpnum{16.118}
\figsetgrptitle{P171-ID177}
\figsetplot{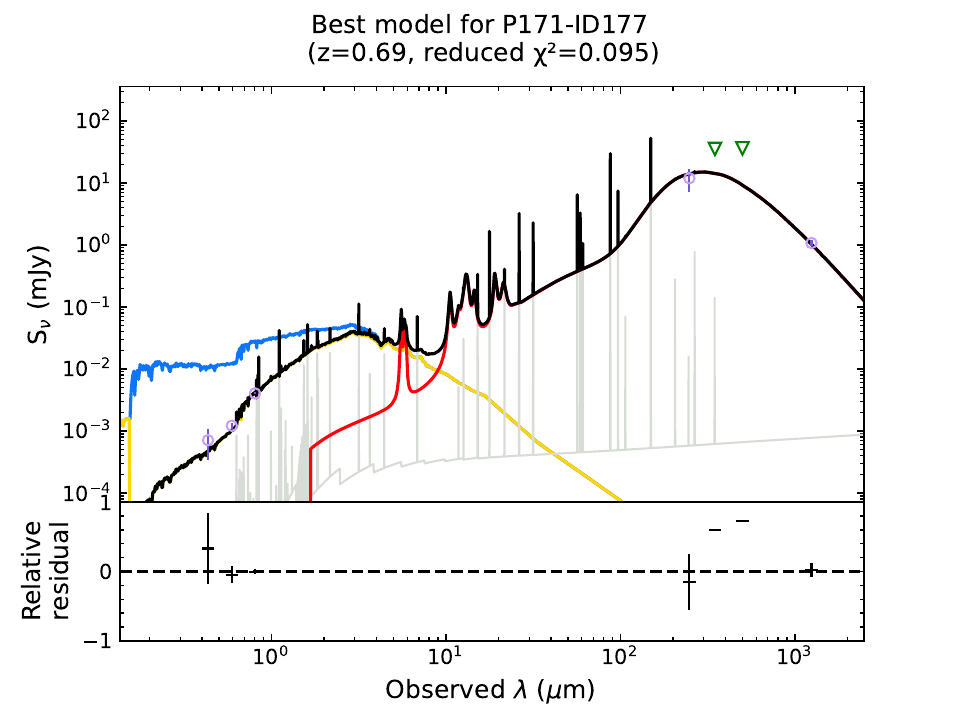}
\figsetgrpnote{The SED and the best-fit model of P171-ID177. Lensing magnification is NOT corrected in this figure. The black solid line represents the composite spectrum of the galaxy. The yellow line illustrates the stellar emission attenuated by interstellar dust. The blue line depicts the unattenuated stellar emission for reference purpose. The orange line corresponds to the emission from an AGN. The red line shows the infrared emission from interstellar dust. The gray line denotes the nebulae emission. The observed data points are represented by purple circles, accompanied by 1$\sigma$ error bars. The bottom panel displays the relative residuals.}
\figsetgrpend

\figsetgrpstart
\figsetgrpnum{16.119}
\figsetgrptitle{P171-ID69}
\figsetplot{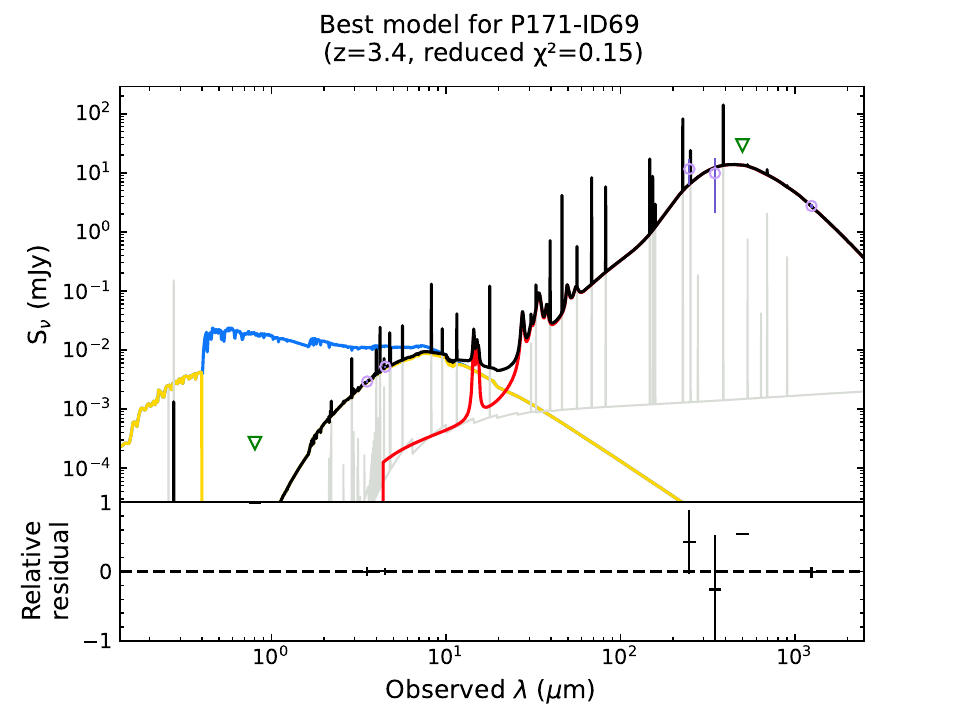}
\figsetgrpnote{The SED and the best-fit model of P171-ID69. Lensing magnification is NOT corrected in this figure. The black solid line represents the composite spectrum of the galaxy. The yellow line illustrates the stellar emission attenuated by interstellar dust. The blue line depicts the unattenuated stellar emission for reference purpose. The orange line corresponds to the emission from an AGN. The red line shows the infrared emission from interstellar dust. The gray line denotes the nebulae emission. The observed data points are represented by purple circles, accompanied by 1$\sigma$ error bars. The bottom panel displays the relative residuals.}
\figsetgrpend

\figsetgrpstart
\figsetgrpnum{16.120}
\figsetgrptitle{R0032-ID127}
\figsetplot{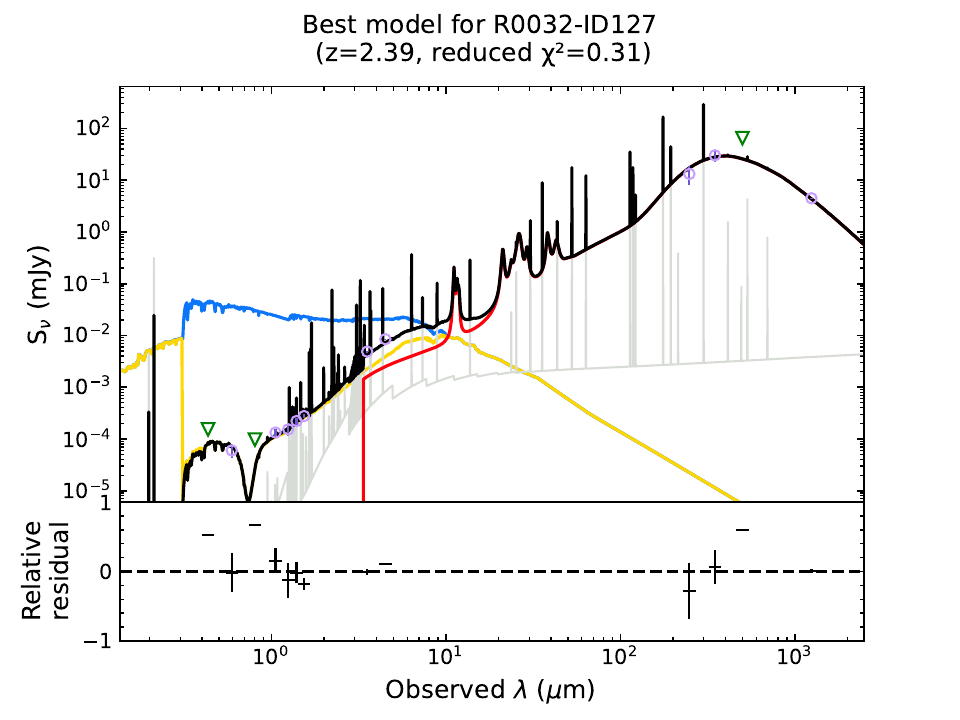}
\figsetgrpnote{The SED and the best-fit model of R0032-ID127. Lensing magnification is NOT corrected in this figure. The black solid line represents the composite spectrum of the galaxy. The yellow line illustrates the stellar emission attenuated by interstellar dust. The blue line depicts the unattenuated stellar emission for reference purpose. The orange line corresponds to the emission from an AGN. The red line shows the infrared emission from interstellar dust. The gray line denotes the nebulae emission. The observed data points are represented by purple circles, accompanied by 1$\sigma$ error bars. The bottom panel displays the relative residuals.}
\figsetgrpend

\figsetgrpstart
\figsetgrpnum{16.121}
\figsetgrptitle{R0032-ID131}
\figsetplot{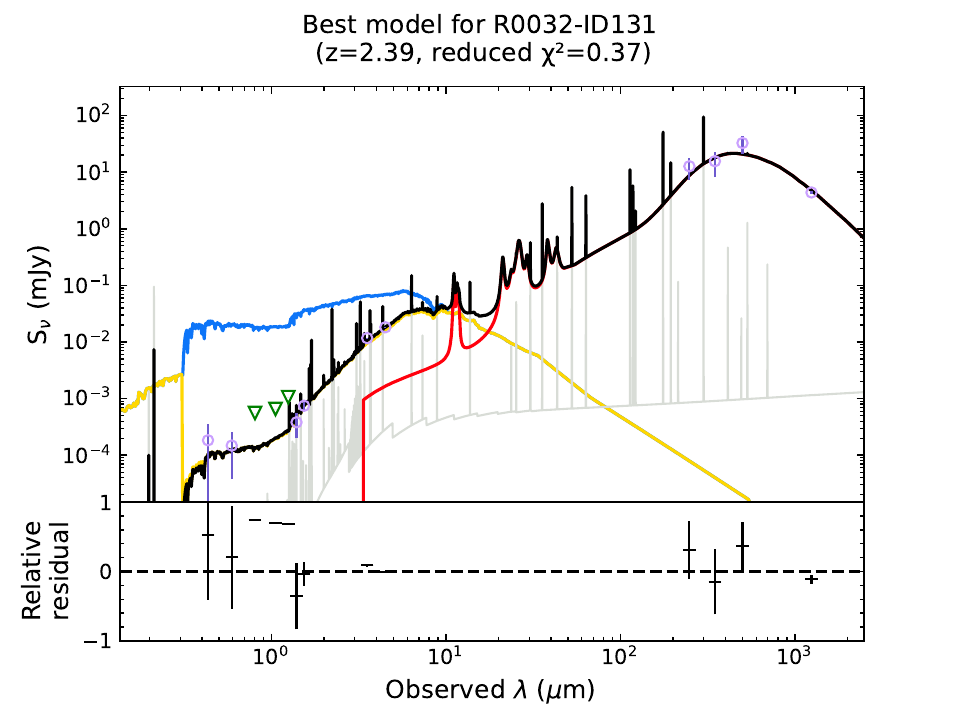}
\figsetgrpnote{The SED and the best-fit model of R0032-ID131. Lensing magnification is NOT corrected in this figure. The black solid line represents the composite spectrum of the galaxy. The yellow line illustrates the stellar emission attenuated by interstellar dust. The blue line depicts the unattenuated stellar emission for reference purpose. The orange line corresponds to the emission from an AGN. The red line shows the infrared emission from interstellar dust. The gray line denotes the nebulae emission. The observed data points are represented by purple circles, accompanied by 1$\sigma$ error bars. The bottom panel displays the relative residuals.}
\figsetgrpend

\figsetgrpstart
\figsetgrpnum{16.122}
\figsetgrptitle{R0032-ID198}
\figsetplot{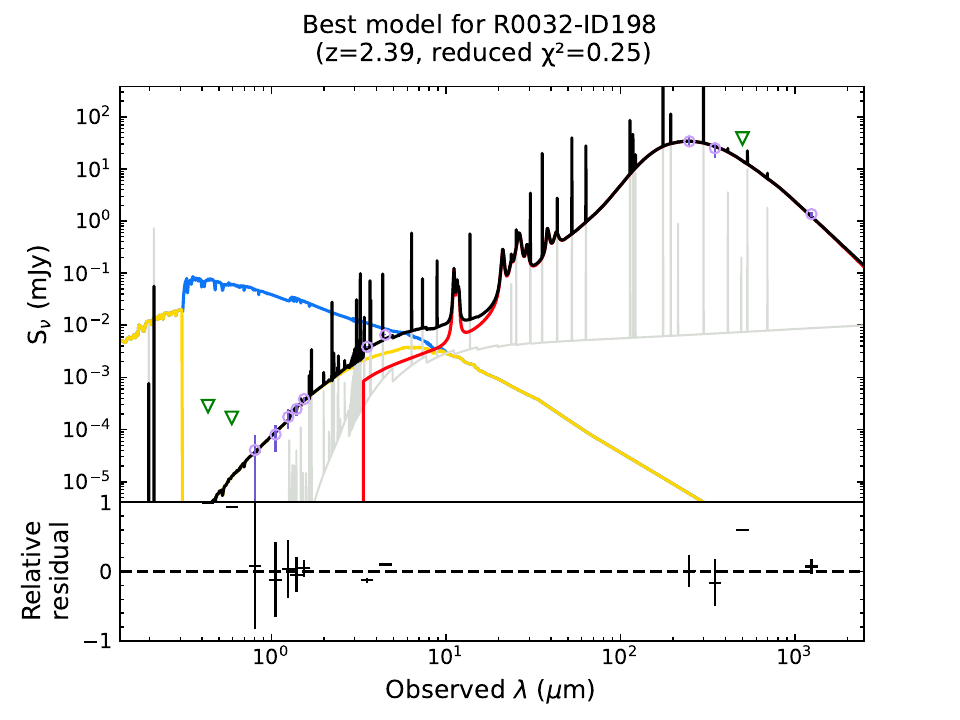}
\figsetgrpnote{The SED and the best-fit model of R0032-ID198. Lensing magnification is NOT corrected in this figure. The black solid line represents the composite spectrum of the galaxy. The yellow line illustrates the stellar emission attenuated by interstellar dust. The blue line depicts the unattenuated stellar emission for reference purpose. The orange line corresponds to the emission from an AGN. The red line shows the infrared emission from interstellar dust. The gray line denotes the nebulae emission. The observed data points are represented by purple circles, accompanied by 1$\sigma$ error bars. The bottom panel displays the relative residuals.}
\figsetgrpend

\figsetgrpstart
\figsetgrpnum{16.123}
\figsetgrptitle{R0032-ID208}
\figsetplot{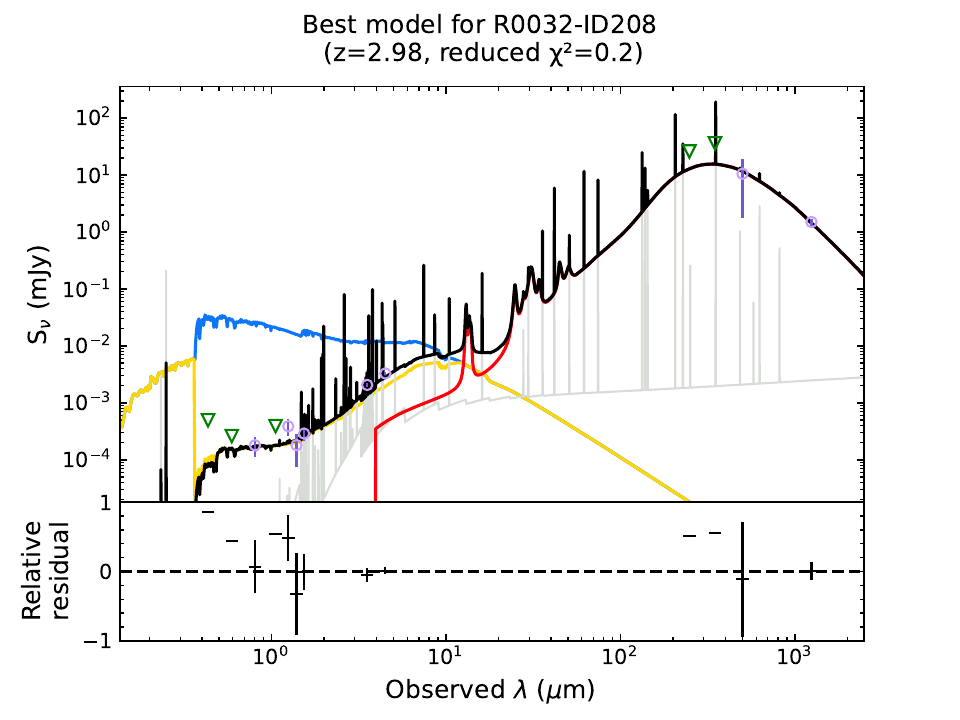}
\figsetgrpnote{The SED and the best-fit model of R0032-ID208. Lensing magnification is NOT corrected in this figure. The black solid line represents the composite spectrum of the galaxy. The yellow line illustrates the stellar emission attenuated by interstellar dust. The blue line depicts the unattenuated stellar emission for reference purpose. The orange line corresponds to the emission from an AGN. The red line shows the infrared emission from interstellar dust. The gray line denotes the nebulae emission. The observed data points are represented by purple circles, accompanied by 1$\sigma$ error bars. The bottom panel displays the relative residuals.}
\figsetgrpend

\figsetgrpstart
\figsetgrpnum{16.124}
\figsetgrptitle{R0032-ID220}
\figsetplot{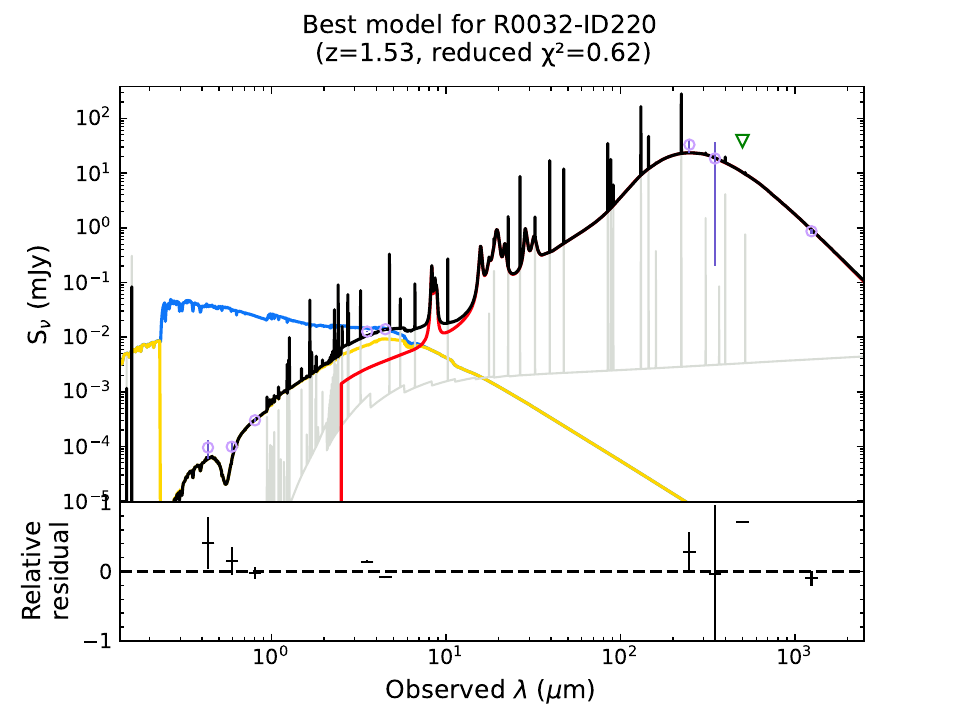}
\figsetgrpnote{The SED and the best-fit model of R0032-ID220. Lensing magnification is NOT corrected in this figure. The black solid line represents the composite spectrum of the galaxy. The yellow line illustrates the stellar emission attenuated by interstellar dust. The blue line depicts the unattenuated stellar emission for reference purpose. The orange line corresponds to the emission from an AGN. The red line shows the infrared emission from interstellar dust. The gray line denotes the nebulae emission. The observed data points are represented by purple circles, accompanied by 1$\sigma$ error bars. The bottom panel displays the relative residuals.}
\figsetgrpend

\figsetgrpstart
\figsetgrpnum{16.125}
\figsetgrptitle{R0032-ID238}
\figsetplot{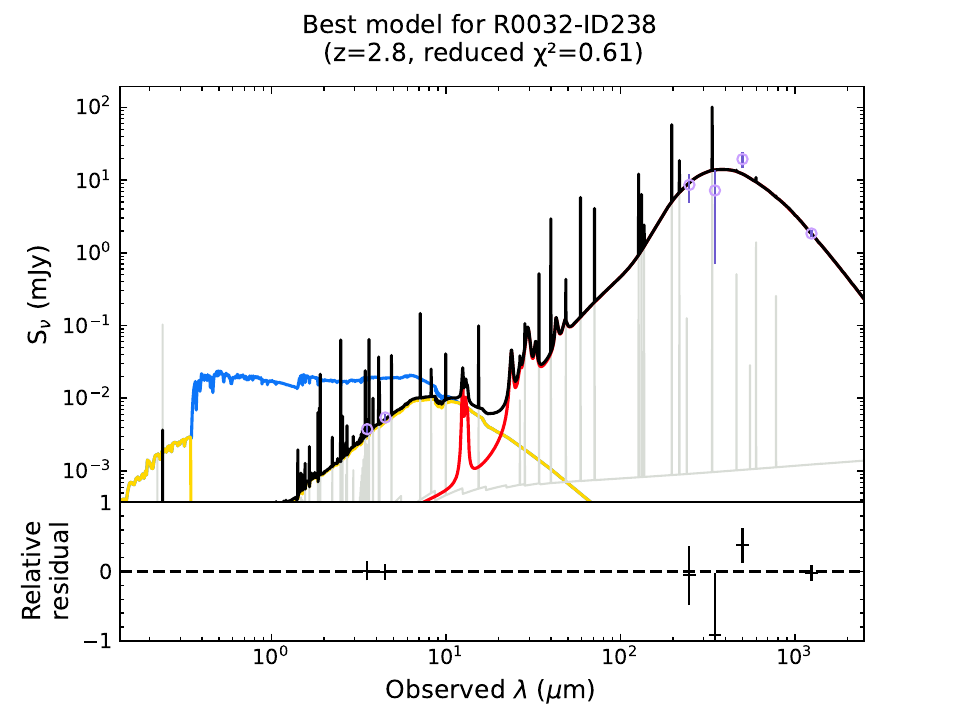}
\figsetgrpnote{The SED and the best-fit model of R0032-ID238. Lensing magnification is NOT corrected in this figure. The black solid line represents the composite spectrum of the galaxy. The yellow line illustrates the stellar emission attenuated by interstellar dust. The blue line depicts the unattenuated stellar emission for reference purpose. The orange line corresponds to the emission from an AGN. The red line shows the infrared emission from interstellar dust. The gray line denotes the nebulae emission. The observed data points are represented by purple circles, accompanied by 1$\sigma$ error bars. The bottom panel displays the relative residuals.}
\figsetgrpend

\figsetgrpstart
\figsetgrpnum{16.126}
\figsetgrptitle{R0032-ID245}
\figsetplot{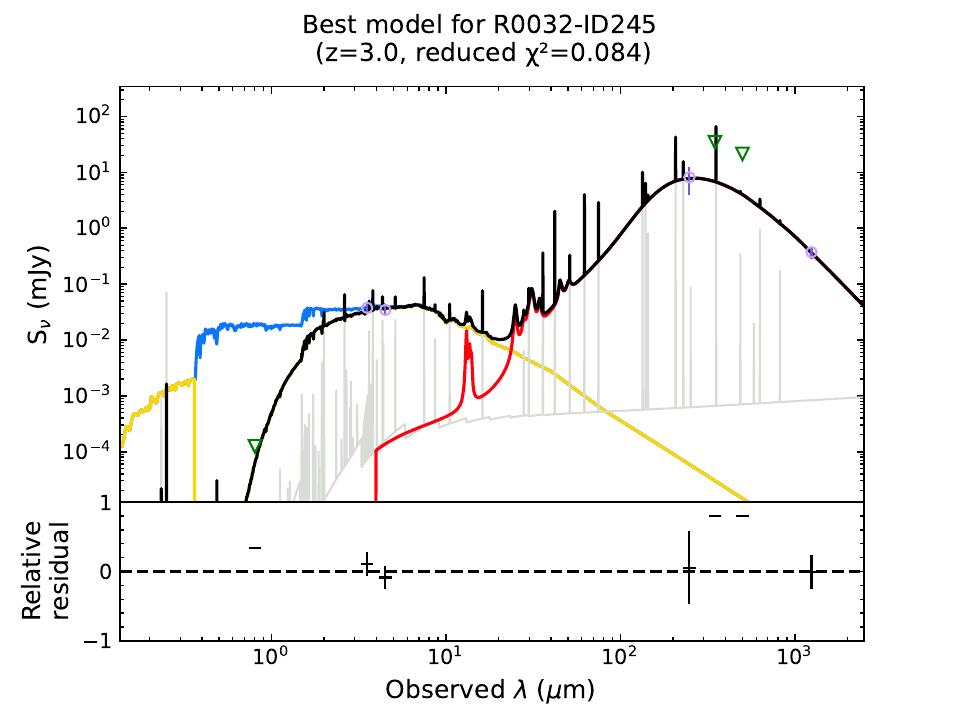}
\figsetgrpnote{The SED and the best-fit model of R0032-ID245. Lensing magnification is NOT corrected in this figure. The black solid line represents the composite spectrum of the galaxy. The yellow line illustrates the stellar emission attenuated by interstellar dust. The blue line depicts the unattenuated stellar emission for reference purpose. The orange line corresponds to the emission from an AGN. The red line shows the infrared emission from interstellar dust. The gray line denotes the nebulae emission. The observed data points are represented by purple circles, accompanied by 1$\sigma$ error bars. The bottom panel displays the relative residuals.}
\figsetgrpend

\figsetgrpstart
\figsetgrpnum{16.127}
\figsetgrptitle{R0032-ID250}
\figsetplot{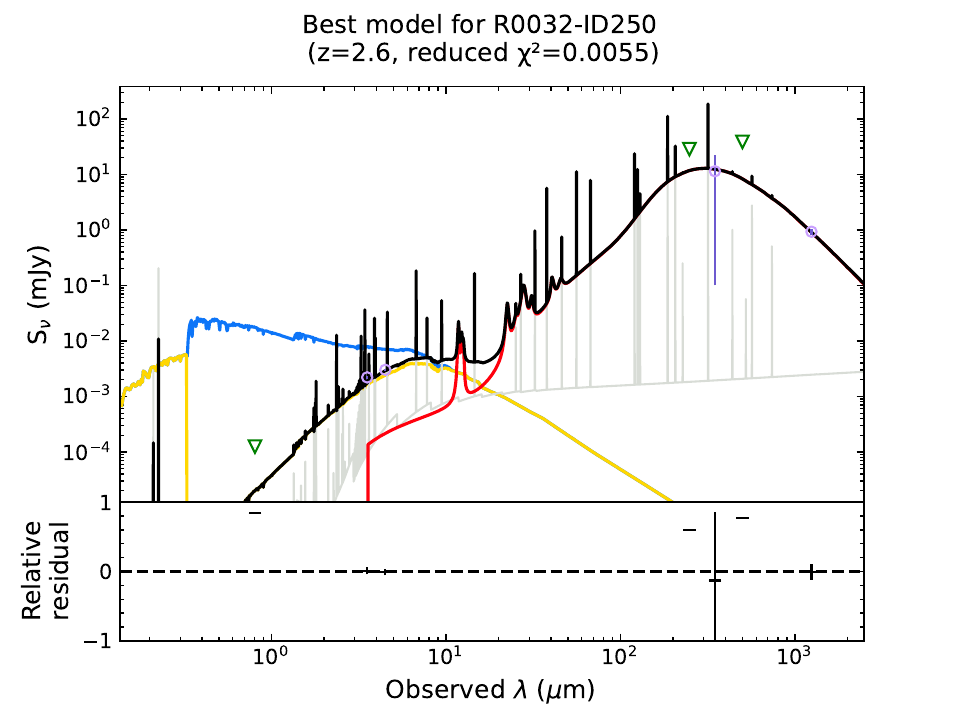}
\figsetgrpnote{The SED and the best-fit model of R0032-ID250. Lensing magnification is NOT corrected in this figure. The black solid line represents the composite spectrum of the galaxy. The yellow line illustrates the stellar emission attenuated by interstellar dust. The blue line depicts the unattenuated stellar emission for reference purpose. The orange line corresponds to the emission from an AGN. The red line shows the infrared emission from interstellar dust. The gray line denotes the nebulae emission. The observed data points are represented by purple circles, accompanied by 1$\sigma$ error bars. The bottom panel displays the relative residuals.}
\figsetgrpend

\figsetgrpstart
\figsetgrpnum{16.128}
\figsetgrptitle{R0032-ID276}
\figsetplot{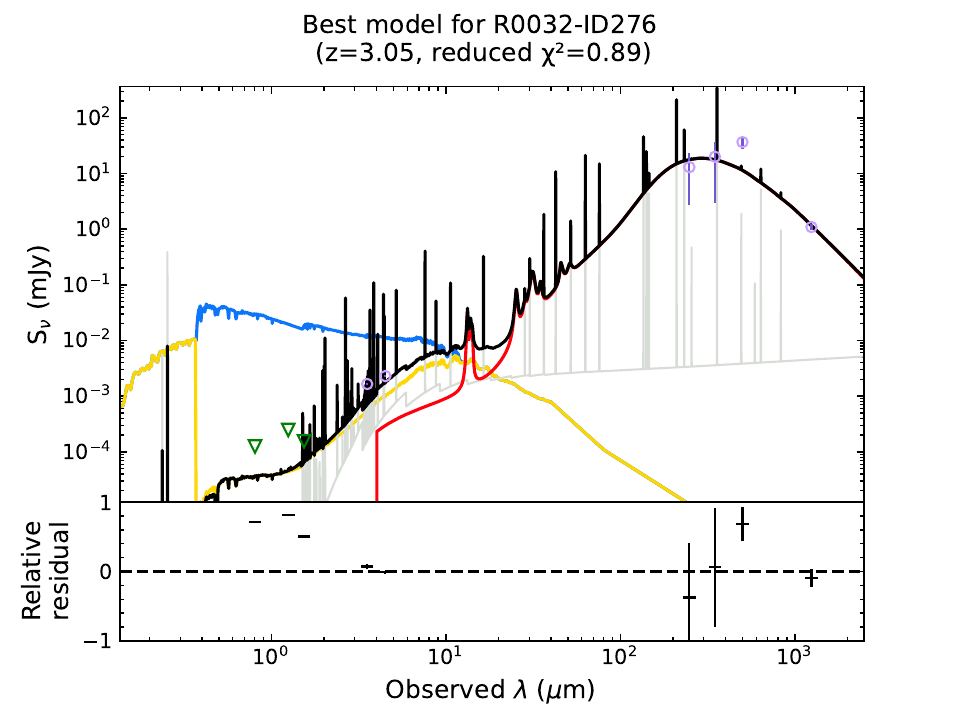}
\figsetgrpnote{The SED and the best-fit model of R0032-ID276. Lensing magnification is NOT corrected in this figure. The black solid line represents the composite spectrum of the galaxy. The yellow line illustrates the stellar emission attenuated by interstellar dust. The blue line depicts the unattenuated stellar emission for reference purpose. The orange line corresponds to the emission from an AGN. The red line shows the infrared emission from interstellar dust. The gray line denotes the nebulae emission. The observed data points are represented by purple circles, accompanied by 1$\sigma$ error bars. The bottom panel displays the relative residuals.}
\figsetgrpend

\figsetgrpstart
\figsetgrpnum{16.129}
\figsetgrptitle{R0032-ID281}
\figsetplot{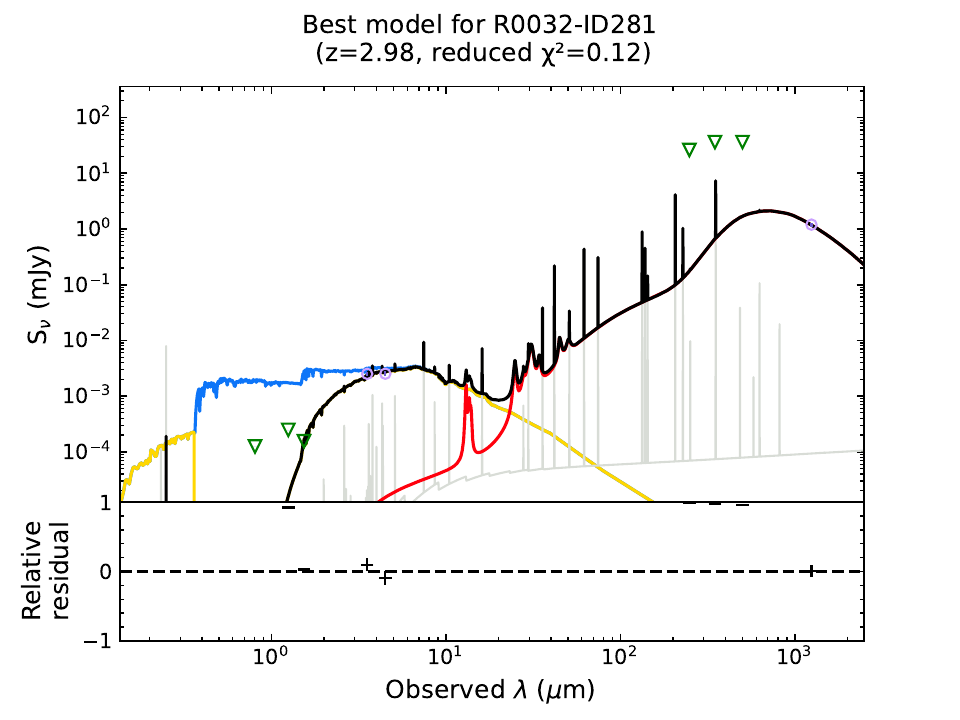}
\figsetgrpnote{The SED and the best-fit model of R0032-ID281. Lensing magnification is NOT corrected in this figure. The black solid line represents the composite spectrum of the galaxy. The yellow line illustrates the stellar emission attenuated by interstellar dust. The blue line depicts the unattenuated stellar emission for reference purpose. The orange line corresponds to the emission from an AGN. The red line shows the infrared emission from interstellar dust. The gray line denotes the nebulae emission. The observed data points are represented by purple circles, accompanied by 1$\sigma$ error bars. The bottom panel displays the relative residuals.}
\figsetgrpend

\figsetgrpstart
\figsetgrpnum{16.130}
\figsetgrptitle{R0032-ID287}
\figsetplot{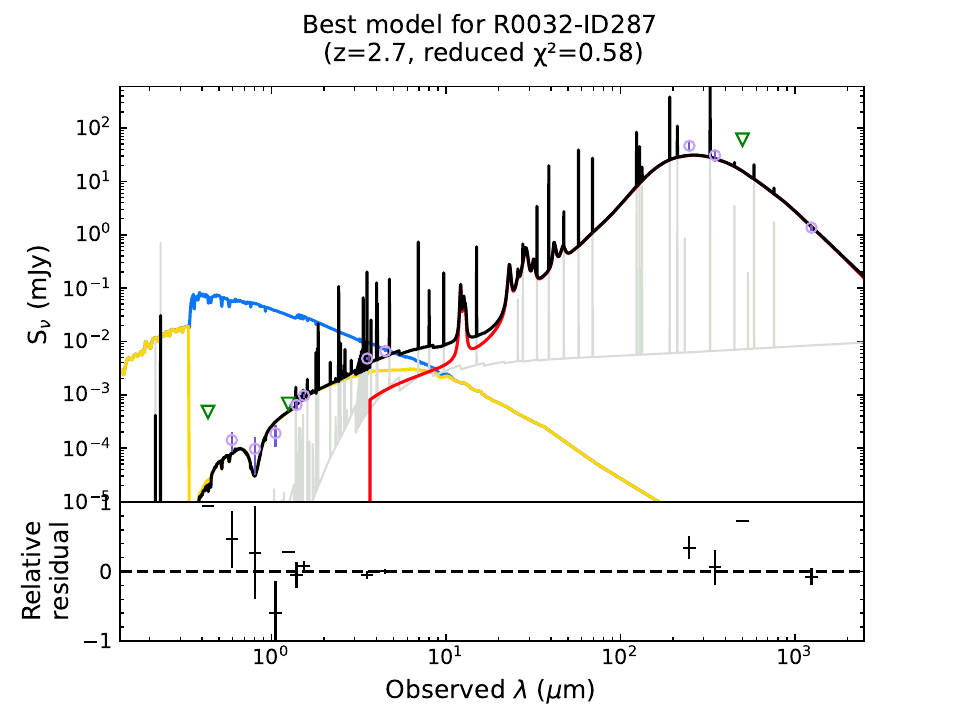}
\figsetgrpnote{The SED and the best-fit model of R0032-ID287. Lensing magnification is NOT corrected in this figure. The black solid line represents the composite spectrum of the galaxy. The yellow line illustrates the stellar emission attenuated by interstellar dust. The blue line depicts the unattenuated stellar emission for reference purpose. The orange line corresponds to the emission from an AGN. The red line shows the infrared emission from interstellar dust. The gray line denotes the nebulae emission. The observed data points are represented by purple circles, accompanied by 1$\sigma$ error bars. The bottom panel displays the relative residuals.}
\figsetgrpend

\figsetgrpstart
\figsetgrpnum{16.131}
\figsetgrptitle{R0032-ID304}
\figsetplot{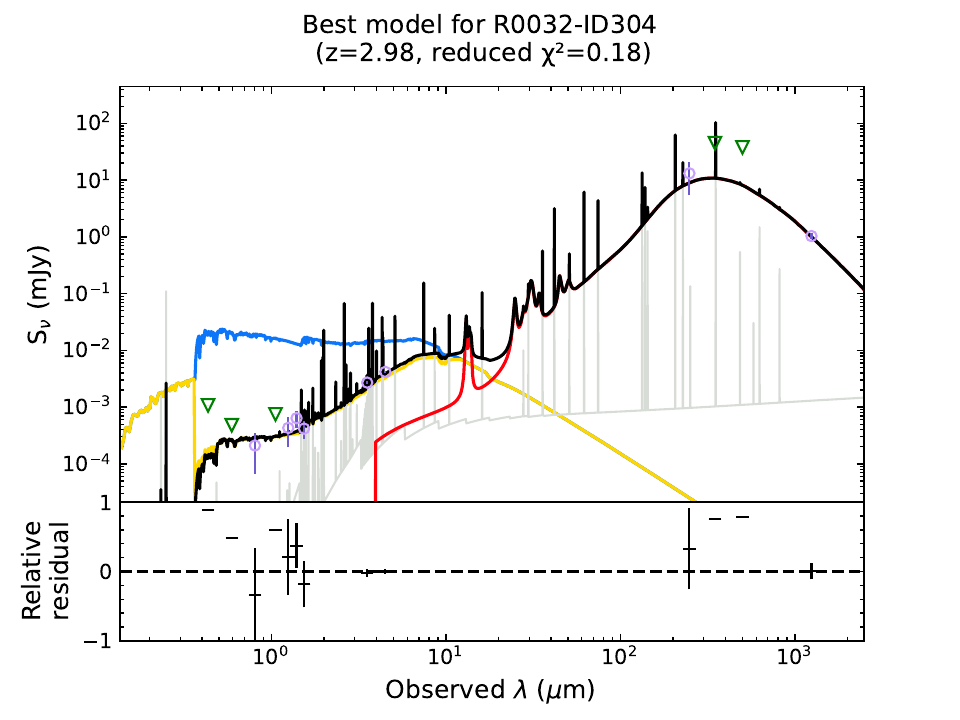}
\figsetgrpnote{The SED and the best-fit model of R0032-ID304. Lensing magnification is NOT corrected in this figure. The black solid line represents the composite spectrum of the galaxy. The yellow line illustrates the stellar emission attenuated by interstellar dust. The blue line depicts the unattenuated stellar emission for reference purpose. The orange line corresponds to the emission from an AGN. The red line shows the infrared emission from interstellar dust. The gray line denotes the nebulae emission. The observed data points are represented by purple circles, accompanied by 1$\sigma$ error bars. The bottom panel displays the relative residuals.}
\figsetgrpend

\figsetgrpstart
\figsetgrpnum{16.132}
\figsetgrptitle{R0032-ID32}
\figsetplot{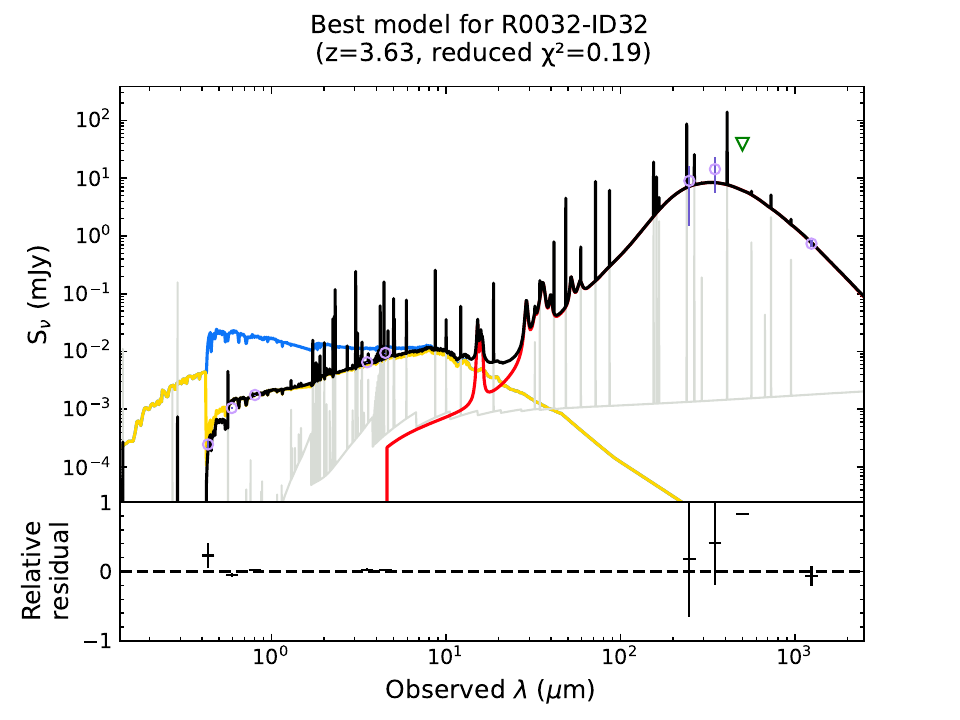}
\figsetgrpnote{The SED and the best-fit model of R0032-ID32. Lensing magnification is NOT corrected in this figure. The black solid line represents the composite spectrum of the galaxy. The yellow line illustrates the stellar emission attenuated by interstellar dust. The blue line depicts the unattenuated stellar emission for reference purpose. The orange line corresponds to the emission from an AGN. The red line shows the infrared emission from interstellar dust. The gray line denotes the nebulae emission. The observed data points are represented by purple circles, accompanied by 1$\sigma$ error bars. The bottom panel displays the relative residuals.}
\figsetgrpend

\figsetgrpstart
\figsetgrpnum{16.133}
\figsetgrptitle{R0032-ID53}
\figsetplot{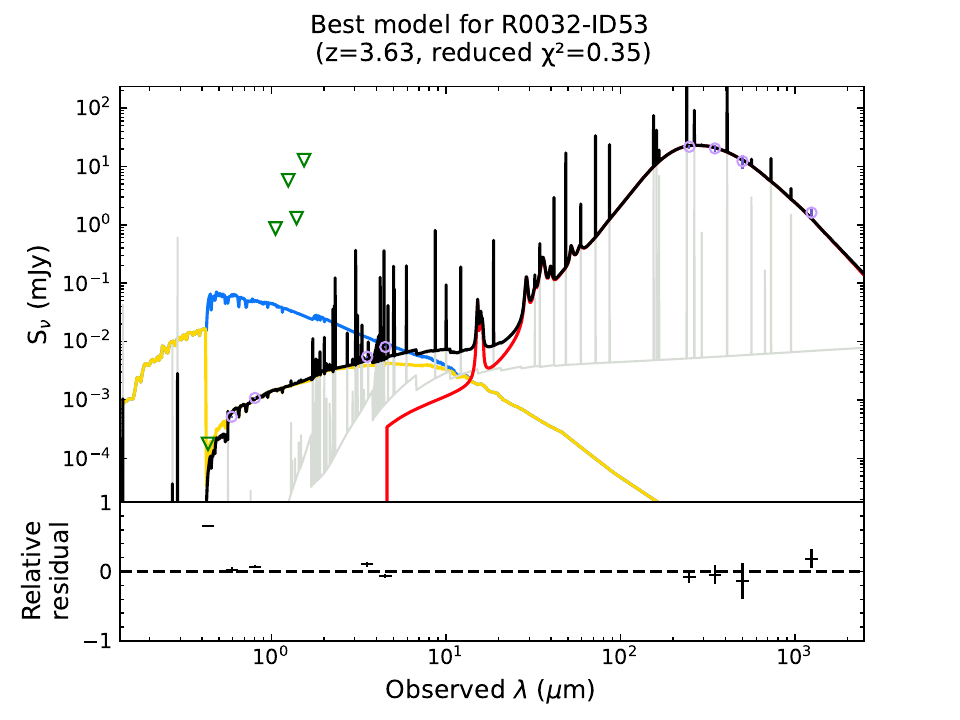}
\figsetgrpnote{The SED and the best-fit model of R0032-ID53. Lensing magnification is NOT corrected in this figure. The black solid line represents the composite spectrum of the galaxy. The yellow line illustrates the stellar emission attenuated by interstellar dust. The blue line depicts the unattenuated stellar emission for reference purpose. The orange line corresponds to the emission from an AGN. The red line shows the infrared emission from interstellar dust. The gray line denotes the nebulae emission. The observed data points are represented by purple circles, accompanied by 1$\sigma$ error bars. The bottom panel displays the relative residuals.}
\figsetgrpend

\figsetgrpstart
\figsetgrpnum{16.134}
\figsetgrptitle{R0032-ID55}
\figsetplot{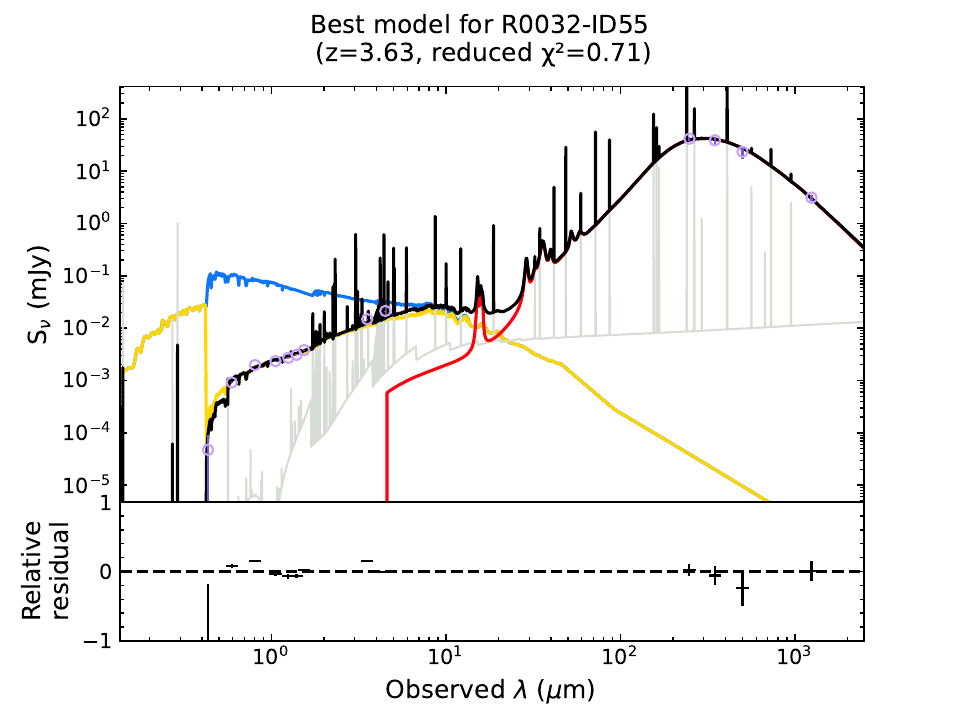}
\figsetgrpnote{The SED and the best-fit model of R0032-ID55. Lensing magnification is NOT corrected in this figure. The black solid line represents the composite spectrum of the galaxy. The yellow line illustrates the stellar emission attenuated by interstellar dust. The blue line depicts the unattenuated stellar emission for reference purpose. The orange line corresponds to the emission from an AGN. The red line shows the infrared emission from interstellar dust. The gray line denotes the nebulae emission. The observed data points are represented by purple circles, accompanied by 1$\sigma$ error bars. The bottom panel displays the relative residuals.}
\figsetgrpend

\figsetgrpstart
\figsetgrpnum{16.135}
\figsetgrptitle{R0032-ID57}
\figsetplot{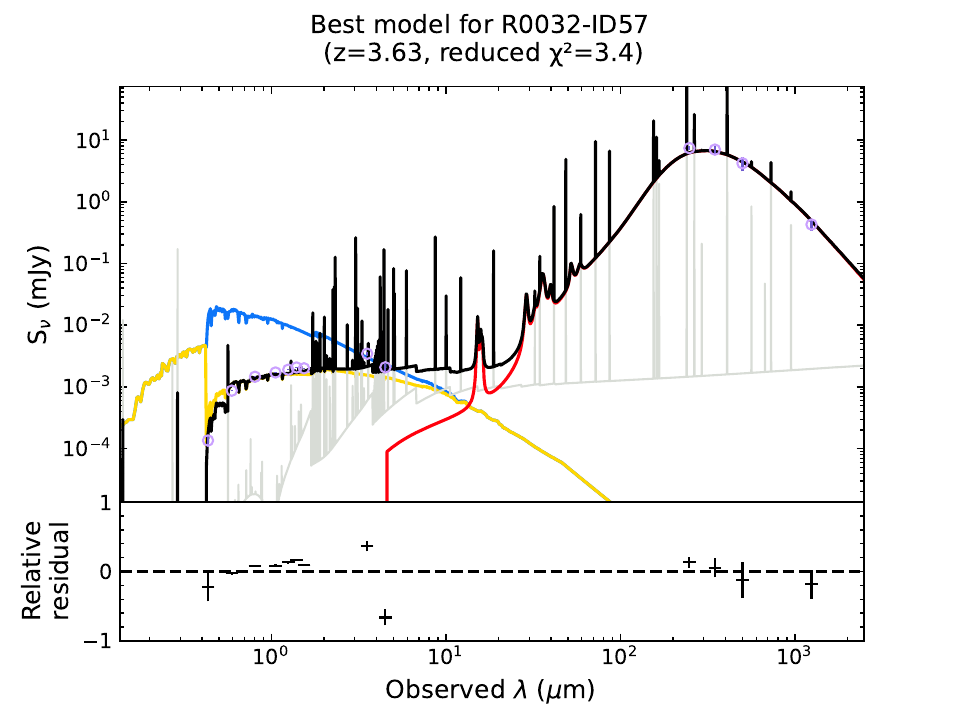}
\figsetgrpnote{The SED and the best-fit model of R0032-ID57. Lensing magnification is NOT corrected in this figure. The black solid line represents the composite spectrum of the galaxy. The yellow line illustrates the stellar emission attenuated by interstellar dust. The blue line depicts the unattenuated stellar emission for reference purpose. The orange line corresponds to the emission from an AGN. The red line shows the infrared emission from interstellar dust. The gray line denotes the nebulae emission. The observed data points are represented by purple circles, accompanied by 1$\sigma$ error bars. The bottom panel displays the relative residuals.}
\figsetgrpend

\figsetgrpstart
\figsetgrpnum{16.136}
\figsetgrptitle{R0032-ID58}
\figsetplot{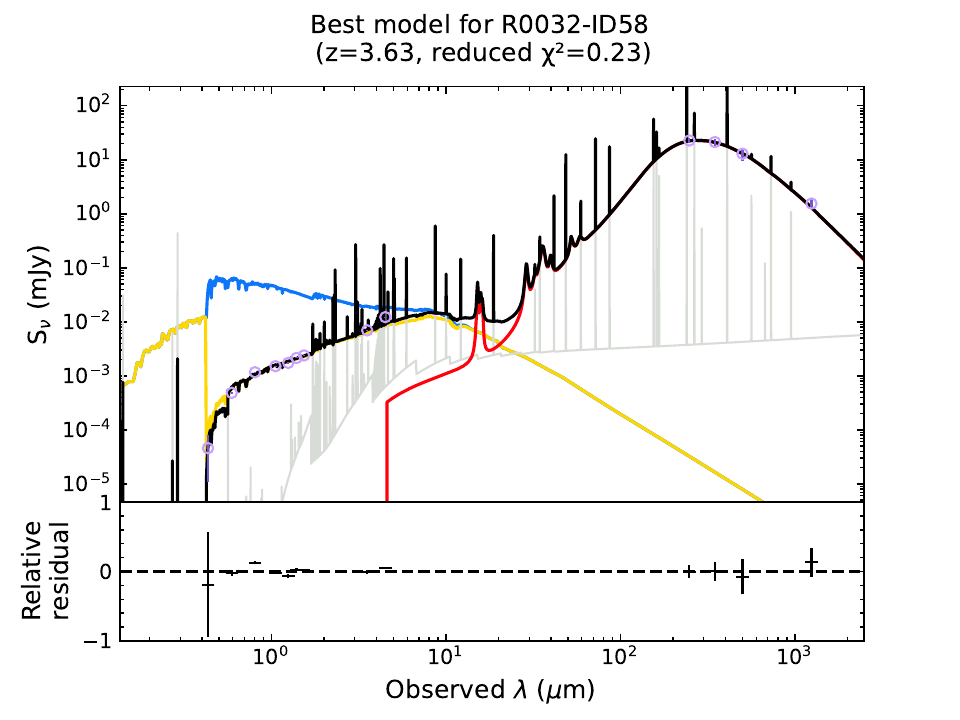}
\figsetgrpnote{The SED and the best-fit model of R0032-ID58. Lensing magnification is NOT corrected in this figure. The black solid line represents the composite spectrum of the galaxy. The yellow line illustrates the stellar emission attenuated by interstellar dust. The blue line depicts the unattenuated stellar emission for reference purpose. The orange line corresponds to the emission from an AGN. The red line shows the infrared emission from interstellar dust. The gray line denotes the nebulae emission. The observed data points are represented by purple circles, accompanied by 1$\sigma$ error bars. The bottom panel displays the relative residuals.}
\figsetgrpend

\figsetgrpstart
\figsetgrpnum{16.137}
\figsetgrptitle{R0032-ID63}
\figsetplot{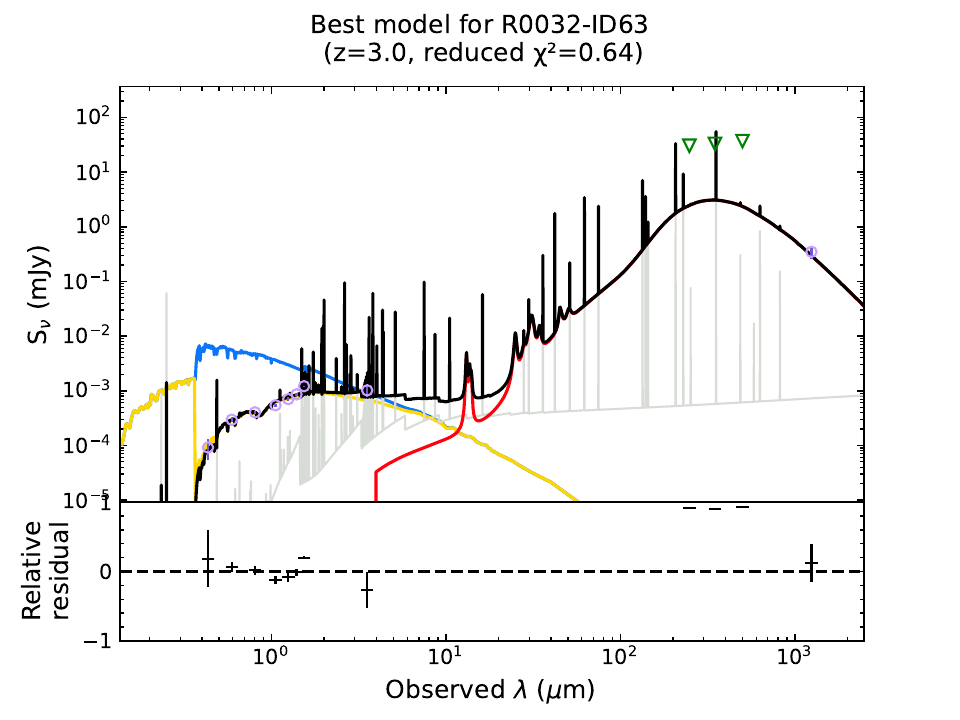}
\figsetgrpnote{The SED and the best-fit model of R0032-ID63. Lensing magnification is NOT corrected in this figure. The black solid line represents the composite spectrum of the galaxy. The yellow line illustrates the stellar emission attenuated by interstellar dust. The blue line depicts the unattenuated stellar emission for reference purpose. The orange line corresponds to the emission from an AGN. The red line shows the infrared emission from interstellar dust. The gray line denotes the nebulae emission. The observed data points are represented by purple circles, accompanied by 1$\sigma$ error bars. The bottom panel displays the relative residuals.}
\figsetgrpend

\figsetgrpstart
\figsetgrpnum{16.138}
\figsetgrptitle{R0032-ID81}
\figsetplot{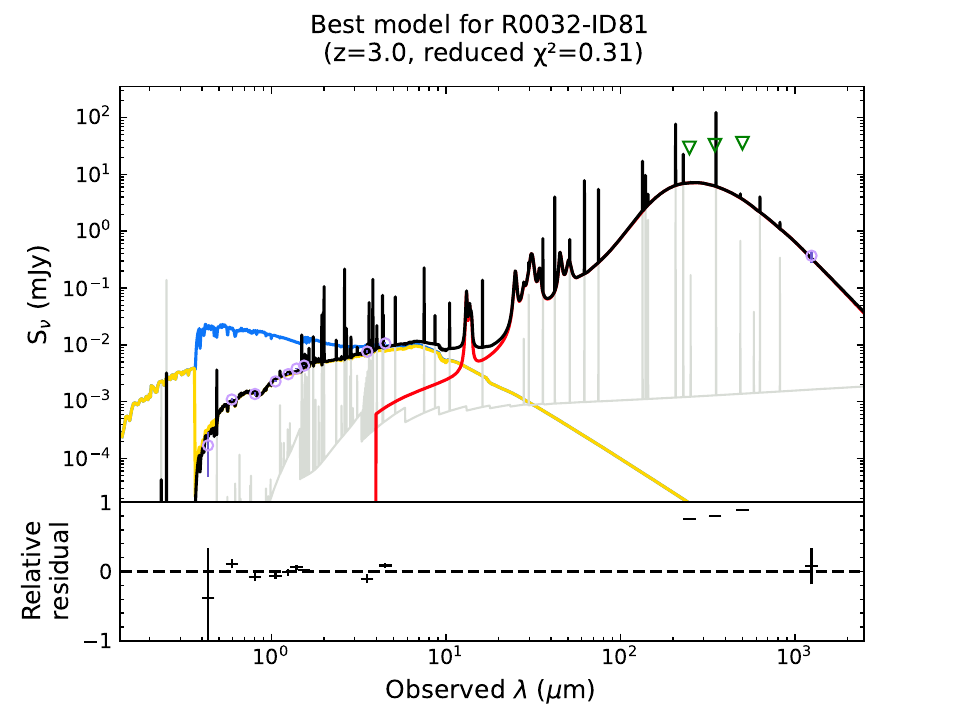}
\figsetgrpnote{The SED and the best-fit model of R0032-ID81. Lensing magnification is NOT corrected in this figure. The black solid line represents the composite spectrum of the galaxy. The yellow line illustrates the stellar emission attenuated by interstellar dust. The blue line depicts the unattenuated stellar emission for reference purpose. The orange line corresponds to the emission from an AGN. The red line shows the infrared emission from interstellar dust. The gray line denotes the nebulae emission. The observed data points are represented by purple circles, accompanied by 1$\sigma$ error bars. The bottom panel displays the relative residuals.}
\figsetgrpend

\figsetgrpstart
\figsetgrpnum{16.139}
\figsetgrptitle{R0600-ID111}
\figsetplot{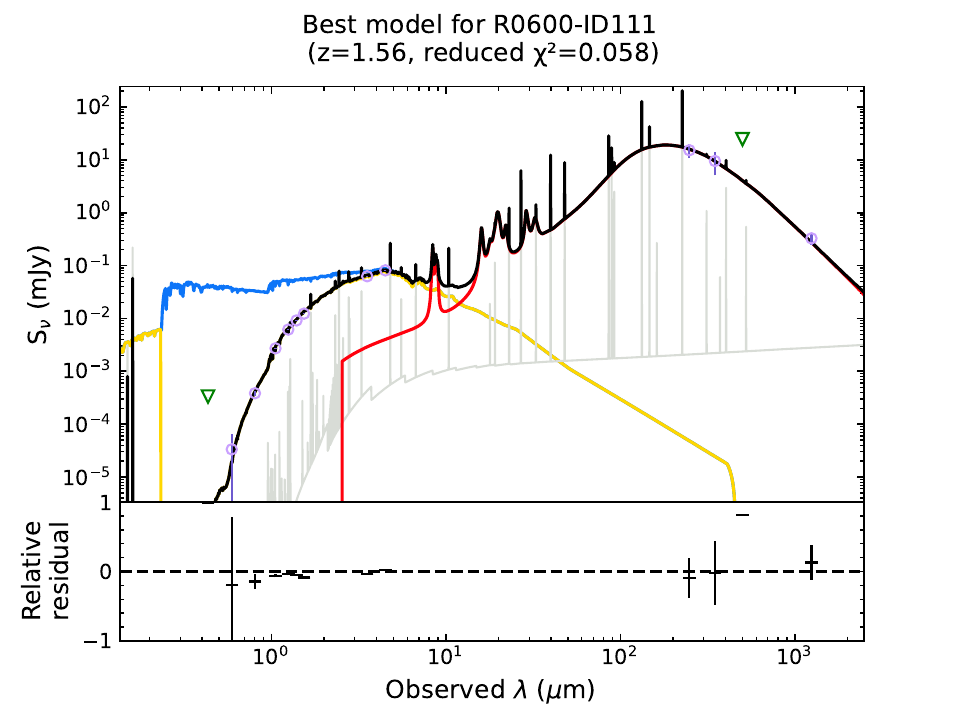}
\figsetgrpnote{The SED and the best-fit model of R0600-ID111. Lensing magnification is NOT corrected in this figure. The black solid line represents the composite spectrum of the galaxy. The yellow line illustrates the stellar emission attenuated by interstellar dust. The blue line depicts the unattenuated stellar emission for reference purpose. The orange line corresponds to the emission from an AGN. The red line shows the infrared emission from interstellar dust. The gray line denotes the nebulae emission. The observed data points are represented by purple circles, accompanied by 1$\sigma$ error bars. The bottom panel displays the relative residuals.}
\figsetgrpend

\figsetgrpstart
\figsetgrpnum{16.140}
\figsetgrptitle{R0600-ID12}
\figsetplot{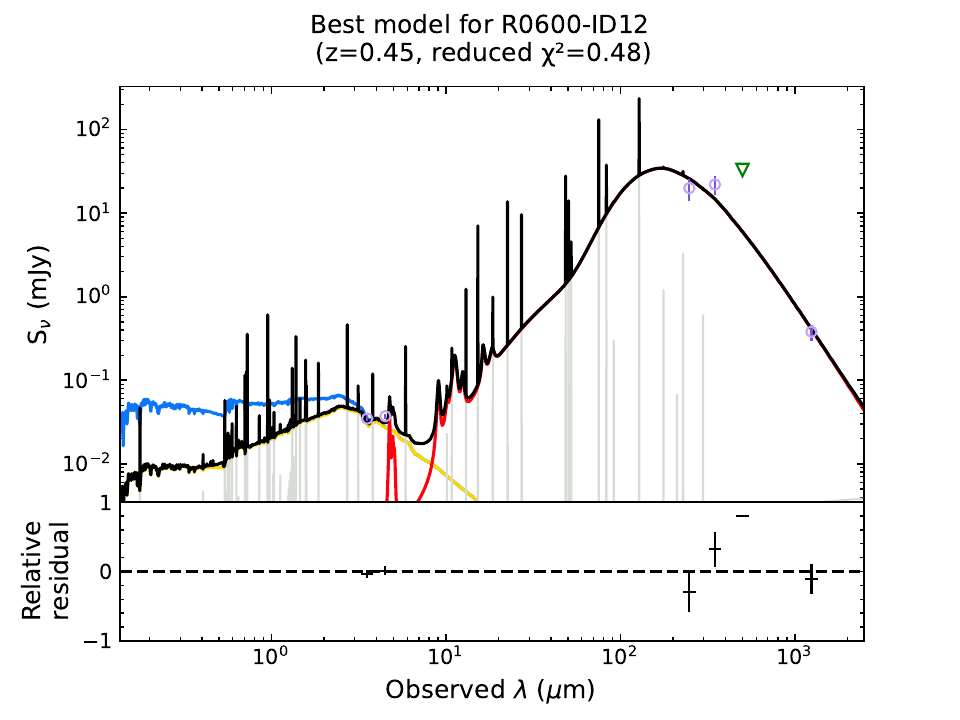}
\figsetgrpnote{The SED and the best-fit model of R0600-ID12. Lensing magnification is NOT corrected in this figure. The black solid line represents the composite spectrum of the galaxy. The yellow line illustrates the stellar emission attenuated by interstellar dust. The blue line depicts the unattenuated stellar emission for reference purpose. The orange line corresponds to the emission from an AGN. The red line shows the infrared emission from interstellar dust. The gray line denotes the nebulae emission. The observed data points are represented by purple circles, accompanied by 1$\sigma$ error bars. The bottom panel displays the relative residuals.}
\figsetgrpend

\figsetgrpstart
\figsetgrpnum{16.141}
\figsetgrptitle{R0600-ID13}
\figsetplot{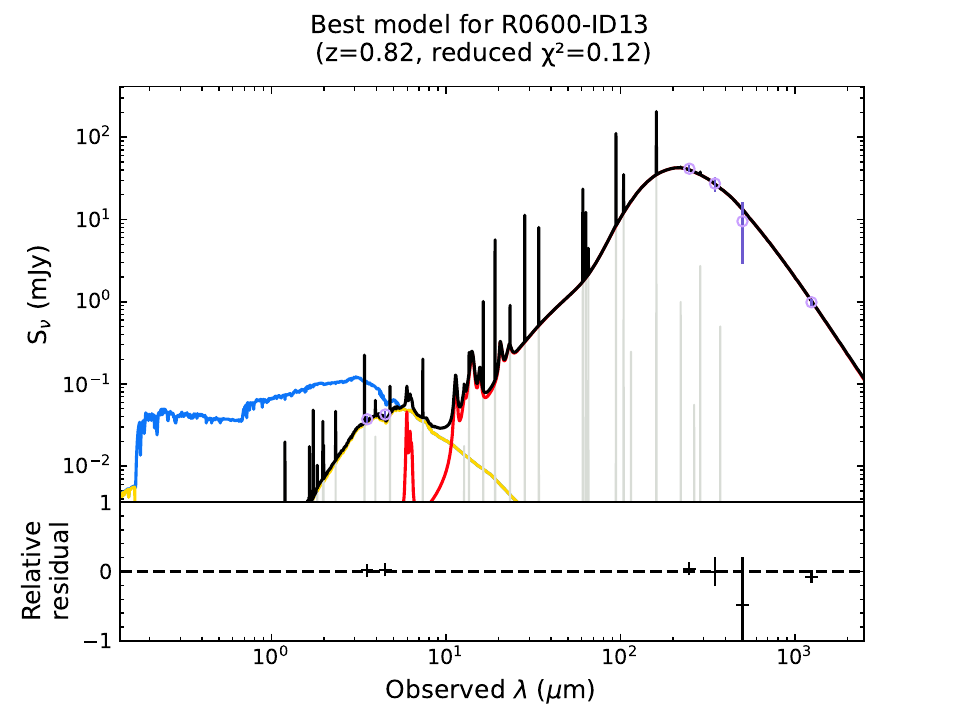}
\figsetgrpnote{The SED and the best-fit model of R0600-ID13. Lensing magnification is NOT corrected in this figure. The black solid line represents the composite spectrum of the galaxy. The yellow line illustrates the stellar emission attenuated by interstellar dust. The blue line depicts the unattenuated stellar emission for reference purpose. The orange line corresponds to the emission from an AGN. The red line shows the infrared emission from interstellar dust. The gray line denotes the nebulae emission. The observed data points are represented by purple circles, accompanied by 1$\sigma$ error bars. The bottom panel displays the relative residuals.}
\figsetgrpend

\figsetgrpstart
\figsetgrpnum{16.142}
\figsetgrptitle{R0949-ID10}
\figsetplot{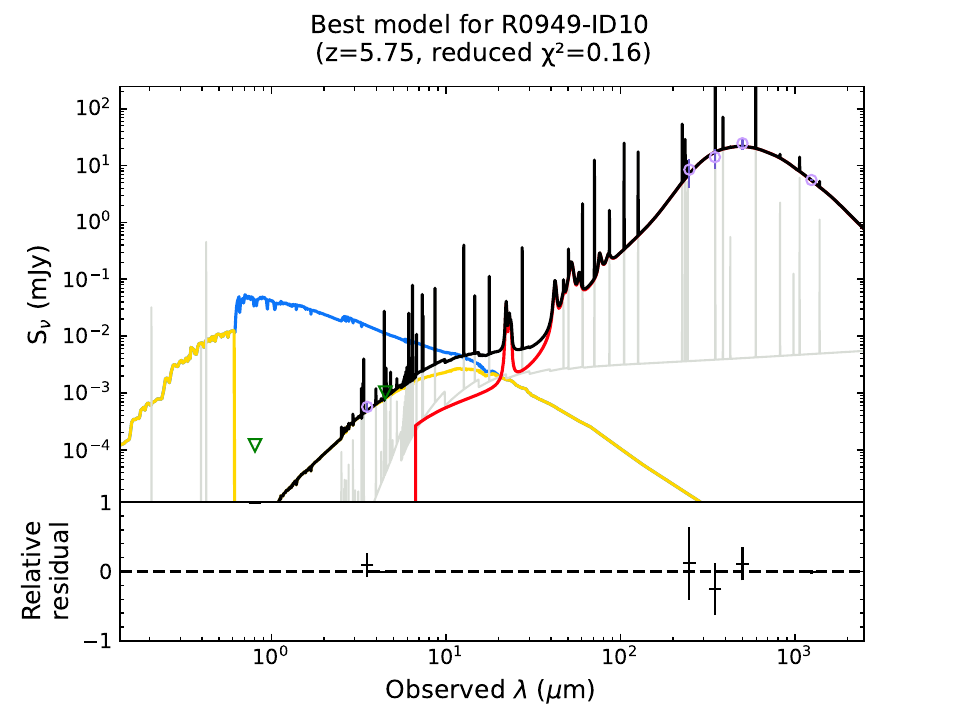}
\figsetgrpnote{The SED and the best-fit model of R0949-ID10. Lensing magnification is NOT corrected in this figure. The black solid line represents the composite spectrum of the galaxy. The yellow line illustrates the stellar emission attenuated by interstellar dust. The blue line depicts the unattenuated stellar emission for reference purpose. The orange line corresponds to the emission from an AGN. The red line shows the infrared emission from interstellar dust. The gray line denotes the nebulae emission. The observed data points are represented by purple circles, accompanied by 1$\sigma$ error bars. The bottom panel displays the relative residuals.}
\figsetgrpend

\figsetgrpstart
\figsetgrpnum{16.143}
\figsetgrptitle{R0949-ID113}
\figsetplot{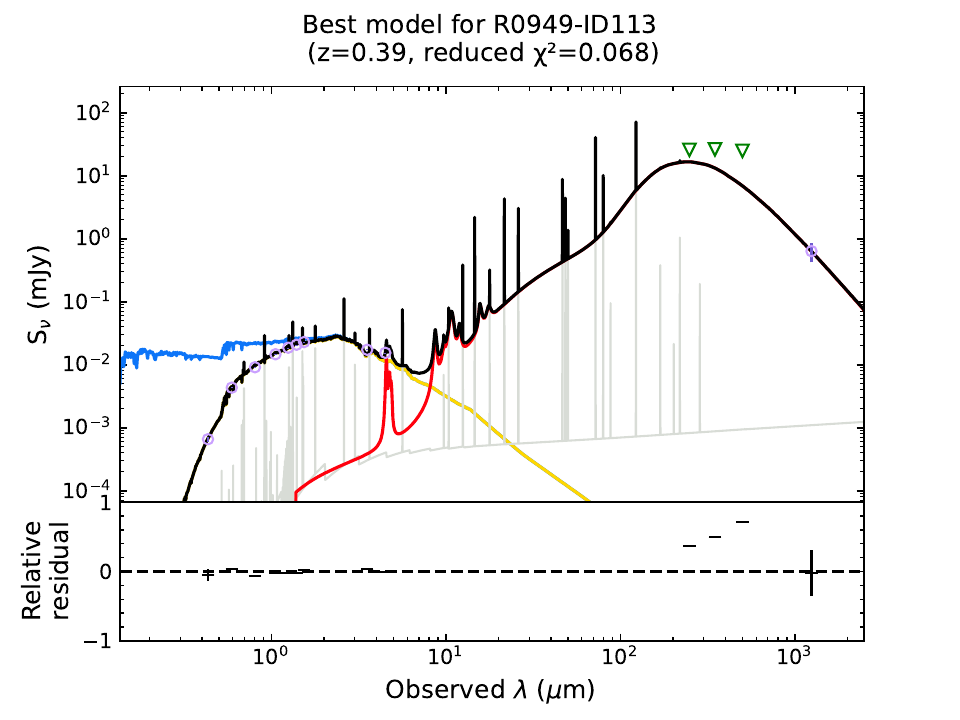}
\figsetgrpnote{The SED and the best-fit model of R0949-ID113. Lensing magnification is NOT corrected in this figure. The black solid line represents the composite spectrum of the galaxy. The yellow line illustrates the stellar emission attenuated by interstellar dust. The blue line depicts the unattenuated stellar emission for reference purpose. The orange line corresponds to the emission from an AGN. The red line shows the infrared emission from interstellar dust. The gray line denotes the nebulae emission. The observed data points are represented by purple circles, accompanied by 1$\sigma$ error bars. The bottom panel displays the relative residuals.}
\figsetgrpend

\figsetgrpstart
\figsetgrpnum{16.144}
\figsetgrptitle{R0949-ID119}
\figsetplot{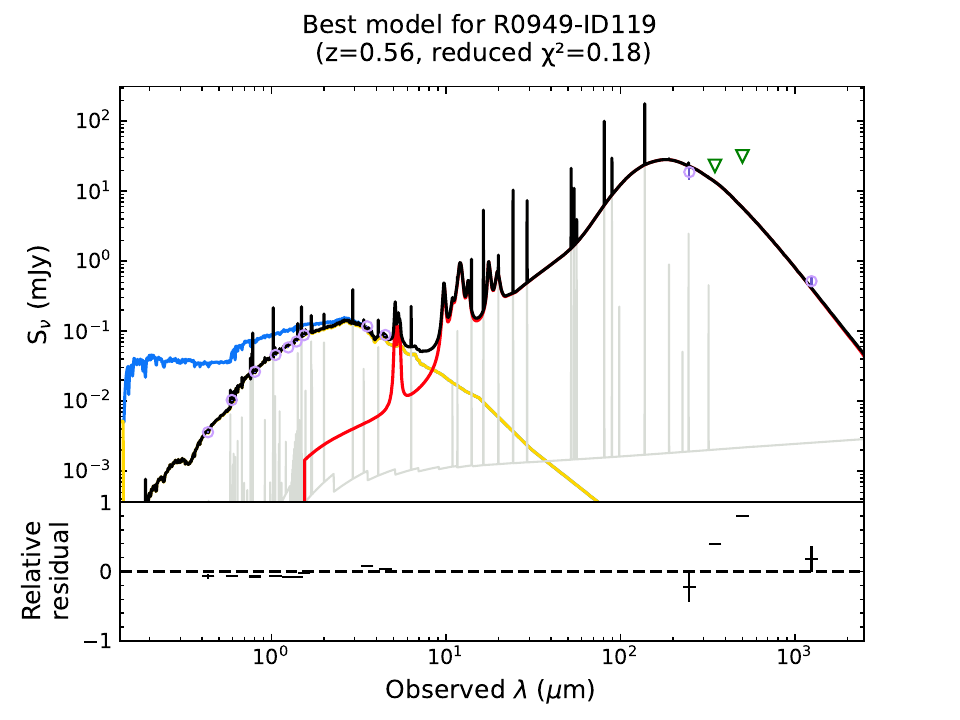}
\figsetgrpnote{The SED and the best-fit model of R0949-ID119. Lensing magnification is NOT corrected in this figure. The black solid line represents the composite spectrum of the galaxy. The yellow line illustrates the stellar emission attenuated by interstellar dust. The blue line depicts the unattenuated stellar emission for reference purpose. The orange line corresponds to the emission from an AGN. The red line shows the infrared emission from interstellar dust. The gray line denotes the nebulae emission. The observed data points are represented by purple circles, accompanied by 1$\sigma$ error bars. The bottom panel displays the relative residuals.}
\figsetgrpend

\figsetgrpstart
\figsetgrpnum{16.145}
\figsetgrptitle{R0949-ID124}
\figsetplot{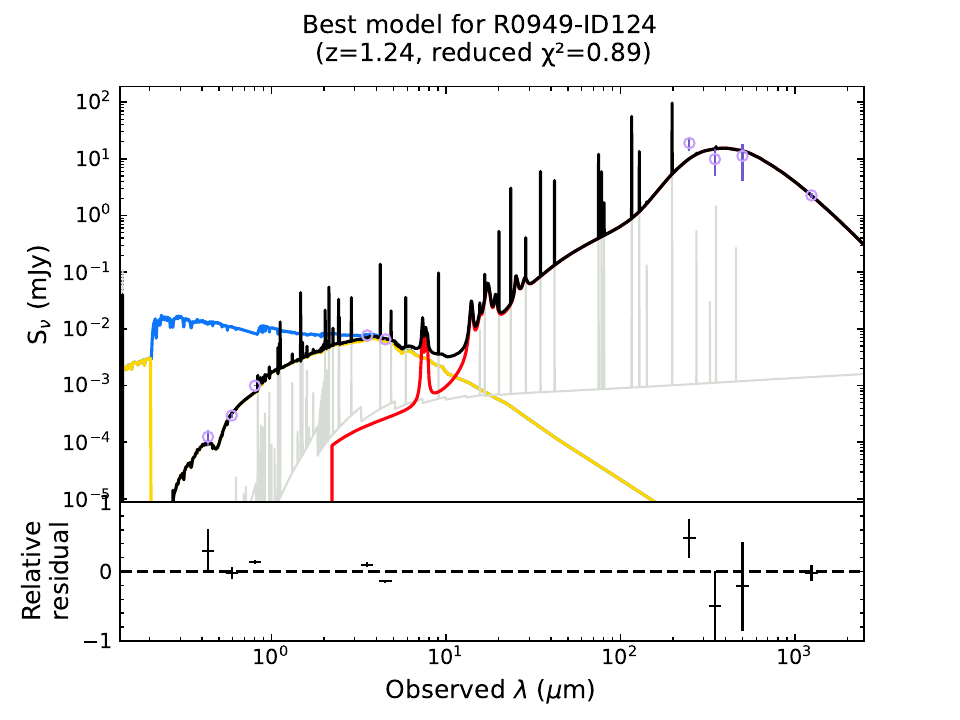}
\figsetgrpnote{The SED and the best-fit model of R0949-ID124. Lensing magnification is NOT corrected in this figure. The black solid line represents the composite spectrum of the galaxy. The yellow line illustrates the stellar emission attenuated by interstellar dust. The blue line depicts the unattenuated stellar emission for reference purpose. The orange line corresponds to the emission from an AGN. The red line shows the infrared emission from interstellar dust. The gray line denotes the nebulae emission. The observed data points are represented by purple circles, accompanied by 1$\sigma$ error bars. The bottom panel displays the relative residuals.}
\figsetgrpend

\figsetgrpstart
\figsetgrpnum{16.146}
\figsetgrptitle{R0949-ID14}
\figsetplot{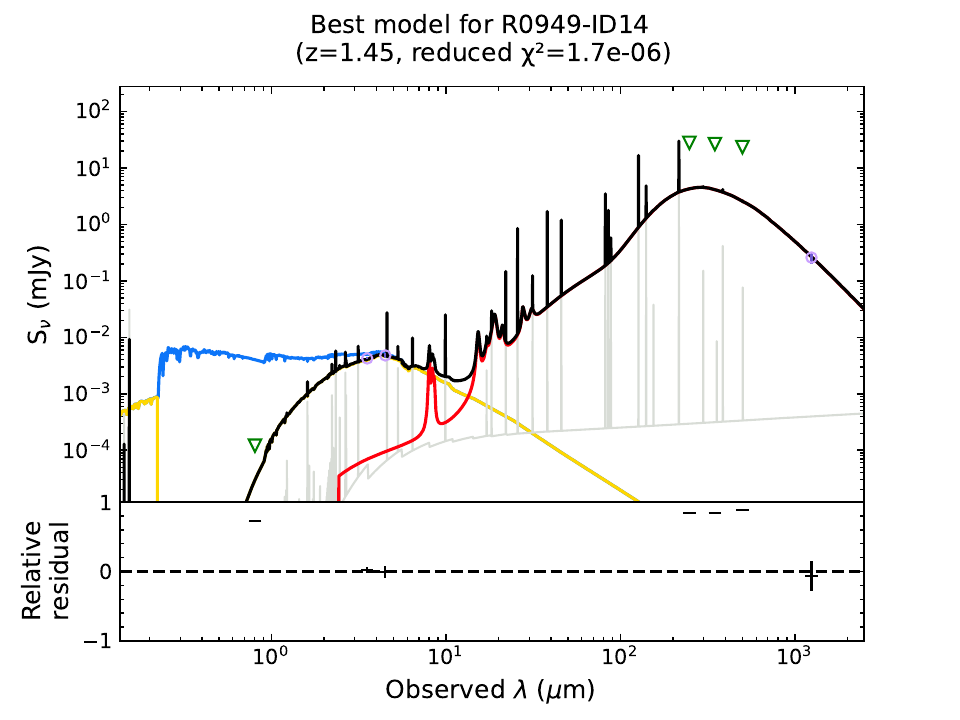}
\figsetgrpnote{The SED and the best-fit model of R0949-ID14. Lensing magnification is NOT corrected in this figure. The black solid line represents the composite spectrum of the galaxy. The yellow line illustrates the stellar emission attenuated by interstellar dust. The blue line depicts the unattenuated stellar emission for reference purpose. The orange line corresponds to the emission from an AGN. The red line shows the infrared emission from interstellar dust. The gray line denotes the nebulae emission. The observed data points are represented by purple circles, accompanied by 1$\sigma$ error bars. The bottom panel displays the relative residuals.}
\figsetgrpend

\figsetgrpstart
\figsetgrpnum{16.147}
\figsetgrptitle{R0949-ID19}
\figsetplot{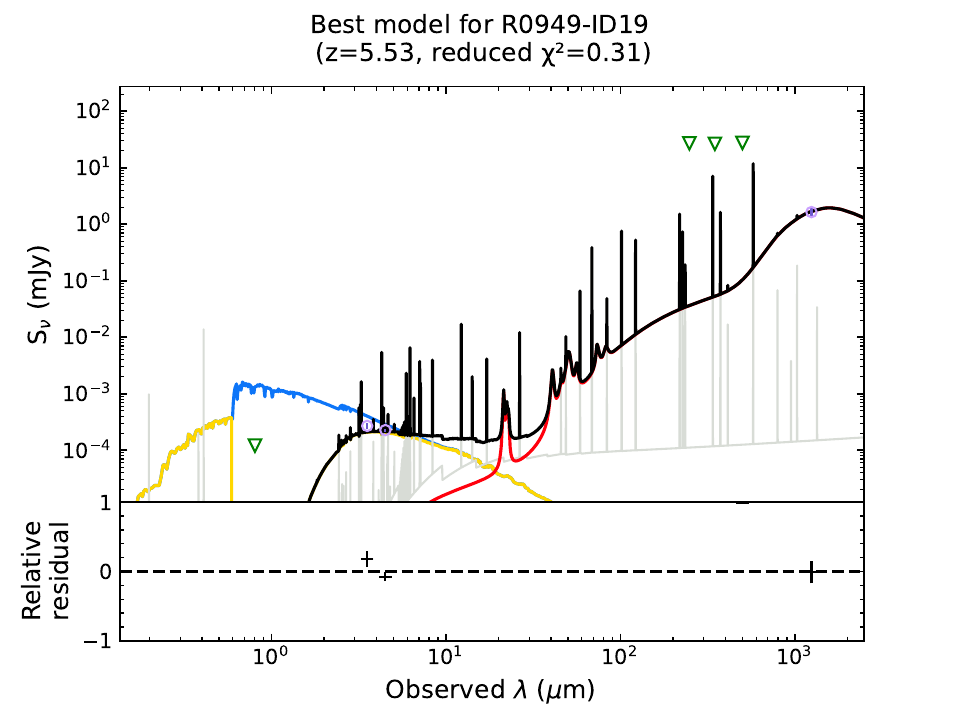}
\figsetgrpnote{The SED and the best-fit model of R0949-ID19. Lensing magnification is NOT corrected in this figure. The black solid line represents the composite spectrum of the galaxy. The yellow line illustrates the stellar emission attenuated by interstellar dust. The blue line depicts the unattenuated stellar emission for reference purpose. The orange line corresponds to the emission from an AGN. The red line shows the infrared emission from interstellar dust. The gray line denotes the nebulae emission. The observed data points are represented by purple circles, accompanied by 1$\sigma$ error bars. The bottom panel displays the relative residuals.}
\figsetgrpend

\figsetgrpstart
\figsetgrpnum{16.148}
\figsetgrptitle{R1347-ID145}
\figsetplot{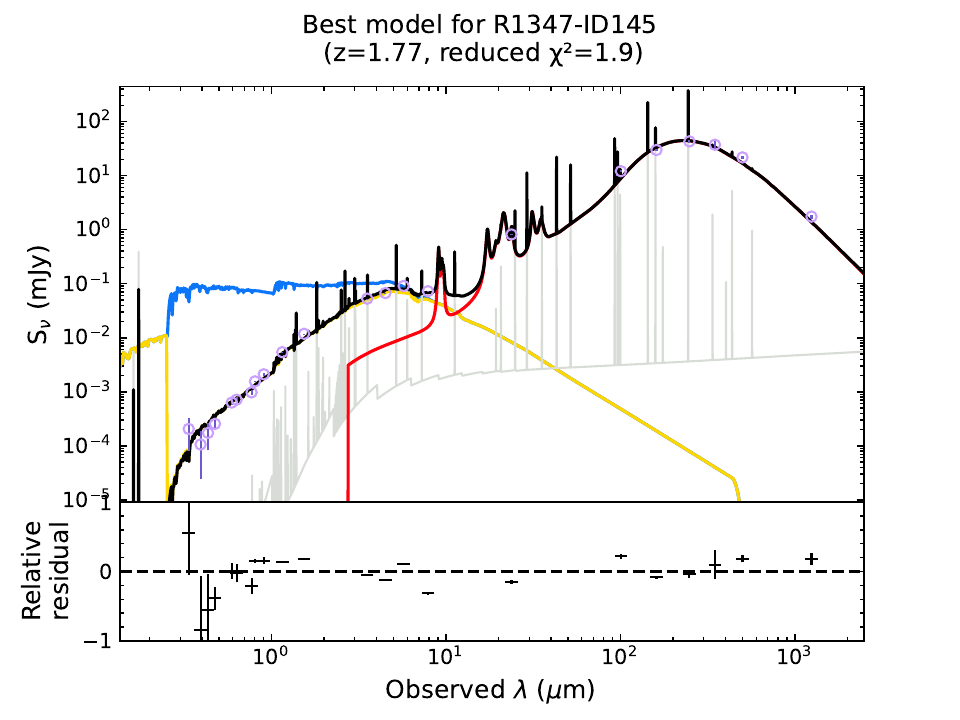}
\figsetgrpnote{The SED and the best-fit model of R1347-ID145. Lensing magnification is NOT corrected in this figure. The black solid line represents the composite spectrum of the galaxy. The yellow line illustrates the stellar emission attenuated by interstellar dust. The blue line depicts the unattenuated stellar emission for reference purpose. The orange line corresponds to the emission from an AGN. The red line shows the infrared emission from interstellar dust. The gray line denotes the nebulae emission. The observed data points are represented by purple circles, accompanied by 1$\sigma$ error bars. The bottom panel displays the relative residuals.}
\figsetgrpend

\figsetgrpstart
\figsetgrpnum{16.149}
\figsetgrptitle{R1347-ID148}
\figsetplot{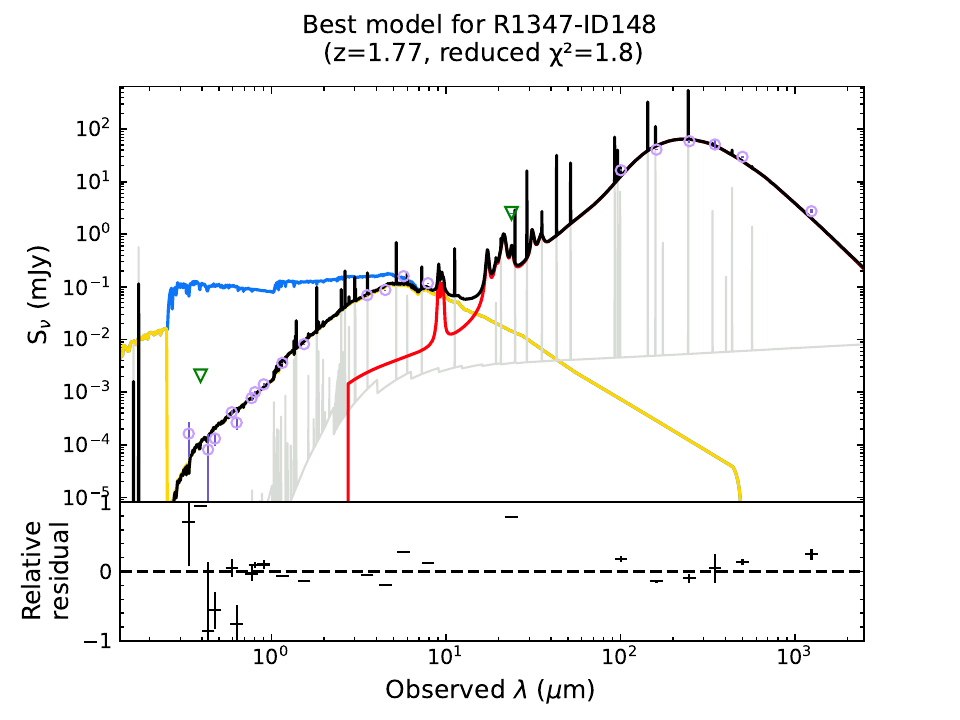}
\figsetgrpnote{The SED and the best-fit model of R1347-ID148. Lensing magnification is NOT corrected in this figure. The black solid line represents the composite spectrum of the galaxy. The yellow line illustrates the stellar emission attenuated by interstellar dust. The blue line depicts the unattenuated stellar emission for reference purpose. The orange line corresponds to the emission from an AGN. The red line shows the infrared emission from interstellar dust. The gray line denotes the nebulae emission. The observed data points are represented by purple circles, accompanied by 1$\sigma$ error bars. The bottom panel displays the relative residuals.}
\figsetgrpend

\figsetgrpstart
\figsetgrpnum{16.150}
\figsetgrptitle{R1347-ID166}
\figsetplot{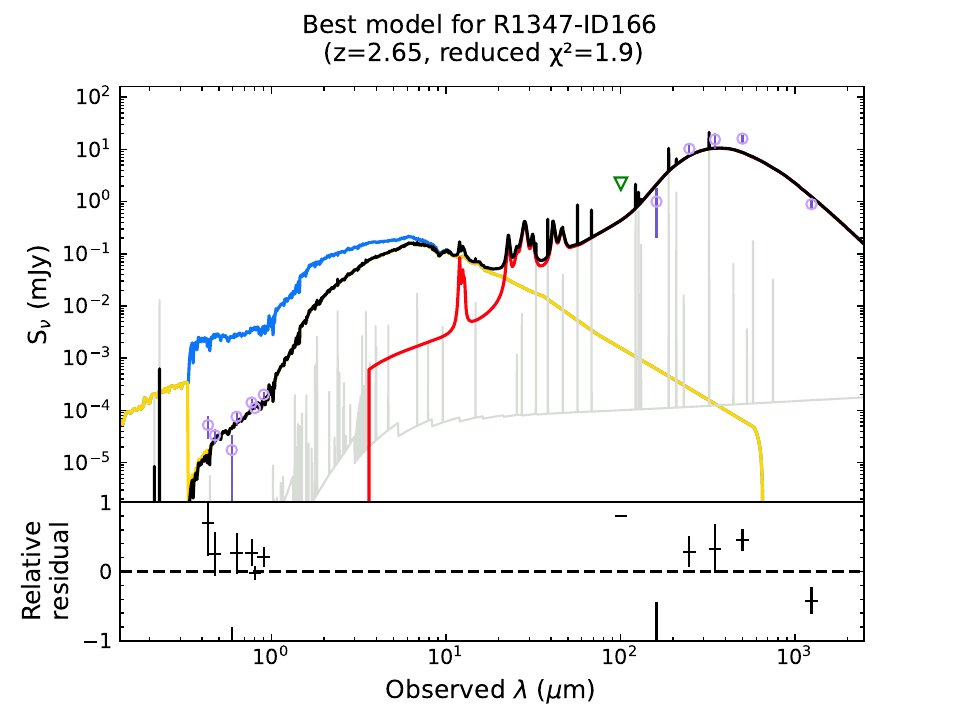}
\figsetgrpnote{The SED and the best-fit model of R1347-ID166. Lensing magnification is NOT corrected in this figure. The black solid line represents the composite spectrum of the galaxy. The yellow line illustrates the stellar emission attenuated by interstellar dust. The blue line depicts the unattenuated stellar emission for reference purpose. The orange line corresponds to the emission from an AGN. The red line shows the infrared emission from interstellar dust. The gray line denotes the nebulae emission. The observed data points are represented by purple circles, accompanied by 1$\sigma$ error bars. The bottom panel displays the relative residuals.}
\figsetgrpend

\figsetgrpstart
\figsetgrpnum{16.151}
\figsetgrptitle{R1347-ID41}
\figsetplot{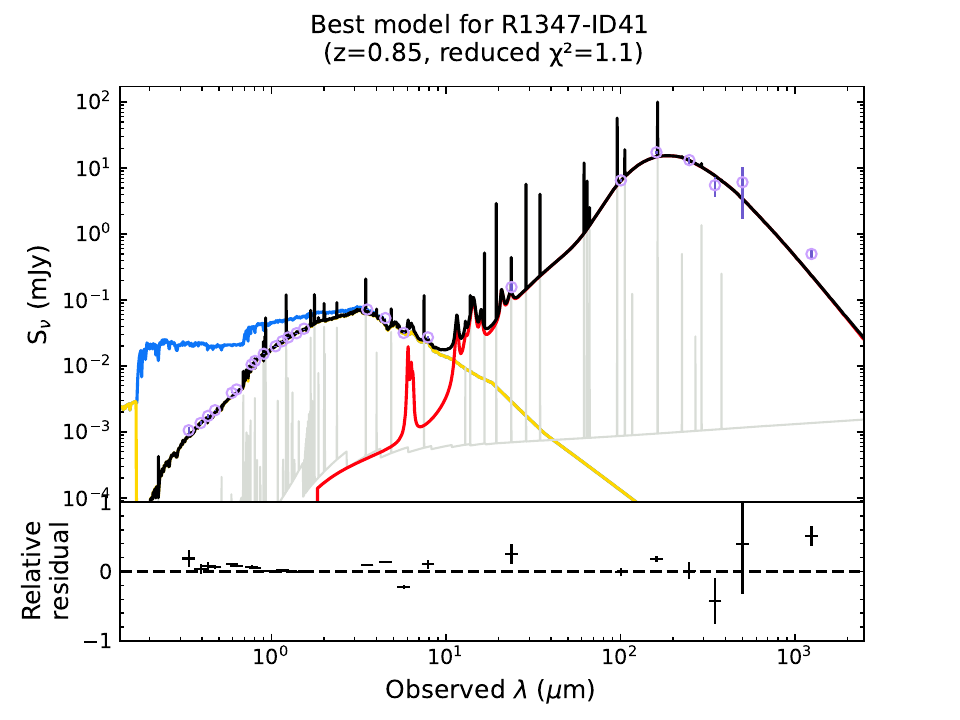}
\figsetgrpnote{The SED and the best-fit model of R1347-ID41. Lensing magnification is NOT corrected in this figure. The black solid line represents the composite spectrum of the galaxy. The yellow line illustrates the stellar emission attenuated by interstellar dust. The blue line depicts the unattenuated stellar emission for reference purpose. The orange line corresponds to the emission from an AGN. The red line shows the infrared emission from interstellar dust. The gray line denotes the nebulae emission. The observed data points are represented by purple circles, accompanied by 1$\sigma$ error bars. The bottom panel displays the relative residuals.}
\figsetgrpend

\figsetgrpstart
\figsetgrpnum{16.152}
\figsetgrptitle{R2129-ID37}
\figsetplot{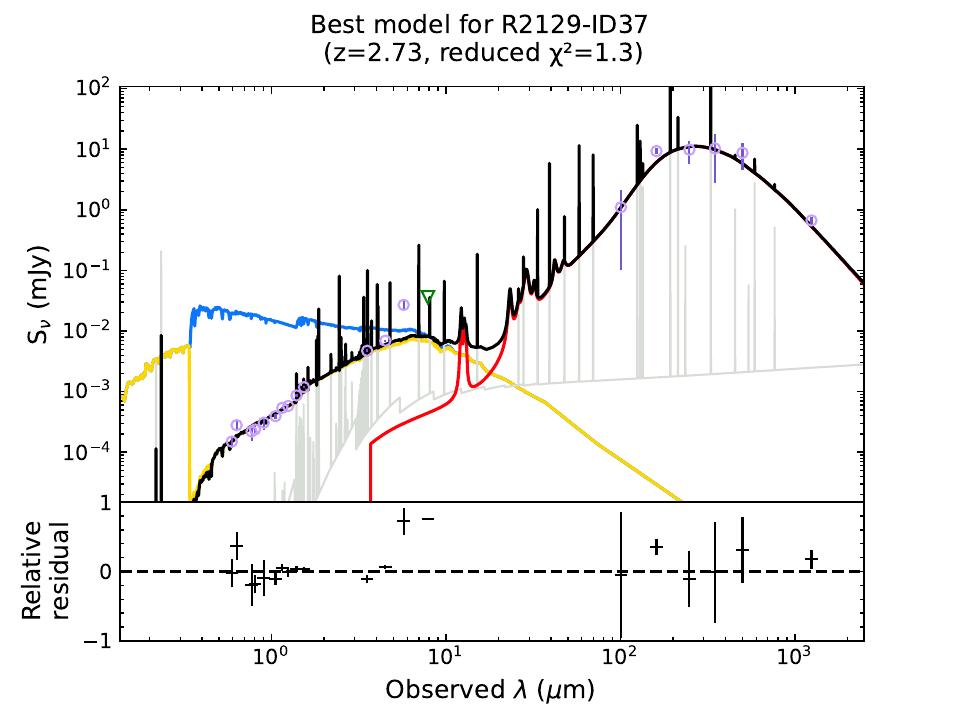}
\figsetgrpnote{The SED and the best-fit model of R2129-ID37. Lensing magnification is NOT corrected in this figure. The black solid line represents the composite spectrum of the galaxy. The yellow line illustrates the stellar emission attenuated by interstellar dust. The blue line depicts the unattenuated stellar emission for reference purpose. The orange line corresponds to the emission from an AGN. The red line shows the infrared emission from interstellar dust. The gray line denotes the nebulae emission. The observed data points are represented by purple circles, accompanied by 1$\sigma$ error bars. The bottom panel displays the relative residuals.}
\figsetgrpend

\figsetgrpstart
\figsetgrpnum{16.153}
\figsetgrptitle{R2211-ID164}
\figsetplot{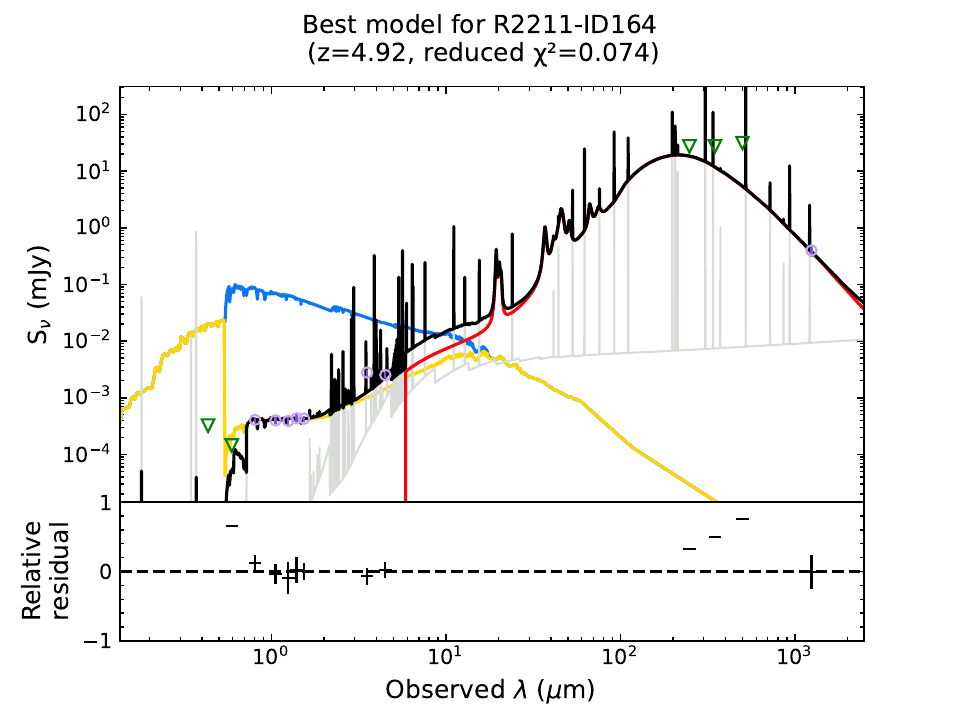}
\figsetgrpnote{The SED and the best-fit model of R2211-ID164. Lensing magnification is NOT corrected in this figure. The black solid line represents the composite spectrum of the galaxy. The yellow line illustrates the stellar emission attenuated by interstellar dust. The blue line depicts the unattenuated stellar emission for reference purpose. The orange line corresponds to the emission from an AGN. The red line shows the infrared emission from interstellar dust. The gray line denotes the nebulae emission. The observed data points are represented by purple circles, accompanied by 1$\sigma$ error bars. The bottom panel displays the relative residuals.}
\figsetgrpend

\figsetgrpstart
\figsetgrpnum{16.154}
\figsetgrptitle{R2211-ID171}
\figsetplot{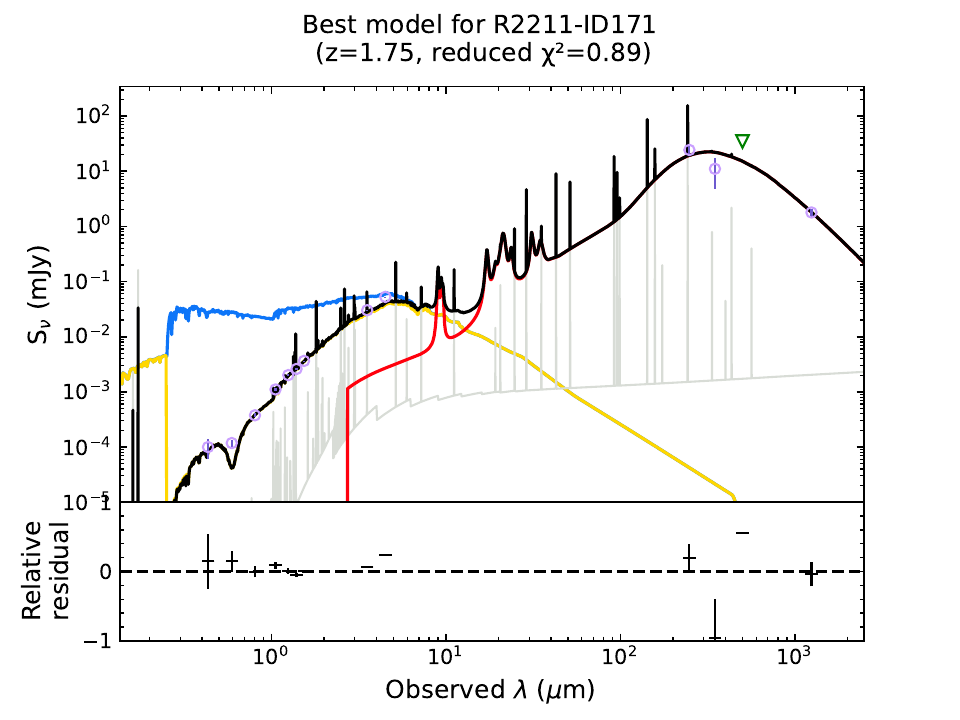}
\figsetgrpnote{The SED and the best-fit model of R2211-ID171. Lensing magnification is NOT corrected in this figure. The black solid line represents the composite spectrum of the galaxy. The yellow line illustrates the stellar emission attenuated by interstellar dust. The blue line depicts the unattenuated stellar emission for reference purpose. The orange line corresponds to the emission from an AGN. The red line shows the infrared emission from interstellar dust. The gray line denotes the nebulae emission. The observed data points are represented by purple circles, accompanied by 1$\sigma$ error bars. The bottom panel displays the relative residuals.}
\figsetgrpend

\figsetgrpstart
\figsetgrpnum{16.155}
\figsetgrptitle{R2211-ID19}
\figsetplot{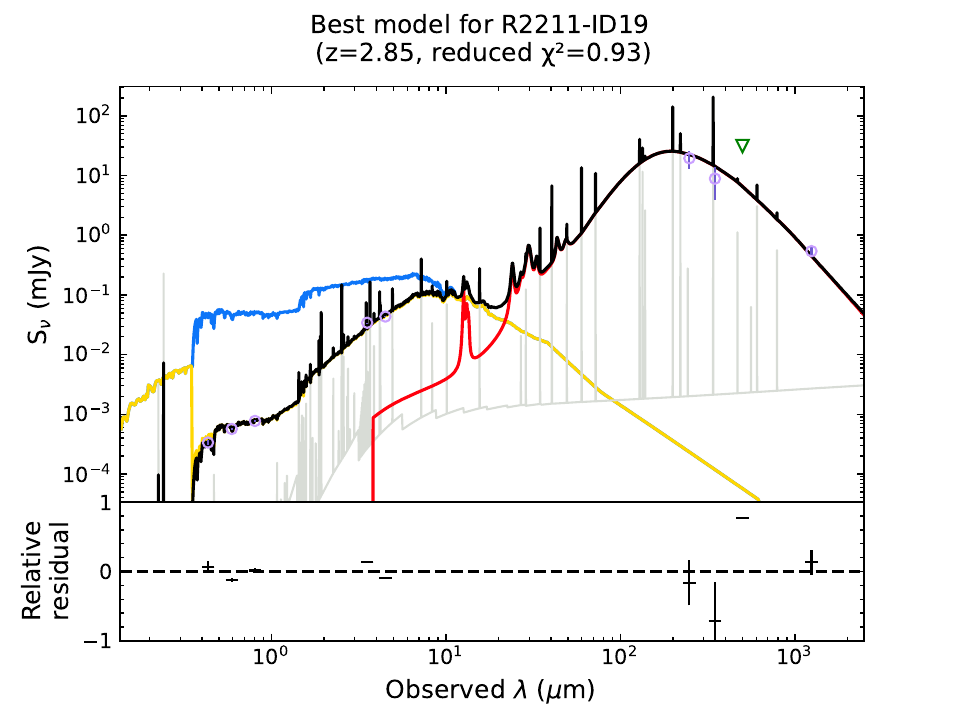}
\figsetgrpnote{The SED and the best-fit model of R2211-ID19. Lensing magnification is NOT corrected in this figure. The black solid line represents the composite spectrum of the galaxy. The yellow line illustrates the stellar emission attenuated by interstellar dust. The blue line depicts the unattenuated stellar emission for reference purpose. The orange line corresponds to the emission from an AGN. The red line shows the infrared emission from interstellar dust. The gray line denotes the nebulae emission. The observed data points are represented by purple circles, accompanied by 1$\sigma$ error bars. The bottom panel displays the relative residuals.}
\figsetgrpend

\figsetgrpstart
\figsetgrpnum{16.156}
\figsetgrptitle{R2211-ID35}
\figsetplot{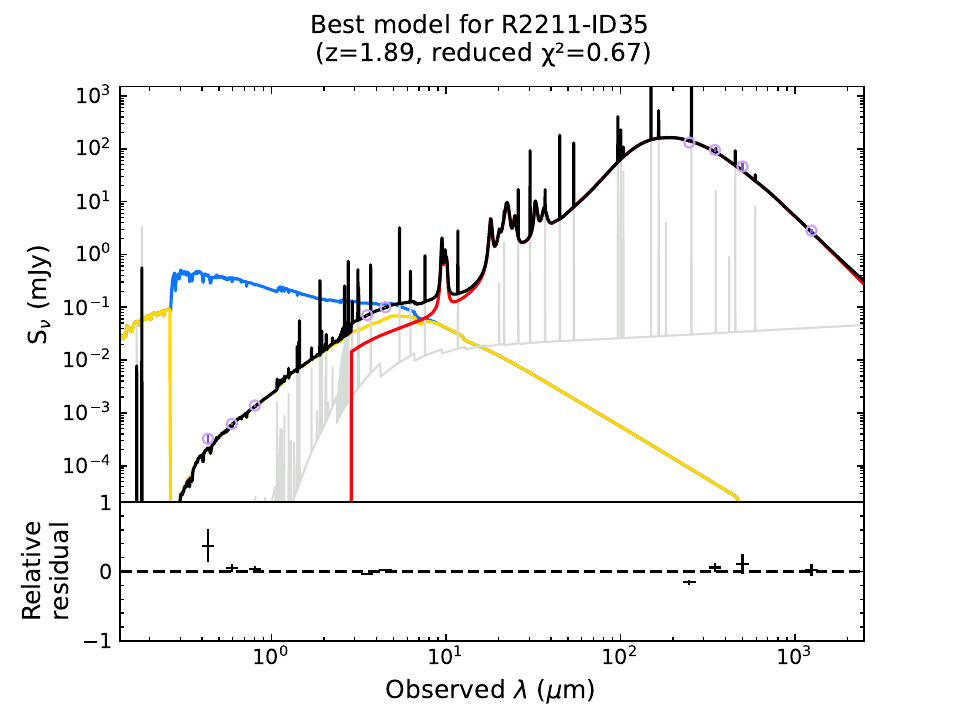}
\figsetgrpnote{The SED and the best-fit model of R2211-ID35. Lensing magnification is NOT corrected in this figure. The black solid line represents the composite spectrum of the galaxy. The yellow line illustrates the stellar emission attenuated by interstellar dust. The blue line depicts the unattenuated stellar emission for reference purpose. The orange line corresponds to the emission from an AGN. The red line shows the infrared emission from interstellar dust. The gray line denotes the nebulae emission. The observed data points are represented by purple circles, accompanied by 1$\sigma$ error bars. The bottom panel displays the relative residuals.}
\figsetgrpend

\figsetgrpstart
\figsetgrpnum{16.157}
\figsetgrptitle{SM0723-ID124}
\figsetplot{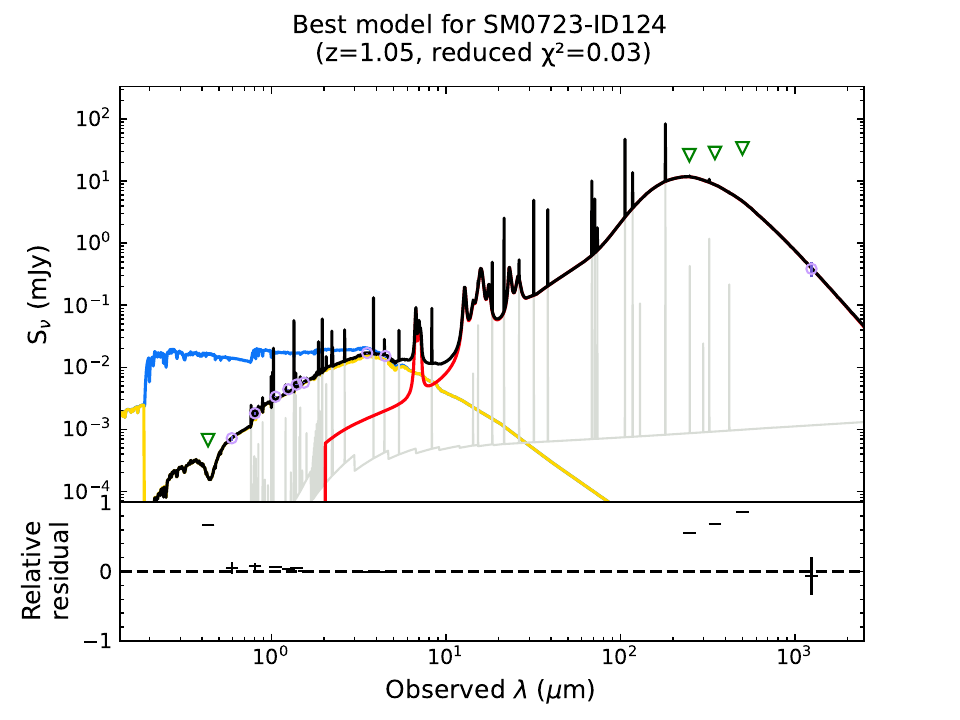}
\figsetgrpnote{The SED and the best-fit model of SM0723-ID124. Lensing magnification is NOT corrected in this figure. The black solid line represents the composite spectrum of the galaxy. The yellow line illustrates the stellar emission attenuated by interstellar dust. The blue line depicts the unattenuated stellar emission for reference purpose. The orange line corresponds to the emission from an AGN. The red line shows the infrared emission from interstellar dust. The gray line denotes the nebulae emission. The observed data points are represented by purple circles, accompanied by 1$\sigma$ error bars. The bottom panel displays the relative residuals.}
\figsetgrpend

\figsetgrpstart
\figsetgrpnum{16.158}
\figsetgrptitle{SM0723-ID61}
\figsetplot{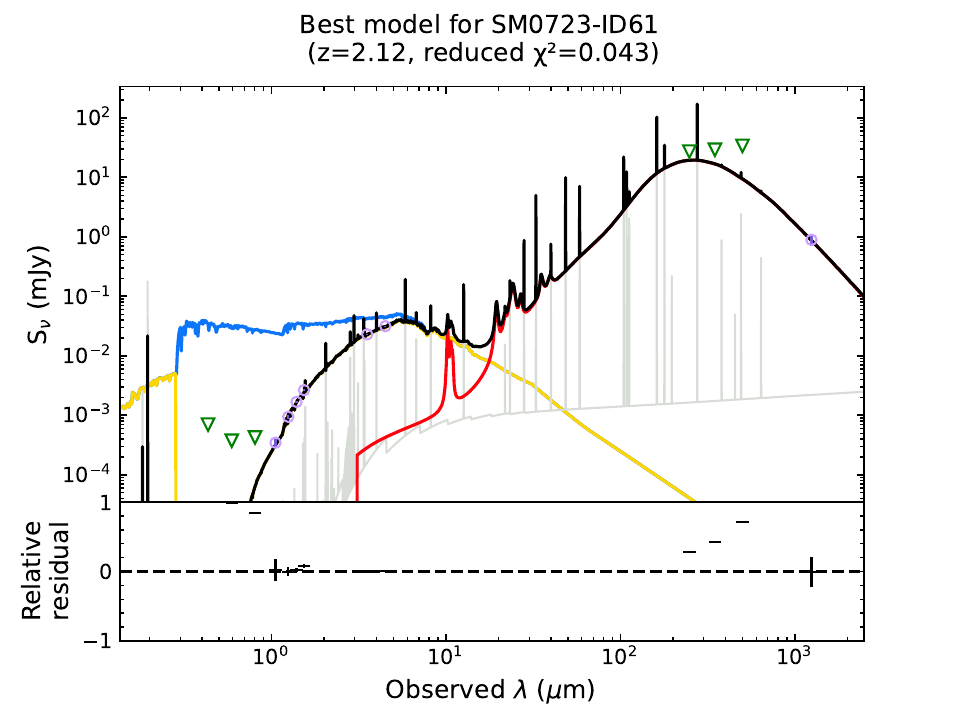}
\figsetgrpnote{The SED and the best-fit model of SM0723-ID61. Lensing magnification is NOT corrected in this figure. The black solid line represents the composite spectrum of the galaxy. The yellow line illustrates the stellar emission attenuated by interstellar dust. The blue line depicts the unattenuated stellar emission for reference purpose. The orange line corresponds to the emission from an AGN. The red line shows the infrared emission from interstellar dust. The gray line denotes the nebulae emission. The observed data points are represented by purple circles, accompanied by 1$\sigma$ error bars. The bottom panel displays the relative residuals.}
\figsetgrpend

\figsetgrpstart
\figsetgrpnum{16.159}
\figsetgrptitle{SM0723-ID93}
\figsetplot{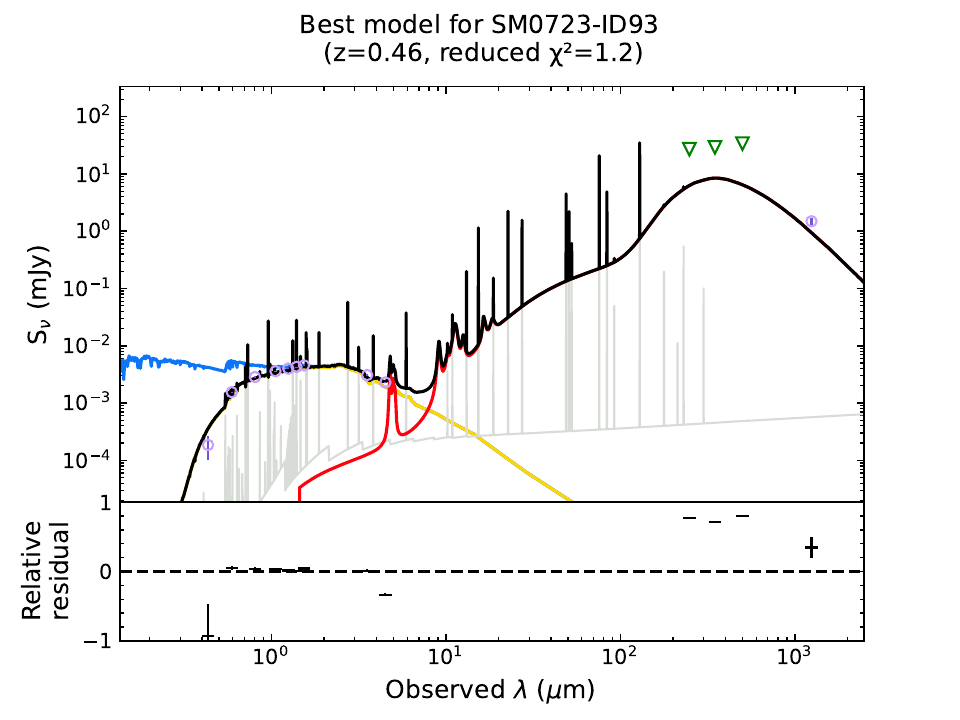}
\figsetgrpnote{The SED and the best-fit model of SM0723-ID93. Lensing magnification is NOT corrected in this figure. The black solid line represents the composite spectrum of the galaxy. The yellow line illustrates the stellar emission attenuated by interstellar dust. The blue line depicts the unattenuated stellar emission for reference purpose. The orange line corresponds to the emission from an AGN. The red line shows the infrared emission from interstellar dust. The gray line denotes the nebulae emission. The observed data points are represented by purple circles, accompanied by 1$\sigma$ error bars. The bottom panel displays the relative residuals.}
\figsetgrpend

\figsetgrpstart
\figsetgrpnum{16.160}
\figsetgrptitle{SM0723-ID98}
\figsetplot{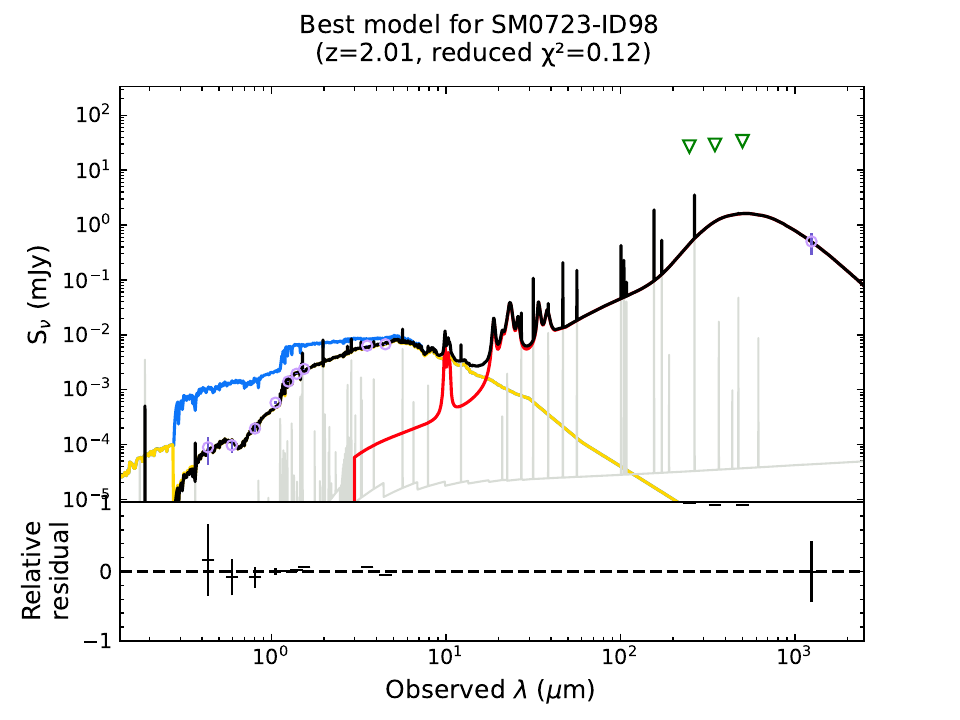}
\figsetgrpnote{The SED and the best-fit model of SM0723-ID98. Lensing magnification is NOT corrected in this figure. The black solid line represents the composite spectrum of the galaxy. The yellow line illustrates the stellar emission attenuated by interstellar dust. The blue line depicts the unattenuated stellar emission for reference purpose. The orange line corresponds to the emission from an AGN. The red line shows the infrared emission from interstellar dust. The gray line denotes the nebulae emission. The observed data points are represented by purple circles, accompanied by 1$\sigma$ error bars. The bottom panel displays the relative residuals.}
\figsetgrpend

\figsetend

\section{Validity of photometric redshift and Impact of redshift uncertainty on physical properties}\label{appendix:redshift}

In this section, we discuss the validity of photometric redshifts and the impact of redshift uncertainty on physical properties. The photometric redshifts of the ALCS sources are estimated by several methods. For example, most of the photometric redshifts of tier-1/2/5/6 sources are estimated by optical to near-infrared SED analysis with \texttt{EAZY} \citep{2022ApJS..263...38K}, whereas those of five objects (A2537-ID24, A2537-ID66, M0035-ID33, R0032-ID63, R0032-ID81) are taken from the literature and those of two objects  (ACT0102-ID223, ACT0102-ID294) are estimated by the sky positions and the lens model. The photometric redshifts of the tier-3/4/7/8 sources are estimated by near-infrared to millimeter SED analysis with a composite SED model obtained from AS2UDS sample \citep{2020MNRAS.494.3828D}. To simplify the discussion, here we only discuss the photometric redshifts estimated by \texttt{EAZY}. 

\begin{figure}[h]
 \epsscale{1.0}
 \plotone{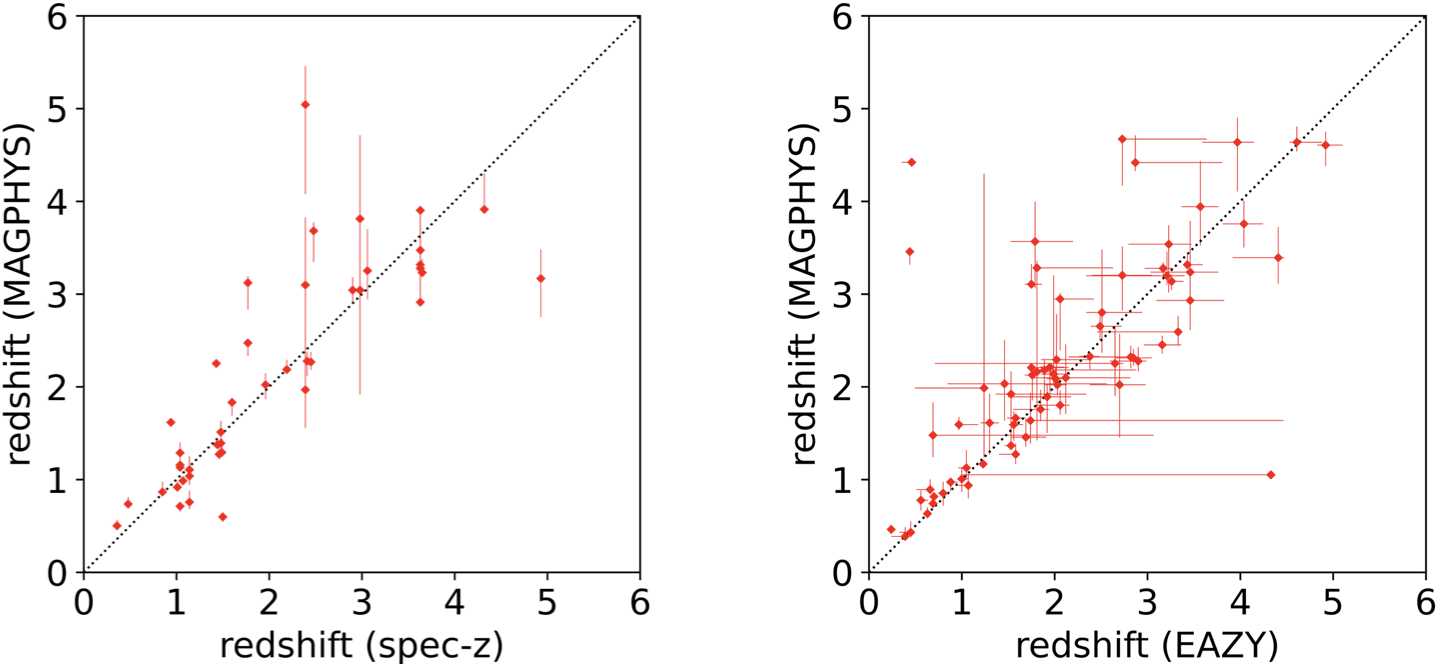}
 \caption{(Left) Comparison between spectroscopic redshift and photometric redshift estimated by \texttt{MAGPHYS}. (Right) Comparison between photometric redshift estimated by \texttt{EAZY} and that estimated by \texttt{MAGPHYS}. The black dotted lines show the identity relations. \label{figure:photoz_redshift}}
\end{figure}

First, we calculate the photometric redshifts of tier-1/2/5/6 sources by optical to millimeter SED analysis with \texttt{MAGPHYS photo-z extension} \citep{2019ApJ...882...61B}. Since the current version of \texttt{MAGPHYS} cannot treat an AGN component, the ALCS-XAGNs and ALCS-nonXAGNs are not analyzed in this section. Figure~\ref{figure:photoz_redshift} shows the results. The left panel plots photometric redshifts estimated by \texttt{MAGPHYS photo-z extension} versus spectroscopic ones. We confirm good agreement in those values, where 34/44 (77\%) sources meet $|z_{\mathrm{spec}}-z_{\mathrm{photo}}|/(1+z_{\mathrm{spec}})<0.2$. The right panel plots photometric redshifts derived by \texttt{MAGPHYS} versus ones obtained by \texttt{EAZY}. We again confirm good agreement in those values, where 59/72 (82\%) sources meet $|z_{\mathrm{MAGPHYS}}-z_{\mathrm{EAZY}}|/(1+z_{\mathrm{EAZY}})<0.2$.

\begin{figure}[h]
 \epsscale{1.17}
 \plotone{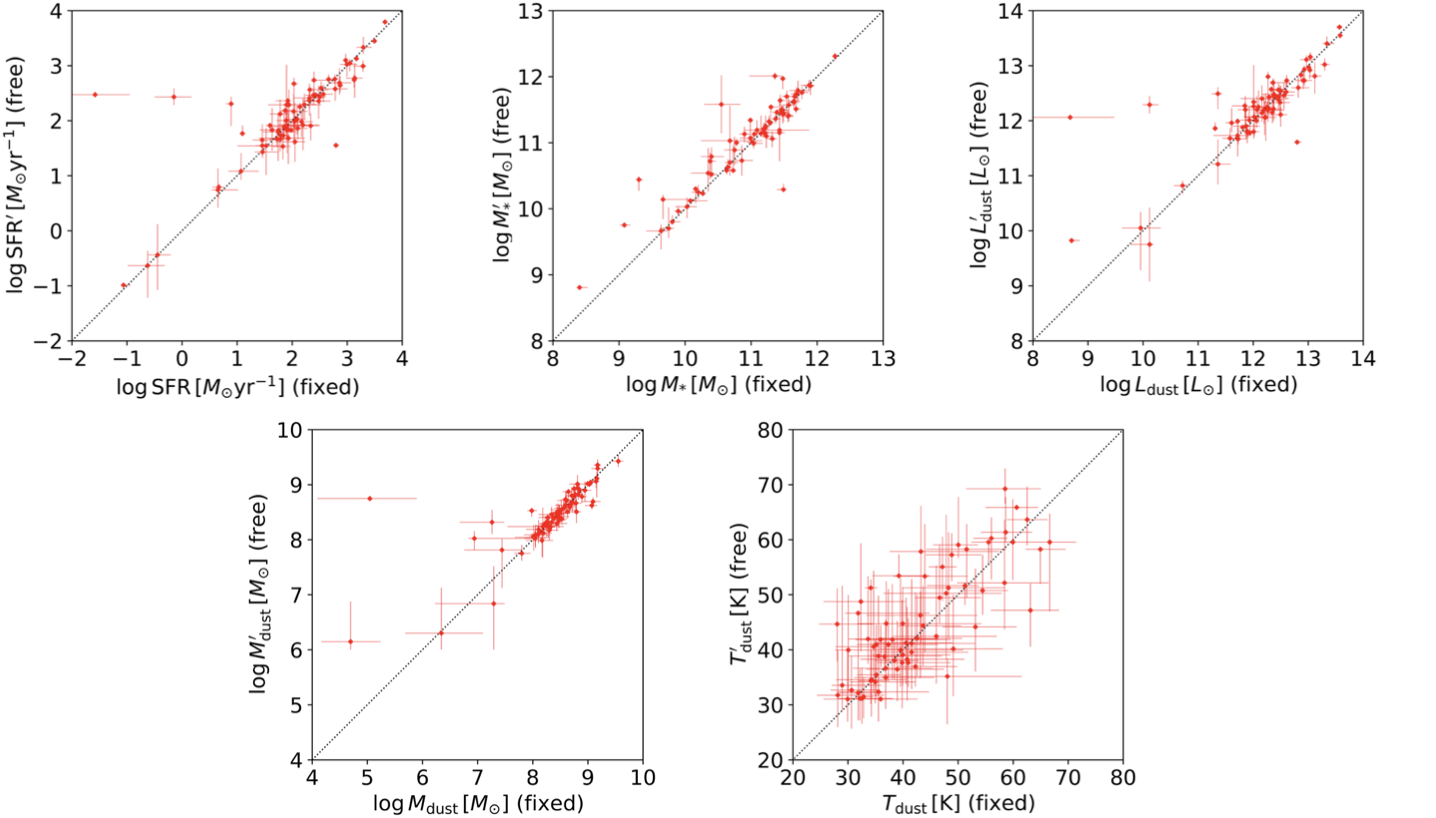}
 \caption{Comparison of the physical properties estimated with redshifts left as a free parameter (vertical axis) versus those estimated at fixed redshifts (horizontal axis). The black dotted lines show the identity relations. The lensing magnification is not corrected for the SFR, stellar mass, dust luminosity, and dust mass. \label{figure:photoz_properties}}
\end{figure}

Second, in order to evaluate the impact of redshift uncertainty on the derived physical properties, we independently estimate the physical parameters by optical to millimeter SED analysis with \texttt{MAGPHYS high-z extension (v2)} \citep{2015ApJ...806..110D,2020ApJ...888..108B}, where the redshifts are fixed at the photometric ones derived by \texttt{EAZY}. Figure~\ref{figure:photoz_properties} plots the physical properties estimated with variable redshifts versus those estimated at fixed redshifts. The standard deviations from the identity relations are 0.54 dex, 0.29 dex, 0.47 dex, 0.42 dex, and 6.1 K for SFR, stellar mass, dust luminosity, dust mass, and dust temperature, respectively. This indicates that the uncertainties in redshift measurement can cause at least these levels of uncertainties in each of the parameters.

\section{Comparison between \texttt{CIGALE} and \texttt{MAGPHYS}}\label{appendix:magphys}

\begin{figure}[htb]
 \epsscale{1.0}
 \plotone{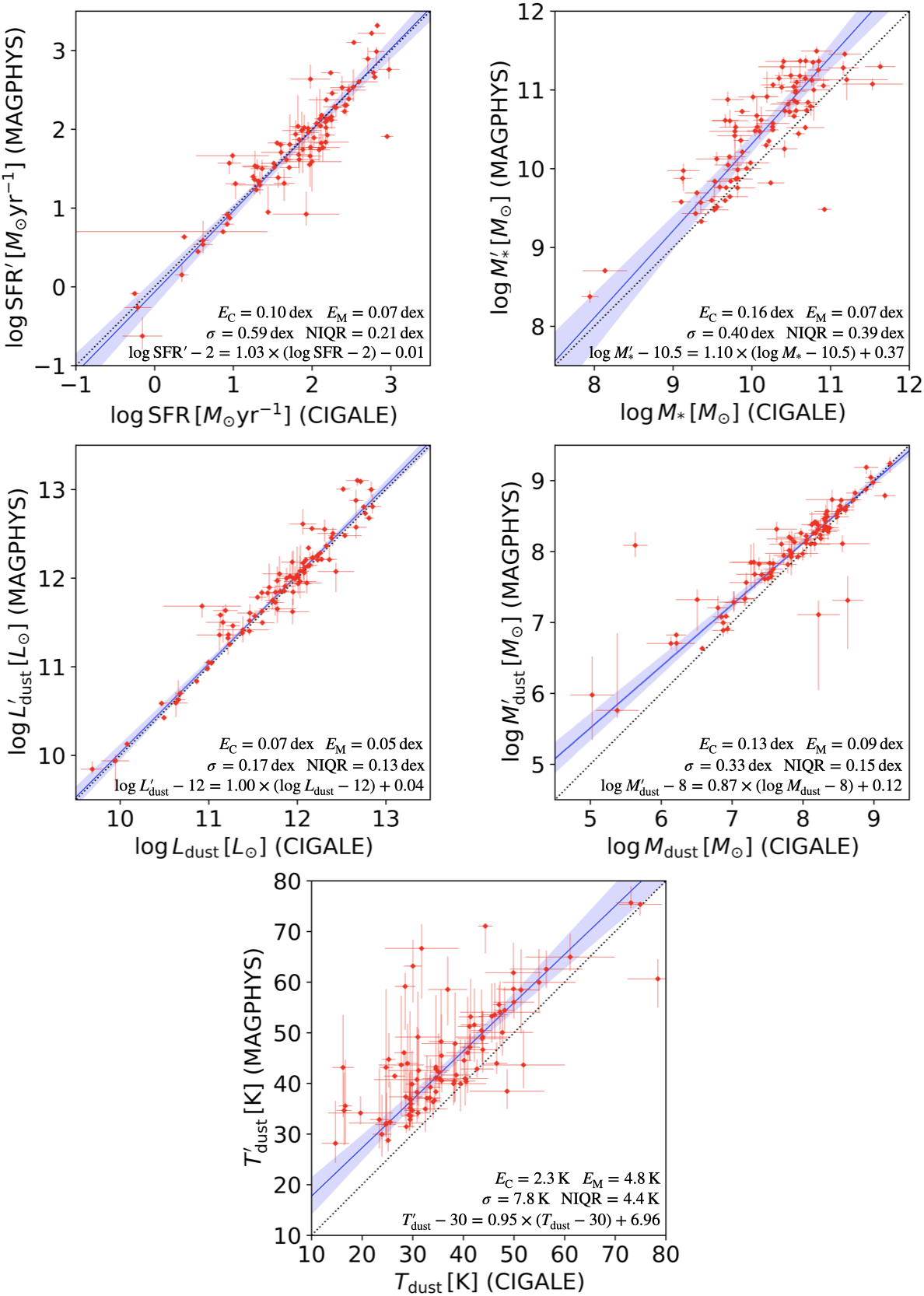}
 \caption{Comparison of SFR, stellar mass, dust luminosity, dust mass, and dust temperature measured by \texttt{CIGALE} and \texttt{MAGPHYS}. The blue solid lines show the regression lines, which are shown in the bottom equations in each panel. The semi-transparent blue bands show the 95\% confidence regions of each correlation. The standard deviations ($\sigma$) and normalized interquartile ranges (NIQR) around the regression lines measured in the vertical axes are shown at the bottom of each panel. We also show the median value of the statistical errors estimated with \text{CIGALE} ($E_{\mathrm{C}}$) and \text{MAGPHYS} ($E_{\mathrm{M}}$) for comparison. The black dotted lines show the identity relations. For a conservative discussion, only the tier-1 samples are plotted for the ALCS sample. The multiply-imaged sources are plotted individually. The lensing magnification is corrected for the SFR, stellar mass, dust luminosity, and dust mass. \label{figure:cigale_magphys}}
\end{figure}

\begin{figure}[htb]
 \epsscale{0.5}
 \plotone{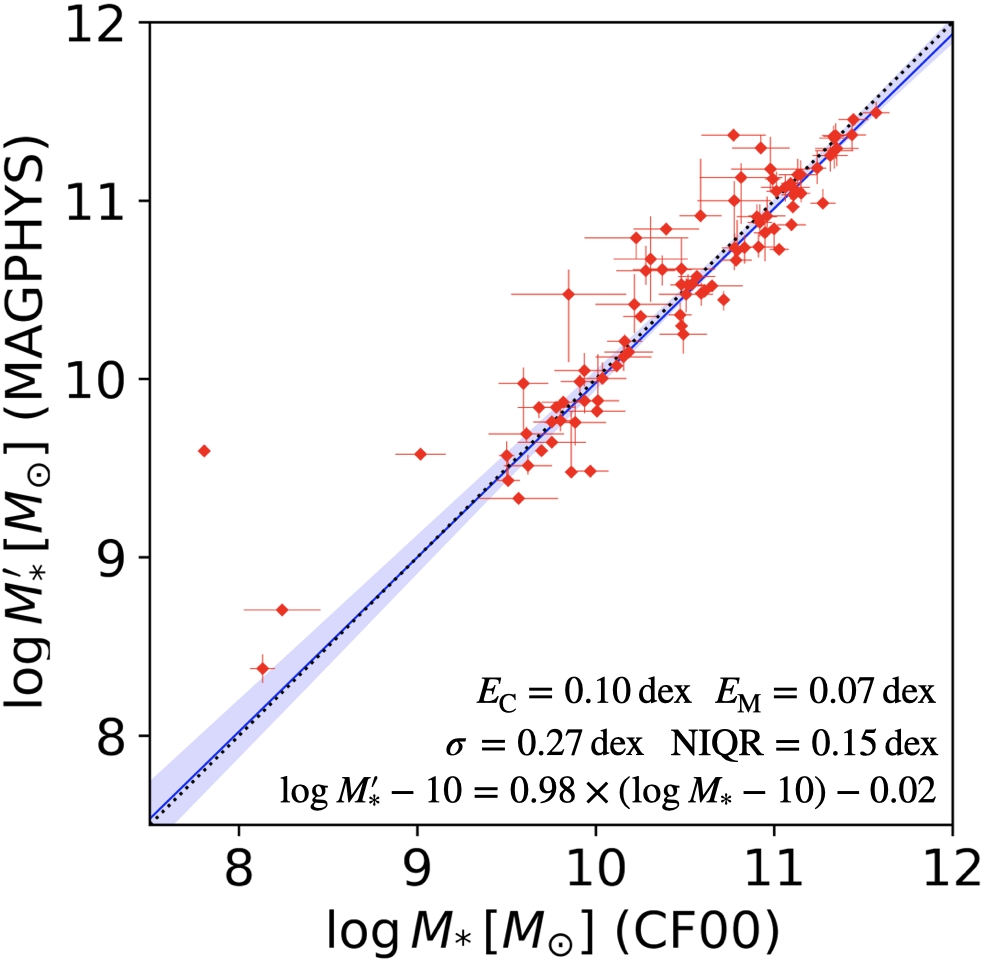}
 \caption{Comparison of stellar mass estimated with \texttt{CIGALE} using the \citet{2000ApJ...539..718C} model with that estimated with \texttt{MAGPHYS}. The symbols are the same as those in Figure~\ref{figure:cigale_magphys}. The lensing magnification is corrected for the stellar masses. \label{figure:CF00_magphys}}
\end{figure}

In this section, we check possible systematic differences in the estimation of physical properties between \texttt{CIGALE} and \texttt{MAGPHYS}. See also \citet{2019A&A...621A..51H} and \citet{2023ApJ...944..141P} for comprehensive studies among some popular SED analysis codes. For a conservative discussion, we only discuss the tier-1 samples, where the physical properties are more reliably constrained than the other samples. In addition, since the current version of \texttt{MAGPHYS} cannot treat an AGN component, ALCS-XAGNs and ALCS-nonXAGNs are excluded. Figure~\ref{figure:cigale_magphys} compares the major physical properties estimated with \texttt{CIGALE} with those with \texttt{MAGPHYS}. We perform Passing-Bablok regression analysis for each parameter, where the conversion relation is shown at the bottom of each panel. We find good agreement for SFR, dust luminosity, and dust mass, while a large discrepancy is found for dust temperature (${\sim}$7 K at 30 K) and stellar mass (${\sim}$0.37 dex at $\log M_*/M_{\odot}=10.5$). These differences can be explained as follows:
\begin{enumerate}
 \item The systematic difference in dust temperature can be attributed to its definition. In the EThemis model (used in \texttt{CIGALE}), dust temperature is defined as the intrinsic temperature of cold interstellar dust illuminated by a radiation field of $U=U_{\mathrm{min}}$. By contrast, in \texttt{MAGPHYS}, dust temperature is defined as the luminosity-weighted average of warm birth clouds and cold interstellar dust. Since dust temperature derived by \texttt{MAGPHYS} is also affected by the warm dust components, this value tends to be larger than that derived by \texttt{CIGALE}.
 \item The systematic difference in stellar mass can be attributed to the difference in dust attenuation models. In this study, we use the modified Calzetti law in \texttt{CIGALE}, whereas the \citet{2000ApJ...539..718C} model is used in \texttt{MAGPHYS}. The modified Calzetti law assumes a single dust component (interstellar dust) with flexible attenuation curves. By contrast, the \citet{2000ApJ...539..718C} model assumes two dust components with fixed power-law attenuation slopes (interstellar dust and birth cloud). In the model of \citet{2000ApJ...539..718C}, even near-infrared emission from young stars can be heavily attenuated by thick birth clouds. This may cause overestimation of intrinsic near-infrared luminosities, and hence stellar masses in \texttt{MAGPHYS}.\footnote{By contrast, stellar masses can be underestimated with \texttt{CIGALE} because of the simplified dust component, which may underestimate the near-infrared emission from young stars.} In \texttt{CIGALE}, we can also choose the \citet{2000ApJ...539..718C} model for the dust attenuation model. In this case, we confirm good agreement between stellar mass estimated with \texttt{CIGALE} and that with \texttt{MAGPHYS} (Figure~\ref{figure:CF00_magphys}).
\end{enumerate}
Moreover, we confirm that the standard deviations around the regression lines are 0.59 dex, 0.40 dex, 0.17 dex, 0.33 dex, and 7.8 K for SFR, stellar mass, dust luminosity, dust mass, and dust temperature, respectively. These values are significantly larger than the median statistical errors estimated with \text{CIGALE} and \text{MAGPHYS}. This indicates that the reported errors in the physical properties are fairly underestimated, considering the uncertainties in the SED models.

\section{Conversion method of dust temperature}\label{appendix:temperature}

In this section, we summarize the conversion method of dust temperature among different definitions. First, for the ALESS sample, \citet{2021ApJ...919...30D} adopted a single optically-thin graybody with various emissivity indices. In this case, characteristic dust temperature is calculated as:
\begin{equation}\label{equation:gray_Tconvert}
 T_{\mathrm{dust}}^{\mathrm{peak}} = \left[W\left(-(\beta+3) e^{-\beta-3}\right)+\beta+3\right]/2.821 \times T_{\mathrm{dust}}
\end{equation}
where $W$ is the Lambert W function (principal branch) and $\beta$ is the emissivity index. Second, for the AS2UDS sample, \citet{2020MNRAS.494.3828D} estimated the dust temperatures by adopting a single optically-thin graybody with an emissivity index of 1.8. This value is converted to the characteristic temperature by applying Equation~\ref{equation:gray_Tconvert} assuming $\beta=1.8$, which is described as:
\begin{equation}\label{equation:AS2UDS_Tconvert}
 T_{\mathrm{dust}}^{\mathrm{peak}}\simeq 1.7\times T_{\mathrm{dust}}
\end{equation}
Third, in the case of the ALCS sample, as shown in \citet{2019A&A...624A..80N}, the far-infrared emission from the EThemis model can be approximated by a graybody with an emissivity index of 1.79. Therefore, we convert the value by applying Equation~\ref{equation:AS2UDS_Tconvert}. Finally, for the ASPECS sample, \citet{2020ApJ...901...79A} used \texttt{MAGPHYS} to estimate the dust temperature. Since \texttt{MAGPHYS} uses a complex dust emission model, it is difficult to obtain the conversion relation from the model itself. Hence, we utilize the conversion law between \texttt{MAGPHYS} and \texttt{CIGALE} denoted in Figure~\ref{figure:cigale_magphys}, and yield the conversion relation using Equation~\ref{equation:AS2UDS_Tconvert}, which is described as:
\begin{eqnarray}
 T_{\mathrm{dust}}^{\mathrm{peak}}
 &\simeq& 1.7 * T_{\mathrm{dust\,[CIGALE]}} \nonumber\\
 &\simeq& 1.7 * (T_{\mathrm{dust\,[MAGPHYS]}}-8.3)/0.95
\end{eqnarray}

\bibliography{citation}{}

\begin{thebibliography}{}
\expandafter\ifx\csname natexlab\endcsname\relax\def\natexlab#1{#1}\fi
\providecommand{\url}[1]{\href{#1}{#1}}
\providecommand{\dodoi}[1]{doi:~\href{http://doi.org/#1}{\nolinkurl{#1}}}
\providecommand{\doeprint}[1]{\href{http://ascl.net/#1}{\nolinkurl{http://ascl.net/#1}}}
\providecommand{\doarXiv}[1]{\href{https://arxiv.org/abs/#1}{\nolinkurl{https://arxiv.org/abs/#1}}}

\bibitem[{{Algera} {et~al.}(2020){Algera}, {Smail}, {Dudzevi{\v{c}}i{\={u}}t{\.{e}}}, {Swinbank}, {Stach}, {Hodge}, {Thomson}, {Almaini}, {Arumugam}, {Blain}, {Calistro-Rivera}, {Chapman}, {Chen}, {da Cunha}, {Farrah}, {Leslie}, {Scott}, {van der Vlugt}, {Wardlow}, \& {van der Werf}}]{2020ApJ...903..138A}
{Algera}, H.~S.~B., {Smail}, I., {Dudzevi{\v{c}}i{\={u}}t{\.{e}}}, U., {et~al.} 2020, \apj, 903, 138, \dodoi{10.3847/1538-4357/abb77b}

\bibitem[{{{\'A}lvarez-M{\'a}rquez} {et~al.}(2016){{\'A}lvarez-M{\'a}rquez}, {Burgarella}, {Heinis}, {Buat}, {Lo Faro}, {B{\'e}thermin}, {L{\'o}pez-Fort{\'\i}n}, {Cooray}, {Farrah}, {Hurley}, {Ibar}, {Ilbert}, {Koekemoer}, {Lemaux}, {P{\'e}rez-Fournon}, {Rodighiero}, {Salvato}, {Scott}, {Taniguchi}, {Vieira}, \& {Wang}}]{2016A&A...587A.122A}
{{\'A}lvarez-M{\'a}rquez}, J., {Burgarella}, D., {Heinis}, S., {et~al.} 2016, \aap, 587, A122, \dodoi{10.1051/0004-6361/201527190}

\bibitem[{{Aravena} {et~al.}(2016{\natexlab{a}}){Aravena}, {Decarli}, {Walter}, {Da Cunha}, {Bauer}, {Carilli}, {Daddi}, {Elbaz}, {Ivison}, {Riechers}, {Smail}, {Swinbank}, {Weiss}, {Anguita}, {Assef}, {Bell}, {Bertoldi}, {Bacon}, {Bouwens}, {Cortes}, {Cox}, {G{\'o}nzalez-L{\'o}pez}, {Hodge}, {Ibar}, {Inami}, {Infante}, {Karim}, {Le Le F{\`e}vre}, {Magnelli}, {Ota}, {Popping}, {Sheth}, {van der Werf}, \& {Wagg}}]{2016ApJ...833...68A}
{Aravena}, M., {Decarli}, R., {Walter}, F., {et~al.} 2016{\natexlab{a}}, \apj, 833, 68, \dodoi{10.3847/1538-4357/833/1/68}

\bibitem[{{Aravena} {et~al.}(2016{\natexlab{b}}){Aravena}, {Decarli}, {Walter}, {Bouwens}, {Oesch}, {Carilli}, {Bauer}, {Da Cunha}, {Daddi}, {G{\'o}nzalez-L{\'o}pez}, {Ivison}, {Riechers}, {Smail}, {Swinbank}, {Weiss}, {Anguita}, {Bacon}, {Bell}, {Bertoldi}, {Cortes}, {Cox}, {Hodge}, {Ibar}, {Inami}, {Infante}, {Karim}, {Magnelli}, {Ota}, {Popping}, {van der Werf}, {Wagg}, \& {Fudamoto}}]{2016ApJ...833...71A}
---. 2016{\natexlab{b}}, \apj, 833, 71, \dodoi{10.3847/1538-4357/833/1/71}

\bibitem[{{Aravena} {et~al.}(2019){Aravena}, {Decarli}, {G{\'o}nzalez-L{\'o}pez}, {Boogaard}, {Walter}, {Carilli}, {Popping}, {Weiss}, {Assef}, {Bacon}, {Bauer}, {Bertoldi}, {Bouwens}, {Contini}, {Cortes}, {Cox}, {da Cunha}, {Daddi}, {D{\'\i}az-Santos}, {Elbaz}, {Hodge}, {Inami}, {Ivison}, {Le F{\`e}vre}, {Magnelli}, {Oesch}, {Riechers}, {Smail}, {Somerville}, {Swinbank}, {Uzgil}, {van der Werf}, {Wagg}, \& {Wisotzki}}]{2019ApJ...882..136A}
{Aravena}, M., {Decarli}, R., {G{\'o}nzalez-L{\'o}pez}, J., {et~al.} 2019, \apj, 882, 136, \dodoi{10.3847/1538-4357/ab30df}

\bibitem[{{Aravena} {et~al.}(2020){Aravena}, {Boogaard}, {G{\'o}nzalez-L{\'o}pez}, {Decarli}, {Walter}, {Carilli}, {Smail}, {Weiss}, {Assef}, {Bauer}, {Bouwens}, {Cortes}, {Cox}, {da Cunha}, {Daddi}, {D{\'\i}az-Santos}, {Inami}, {Ivison}, {Novak}, {Popping}, {Riechers}, {van der Werf}, \& {Wagg}}]{2020ApJ...901...79A}
{Aravena}, M., {Boogaard}, L., {G{\'o}nzalez-L{\'o}pez}, J., {et~al.} 2020, \apj, 901, 79, \dodoi{10.3847/1538-4357/ab99a2}

\bibitem[{{Battisti} {et~al.}(2020){Battisti}, {Cunha}, {Shivaei}, \& {Calzetti}}]{2020ApJ...888..108B}
{Battisti}, A.~J., {Cunha}, E.~d., {Shivaei}, I., \& {Calzetti}, D. 2020, \apj, 888, 108, \dodoi{10.3847/1538-4357/ab5fdd}

\bibitem[{{Battisti} {et~al.}(2019){Battisti}, {da Cunha}, {Grasha}, {Salvato}, {Daddi}, {Davies}, {Jin}, {Liu}, {Schinnerer}, {Vaccari}, \& {COSMOS Collaboration}}]{2019ApJ...882...61B}
{Battisti}, A.~J., {da Cunha}, E., {Grasha}, K., {et~al.} 2019, \apj, 882, 61, \dodoi{10.3847/1538-4357/ab345d}

\bibitem[{{Beckwith} {et~al.}(2006){Beckwith}, {Stiavelli}, {Koekemoer}, {Caldwell}, {Ferguson}, {Hook}, {Lucas}, {Bergeron}, {Corbin}, {Jogee}, {Panagia}, {Robberto}, {Royle}, {Somerville}, \& {Sosey}}]{2006AJ....132.1729B}
{Beckwith}, S. V.~W., {Stiavelli}, M., {Koekemoer}, A.~M., {et~al.} 2006, \aj, 132, 1729, \dodoi{10.1086/507302}

\bibitem[{{Bertin} \& {Arnouts}(1996)}]{1996A&AS..117..393B}
{Bertin}, E., \& {Arnouts}, S. 1996, \aaps, 117, 393, \dodoi{10.1051/aas:1996164}

\bibitem[{{Blain} {et~al.}(2002){Blain}, {Smail}, {Ivison}, {Kneib}, \& {Frayer}}]{2002PhR...369..111B}
{Blain}, A.~W., {Smail}, I., {Ivison}, R.~J., {Kneib}, J.~P., \& {Frayer}, D.~T. 2002, \physrep, 369, 111, \dodoi{10.1016/S0370-1573(02)00134-5}

\bibitem[{{Boogaard} {et~al.}(2019){Boogaard}, {Decarli}, {Gonz{\'a}lez-L{\'o}pez}, {van der Werf}, {Walter}, {Bouwens}, {Aravena}, {Carilli}, {Bauer}, {Brinchmann}, {Contini}, {Cox}, {da Cunha}, {Daddi}, {D{\'\i}az-Santos}, {Hodge}, {Inami}, {Ivison}, {Maseda}, {Matthee}, {Oesch}, {Popping}, {Riechers}, {Schaye}, {Schouws}, {Smail}, {Weiss}, {Wisotzki}, {Bacon}, {Cortes}, {Rix}, {Somerville}, {Swinbank}, \& {Wagg}}]{2019ApJ...882..140B}
{Boogaard}, L.~A., {Decarli}, R., {Gonz{\'a}lez-L{\'o}pez}, J., {et~al.} 2019, \apj, 882, 140, \dodoi{10.3847/1538-4357/ab3102}

\bibitem[{{Boogaard} {et~al.}(2020){Boogaard}, {van der Werf}, {Weiss}, {Popping}, {Decarli}, {Walter}, {Aravena}, {Bouwens}, {Riechers}, {Gonz{\'a}lez-L{\'o}pez}, {Smail}, {Carilli}, {Kaasinen}, {Daddi}, {Cox}, {D{\'\i}az-Santos}, {Inami}, {Cortes}, \& {Wagg}}]{2020ApJ...902..109B}
{Boogaard}, L.~A., {van der Werf}, P., {Weiss}, A., {et~al.} 2020, \apj, 902, 109, \dodoi{10.3847/1538-4357/abb82f}

\bibitem[{{Boquien} {et~al.}(2019){Boquien}, {Burgarella}, {Roehlly}, {Buat}, {Ciesla}, {Corre}, {Inoue}, \& {Salas}}]{2019A&A...622A.103B}
{Boquien}, M., {Burgarella}, D., {Roehlly}, Y., {et~al.} 2019, \aap, 622, A103, \dodoi{10.1051/0004-6361/201834156}

\bibitem[{{Boquien} {et~al.}(2022){Boquien}, {Buat}, {Burgarella}, {Bardelli}, {B{\'e}thermin}, {Faisst}, {Ginolfi}, {Hathi}, {Jones}, {Koekemoer}, {Lemaux}, {Narayanan}, {Romano}, {Schaerer}, {Vergani}, {Zamorani}, \& {Zucca}}]{2022A&A...663A..50B}
{Boquien}, M., {Buat}, V., {Burgarella}, D., {et~al.} 2022, \aap, 663, A50, \dodoi{10.1051/0004-6361/202142537}

\bibitem[{{Bouchet} {et~al.}(1985){Bouchet}, {Lequeux}, {Maurice}, {Prevot}, \& {Prevot-Burnichon}}]{1985A&A...149..330B}
{Bouchet}, P., {Lequeux}, J., {Maurice}, E., {Prevot}, L., \& {Prevot-Burnichon}, M.~L. 1985, \aap, 149, 330

\bibitem[{{Bouwens} {et~al.}(2020){Bouwens}, {Gonz{\'a}lez-L{\'o}pez}, {Aravena}, {Decarli}, {Novak}, {Stefanon}, {Walter}, {Boogaard}, {Carilli}, {Dudzevi{\v{c}}i{\={u}}t{\.{e}}}, {Smail}, {Daddi}, {da Cunha}, {Ivison}, {Nanayakkara}, {Cortes}, {Cox}, {Inami}, {Oesch}, {Popping}, {Riechers}, {van der Werf}, {Weiss}, {Fudamoto}, \& {Wagg}}]{2020ApJ...902..112B}
{Bouwens}, R., {Gonz{\'a}lez-L{\'o}pez}, J., {Aravena}, M., {et~al.} 2020, \apj, 902, 112, \dodoi{10.3847/1538-4357/abb830}

\bibitem[{{Bouwens} {et~al.}(2016){Bouwens}, {Aravena}, {Decarli}, {Walter}, {da Cunha}, {Labb{\'e}}, {Bauer}, {Bertoldi}, {Carilli}, {Chapman}, {Daddi}, {Hodge}, {Ivison}, {Karim}, {Le Fevre}, {Magnelli}, {Ota}, {Riechers}, {Smail}, {van der Werf}, {Weiss}, {Cox}, {Elbaz}, {Gonzalez-Lopez}, {Infante}, {Oesch}, {Wagg}, \& {Wilkins}}]{2016ApJ...833...72B}
{Bouwens}, R.~J., {Aravena}, M., {Decarli}, R., {et~al.} 2016, \apj, 833, 72, \dodoi{10.3847/1538-4357/833/1/72}

\bibitem[{{Brightman} {et~al.}(2017){Brightman}, {Balokovi{\'c}}, {Ballantyne}, {Bauer}, {Boorman}, {Buchner}, {Brandt}, {Comastri}, {Del Moro}, {Farrah}, {Gandhi}, {Harrison}, {Koss}, {Lanz}, {Masini}, {Ricci}, {Stern}, {Vasudevan}, \& {Walton}}]{2017ApJ...844...10B}
{Brightman}, M., {Balokovi{\'c}}, M., {Ballantyne}, D.~R., {et~al.} 2017, \apj, 844, 10, \dodoi{10.3847/1538-4357/aa75c9}

\bibitem[{{Bruzual} \& {Charlot}(2003)}]{2003MNRAS.344.1000B}
{Bruzual}, G., \& {Charlot}, S. 2003, \mnras, 344, 1000, \dodoi{10.1046/j.1365-8711.2003.06897.x}

\bibitem[{{Burnham} {et~al.}(2021){Burnham}, {Casey}, {Zavala}, {Manning}, {Spilker}, {Chapman}, {Chen}, {Cooray}, {Sanders}, \& {Scoville}}]{2021ApJ...910...89B}
{Burnham}, A.~D., {Casey}, C.~M., {Zavala}, J.~A., {et~al.} 2021, \apj, 910, 89, \dodoi{10.3847/1538-4357/abe401}

\bibitem[{{Calzetti} {et~al.}(2000){Calzetti}, {Armus}, {Bohlin}, {Kinney}, {Koornneef}, \& {Storchi-Bergmann}}]{2000ApJ...533..682C}
{Calzetti}, D., {Armus}, L., {Bohlin}, R.~C., {et~al.} 2000, \apj, 533, 682, \dodoi{10.1086/308692}

\bibitem[{{Calzetti} {et~al.}(1994){Calzetti}, {Kinney}, \& {Storchi-Bergmann}}]{1994ApJ...429..582C}
{Calzetti}, D., {Kinney}, A.~L., \& {Storchi-Bergmann}, T. 1994, \apj, 429, 582, \dodoi{10.1086/174346}

\bibitem[{{Caputi} {et~al.}(2021){Caputi}, {Caminha}, {Fujimoto}, {Kohno}, {Sun}, {Egami}, {Deshmukh}, {Tang}, {Ao}, {Bradley}, {Coe}, {Espada}, {Grillo}, {Hatsukade}, {Knudsen}, {Lee}, {Magdis}, {Morokuma-Matsui}, {Oesch}, {Ouchi}, {Rosati}, {Umehata}, {Valentino}, {Vanzella}, {Wang}, {Wu}, \& {Zitrin}}]{2021ApJ...908..146C}
{Caputi}, K.~I., {Caminha}, G.~B., {Fujimoto}, S., {et~al.} 2021, \apj, 908, 146, \dodoi{10.3847/1538-4357/abd4d0}

\bibitem[{{Carilli} {et~al.}(2016){Carilli}, {Chluba}, {Decarli}, {Walter}, {Aravena}, {Wagg}, {Popping}, {Cortes}, {Hodge}, {Weiss}, {Bertoldi}, \& {Riechers}}]{2016ApJ...833...73C}
{Carilli}, C.~L., {Chluba}, J., {Decarli}, R., {et~al.} 2016, \apj, 833, 73, \dodoi{10.3847/1538-4357/833/1/73}

\bibitem[{{Casey}(2012)}]{2012MNRAS.425.3094C}
{Casey}, C.~M. 2012, \mnras, 425, 3094, \dodoi{10.1111/j.1365-2966.2012.21455.x}

\bibitem[{{Casey} {et~al.}(2014){Casey}, {Narayanan}, \& {Cooray}}]{2014PhR...541...45C}
{Casey}, C.~M., {Narayanan}, D., \& {Cooray}, A. 2014, \physrep, 541, 45, \dodoi{10.1016/j.physrep.2014.02.009}

\bibitem[{{Cazaux} \& {Spaans}(2009)}]{2009A&A...496..365C}
{Cazaux}, S., \& {Spaans}, M. 2009, \aap, 496, 365, \dodoi{10.1051/0004-6361:200811302}

\bibitem[{{Cazaux} \& {Tielens}(2004)}]{2004ApJ...604..222C}
{Cazaux}, S., \& {Tielens}, A.~G.~G.~M. 2004, \apj, 604, 222, \dodoi{10.1086/381775}

\bibitem[{{Chabrier}(2003)}]{2003PASP..115..763C}
{Chabrier}, G. 2003, \pasp, 115, 763, \dodoi{10.1086/376392}

\bibitem[{{Charlot} \& {Fall}(2000)}]{2000ApJ...539..718C}
{Charlot}, S., \& {Fall}, S.~M. 2000, \apj, 539, 718, \dodoi{10.1086/309250}

\bibitem[{{Chen} {et~al.}(2015){Chen}, {Smail}, {Swinbank}, {Simpson}, {Ma}, {Alexander}, {Biggs}, {Brandt}, {Chapman}, {Coppin}, {Danielson}, {Dannerbauer}, {Edge}, {Greve}, {Ivison}, {Karim}, {Menten}, {Schinnerer}, {Walter}, {Wardlow}, {Wei{\ss}}, \& {van der Werf}}]{2015ApJ...799..194C}
{Chen}, C.-C., {Smail}, I., {Swinbank}, A.~M., {et~al.} 2015, \apj, 799, 194, \dodoi{10.1088/0004-637X/799/2/194}

\bibitem[{{Ciesla} {et~al.}(2023){Ciesla}, {G{\'o}mez-Guijarro}, {Buat}, {Elbaz}, {Jin}, {B{\'e}thermin}, {Daddi}, {Franco}, {Inami}, {Magdis}, {Magnelli}, \& {Xiao}}]{2023A&A...672A.191C}
{Ciesla}, L., {G{\'o}mez-Guijarro}, C., {Buat}, V., {et~al.} 2023, \aap, 672, A191, \dodoi{10.1051/0004-6361/202245376}

\bibitem[{{Clements} {et~al.}(2018){Clements}, {Pearson}, {Farrah}, {Greenslade}, {Bernard-Salas}, {Gonz{\'a}lez-Alfonso}, {Afonso}, {Efstathiou}, {Rigopoulou}, {Lebouteiller}, {Hurley}, \& {Spoon}}]{2018MNRAS.475.2097C}
{Clements}, D.~L., {Pearson}, C., {Farrah}, D., {et~al.} 2018, \mnras, 475, 2097, \dodoi{10.1093/mnras/stx3227}

\bibitem[{{Coe} {et~al.}(2019){Coe}, {Salmon}, {Brada{\v{c}}}, {Bradley}, {Sharon}, {Zitrin}, {Acebron}, {Cerny}, {Cibirka}, {Strait}, {Paterno-Mahler}, {Mahler}, {Avila}, {Ogaz}, {Huang}, {Pelliccia}, {Stark}, {Mainali}, {Oesch}, {Trenti}, {Carrasco}, {Dawson}, {Rodney}, {Strolger}, {Riess}, {Jones}, {Frye}, {Czakon}, {Umetsu}, {Vulcani}, {Graur}, {Jha}, {Graham}, {Molino}, {Nonino}, {Hjorth}, {Selsing}, {Christensen}, {Kikuchihara}, {Ouchi}, {Oguri}, {Welch}, {Lemaux}, {Andrade-Santos}, {Hoag}, {Johnson}, {Peterson}, {Past}, {Fox}, {Agulli}, {Livermore}, {Ryan}, {Lam}, {Sendra-Server}, {Toft}, {Lovisari}, \& {Su}}]{2019ApJ...884...85C}
{Coe}, D., {Salmon}, B., {Brada{\v{c}}}, M., {et~al.} 2019, \apj, 884, 85, \dodoi{10.3847/1538-4357/ab412b}

\bibitem[{{Compi{\`e}gne} {et~al.}(2011){Compi{\`e}gne}, {Verstraete}, {Jones}, {Bernard}, {Boulanger}, {Flagey}, {Le Bourlot}, {Paradis}, \& {Ysard}}]{2011A&A...525A.103C}
{Compi{\`e}gne}, M., {Verstraete}, L., {Jones}, A., {et~al.} 2011, \aap, 525, A103, \dodoi{10.1051/0004-6361/201015292}

\bibitem[{{Cooke} {et~al.}(2018){Cooke}, {Smail}, {Swinbank}, {Stach}, {An}, {Gullberg}, {Almaini}, {Simpson}, {Wardlow}, {Blain}, {Chapman}, {Chen}, {Conselice}, {Coppin}, {Farrah}, {Maltby}, {Micha{\l}owski}, {Scott}, {Simpson}, {Thomson}, \& {van der Werf}}]{2018ApJ...861..100C}
{Cooke}, E.~A., {Smail}, I., {Swinbank}, A.~M., {et~al.} 2018, \apj, 861, 100, \dodoi{10.3847/1538-4357/aac6ba}

\bibitem[{{Coppin} {et~al.}(2008){Coppin}, {Halpern}, {Scott}, {Borys}, {Dunlop}, {Dunne}, {Ivison}, {Wagg}, {Aretxaga}, {Battistelli}, {Benson}, {Blain}, {Chapman}, {Clements}, {Dye}, {Farrah}, {Hughes}, {Jenness}, {van Kampen}, {Lacey}, {Mortier}, {Pope}, {Priddey}, {Serjeant}, {Smail}, {Stevens}, \& {Vaccari}}]{2008MNRAS.384.1597C}
{Coppin}, K., {Halpern}, M., {Scott}, D., {et~al.} 2008, \mnras, 384, 1597, \dodoi{10.1111/j.1365-2966.2007.12808.x}

\bibitem[{{Cortese} {et~al.}(2014){Cortese}, {Fritz}, {Bianchi}, {Boselli}, {Ciesla}, {Bendo}, {Boquien}, {Roussel}, {Baes}, {Buat}, {Clemens}, {Cooray}, {Cormier}, {Davies}, {De Looze}, {Eales}, {Fuller}, {Hunt}, {Madden}, {Munoz-Mateos}, {Pappalardo}, {Pierini}, {R{\'e}my-Ruyer}, {Sauvage}, {di Serego Alighieri}, {Smith}, {Spinoglio}, {Vaccari}, \& {Vlahakis}}]{2014MNRAS.440..942C}
{Cortese}, L., {Fritz}, J., {Bianchi}, S., {et~al.} 2014, \mnras, 440, 942, \dodoi{10.1093/mnras/stu175}

\bibitem[{{da Cunha} {et~al.}(2008){da Cunha}, {Charlot}, \& {Elbaz}}]{2008MNRAS.388.1595D}
{da Cunha}, E., {Charlot}, S., \& {Elbaz}, D. 2008, \mnras, 388, 1595, \dodoi{10.1111/j.1365-2966.2008.13535.x}

\bibitem[{{da Cunha} {et~al.}(2015){da Cunha}, {Walter}, {Smail}, {Swinbank}, {Simpson}, {Decarli}, {Hodge}, {Weiss}, {van der Werf}, {Bertoldi}, {Chapman}, {Cox}, {Danielson}, {Dannerbauer}, {Greve}, {Ivison}, {Karim}, \& {Thomson}}]{2015ApJ...806..110D}
{da Cunha}, E., {Walter}, F., {Smail}, I.~R., {et~al.} 2015, \apj, 806, 110, \dodoi{10.1088/0004-637X/806/1/110}

\bibitem[{{da Cunha} {et~al.}(2021){da Cunha}, {Hodge}, {Casey}, {Algera}, {Kaasinen}, {Smail}, {Walter}, {Brandt}, {Dannerbauer}, {Decarli}, {Groves}, {Knudsen}, {Swinbank}, {Weiss}, {van der Werf}, \& {Zavala}}]{2021ApJ...919...30D}
{da Cunha}, E., {Hodge}, J.~A., {Casey}, C.~M., {et~al.} 2021, \apj, 919, 30, \dodoi{10.3847/1538-4357/ac0ae0}

\bibitem[{{Danielson} {et~al.}(2017){Danielson}, {Swinbank}, {Smail}, {Simpson}, {Casey}, {Chapman}, {da Cunha}, {Hodge}, {Walter}, {Wardlow}, {Alexander}, {Brandt}, {de Breuck}, {Coppin}, {Dannerbauer}, {Dickinson}, {Edge}, {Gawiser}, {Ivison}, {Karim}, {Kovacs}, {Lutz}, {Menten}, {Schinnerer}, {Wei{\ss}}, \& {van der Werf}}]{2017ApJ...840...78D}
{Danielson}, A.~L.~R., {Swinbank}, A.~M., {Smail}, I., {et~al.} 2017, \apj, 840, 78, \dodoi{10.3847/1538-4357/aa6caf}

\bibitem[{{Decarli} {et~al.}(2014){Decarli}, {Smail}, {Walter}, {Swinbank}, {Chapman}, {Coppin}, {Cox}, {Dannerbauer}, {Greve}, {Hodge}, {Ivison}, {Karim}, {Knudsen}, {Lindroos}, {Rix}, {Schinnerer}, {Simpson}, {van der Werf}, \& {Wei{\ss}}}]{2014ApJ...780..115D}
{Decarli}, R., {Smail}, I., {Walter}, F., {et~al.} 2014, \apj, 780, 115, \dodoi{10.1088/0004-637X/780/2/115}

\bibitem[{{Decarli} {et~al.}(2016{\natexlab{a}}){Decarli}, {Walter}, {Aravena}, {Carilli}, {Bouwens}, {da Cunha}, {Daddi}, {Ivison}, {Popping}, {Riechers}, {Smail}, {Swinbank}, {Weiss}, {Anguita}, {Assef}, {Bauer}, {Bell}, {Bertoldi}, {Chapman}, {Colina}, {Cortes}, {Cox}, {Dickinson}, {Elbaz}, {G{\'o}nzalez-L{\'o}pez}, {Ibar}, {Infante}, {Hodge}, {Karim}, {Le Fevre}, {Magnelli}, {Neri}, {Oesch}, {Ota}, {Rix}, {Sargent}, {Sheth}, {van der Wel}, {van der Werf}, \& {Wagg}}]{2016ApJ...833...69D}
{Decarli}, R., {Walter}, F., {Aravena}, M., {et~al.} 2016{\natexlab{a}}, \apj, 833, 69, \dodoi{10.3847/1538-4357/833/1/69}

\bibitem[{{Decarli} {et~al.}(2016{\natexlab{b}}){Decarli}, {Walter}, {Aravena}, {Carilli}, {Bouwens}, {da Cunha}, {Daddi}, {Elbaz}, {Riechers}, {Smail}, {Swinbank}, {Weiss}, {Bacon}, {Bauer}, {Bell}, {Bertoldi}, {Chapman}, {Colina}, {Cortes}, {Cox}, {G{\'o}nzalez-L{\'o}pez}, {Inami}, {Ivison}, {Hodge}, {Karim}, {Magnelli}, {Ota}, {Popping}, {Rix}, {Sargent}, {van der Wel}, \& {van der Werf}}]{2016ApJ...833...70D}
---. 2016{\natexlab{b}}, \apj, 833, 70, \dodoi{10.3847/1538-4357/833/1/70}

\bibitem[{{Decarli} {et~al.}(2019){Decarli}, {Walter}, {G{\'o}nzalez-L{\'o}pez}, {Aravena}, {Boogaard}, {Carilli}, {Cox}, {Daddi}, {Popping}, {Riechers}, {Uzgil}, {Weiss}, {Assef}, {Bacon}, {Bauer}, {Bertoldi}, {Bouwens}, {Contini}, {Cortes}, {da Cunha}, {D{\'\i}az-Santos}, {Elbaz}, {Inami}, {Hodge}, {Ivison}, {Le F{\`e}vre}, {Magnelli}, {Novak}, {Oesch}, {Rix}, {Sargent}, {Smail}, {Swinbank}, {Somerville}, {van der Werf}, {Wagg}, \& {Wisotzki}}]{2019ApJ...882..138D}
{Decarli}, R., {Walter}, F., {G{\'o}nzalez-L{\'o}pez}, J., {et~al.} 2019, \apj, 882, 138, \dodoi{10.3847/1538-4357/ab30fe}

\bibitem[{{Decarli} {et~al.}(2020){Decarli}, {Aravena}, {Boogaard}, {Carilli}, {Gonz{\'a}lez-L{\'o}pez}, {Walter}, {Cortes}, {Cox}, {da Cunha}, {Daddi}, {D{\'\i}az-Santos}, {Hodge}, {Inami}, {Neeleman}, {Novak}, {Oesch}, {Popping}, {Riechers}, {Smail}, {Uzgil}, {van der Werf}, {Wagg}, \& {Weiss}}]{2020ApJ...902..110D}
{Decarli}, R., {Aravena}, M., {Boogaard}, L., {et~al.} 2020, \apj, 902, 110, \dodoi{10.3847/1538-4357/abaa3b}

\bibitem[{{Drew} \& {Casey}(2022)}]{2022ApJ...930..142D}
{Drew}, P.~M., \& {Casey}, C.~M. 2022, \apj, 930, 142, \dodoi{10.3847/1538-4357/ac6270}

\bibitem[{{Dudzevi{\v{c}}i{\={u}}t{\.{e}}} {et~al.}(2020){Dudzevi{\v{c}}i{\={u}}t{\.{e}}}, {Smail}, {Swinbank}, {Stach}, {Almaini}, {da Cunha}, {An}, {Arumugam}, {Birkin}, {Blain}, {Chapman}, {Chen}, {Conselice}, {Coppin}, {Dunlop}, {Farrah}, {Geach}, {Gullberg}, {Hartley}, {Hodge}, {Ivison}, {Maltby}, {Scott}, {Simpson}, {Simpson}, {Thomson}, {Walter}, {Wardlow}, {Weiss}, \& {van der Werf}}]{2020MNRAS.494.3828D}
{Dudzevi{\v{c}}i{\={u}}t{\.{e}}}, U., {Smail}, I., {Swinbank}, A.~M., {et~al.} 2020, \mnras, 494, 3828, \dodoi{10.1093/mnras/staa769}

\bibitem[{{Dudzevi{\v{c}}i{\={u}}t{\.{e}}} {et~al.}(2021){Dudzevi{\v{c}}i{\={u}}t{\.{e}}}, {Smail}, {Swinbank}, {Lim}, {Wang}, {Simpson}, {Ao}, {Chapman}, {Chen}, {Clements}, {Dannerbauer}, {Ho}, {Hwang}, {Koprowski}, {Lee}, {Scott}, {Shim}, {Shirley}, \& {Toba}}]{2021MNRAS.500..942D}
---. 2021, \mnras, 500, 942, \dodoi{10.1093/mnras/staa3285}

\bibitem[{{Dunlop} {et~al.}(2017){Dunlop}, {McLure}, {Biggs}, {Geach}, {Micha{\l}owski}, {Ivison}, {Rujopakarn}, {van Kampen}, {Kirkpatrick}, {Pope}, {Scott}, {Swinbank}, {Targett}, {Aretxaga}, {Austermann}, {Best}, {Bruce}, {Chapin}, {Charlot}, {Cirasuolo}, {Coppin}, {Ellis}, {Finkelstein}, {Hayward}, {Hughes}, {Ibar}, {Jagannathan}, {Khochfar}, {Koprowski}, {Narayanan}, {Nyland}, {Papovich}, {Peacock}, {Rieke}, {Robertson}, {Vernstrom}, {Werf}, {Wilson}, \& {Yun}}]{2017MNRAS.466..861D}
{Dunlop}, J.~S., {McLure}, R.~J., {Biggs}, A.~D., {et~al.} 2017, \mnras, 466, 861, \dodoi{10.1093/mnras/stw3088}

\bibitem[{{Franco} {et~al.}(2018){Franco}, {Elbaz}, {B{\'e}thermin}, {Magnelli}, {Schreiber}, {Ciesla}, {Dickinson}, {Nagar}, {Silverman}, {Daddi}, {Alexander}, {Wang}, {Pannella}, {Le Floc'h}, {Pope}, {Giavalisco}, {Maury}, {Bournaud}, {Chary}, {Demarco}, {Ferguson}, {Finkelstein}, {Inami}, {Iono}, {Juneau}, {Lagache}, {Leiton}, {Lin}, {Magdis}, {Messias}, {Motohara}, {Mullaney}, {Okumura}, {Papovich}, {Pforr}, {Rujopakarn}, {Sargent}, {Shu}, \& {Zhou}}]{2018A&A...620A.152F}
{Franco}, M., {Elbaz}, D., {B{\'e}thermin}, M., {et~al.} 2018, \aap, 620, A152, \dodoi{10.1051/0004-6361/201832928}

\bibitem[{{Franco} {et~al.}(2020{\natexlab{a}}){Franco}, {Elbaz}, {Zhou}, {Magnelli}, {Schreiber}, {Ciesla}, {Dickinson}, {Nagar}, {Magdis}, {Alexander}, {B{\'e}thermin}, {Demarco}, {Daddi}, {Wang}, {Mullaney}, {Sargent}, {Inami}, {Shu}, {Bournaud}, {Chary}, {Coogan}, {Ferguson}, {Finkelstein}, {Giavalisco}, {G{\'o}mez-Guijarro}, {Iono}, {Juneau}, {Lagache}, {Lin}, {Motohara}, {Okumura}, {Pannella}, {Papovich}, {Pope}, {Rujopakarn}, {Silverman}, \& {Xiao}}]{2020A&A...643A..30F}
{Franco}, M., {Elbaz}, D., {Zhou}, L., {et~al.} 2020{\natexlab{a}}, \aap, 643, A30, \dodoi{10.1051/0004-6361/202038312}

\bibitem[{{Franco} {et~al.}(2020{\natexlab{b}}){Franco}, {Elbaz}, {Zhou}, {Magnelli}, {Schreiber}, {Ciesla}, {Dickinson}, {Nagar}, {Magdis}, {Alexander}, {B{\'e}thermin}, {Demarco}, {Daddi}, {Wang}, {Mullaney}, {Inami}, {Shu}, {Bournaud}, {Chary}, {Coogan}, {Ferguson}, {Finkelstein}, {Giavalisco}, {G{\'o}mez-Guijarro}, {Iono}, {Juneau}, {Lagache}, {Lin}, {Motohara}, {Okumura}, {Pannella}, {Papovich}, {Pope}, {Rujopakarn}, {Silverman}, \& {Xiao}}]{2020A&A...643A..53F}
---. 2020{\natexlab{b}}, \aap, 643, A53, \dodoi{10.1051/0004-6361/202038310}

\bibitem[{{Fruscione} {et~al.}(2006){Fruscione}, {McDowell}, {Allen}, {Brickhouse}, {Burke}, {Davis}, {Durham}, {Elvis}, {Galle}, {Harris}, {Huenemoerder}, {Houck}, {Ishibashi}, {Karovska}, {Nicastro}, {Noble}, {Nowak}, {Primini}, {Siemiginowska}, {Smith}, \& {Wise}}]{2006SPIE.6270E..1VF}
{Fruscione}, A., {McDowell}, J.~C., {Allen}, G.~E., {et~al.} 2006, in Society of Photo-Optical Instrumentation Engineers (SPIE) Conference Series, Vol. 6270, Society of Photo-Optical Instrumentation Engineers (SPIE) Conference Series, ed. D.~R. {Silva} \& R.~E. {Doxsey}, 62701V, \dodoi{10.1117/12.671760}

\bibitem[{{Fudamoto} {et~al.}(2020){Fudamoto}, {Oesch}, {Magnelli}, {Schinnerer}, {Liu}, {Lang}, {Jim{\'e}nez-Andrade}, {Groves}, {Leslie}, \& {Sargent}}]{2020MNRAS.491.4724F}
{Fudamoto}, Y., {Oesch}, P.~A., {Magnelli}, B., {et~al.} 2020, \mnras, 491, 4724, \dodoi{10.1093/mnras/stz3248}

\bibitem[{{Fujimoto} {et~al.}(2018){Fujimoto}, {Ouchi}, {Kohno}, {Yamaguchi}, {Hatsukade}, {Ueda}, {Shibuya}, {Inoue}, {Oogi}, {Toft}, {G{\'o}mez-Guijarro}, {Wang}, {Espada}, {Nagao}, {Tanaka}, {Ao}, {Umehata}, {Taniguchi}, {Nakanishi}, {Rujopakarn}, {Ivison}, {Wang}, {Lee}, {Tadaki}, {Tamura}, \& {Dunlop}}]{2018ApJ...861....7F}
{Fujimoto}, S., {Ouchi}, M., {Kohno}, K., {et~al.} 2018, \apj, 861, 7, \dodoi{10.3847/1538-4357/aac6c4}

\bibitem[{{Fujimoto} {et~al.}(2021){Fujimoto}, {Oguri}, {Brammer}, {Yoshimura}, {Laporte}, {Gonz{\'a}lez-L{\'o}pez}, {Caminha}, {Kohno}, {Zitrin}, {Richard}, {Ouchi}, {Bauer}, {Smail}, {Hatsukade}, {Ono}, {Kokorev}, {Umehata}, {Schaerer}, {Knudsen}, {Sun}, {Magdis}, {Valentino}, {Ao}, {Toft}, {Dessauges-Zavadsky}, {Shimasaku}, {Caputi}, {Kusakabe}, {Morokuma-Matsui}, {Shotaro}, {Egami}, {Lee}, {Rawle}, \& {Espada}}]{2021ApJ...911...99F}
{Fujimoto}, S., {Oguri}, M., {Brammer}, G., {et~al.} 2021, \apj, 911, 99, \dodoi{10.3847/1538-4357/abd7ec}

\bibitem[{{Fujimoto} {et~al.}(2023){Fujimoto}, {Kohno}, {Ouchi}, {Oguri}, {Kokorev}, {Brammer}, {Sun}, {Gonzalez-Lopez}, {Bauer}, {Caminha}, {Hatsukade}, {Richard}, {Smail}, {Tsujita}, {Ueda}, {Uematsu}, {Zitrin}, {Coe}, {Kneib}, {Postman}, {Umetsu}, {Lagos}, {Popping}, {Ao}, {Bradley}, {Caputi}, {Dessauges-Zavadsky}, {Egami}, {Espada}, {Ivison}, {Jauzac}, {Knudsen}, {Koekemoer}, {Magdis}, {Mahler}, {Munoz Arancibia}, {Rawle}, {Shimasaku}, {Toft}, {Umehata}, {Valentino}, {Wang}, \& {Wang}}]{2023arXiv230301658F}
{Fujimoto}, S., {Kohno}, K., {Ouchi}, M., {et~al.} 2023, arXiv e-prints, arXiv:2303.01658, \dodoi{10.48550/arXiv.2303.01658}

\bibitem[{{Geach} {et~al.}(2017){Geach}, {Dunlop}, {Halpern}, {Smail}, {van der Werf}, {Alexander}, {Almaini}, {Aretxaga}, {Arumugam}, {Asboth}, {Banerji}, {Beanlands}, {Best}, {Blain}, {Birkinshaw}, {Chapin}, {Chapman}, {Chen}, {Chrysostomou}, {Clarke}, {Clements}, {Conselice}, {Coppin}, {Cowley}, {Danielson}, {Eales}, {Edge}, {Farrah}, {Gibb}, {Harrison}, {Hine}, {Hughes}, {Ivison}, {Jarvis}, {Jenness}, {Jones}, {Karim}, {Koprowski}, {Knudsen}, {Lacey}, {Mackenzie}, {Marsden}, {McAlpine}, {McMahon}, {Meijerink}, {Micha{\l}owski}, {Oliver}, {Page}, {Peacock}, {Rigopoulou}, {Robson}, {Roseboom}, {Rotermund}, {Scott}, {Serjeant}, {Simpson}, {Simpson}, {Smith}, {Spaans}, {Stanley}, {Stevens}, {Swinbank}, {Targett}, {Thomson}, {Valiante}, {Wake}, {Webb}, {Willott}, {Zavala}, \& {Zemcov}}]{2017MNRAS.465.1789G}
{Geach}, J.~E., {Dunlop}, J.~S., {Halpern}, M., {et~al.} 2017, \mnras, 465, 1789, \dodoi{10.1093/mnras/stw2721}

\bibitem[{{G{\'o}mez-Guijarro} {et~al.}(2022{\natexlab{a}}){G{\'o}mez-Guijarro}, {Elbaz}, {Xiao}, {B{\'e}thermin}, {Franco}, {Magnelli}, {Daddi}, {Dickinson}, {Demarco}, {Inami}, {Rujopakarn}, {Magdis}, {Shu}, {Chary}, {Zhou}, {Alexander}, {Bournaud}, {Ciesla}, {Ferguson}, {Finkelstein}, {Giavalisco}, {Iono}, {Juneau}, {Kartaltepe}, {Lagache}, {Le Floc'h}, {Leiton}, {Lin}, {Motohara}, {Mullaney}, {Okumura}, {Pannella}, {Papovich}, {Pope}, {Sargent}, {Silverman}, {Treister}, \& {Wang}}]{2022A&A...658A..43G}
{G{\'o}mez-Guijarro}, C., {Elbaz}, D., {Xiao}, M., {et~al.} 2022{\natexlab{a}}, \aap, 658, A43, \dodoi{10.1051/0004-6361/202141615}

\bibitem[{{G{\'o}mez-Guijarro} {et~al.}(2022{\natexlab{b}}){G{\'o}mez-Guijarro}, {Elbaz}, {Xiao}, {Kokorev}, {Magdis}, {Magnelli}, {Daddi}, {Valentino}, {Sargent}, {Dickinson}, {B{\'e}thermin}, {Franco}, {Pope}, {Kalita}, {Ciesla}, {Demarco}, {Inami}, {Rujopakarn}, {Shu}, {Wang}, {Zhou}, {Alexander}, {Bournaud}, {Chary}, {Ferguson}, {Finkelstein}, {Giavalisco}, {Iono}, {Juneau}, {Kartaltepe}, {Lagache}, {Le Floc'h}, {Leiton}, {Leroy}, {Lin}, {Motohara}, {Mullaney}, {Okumura}, {Pannella}, {Papovich}, \& {Treister}}]{2022A&A...659A.196G}
---. 2022{\natexlab{b}}, \aap, 659, A196, \dodoi{10.1051/0004-6361/202142352}

\bibitem[{{Gonz{\'a}lez-L{\'o}pez} {et~al.}(2017){Gonz{\'a}lez-L{\'o}pez}, {Bauer}, {Romero-Ca{\~n}izales}, {Kneissl}, {Villard}, {Carvajal}, {Kim}, {Laporte}, {Anguita}, {Aravena}, {Bouwens}, {Bradley}, {Carrasco}, {Demarco}, {Ford}, {Ibar}, {Infante}, {Messias}, {Mu{\~n}oz Arancibia}, {Nagar}, {Padilla}, {Treister}, {Troncoso}, \& {Zitrin}}]{2017A&A...597A..41G}
{Gonz{\'a}lez-L{\'o}pez}, J., {Bauer}, F.~E., {Romero-Ca{\~n}izales}, C., {et~al.} 2017, \aap, 597, A41, \dodoi{10.1051/0004-6361/201628806}

\bibitem[{{Gonz{\'a}lez-L{\'o}pez} {et~al.}(2020){Gonz{\'a}lez-L{\'o}pez}, {Novak}, {Decarli}, {Walter}, {Aravena}, {Carilli}, {Boogaard}, {Popping}, {Weiss}, {Assef}, {Bauer}, {Bouwens}, {Cortes}, {Cox}, {Daddi}, {Cunha}, {D{\'\i}az-Santos}, {Ivison}, {Magnelli}, {Riechers}, {Smail}, {van der Werf}, \& {Wagg}}]{2020ApJ...897...91G}
{Gonz{\'a}lez-L{\'o}pez}, J., {Novak}, M., {Decarli}, R., {et~al.} 2020, \apj, 897, 91, \dodoi{10.3847/1538-4357/ab765b}

\bibitem[{{Gould} \& {Salpeter}(1963)}]{1963ApJ...138..393G}
{Gould}, R.~J., \& {Salpeter}, E.~E. 1963, \apj, 138, 393, \dodoi{10.1086/147654}

\bibitem[{{Gullberg} {et~al.}(2019){Gullberg}, {Smail}, {Swinbank}, {Dudzevi{\v{c}}i{\={u}}t{\.{e}}}, {Stach}, {Thomson}, {Almaini}, {Chen}, {Conselice}, {Cooke}, {Farrah}, {Ivison}, {Maltby}, {Micha{\l}owski}, {Simpson}, {Scott}, {Wardlow}, \& {Weiss}}]{2019MNRAS.490.4956G}
{Gullberg}, B., {Smail}, I., {Swinbank}, A.~M., {et~al.} 2019, \mnras, 490, 4956, \dodoi{10.1093/mnras/stz2835}

\bibitem[{{Gupta} {et~al.}(2021){Gupta}, {Ricci}, {Tortosa}, {Ueda}, {Kawamuro}, {Koss}, {Trakhtenbrot}, {Oh}, {Bauer}, {Ricci}, {Privon}, {Zappacosta}, {Stern}, {Kakkad}, {Piconcelli}, {Veilleux}, {Mushotzky}, {Caglar}, {Ichikawa}, {Elagali}, {Powell}, {Urry}, \& {Harrison}}]{2021MNRAS.504..428G}
{Gupta}, K.~K., {Ricci}, C., {Tortosa}, A., {et~al.} 2021, \mnras, 504, 428, \dodoi{10.1093/mnras/stab839}

\bibitem[{{Hatsukade} {et~al.}(2018){Hatsukade}, {Kohno}, {Yamaguchi}, {Umehata}, {Ao}, {Aretxaga}, {Caputi}, {Dunlop}, {Egami}, {Espada}, {Fujimoto}, {Hayatsu}, {Hughes}, {Ikarashi}, {Iono}, {Ivison}, {Kawabe}, {Kodama}, {Lee}, {Matsuda}, {Nakanishi}, {Ohta}, {Ouchi}, {Rujopakarn}, {Suzuki}, {Tamura}, {Ueda}, {Wang}, {Wang}, {Wilson}, {Yoshimura}, \& {Yun}}]{2018PASJ...70..105H}
{Hatsukade}, B., {Kohno}, K., {Yamaguchi}, Y., {et~al.} 2018, \pasj, 70, 105, \dodoi{10.1093/pasj/psy104}

\bibitem[{{Hatsukade} {et~al.}(2020){Hatsukade}, {Kohno}, {Yamaguchi}, {Umehata}, {Ao}, {Aretxaga}, {Caputi}, {Dunlop}, {Egami}, {Espada}, {Fujimoto}, {Hayatsu}, {Hughes}, {Ikarashi}, {Iono}, {Ivison}, {Kawabe}, {Kodama}, {Lee}, {Matsuda}, {Nakanishi}, {Ohta}, {Ouchi}, {Rujopakarn}, {Suzuki}, {Tamura}, {Ueda}, {Wang}, {Wang}, {Wilson}, {Yoshimura}, {Yun}, \& {Asagao Team}}]{2020IAUS..352..239H}
{Hatsukade}, B., {Kohno}, K., {Yamaguchi}, Y., {et~al.} 2020, in Uncovering Early Galaxy Evolution in the ALMA and JWST Era, ed. E.~{da Cunha}, J.~{Hodge}, J.~{Afonso}, L.~{Pentericci}, \& D.~{Sobral}, Vol. 352, 239--240, \dodoi{10.1017/S1743921319009542}

\bibitem[{{Hirashita} \& {Ferrara}(2002)}]{2002MNRAS.337..921H}
{Hirashita}, H., \& {Ferrara}, A. 2002, \mnras, 337, 921, \dodoi{10.1046/j.1365-8711.2002.05968.x}

\bibitem[{{Hodge} \& {da Cunha}(2020)}]{2020RSOS....700556H}
{Hodge}, J.~A., \& {da Cunha}, E. 2020, Royal Society Open Science, 7, 200556, \dodoi{10.1098/rsos.200556}

\bibitem[{{Hodge} {et~al.}(2013){Hodge}, {Karim}, {Smail}, {Swinbank}, {Walter}, {Biggs}, {Ivison}, {Weiss}, {Alexander}, {Bertoldi}, {Brandt}, {Chapman}, {Coppin}, {Cox}, {Danielson}, {Dannerbauer}, {De Breuck}, {Decarli}, {Edge}, {Greve}, {Knudsen}, {Menten}, {Rix}, {Schinnerer}, {Simpson}, {Wardlow}, \& {van der Werf}}]{2013ApJ...768...91H}
{Hodge}, J.~A., {Karim}, A., {Smail}, I., {et~al.} 2013, \apj, 768, 91, \dodoi{10.1088/0004-637X/768/1/91}

\bibitem[{{Hunt} {et~al.}(2019){Hunt}, {De Looze}, {Boquien}, {Nikutta}, {Rossi}, {Bianchi}, {Dale}, {Granato}, {Kennicutt}, {Silva}, {Ciesla}, {Rela{\~n}o}, {Viaene}, {Brandl}, {Calzetti}, {Croxall}, {Draine}, {Galametz}, {Gordon}, {Groves}, {Helou}, {Herrera-Camus}, {Hinz}, {Koda}, {Salim}, {Sandstrom}, {Smith}, {Wilson}, \& {Zibetti}}]{2019A&A...621A..51H}
{Hunt}, L.~K., {De Looze}, I., {Boquien}, M., {et~al.} 2019, \aap, 621, A51, \dodoi{10.1051/0004-6361/201834212}

\bibitem[{{Inami} {et~al.}(2020){Inami}, {Decarli}, {Walter}, {Weiss}, {Carilli}, {Aravena}, {Boogaard}, {Gonza{\'l}ez-L{\'o}pez}, {Popping}, {da Cunha}, {Bacon}, {Bauer}, {Contini}, {Cortes}, {Cox}, {Daddi}, {D{\'\i}az-Santos}, {Kaasinen}, {Riechers}, {Wagg}, {van der Werf}, \& {Wisotzki}}]{2020ApJ...902..113I}
{Inami}, H., {Decarli}, R., {Walter}, F., {et~al.} 2020, \apj, 902, 113, \dodoi{10.3847/1538-4357/abba2f}

\bibitem[{{Inoue}(2011)}]{2011MNRAS.415.2920I}
{Inoue}, A.~K. 2011, \mnras, 415, 2920, \dodoi{10.1111/j.1365-2966.2011.18906.x}

\bibitem[{{Jolly} {et~al.}(2021){Jolly}, {Knudsen}, {Laporte}, {Richard}, {Fujimoto}, {Kohno}, {Ao}, {Bauer}, {Egami}, {Espada}, {Dessauges-Zavadsky}, {Magdis}, {Schaerer}, {Sun}, {Valentino}, {Wang}, \& {Zitrin}}]{2021A&A...652A.128J}
{Jolly}, J.-B., {Knudsen}, K., {Laporte}, N., {et~al.} 2021, \aap, 652, A128, \dodoi{10.1051/0004-6361/202140878}

\bibitem[{{Jones} {et~al.}(2017){Jones}, {K{\"o}hler}, {Ysard}, {Bocchio}, \& {Verstraete}}]{2017A&A...602A..46J}
{Jones}, A.~P., {K{\"o}hler}, M., {Ysard}, N., {Bocchio}, M., \& {Verstraete}, L. 2017, \aap, 602, A46, \dodoi{10.1051/0004-6361/201630225}

\bibitem[{{Karim} {et~al.}(2013){Karim}, {Swinbank}, {Hodge}, {Smail}, {Walter}, {Biggs}, {Simpson}, {Danielson}, {Alexander}, {Bertoldi}, {de Breuck}, {Chapman}, {Coppin}, {Dannerbauer}, {Edge}, {Greve}, {Ivison}, {Knudsen}, {Menten}, {Schinnerer}, {Wardlow}, {Wei{\ss}}, \& {van der Werf}}]{2013MNRAS.432....2K}
{Karim}, A., {Swinbank}, A.~M., {Hodge}, J.~A., {et~al.} 2013, \mnras, 432, 2, \dodoi{10.1093/mnras/stt196}

\bibitem[{{Kashyap} {et~al.}(2010){Kashyap}, {van Dyk}, {Connors}, {Freeman}, {Siemiginowska}, {Xu}, \& {Zezas}}]{2010ApJ...719..900K}
{Kashyap}, V.~L., {van Dyk}, D.~A., {Connors}, A., {et~al.} 2010, \apj, 719, 900, \dodoi{10.1088/0004-637X/719/1/900}

\bibitem[{{Kohno} {et~al.}(2023){Kohno}, {Fujimoto}, {Tsujita}, {Kokorev}, {Brammer}, {Magdis}, {Valentino}, {Laporte}, {Sun}, {Egami}, {Bauer}, {Guerrero}, {Nagar}, {Caputi}, {Caminha}, {Jolly}, {Knudsen}, {Uematsu}, {Ueda}, {Oguri}, {Zitrin}, {Ouchi}, {Ono}, {Gonzalez-Lopez}, {Richard}, {Smail}, {Coe}, {Postman}, {Bradley}, {Koekemoer}, {Munoz Arancibia}, {Dessauges-Zavadsky}, {Espada}, {Umehata}, {Hatsukade}, {Egusa}, {Shimasaku}, {Matsui-Morokuma}, {Wang}, {Wang}, {Ao}, {Baker}, {Lee}, {Lagos}, {Hughes}, \& {ALCS collaboration}}]{2023arXiv230515126K}
{Kohno}, K., {Fujimoto}, S., {Tsujita}, A., {et~al.} 2023, arXiv e-prints, arXiv:2305.15126, \dodoi{10.48550/arXiv.2305.15126}

\bibitem[{{Kokorev} {et~al.}(2022){Kokorev}, {Brammer}, {Fujimoto}, {Kohno}, {Magdis}, {Valentino}, {Toft}, {Oesch}, {Davidzon}, {Bauer}, {Coe}, {Egami}, {Oguri}, {Ouchi}, {Postman}, {Richard}, {Jolly}, {Knudsen}, {Sun}, {Weaver}, {Ao}, {Baker}, {Bradley}, {Caputi}, {Dessauges-Zavadsky}, {Espada}, {Hatsukade}, {Koekemoer}, {Mu{\~n}oz Arancibia}, {Shimasaku}, {Umehata}, {Wang}, \& {Wang}}]{2022ApJS..263...38K}
{Kokorev}, V., {Brammer}, G., {Fujimoto}, S., {et~al.} 2022, \apjs, 263, 38, \dodoi{10.3847/1538-4365/ac9909}

\bibitem[{{Kokorev} {et~al.}(2023){Kokorev}, {Jin}, {Magdis}, {Caputi}, {Valentino}, {Dayal}, {Trebitsch}, {Brammer}, {Fujimoto}, {Bauer}, {Iani}, {Kohno}, {Bl{\'a}nquez Ses{\'e}}, {G{\'o}mez-Guijarro}, {Rinaldi}, \& {Navarro-Carrera}}]{2023ApJ...945L..25K}
{Kokorev}, V., {Jin}, S., {Magdis}, G.~E., {et~al.} 2023, \apjl, 945, L25, \dodoi{10.3847/2041-8213/acbd9d}

\bibitem[{{Komatsu} {et~al.}(2011){Komatsu}, {Smith}, {Dunkley}, {Bennett}, {Gold}, {Hinshaw}, {Jarosik}, {Larson}, {Nolta}, {Page}, {Spergel}, {Halpern}, {Hill}, {Kogut}, {Limon}, {Meyer}, {Odegard}, {Tucker}, {Weiland}, {Wollack}, \& {Wright}}]{2011ApJS..192...18K}
{Komatsu}, E., {Smith}, K.~M., {Dunkley}, J., {et~al.} 2011, \apjs, 192, 18, \dodoi{10.1088/0067-0049/192/2/18}

\bibitem[{{Koprowski} {et~al.}(2020){Koprowski}, {Coppin}, {Geach}, {Dudzevi{\v{c}}i{\={u}}t{\.{e}}}, {Smail}, {Almaini}, {An}, {Blain}, {Chapman}, {Chen}, {Conselice}, {Dunlop}, {Farrah}, {Gullberg}, {Hartley}, {Ivison}, {Karska}, {Maltby}, {Malek}, {Micha{\l}owski}, {Pope}, {Salim}, {Scott}, {Simpson}, {Simpson}, {Swinbank}, {Thomson}, {Wardlow}, {van der Werf}, \& {Whitaker}}]{2020MNRAS.492.4927K}
{Koprowski}, M.~P., {Coppin}, K.~E.~K., {Geach}, J.~E., {et~al.} 2020, \mnras, 492, 4927, \dodoi{10.1093/mnras/staa160}

\bibitem[{{Kormendy} \& {Ho}(2013)}]{2013ARA&A..51..511K}
{Kormendy}, J., \& {Ho}, L.~C. 2013, \araa, 51, 511, \dodoi{10.1146/annurev-astro-082708-101811}

\bibitem[{{Laporte} {et~al.}(2021){Laporte}, {Zitrin}, {Ellis}, {Fujimoto}, {Brammer}, {Richard}, {Oguri}, {Caminha}, {Kohno}, {Yoshimura}, {Ao}, {Bauer}, {Caputi}, {Egami}, {Espada}, {Gonz{\'a}lez-L{\'o}pez}, {Hatsukade}, {Knudsen}, {Lee}, {Magdis}, {Ouchi}, {Valentino}, \& {Wang}}]{2021MNRAS.505.4838L}
{Laporte}, N., {Zitrin}, A., {Ellis}, R.~S., {et~al.} 2021, \mnras, 505, 4838, \dodoi{10.1093/mnras/stab191}

\bibitem[{{Lequeux} {et~al.}(1982){Lequeux}, {Maurice}, {Prevot-Burnichon}, {Prevot}, \& {Rocca-Volmerange}}]{1982A&A...113L..15L}
{Lequeux}, J., {Maurice}, E., {Prevot-Burnichon}, M.~L., {Prevot}, L., \& {Rocca-Volmerange}, B. 1982, \aap, 113, L15

\bibitem[{{Lotz} {et~al.}(2017){Lotz}, {Koekemoer}, {Coe}, {Grogin}, {Capak}, {Mack}, {Anderson}, {Avila}, {Barker}, {Borncamp}, {Brammer}, {Durbin}, {Gunning}, {Hilbert}, {Jenkner}, {Khandrika}, {Levay}, {Lucas}, {MacKenty}, {Ogaz}, {Porterfield}, {Reid}, {Robberto}, {Royle}, {Smith}, {Storrie-Lombardi}, {Sunnquist}, {Surace}, {Taylor}, {Williams}, {Bullock}, {Dickinson}, {Finkelstein}, {Natarajan}, {Richard}, {Robertson}, {Tumlinson}, {Zitrin}, {Flanagan}, {Sembach}, {Soifer}, \& {Mountain}}]{2017ApJ...837...97L}
{Lotz}, J.~M., {Koekemoer}, A., {Coe}, D., {et~al.} 2017, \apj, 837, 97, \dodoi{10.3847/1538-4357/837/1/97}

\bibitem[{{Lusso} {et~al.}(2012){Lusso}, {Comastri}, {Simmons}, {Mignoli}, {Zamorani}, {Vignali}, {Brusa}, {Shankar}, {Lutz}, {Trump}, {Maiolino}, {Gilli}, {Bolzonella}, {Puccetti}, {Salvato}, {Impey}, {Civano}, {Elvis}, {Mainieri}, {Silverman}, {Koekemoer}, {Bongiorno}, {Merloni}, {Berta}, {Le Floc'h}, {Magnelli}, {Pozzi}, \& {Riguccini}}]{2012MNRAS.425..623L}
{Lusso}, E., {Comastri}, A., {Simmons}, B.~D., {et~al.} 2012, \mnras, 425, 623, \dodoi{10.1111/j.1365-2966.2012.21513.x}

\bibitem[{{Madau} \& {Dickinson}(2014)}]{2014ARA&A..52..415M}
{Madau}, P., \& {Dickinson}, M. 2014, \araa, 52, 415, \dodoi{10.1146/annurev-astro-081811-125615}

\bibitem[{{Magnelli} {et~al.}(2020){Magnelli}, {Boogaard}, {Decarli}, {G{\'o}nzalez-L{\'o}pez}, {Novak}, {Popping}, {Smail}, {Walter}, {Aravena}, {Assef}, {Bauer}, {Bertoldi}, {Carilli}, {Cortes}, {Cunha}, {Daddi}, {D{\'\i}az-Santos}, {Inami}, {Ivison}, {F{\`e}vre}, {Oesch}, {Riechers}, {Rix}, {Sargent}, {Werf}, {Wagg}, \& {Weiss}}]{2020ApJ...892...66M}
{Magnelli}, B., {Boogaard}, L., {Decarli}, R., {et~al.} 2020, \apj, 892, 66, \dodoi{10.3847/1538-4357/ab7897}

\bibitem[{{Magorrian} {et~al.}(1998){Magorrian}, {Tremaine}, {Richstone}, {Bender}, {Bower}, {Dressler}, {Faber}, {Gebhardt}, {Green}, {Grillmair}, {Kormendy}, \& {Lauer}}]{1998AJ....115.2285M}
{Magorrian}, J., {Tremaine}, S., {Richstone}, D., {et~al.} 1998, \aj, 115, 2285, \dodoi{10.1086/300353}

\bibitem[{{Marconi} \& {Hunt}(2003)}]{2003ApJ...589L..21M}
{Marconi}, A., \& {Hunt}, L.~K. 2003, \apjl, 589, L21, \dodoi{10.1086/375804}

\bibitem[{{McMullin} {et~al.}(2007){McMullin}, {Waters}, {Schiebel}, {Young}, \& {Golap}}]{2007ASPC..376..127M}
{McMullin}, J.~P., {Waters}, B., {Schiebel}, D., {Young}, W., \& {Golap}, K. 2007, in Astronomical Society of the Pacific Conference Series, Vol. 376, Astronomical Data Analysis Software and Systems XVI, ed. R.~A. {Shaw}, F.~{Hill}, \& D.~J. {Bell}, 127

\bibitem[{{Nandy} {et~al.}(1980){Nandy}, {Morgan}, {Willis}, {Wilson}, {Gondhalekar}, \& {Houziaux}}]{1980Natur.283..725N}
{Nandy}, K., {Morgan}, D.~H., {Willis}, A.~J., {et~al.} 1980, \nat, 283, 725, \dodoi{10.1038/283725a0}

\bibitem[{{Nandy} {et~al.}(1975){Nandy}, {Thompson}, {Jamar}, {Monfils}, \& {Wilson}}]{1975A&A....44..195N}
{Nandy}, K., {Thompson}, G.~I., {Jamar}, C., {Monfils}, A., \& {Wilson}, R. 1975, \aap, 44, 195

\bibitem[{{Narayanan} {et~al.}(2018){Narayanan}, {Dav{\'e}}, {Johnson}, {Thompson}, {Conroy}, \& {Geach}}]{2018MNRAS.474.1718N}
{Narayanan}, D., {Dav{\'e}}, R., {Johnson}, B.~D., {et~al.} 2018, \mnras, 474, 1718, \dodoi{10.1093/mnras/stx2860}

\bibitem[{{Nasa High Energy Astrophysics Science Archive Research Center (Heasarc)}(2014)}]{2014ascl.soft08004N}
{Nasa High Energy Astrophysics Science Archive Research Center (Heasarc)}. 2014, {HEAsoft: Unified Release of FTOOLS and XANADU}, Astrophysics Source Code Library, record ascl:1408.004.
\newblock \doeprint{1408.004}

\bibitem[{{Nersesian} {et~al.}(2019){Nersesian}, {Xilouris}, {Bianchi}, {Galliano}, {Jones}, {Baes}, {Casasola}, {Cassar{\`a}}, {Clark}, {Davies}, {Decleir}, {Dobbels}, {De Looze}, {De Vis}, {Fritz}, {Galametz}, {Madden}, {Mosenkov}, {Tr{\v{c}}ka}, {Verstocken}, {Viaene}, \& {Lianou}}]{2019A&A...624A..80N}
{Nersesian}, A., {Xilouris}, E.~M., {Bianchi}, S., {et~al.} 2019, \aap, 624, A80, \dodoi{10.1051/0004-6361/201935118}

\bibitem[{{Ogawa} {et~al.}(2019){Ogawa}, {Ueda}, {Yamada}, {Tanimoto}, \& {Kawaguchi}}]{2019ApJ...875..115O}
{Ogawa}, S., {Ueda}, Y., {Yamada}, S., {Tanimoto}, A., \& {Kawaguchi}, T. 2019, \apj, 875, 115, \dodoi{10.3847/1538-4357/ab0e08}

\bibitem[{{Oguri}(2010)}]{2010PASJ...62.1017O}
{Oguri}, M. 2010, \pasj, 62, 1017, \dodoi{10.1093/pasj/62.4.1017}

\bibitem[{{Pacifici} {et~al.}(2023){Pacifici}, {Iyer}, {Mobasher}, {da Cunha}, {Acquaviva}, {Burgarella}, {Calistro Rivera}, {Carnall}, {Chang}, {Chartab}, {Cooke}, {Fairhurst}, {Kartaltepe}, {Leja}, {Ma{\l}ek}, {Salmon}, {Torelli}, {Vidal-Garc{\'\i}a}, {Boquien}, {Brammer}, {Brown}, {Capak}, {Chevallard}, {Circosta}, {Croton}, {Davidzon}, {Dickinson}, {Duncan}, {Faber}, {Ferguson}, {Fontana}, {Guo}, {Haeussler}, {Hemmati}, {Jafariyazani}, {Kassin}, {Larson}, {Lee}, {Mantha}, {Marchi}, {Nayyeri}, {Newman}, {Pandya}, {Pforr}, {Reddy}, {Sanders}, {Shah}, {Shahidi}, {Stevans}, {Triani}, {Tyler}, {Vanderhoof}, {de la Vega}, {Wang}, \& {Weston}}]{2023ApJ...944..141P}
{Pacifici}, C., {Iyer}, K.~G., {Mobasher}, B., {et~al.} 2023, \apj, 944, 141, \dodoi{10.3847/1538-4357/acacff}

\bibitem[{{Pei}(1992)}]{1992ApJ...395..130P}
{Pei}, Y.~C. 1992, \apj, 395, 130, \dodoi{10.1086/171637}

\bibitem[{{Pettini} {et~al.}(1998){Pettini}, {Kellogg}, {Steidel}, {Dickinson}, {Adelberger}, \& {Giavalisco}}]{1998ApJ...508..539P}
{Pettini}, M., {Kellogg}, M., {Steidel}, C.~C., {et~al.} 1998, \apj, 508, 539, \dodoi{10.1086/306431}

\bibitem[{{Planck Collaboration} {et~al.}(2011){Planck Collaboration}, {Abergel}, {Ade}, {Aghanim}, {Arnaud}, {Ashdown}, {Aumont}, {Baccigalupi}, {Balbi}, {Banday}, {Barreiro}, {Bartlett}, {Battaner}, {Benabed}, {Beno{\^\i}t}, {Bernard}, {Bersanelli}, {Bhatia}, {Bock}, {Bonaldi}, {Bond}, {Borrill}, {Bouchet}, {Boulanger}, {Bucher}, {Burigana}, {Cabella}, {Cardoso}, {Catalano}, {Cay{\'o}n}, {Challinor}, {Chamballu}, {Chiang}, {Chiang}, {Christensen}, {Clements}, {Colombi}, {Couchot}, {Coulais}, {Crill}, {Cuttaia}, {Danese}, {Davies}, {Davis}, {de Bernardis}, {de Gasperis}, {de Rosa}, {de Zotti}, {Delabrouille}, {Delouis}, {D{\'e}sert}, {Dickinson}, {Dobashi}, {Donzelli}, {Dor{\'e}}, {D{\"o}rl}, {Douspis}, {Dupac}, {Efstathiou}, {En{\ss}lin}, {Eriksen}, {Finelli}, {Forni}, {Frailis}, {Franceschi}, {Galeotta}, {Ganga}, {Giard}, {Giardino}, {Giraud-H{\'e}raud}, {Gonz{\'a}lez-Nuevo}, {G{\'o}rski}, {Gratton}, {Gregorio}, {Gruppuso}, {Guillet}, {Hansen}, {Harrison}, {Henrot-Versill{\'e}}, {Herranz}, {Hildebrandt}, {Hivon}, {Hobson}, {Holmes}, {Hovest}, {Hoyland}, {Huffenberger}, {Jaffe}, {Jones}, {Jones}, {Juvela}, {Keih{\"a}nen}, {Keskitalo}, {Kisner}, {Kneissl}, {Knox}, {Kurki-Suonio}, {Lagache}, {Lamarre}, {Lasenby}, {Laureijs}, {Lawrence}, {Leach}, {Leonardi}, {Leroy}, {Linden-V{\o}rnle}, {L{\'o}pez-Caniego}, {Lubin}, {Mac{\'\i}as-P{\'e}rez}, {MacTavish}, {Maffei}, {Mandolesi}, {Mann}, {Maris}, {Marshall}, {Martin}, {Mart{\'\i}nez-Gonz{\'a}lez}, {Masi}, {Matarrese}, {Matthai}, {Mazzotta}, {McGehee}, {Meinhold}, {Melchiorri}, {Mendes}, {Mennella}, {Mitra}, {Miville-Desch{\^e}nes}, {Moneti}, {Montier}, {Morgante}, {Mortlock}, {Munshi}, {Murphy}, {Naselsky}, {Natoli}, {Netterfield}, {N{\o}rgaard-Nielsen}, {Noviello}, {Novikov}, {Novikov}, {Osborne}, {Pajot}, {Paladini}, {Pasian}, {Patanchon}, {Perdereau}, {Perotto}, {Perrotta}, {Piacentini}, {Piat}, {Plaszczynski}, {Pointecouteau}, {Polenta}, {Ponthieu}, {Poutanen}, {Pr{\'e}zeau}, {Prunet}, {Puget}, {Reach}, {Rebolo}, {Reinecke}, {Renault}, {Ricciardi}, {Riller}, {Ristorcelli}, {Rocha}, {Rosset}, {Rubi{\~n}o-Mart{\'\i}n}, {Rusholme}, {Sandri}, {Santos}, {Savini}, {Scott}, {Seiffert}, {Shellard}, {Smoot}, {Starck}, {Stivoli}, {Stolyarov}, {Sudiwala}, {Sygnet}, {Tauber}, {Terenzi}, {Toffolatti}, {Tomasi}, {Torre}, {Tristram}, {Tuovinen}, {Umana}, {Valenziano}, {Verstraete}, {Vielva}, {Villa}, {Vittorio}, {Wade}, {Wandelt}, {Yvon}, {Zacchei}, \& {Zonca}}]{2011A&A...536A..25P}
{Planck Collaboration}, {Abergel}, A., {Ade}, P.~A.~R., {et~al.} 2011, \aap, 536, A25, \dodoi{10.1051/0004-6361/201116483}

\bibitem[{{Popping} {et~al.}(2017){Popping}, {Puglisi}, \& {Norman}}]{2017MNRAS.472.2315P}
{Popping}, G., {Puglisi}, A., \& {Norman}, C.~A. 2017, \mnras, 472, 2315, \dodoi{10.1093/mnras/stx2202}

\bibitem[{{Popping} {et~al.}(2019){Popping}, {Pillepich}, {Somerville}, {Decarli}, {Walter}, {Aravena}, {Carilli}, {Cox}, {Nelson}, {Riechers}, {Weiss}, {Boogaard}, {Bouwens}, {Contini}, {Cortes}, {da Cunha}, {Daddi}, {D{\'\i}az-Santos}, {Diemer}, {Gonz{\'a}lez-L{\'o}pez}, {Hernquist}, {Ivison}, {Le F{\`e}vre}, {Marinacci}, {Rix}, {Swinbank}, {Vogelsberger}, {van der Werf}, {Wagg}, \& {Yung}}]{2019ApJ...882..137P}
{Popping}, G., {Pillepich}, A., {Somerville}, R.~S., {et~al.} 2019, \apj, 882, 137, \dodoi{10.3847/1538-4357/ab30f2}

\bibitem[{{Popping} {et~al.}(2020){Popping}, {Walter}, {Behroozi}, {Gonz{\'a}lez-L{\'o}pez}, {Hayward}, {Somerville}, {van der Werf}, {Aravena}, {Assef}, {Boogaard}, {Bauer}, {Cortes}, {Cox}, {D{\'\i}az-Santos}, {Decarli}, {Franco}, {Ivison}, {Riechers}, {Rix}, \& {Weiss}}]{2020ApJ...891..135P}
{Popping}, G., {Walter}, F., {Behroozi}, P., {et~al.} 2020, \apj, 891, 135, \dodoi{10.3847/1538-4357/ab76c0}

\bibitem[{{Postman} {et~al.}(2012){Postman}, {Coe}, {Ben{\'\i}tez}, {Bradley}, {Broadhurst}, {Donahue}, {Ford}, {Graur}, {Graves}, {Jouvel}, {Koekemoer}, {Lemze}, {Medezinski}, {Molino}, {Moustakas}, {Ogaz}, {Riess}, {Rodney}, {Rosati}, {Umetsu}, {Zheng}, {Zitrin}, {Bartelmann}, {Bouwens}, {Czakon}, {Golwala}, {Host}, {Infante}, {Jha}, {Jimenez-Teja}, {Kelson}, {Lahav}, {Lazkoz}, {Maoz}, {McCully}, {Melchior}, {Meneghetti}, {Merten}, {Moustakas}, {Nonino}, {Patel}, {Reg{\"o}s}, {Sayers}, {Seitz}, \& {Van der Wel}}]{2012ApJS..199...25P}
{Postman}, M., {Coe}, D., {Ben{\'\i}tez}, N., {et~al.} 2012, \apjs, 199, 25, \dodoi{10.1088/0067-0049/199/2/25}

\bibitem[{{Prevot} {et~al.}(1984){Prevot}, {Lequeux}, {Maurice}, {Prevot}, \& {Rocca-Volmerange}}]{1984A&A...132..389P}
{Prevot}, M.~L., {Lequeux}, J., {Maurice}, E., {Prevot}, L., \& {Rocca-Volmerange}, B. 1984, \aap, 132, 389

\bibitem[{{Reddy} {et~al.}(2010){Reddy}, {Erb}, {Pettini}, {Steidel}, \& {Shapley}}]{2010ApJ...712.1070R}
{Reddy}, N.~A., {Erb}, D.~K., {Pettini}, M., {Steidel}, C.~C., \& {Shapley}, A.~E. 2010, \apj, 712, 1070, \dodoi{10.1088/0004-637X/712/2/1070}

\bibitem[{{Ricci} {et~al.}(2017){Ricci}, {Trakhtenbrot}, {Koss}, {Ueda}, {Del Vecchio}, {Treister}, {Schawinski}, {Paltani}, {Oh}, {Lamperti}, {Berney}, {Gandhi}, {Ichikawa}, {Bauer}, {Ho}, {Asmus}, {Beckmann}, {Soldi}, {Balokovi{\'c}}, {Gehrels}, \& {Markwardt}}]{2017ApJS..233...17R}
{Ricci}, C., {Trakhtenbrot}, B., {Koss}, M.~J., {et~al.} 2017, \apjs, 233, 17, \dodoi{10.3847/1538-4365/aa96ad}

\bibitem[{{Rocca-Volmerange} {et~al.}(1981){Rocca-Volmerange}, {Prevot}, {Ferlet}, {Lequeux}, \& {Prevot-Burnichon}}]{1981A&A....99L...5R}
{Rocca-Volmerange}, B., {Prevot}, L., {Ferlet}, R., {Lequeux}, J., \& {Prevot-Burnichon}, M.~L. 1981, \aap, 99, L5

\bibitem[{{Salim} {et~al.}(2018){Salim}, {Boquien}, \& {Lee}}]{2018ApJ...859...11S}
{Salim}, S., {Boquien}, M., \& {Lee}, J.~C. 2018, \apj, 859, 11, \dodoi{10.3847/1538-4357/aabf3c}

\bibitem[{{Schreiber} {et~al.}(2018){Schreiber}, {Elbaz}, {Pannella}, {Ciesla}, {Wang}, \& {Franco}}]{2018A&A...609A..30S}
{Schreiber}, C., {Elbaz}, D., {Pannella}, M., {et~al.} 2018, \aap, 609, A30, \dodoi{10.1051/0004-6361/201731506}

\bibitem[{{Schreiber} {et~al.}(2015){Schreiber}, {Pannella}, {Elbaz}, {B{\'e}thermin}, {Inami}, {Dickinson}, {Magnelli}, {Wang}, {Aussel}, {Daddi}, {Juneau}, {Shu}, {Sargent}, {Buat}, {Faber}, {Ferguson}, {Giavalisco}, {Koekemoer}, {Magdis}, {Morrison}, {Papovich}, {Santini}, \& {Scott}}]{2015A&A...575A..74S}
{Schreiber}, C., {Pannella}, M., {Elbaz}, D., {et~al.} 2015, \aap, 575, A74, \dodoi{10.1051/0004-6361/201425017}

\bibitem[{{Schwarz}(1978)}]{1978AnSta...6..461S}
{Schwarz}, G. 1978, Annals of Statistics, 6, 461

\bibitem[{{Simpson} {et~al.}(2014){Simpson}, {Swinbank}, {Smail}, {Alexander}, {Brandt}, {Bertoldi}, {de Breuck}, {Chapman}, {Coppin}, {da Cunha}, {Danielson}, {Dannerbauer}, {Greve}, {Hodge}, {Ivison}, {Karim}, {Knudsen}, {Poggianti}, {Schinnerer}, {Thomson}, {Walter}, {Wardlow}, {Wei{\ss}}, \& {van der Werf}}]{2014ApJ...788..125S}
{Simpson}, J.~M., {Swinbank}, A.~M., {Smail}, I., {et~al.} 2014, \apj, 788, 125, \dodoi{10.1088/0004-637X/788/2/125}

\bibitem[{{Simpson} {et~al.}(2017){Simpson}, {Smail}, {Swinbank}, {Ivison}, {Dunlop}, {Geach}, {Almaini}, {Arumugam}, {Bremer}, {Chen}, {Conselice}, {Coppin}, {Farrah}, {Ibar}, {Hartley}, {Ma}, {Micha{\l}owski}, {Scott}, {Spaans}, {Thomson}, \& {van der Werf}}]{2017ApJ...839...58S}
{Simpson}, J.~M., {Smail}, I., {Swinbank}, A.~M., {et~al.} 2017, \apj, 839, 58, \dodoi{10.3847/1538-4357/aa65d0}

\bibitem[{{Smail} {et~al.}(2021){Smail}, {Dudzevi{\v{c}}i{\={u}}t{\.{e}}}, {Stach}, {Almaini}, {Birkin}, {Chapman}, {Chen}, {Geach}, {Gullberg}, {Hodge}, {Ikarashi}, {Ivison}, {Scott}, {Simpson}, {Swinbank}, {Thomson}, {Walter}, {Wardlow}, \& {van der Werf}}]{2021MNRAS.502.3426S}
{Smail}, I., {Dudzevi{\v{c}}i{\={u}}t{\.{e}}}, U., {Stach}, S.~M., {et~al.} 2021, \mnras, 502, 3426, \dodoi{10.1093/mnras/stab283}

\bibitem[{{Smit} {et~al.}(2016){Smit}, {Bouwens}, {Labb{\'e}}, {Franx}, {Wilkins}, \& {Oesch}}]{2016ApJ...833..254S}
{Smit}, R., {Bouwens}, R.~J., {Labb{\'e}}, I., {et~al.} 2016, \apj, 833, 254, \dodoi{10.3847/1538-4357/833/2/254}

\bibitem[{{SSC And IRSA}(2020)}]{https://doi.org/10.26131/irsa433}
{SSC And IRSA}. 2020, Spitzer Enhanced Imaging Products,  IPAC, \dodoi{10.26131/IRSA433}

\bibitem[{{Stach} {et~al.}(2018){Stach}, {Smail}, {Swinbank}, {Simpson}, {Geach}, {An}, {Almaini}, {Arumugam}, {Blain}, {Chapman}, {Chen}, {Conselice}, {Cooke}, {Coppin}, {Dunlop}, {Farrah}, {Gullberg}, {Hartley}, {Ivison}, {Maltby}, {Micha{\l}owski}, {Scott}, {Simpson}, {Thomson}, {Wardlow}, \& {van der Werf}}]{2018ApJ...860..161S}
{Stach}, S.~M., {Smail}, I., {Swinbank}, A.~M., {et~al.} 2018, \apj, 860, 161, \dodoi{10.3847/1538-4357/aac5e5}

\bibitem[{{Stach} {et~al.}(2019){Stach}, {Dudzevi{\v{c}}i{\={u}}t{\.{e}}}, {Smail}, {Swinbank}, {Geach}, {Simpson}, {An}, {Almaini}, {Arumugam}, {Blain}, {Chapman}, {Chen}, {Conselice}, {Cooke}, {Coppin}, {da Cunha}, {Dunlop}, {Farrah}, {Gullberg}, {Hodge}, {Ivison}, {Kocevski}, {Micha{\l}owski}, {Miyaji}, {Scott}, {Thomson}, {Wardlow}, {Weiss}, \& {van der Werf}}]{2019MNRAS.487.4648S}
{Stach}, S.~M., {Dudzevi{\v{c}}i{\={u}}t{\.{e}}}, U., {Smail}, I., {et~al.} 2019, \mnras, 487, 4648, \dodoi{10.1093/mnras/stz1536}

\bibitem[{{Stalevski} {et~al.}(2012){Stalevski}, {Fritz}, {Baes}, {Nakos}, \& {Popovi{\'c}}}]{2012MNRAS.420.2756S}
{Stalevski}, M., {Fritz}, J., {Baes}, M., {Nakos}, T., \& {Popovi{\'c}}, L.~{\v{C}}. 2012, \mnras, 420, 2756, \dodoi{10.1111/j.1365-2966.2011.19775.x}

\bibitem[{{Stalevski} {et~al.}(2016){Stalevski}, {Ricci}, {Ueda}, {Lira}, {Fritz}, \& {Baes}}]{2016MNRAS.458.2288S}
{Stalevski}, M., {Ricci}, C., {Ueda}, Y., {et~al.} 2016, \mnras, 458, 2288, \dodoi{10.1093/mnras/stw444}

\bibitem[{{Steinhardt} {et~al.}(2020){Steinhardt}, {Jauzac}, {Acebron}, {Atek}, {Capak}, {Davidzon}, {Eckert}, {Harvey}, {Koekemoer}, {Lagos}, {Mahler}, {Montes}, {Niemiec}, {Nonino}, {Oesch}, {Richard}, {Rodney}, {Schaller}, {Sharon}, {Strolger}, {Allingham}, {Amara}, {Bah{\'e}}, {B{\oe}hm}, {Bose}, {Bouwens}, {Bradley}, {Brammer}, {Broadhurst}, {Ca{\~n}as}, {Cen}, {Cl{\'e}ment}, {Clowe}, {Coe}, {Connor}, {Darvish}, {Diego}, {Ebeling}, {Edge}, {Egami}, {Ettori}, {Faisst}, {Frye}, {Furtak}, {G{\'o}mez-Guijarro}, {Remolina Gonz{\'a}lez}, {Gonzalez}, {Graur}, {Gruen}, {Harvey}, {Hensley}, {Hovis-Afflerbach}, {Jablonka}, {Jha}, {Jullo}, {Kneib}, {Kokorev}, {Lagattuta}, {Limousin}, {von der Linden}, {Linzer}, {Lopez}, {Magdis}, {Massey}, {Masters}, {Maturi}, {McCully}, {McGee}, {Meneghetti}, {Mobasher}, {Moustakas}, {Murphy}, {Natarajan}, {Neyrinck}, {O'Connor}, {Oguri}, {Pagul}, {Rhodes}, {Rich}, {Robertson}, {Sereno}, {Shan}, {Smith}, {Sneppen}, {Squires}, {Tam}, {Tchernin}, {Toft}, {Umetsu}, {Weaver}, {van Weeren}, {Williams}, {Wilson}, {Yan}, \& {Zitrin}}]{2020ApJS..247...64S}
{Steinhardt}, C.~L., {Jauzac}, M., {Acebron}, A., {et~al.} 2020, \apjs, 247, 64, \dodoi{10.3847/1538-4365/ab75ed}

\bibitem[{{Sun} {et~al.}(2022){Sun}, {Egami}, {Fujimoto}, {Rawle}, {Bauer}, {Kohno}, {Smail}, {P{\'e}rez-Gonz{\'a}lez}, {Ao}, {Chapman}, {Combes}, {Dessauges-Zavadsky}, {Espada}, {Gonz{\'a}lez-L{\'o}pez}, {Koekemoer}, {Kokorev}, {Lee}, {Morokuma-Matsui}, {Mu{\~n}oz Arancibia}, {Oguri}, {Pell{\'o}}, {Ueda}, {Uematsu}, {Valentino}, {Van der Werf}, {Walth}, {Zemcov}, \& {Zitrin}}]{2022ApJ...932...77S}
{Sun}, F., {Egami}, E., {Fujimoto}, S., {et~al.} 2022, \apj, 932, 77, \dodoi{10.3847/1538-4357/ac6e3f}

\bibitem[{{Swinbank} {et~al.}(2012{\natexlab{a}}){Swinbank}, {Karim}, {Smail}, {Hodge}, {Walter}, {Bertoldi}, {Biggs}, {de Breuck}, {Chapman}, {Coppin}, {Cox}, {Danielson}, {Dannerbauer}, {Ivison}, {Greve}, {Knudsen}, {Menten}, {Simpson}, {Schinnerer}, {Wardlow}, {Wei{\ss}}, \& {van der Werf}}]{2012MNRAS.427.1066S}
{Swinbank}, A.~M., {Karim}, A., {Smail}, I., {et~al.} 2012{\natexlab{a}}, \mnras, 427, 1066, \dodoi{10.1111/j.1365-2966.2012.22048.x}

\bibitem[{{Swinbank} {et~al.}(2014){Swinbank}, {Simpson}, {Smail}, {Harrison}, {Hodge}, {Karim}, {Walter}, {Alexander}, {Brandt}, {de Breuck}, {da Cunha}, {Chapman}, {Coppin}, {Danielson}, {Dannerbauer}, {Decarli}, {Greve}, {Ivison}, {Knudsen}, {Lagos}, {Schinnerer}, {Thomson}, {Wardlow}, {Wei{\ss}}, \& {van der Werf}}]{2014MNRAS.438.1267S}
{Swinbank}, A.~M., {Simpson}, J.~M., {Smail}, I., {et~al.} 2014, \mnras, 438, 1267, \dodoi{10.1093/mnras/stt2273}

\bibitem[{{Swinbank} {et~al.}(2012{\natexlab{b}}){Swinbank}, {Smail}, {Karim}, {Hodge}, {Walter}, {Alexander}, {Bertoldi}, {Biggs}, {Brandt}, {De Breuck}, {Chapman}, {Coppin}, {Cox}, {Danielson}, {Dannerbauer}, {Edge}, {Ivison}, {Greve}, {Knudsen}, {Menten}, {Simpson}, {Schinnerer}, {Wardlow}, {Weiss}, \& {van der Werf}}]{2012Msngr.149...40S}
{Swinbank}, M., {Smail}, I., {Karim}, A., {et~al.} 2012{\natexlab{b}}, The Messenger, 149, 40, \dodoi{10.48550/arXiv.1209.1390}

\bibitem[{{Takeuchi} {et~al.}(2012){Takeuchi}, {Yuan}, {Ikeyama}, {Murata}, \& {Inoue}}]{2012ApJ...755..144T}
{Takeuchi}, T.~T., {Yuan}, F.-T., {Ikeyama}, A., {Murata}, K.~L., \& {Inoue}, A.~K. 2012, \apj, 755, 144, \dodoi{10.1088/0004-637X/755/2/144}

\bibitem[{{Tanimoto} {et~al.}(2019){Tanimoto}, {Ueda}, {Odaka}, {Kawaguchi}, {Fukazawa}, \& {Kawamuro}}]{2019ApJ...877...95T}
{Tanimoto}, A., {Ueda}, Y., {Odaka}, H., {et~al.} 2019, \apj, 877, 95, \dodoi{10.3847/1538-4357/ab1b20}

\bibitem[{{Teng} {et~al.}(2015){Teng}, {Rigby}, {Stern}, {Ptak}, {Alexander}, {Bauer}, {Boggs}, {Brandt}, {Christensen}, {Comastri}, {Craig}, {Farrah}, {Gandhi}, {Hailey}, {Harrison}, {Hickox}, {Koss}, {Luo}, {Treister}, \& {Zhang}}]{2015ApJ...814...56T}
{Teng}, S.~H., {Rigby}, J.~R., {Stern}, D., {et~al.} 2015, \apj, 814, 56, \dodoi{10.1088/0004-637X/814/1/56}

\bibitem[{{Thomson} {et~al.}(2014){Thomson}, {Ivison}, {Simpson}, {Swinbank}, {Smail}, {Arumugam}, {Alexander}, {Beelen}, {Brandt}, {Chandra}, {Dannerbauer}, {Greve}, {Hodge}, {Ibar}, {Karim}, {Murphy}, {Schinnerer}, {Sirothia}, {Walter}, {Wardlow}, \& {van der Werf}}]{2014MNRAS.442..577T}
{Thomson}, A.~P., {Ivison}, R.~J., {Simpson}, J.~M., {et~al.} 2014, \mnras, 442, 577, \dodoi{10.1093/mnras/stu839}

\bibitem[{{Toba} {et~al.}(2020){Toba}, {Goto}, {Oi}, {Wang}, {Kim}, {Ho}, {Burgarella}, {Hashimoto}, {Hsieh}, {Huang}, {Hwang}, {Ikeda}, {Kim}, {Kim}, {Lee}, {Malkan}, {Matsuhara}, {Miyaji}, {Momose}, {Ohyama}, {Oyabu}, {Pearson}, {Santos}, {Shim}, {Takagi}, {Ueda}, {Utsumi}, \& {Wada}}]{2020ApJ...899...35T}
{Toba}, Y., {Goto}, T., {Oi}, N., {et~al.} 2020, \apj, 899, 35, \dodoi{10.3847/1538-4357/ab9cb7}

\bibitem[{{Ueda} {et~al.}(2014){Ueda}, {Akiyama}, {Hasinger}, {Miyaji}, \& {Watson}}]{2014ApJ...786..104U}
{Ueda}, Y., {Akiyama}, M., {Hasinger}, G., {Miyaji}, T., \& {Watson}, M.~G. 2014, \apj, 786, 104, \dodoi{10.1088/0004-637X/786/2/104}

\bibitem[{{Ueda} {et~al.}(2003){Ueda}, {Akiyama}, {Ohta}, \& {Miyaji}}]{2003ApJ...598..886U}
{Ueda}, Y., {Akiyama}, M., {Ohta}, K., \& {Miyaji}, T. 2003, \apj, 598, 886, \dodoi{10.1086/378940}

\bibitem[{{Ueda} {et~al.}(2018){Ueda}, {Hatsukade}, {Kohno}, {Yamaguchi}, {Tamura}, {Umehata}, {Akiyama}, {Ao}, {Aretxaga}, {Caputi}, {Dunlop}, {Espada}, {Fujimoto}, {Hayatsu}, {Imanishi}, {Inoue}, {Ivison}, {Kodama}, {Lee}, {Matsuoka}, {Miyaji}, {Morokuma-Matsui}, {Nagao}, {Nakanishi}, {Nyland}, {Ohta}, {Ouchi}, {Rujopakarn}, {Saito}, {Tadaki}, {Tanaka}, {Taniguchi}, {Wang}, {Wang}, {Yoshimura}, \& {Yun}}]{2018ApJ...853...24U}
{Ueda}, Y., {Hatsukade}, B., {Kohno}, K., {et~al.} 2018, \apj, 853, 24, \dodoi{10.3847/1538-4357/aa9f10}

\bibitem[{{Uematsu} {et~al.}(2023){Uematsu}, {Ueda}, {Kohno}, {Yamada}, {Toba}, {Fujimoto}, {Hatsukade}, {Umehata}, {Espada}, {Sun}, {Magdis}, {Kokorev}, \& {Ao}}]{2023ApJ...945..121U}
{Uematsu}, R., {Ueda}, Y., {Kohno}, K., {et~al.} 2023, \apj, 945, 121, \dodoi{10.3847/1538-4357/acb4e9}

\bibitem[{{Uzgil} {et~al.}(2019){Uzgil}, {Carilli}, {Lidz}, {Walter}, {Thyagarajan}, {Decarli}, {Aravena}, {Bertoldi}, {Cortes}, {Gonz{\'a}lez-L{\'o}pez}, {Inami}, {Popping}, {Riechers}, {Van der Werf}, {Wagg}, \& {Weiss}}]{2019ApJ...887...37U}
{Uzgil}, B.~D., {Carilli}, C., {Lidz}, A., {et~al.} 2019, \apj, 887, 37, \dodoi{10.3847/1538-4357/ab517f}

\bibitem[{{Uzgil} {et~al.}(2021){Uzgil}, {Oesch}, {Walter}, {Aravena}, {Boogaard}, {Carilli}, {Decarli}, {D{\'\i}az-Santos}, {Fudamoto}, {Inami}, {Bouwens}, {Cortes}, {Cox}, {Daddi}, {Gonz{\'a}lez-L{\'o}pez}, {Labbe}, {Popping}, {Riechers}, {Stefanon}, {Van der Werf}, \& {Weiss}}]{2021ApJ...912...67U}
{Uzgil}, B.~D., {Oesch}, P.~A., {Walter}, F., {et~al.} 2021, \apj, 912, 67, \dodoi{10.3847/1538-4357/abe86b}

\bibitem[{{Vasudevan} \& {Fabian}(2007)}]{2007MNRAS.381.1235V}
{Vasudevan}, R.~V., \& {Fabian}, A.~C. 2007, \mnras, 381, 1235, \dodoi{10.1111/j.1365-2966.2007.12328.x}

\bibitem[{{Viero} {et~al.}(2022){Viero}, {Sun}, {Chung}, {Moncelsi}, \& {Condon}}]{2022MNRAS.516L..30V}
{Viero}, M.~P., {Sun}, G., {Chung}, D.~T., {Moncelsi}, L., \& {Condon}, S.~S. 2022, \mnras, 516, L30, \dodoi{10.1093/mnrasl/slac075}

\bibitem[{{Walter} {et~al.}(2016){Walter}, {Decarli}, {Aravena}, {Carilli}, {Bouwens}, {da Cunha}, {Daddi}, {Ivison}, {Riechers}, {Smail}, {Swinbank}, {Weiss}, {Anguita}, {Assef}, {Bacon}, {Bauer}, {Bell}, {Bertoldi}, {Chapman}, {Colina}, {Cortes}, {Cox}, {Dickinson}, {Elbaz}, {G{\'o}nzalez-L{\'o}pez}, {Ibar}, {Inami}, {Infante}, {Hodge}, {Karim}, {Le Fevre}, {Magnelli}, {Neri}, {Oesch}, {Ota}, {Popping}, {Rix}, {Sargent}, {Sheth}, {van der Wel}, {van der Werf}, \& {Wagg}}]{2016ApJ...833...67W}
{Walter}, F., {Decarli}, R., {Aravena}, M., {et~al.} 2016, \apj, 833, 67, \dodoi{10.3847/1538-4357/833/1/67}

\bibitem[{{Wang} {et~al.}(2013){Wang}, {Brandt}, {Luo}, {Smail}, {Alexander}, {Danielson}, {Hodge}, {Karim}, {Lehmer}, {Simpson}, {Swinbank}, {Walter}, {Wardlow}, {Xue}, {Chapman}, {Coppin}, {Dannerbauer}, {De Breuck}, {Menten}, \& {van der Werf}}]{2013ApJ...778..179W}
{Wang}, S.~X., {Brandt}, W.~N., {Luo}, B., {et~al.} 2013, \apj, 778, 179, \dodoi{10.1088/0004-637X/778/2/179}

\bibitem[{{Wang} {et~al.}(2017){Wang}, {Lin}, {Lim}, {Smail}, {Chapman}, {Zheng}, {Shim}, {Kodama}, {Almaini}, {Ao}, {Blain}, {Bourne}, {Bunker}, {Chang}, {Chao}, {Chen}, {Clements}, {Conselice}, {Cowley}, {Dannerbauer}, {Dunlop}, {Geach}, {Goto}, {Jiang}, {Ivison}, {Jeong}, {Kohno}, {Kong}, {Lee}, {Lee}, {Lee}, {Micha{\l}owski}, {Oteo}, {Sawicki}, {Scott}, {Shu}, {Simpson}, {Tee}, {Toba}, {Valiante}, {Wang}, {Wang}, \& {Wardlow}}]{2017ApJ...850...37W}
{Wang}, W.-H., {Lin}, W.-C., {Lim}, C.-F., {et~al.} 2017, \apj, 850, 37, \dodoi{10.3847/1538-4357/aa911b}

\bibitem[{{Wei{\ss}} {et~al.}(2009){Wei{\ss}}, {Kov{\'a}cs}, {Coppin}, {Greve}, {Walter}, {Smail}, {Dunlop}, {Knudsen}, {Alexander}, {Bertoldi}, {Brandt}, {Chapman}, {Cox}, {Dannerbauer}, {De Breuck}, {Gawiser}, {Ivison}, {Lutz}, {Menten}, {Koekemoer}, {Kreysa}, {Kurczynski}, {Rix}, {Schinnerer}, \& {van der Werf}}]{2009ApJ...707.1201W}
{Wei{\ss}}, A., {Kov{\'a}cs}, A., {Coppin}, K., {et~al.} 2009, \apj, 707, 1201, \dodoi{10.1088/0004-637X/707/2/1201}

\bibitem[{{Whitaker} {et~al.}(2014){Whitaker}, {Franx}, {Leja}, {van Dokkum}, {Henry}, {Skelton}, {Fumagalli}, {Momcheva}, {Brammer}, {Labb{\'e}}, {Nelson}, \& {Rigby}}]{2014ApJ...795..104W}
{Whitaker}, K.~E., {Franx}, M., {Leja}, J., {et~al.} 2014, \apj, 795, 104, \dodoi{10.1088/0004-637X/795/2/104}

\bibitem[{{Willingale} {et~al.}(2013){Willingale}, {Starling}, {Beardmore}, {Tanvir}, \& {O'Brien}}]{2013MNRAS.431..394W}
{Willingale}, R., {Starling}, R.~L.~C., {Beardmore}, A.~P., {Tanvir}, N.~R., \& {O'Brien}, P.~T. 2013, \mnras, 431, 394, \dodoi{10.1093/mnras/stt175}

\bibitem[{{Yamada} {et~al.}(2021){Yamada}, {Ueda}, {Tanimoto}, {Imanishi}, {Toba}, {Ricci}, \& {Privon}}]{2021ApJS..257...61Y}
{Yamada}, S., {Ueda}, Y., {Tanimoto}, A., {et~al.} 2021, \apjs, 257, 61, \dodoi{10.3847/1538-4365/ac17f5}

\bibitem[{{Yamaguchi} {et~al.}(2019){Yamaguchi}, {Kohno}, {Hatsukade}, {Wang}, {Yoshimura}, {Ao}, {Caputi}, {Dunlop}, {Egami}, {Espada}, {Fujimoto}, {Hayatsu}, {Ivison}, {Kodama}, {Kusakabe}, {Nagao}, {Ouchi}, {Rujopakarn}, {Tadaki}, {Tamura}, {Ueda}, {Umehata}, {Wang}, \& {Yun}}]{2019ApJ...878...73Y}
{Yamaguchi}, Y., {Kohno}, K., {Hatsukade}, B., {et~al.} 2019, \apj, 878, 73, \dodoi{10.3847/1538-4357/ab0d22}

\bibitem[{{Yamaguchi} {et~al.}(2020){Yamaguchi}, {Kohno}, {Hatsukade}, {Wang}, {Yoshimura}, {Ao}, {Dunlop}, {Egami}, {Espada}, {Fujimoto}, {Hayatsu}, {Ivison}, {Kodama}, {Kusakabe}, {Nagao}, {Ouchi}, {Rujopakarn}, {Tadaki}, {Tamura}, {Ueda}, {Umehata}, \& {Wang}}]{2020PASJ...72...69Y}
---. 2020, \pasj, 72, 69, \dodoi{10.1093/pasj/psaa057}

\bibitem[{{Yang} {et~al.}(2022){Yang}, {Boquien}, {Brandt}, {Buat}, {Burgarella}, {Ciesla}, {Lehmer}, {Ma{\l}ek}, {Mountrichas}, {Papovich}, {Pons}, {Stalevski}, {Theul{\'e}}, \& {Zhu}}]{2022ApJ...927..192Y}
{Yang}, G., {Boquien}, M., {Brandt}, W.~N., {et~al.} 2022, \apj, 927, 192, \dodoi{10.3847/1538-4357/ac4971}

\bibitem[{{Zhou} {et~al.}(2020){Zhou}, {Elbaz}, {Franco}, {Magnelli}, {Schreiber}, {Wang}, {Ciesla}, {Daddi}, {Dickinson}, {Nagar}, {Magdis}, {Alexander}, {B{\'e}thermin}, {Demarco}, {Mullaney}, {Bournaud}, {Ferguson}, {Finkelstein}, {Giavalisco}, {Inami}, {Iono}, {Juneau}, {Lagache}, {Messias}, {Motohara}, {Okumura}, {Pannella}, {Papovich}, {Pope}, {Rujopakarn}, {Shi}, {Shu}, \& {Silverman}}]{2020A&A...642A.155Z}
{Zhou}, L., {Elbaz}, D., {Franco}, M., {et~al.} 2020, \aap, 642, A155, \dodoi{10.1051/0004-6361/202038059}

\end{thebibliography}
\bibliographystyle{aasjournal}

\end{document}